%% file: Main.tex
\documentclass{book}
\usepackage{geometry}
\usepackage{tensor}
\date{}
\usepackage{amsthm} 

\usepackage{amsmath}
\usepackage{amssymb}
\usepackage{chngcntr}
\usepackage{subcaption}
\usepackage{graphicx}
\usepackage[T1]{fontenc}
\usepackage[english]{babel}
\usepackage{newtxtext, newtxmath}
\usepackage[T1]{fontenc}
\usepackage{natbib}
\setcitestyle{year,open={(},close={)}}
\usepackage{array}
\usepackage{physics}
\usepackage[bookmarks=True]{hyperref}
\usepackage{bookmark}
\usepackage{orcidlink}
\usepackage{xcolor}
\usepackage{listings}
\definecolor{notebookbackground}{RGB}{220,220,220}

\usepackage{amsthm}
\theoremstyle{definition}
\newtheoremstyle{mycolonstyle}
  {\topsep}{\topsep}
  {\itshape}{0pt}
  {\bfseries}{:\,}{0pt}{}

\theoremstyle{mycolonstyle}
\newtheorem*{definition}{Definition}
\newtheorem*{example}{Example}

\newtheorem*{exercise}{Exercise}

\lstdefinestyle{in}{
    language=[1.0]Mathematica,
    basicstyle=\ttfamily\bfseries,
    keywordstyle=\bfseries,
    commentstyle=\itshape,
    numbers=left,
    numberstyle=\tiny,
    numbersep=10pt,
    frame=single,
    framesep=5pt,
    framerule=0.5pt,
    rulecolor=\color{black},
    xleftmargin=10pt,
    xrightmargin=10pt,
    backgroundcolor=\color{notebookbackground},
    breaklines=true,
    showstringspaces=false,
    escapeinside={(*@}{@*)},
    escapechar=|,
    mathescape 
}
\lstdefinestyle{out}{
    language=[1.0]Mathematica,
    basicstyle=\ttfamily,
    keywordstyle=\bfseries,
    commentstyle=\itshape,
    numbers=none,
    frame=single,
    framesep=5pt,
    framerule=0.5pt,
    rulecolor=\color{gray},
    xleftmargin=10pt,
    xrightmargin=10pt,
    backgroundcolor=\color{white},
    breaklines=true,
    showstringspaces=false,
    escapeinside={(*@}{@*)},
    escapechar=|,
    mathescape 
}

\newcommand{\Z}{\mathbb{Z}}
\newcommand{\R}{\mathbb{R}}

\newcommand{\half}{\frac{1}{2}}

\newcommand{\code}[1]{\texttt{\textbf{#1}}}

\def\implies{\Rightarrow}
\def\inv{{^{-1}}}
\def\homotopy{\s\dot{\sim}\s}
\def\s{\text{  }}

\def\scr{\mathscr}

\def\cs{{\cal S}}
\def\so{{\scr O}}
\def\cm{{\cal M}}
\def\cn{{\cal N}}
\def\cu{{\cal U}}

\def\lie{\hbox{\pounds}}

\font\defbf=eurb10 
\def\db#1{{\defbf #1}}
\def\md{\noindent$\bullet$~}

\def\open#1#2{\left]#1, #2\right[}
\def\rn{\mathbb{R}^n}

\def\d{\text{d}}
\def\I{\text{i}}

\def\bull{{\,\raise0.3ex\hbox{\ensuremath{\scriptscriptstyle\bullet}}\,}}

\def\comm#1#2{{\left[#1\,, #2 \right]}}

\def\implies{\Rightarrow}
\def\inv{^{-1}}

\def\trans{\top}

\def\next{\\}
\def\inprod{\braket}

\def\comm#1#2{{\left[#1\,, #2 \right]}}

\title{\Huge \textbf{Elements of Topology, Differential Geometry and General Relativity for Physicists:  A Mathematica-based Tutorial Approach}}

\author{\Large{Balasubramanian {Ananthanarayan} \textsuperscript{\orcidlink{0000-0001-5955-2123}}, Souvik {Bera}\textsuperscript{\orcidlink{0000-0002-1784-7051}}, Subhasish {Chakrabarty} \textsuperscript{\orcidlink{0000-0002-7564-4607}}},\\ \vspace{0.1cm} \\ \Large{Souradeep {Das} \textsuperscript{\orcidlink{0009-0006-3276-551X}}, Amitabha {Lahiri}\textsuperscript{\orcidlink{0000-0001-8113-6345}}}, \Large{Suhas {Sheikh}\textsuperscript{\orcidlink{0000-0002-8290-9968}}, Sarthak {Talukdar}\textsuperscript{\orcidlink{0000-0002-7127-2922}}}}

\begin{document}

\maketitle

\include{FrontMatter/FM_preface}

\tableofcontents
\include{FrontMatter/FM_AboutAuthors}

\include{FrontMatter/FM_Abstract}

\include{FrontMatter/FM_HowToRead}

\include{FrontMatter/FM_Acknowledgements}

\include{FrontMatter/FM_ListofNotebooks}

\include{FrontMatter/FM_math-info}

\part{Topology}
\include{DiffGeom/DG_topology}		
 
\include{DiffGeom/DG_maps}
\include{Topology/TP_homotopy}
\include{Topology/TP_topophysics}

 \include{DiffGeom/DG_manifolds}		
 \include{DiffGeom/DG_TgtVec}		

  \include{DiffGeom/DG_TgtSpc}			
  \include{DiffGeom/DG_dual} 		
 \include{DiffGeom/DG_vfld}		
 \include{DiffGeom/DG_pbpf}		
 \include{DiffGeom/DG_lie}		

 \include{DiffGeom/DG_flow}		
 \include{DiffGeom/DG_lieder}	
 \include{DiffGeom/DG_tensor}	
 \include{DiffGeom/DG_forms}		
 \include{DiffGeom/DG_extder}	
 
\include{DiffGeom/DG_vol}		
\include{DiffGeom/DG_metric}	
\include{DiffGeom/DG_hodge}		
\include{DiffGeom/DG_maxwell}	
 \include{DiffGeom/DG_stokes}		
 \include{DiffGeom/DG_liegroup}	
 \include{DiffGeom/DG_identity}		
\include{DiffGeom/DG_1ps}	

\include{DiffGeom/DG_FB_Conn_Curv} 

\include{DiffGeom/GR}


\include{data}
\bibliography{topo}

\begin{small}
\input{book_index.tex}
\end{small}

\end{document}

%% file: FrontMatter/FM_preface.tex
\chapter*{Preface}
\addcontentsline{toc}{chapter}{Preface}

Over the years, topology and differential geometry have acquired immense importance within the physics community. Although these subjects are more than a century old, their development has benefited greatly from major contributions by theoretical physicists and from sustained interaction between physicists and mathematicians. In recent decades, numerous books introducing these subjects have appeared. Nevertheless, the material often remains challenging and inaccessible to students encountering it for the first time.

The lecture notes prepared by one of us (AL) on these topics have been available online for several years and have proved popular among students. These notes form the foundation of the present work.

Mathematica has also become a widely used tool in both research and education. Although several packages and built-in functionalities for differential geometry have been developed within Mathematica, these resources remain, in our experience, too scattered to serve as a unified basis for instruction. Our aim has therefore been to develop a concrete and reasonably comprehensive resource that helps beginning students understand the fundamental concepts of topology and differential geometry, appreciate their applications from a physicist's perspective, and develop familiarity with symbolic computation.

To this end, we have assembled a collection of example notebooks based on well-established theoretical material. The relevant theory is reviewed briefly, while the accompanying notebooks demonstrate specific concepts through explicit computations. With minor modifications, many of these examples may also serve as starting points for further exploration by interested students. The examples are intended to complement and clarify the formal development, making the underlying concepts and ideas more accessible.

This work is in keeping with recent initiatives by EPJ publishers and their collaborating European physical societies to make educational material available in a comprehensive and useful form. We also acknowledge the use of artificial-intelligence tools at various stages of the preparation of this work. We hope that the combination of the original lecture notes with extensive Mathematica support will meet the needs of both students and instructors who wish to use the material as a pedagogical resource.

In addition to the built-in commands of Mathematica, we have employed a small number of well-known packages for illustrative purposes, particularly in situations where the built-in functionality is insufficient. These include \texttt{DifferentialForms}, \texttt{Difform}, \texttt{RGTC}, and \texttt{Ricci}. As these packages are open source, they can be used and explored freely for educational purposes.

This work began to take shape when several of the authors were students attending courses at the Indian Institute of Science and contributed to the development of the Mathematica programs. It has evolved through a deeply collaborative, interactive, and innovative effort. We hope that the material presented here will prove valuable to students, teachers, and researchers working in this field.

%% file: FrontMatter/FM_AboutAuthors.tex
\chapter*{About the Authors} 

\begin{description}
    \item[\textbf{\large Balasubramanian Ananthanarayan}]
    \item \vspace{0.2cm} Ananthanarayan is a professor at the Centre for High Energy Physics, Indian Institute of Science, and an experienced teacher and researcher.
    
    \vspace{0.5cm}
    \item[\textbf{\large Souvik Bera}]
    \item \vspace{0.2cm}
    Souvik is a postdoctoral researcher at the Asia Pacific Center for Theoretical Physics, Pohang.  He works on high energy physics and mathematical physics.
    
    \vspace{0.5cm}
    \item[\textbf{\large Subhasish Chakrabarty}]
    \item \vspace{0.2cm} Subhasish is an assistant professor in the department of Physics of Ramanuj Gupta Degree College, Silchar, India. His area of interest is Theoretical High Energy Physics.
    
    \vspace{0.5cm}
    \item[\textbf{\large Souradeep Das}]
    \item \vspace{0.2cm} Souradeep is a first-year graduate student at the Ohio State University. He works on astroparticle physics and cosmology.
    
    \vspace{0.5cm}
    \item[\textbf{\large Amitabha Lahiri}]
    \item \vspace{0.2cm} Amitabha is currently a Visiting (Honorary) Fellow  at the S. N. Bose National Centre for Basic Sciences, from where he retired as Senior Professor. He does research on theoretical physics and has nearly 30 years of graduate teaching experience.

    \vspace{0.5cm}
    \item[\textbf{\large Suhas Sheikh}]
    \item \vspace{0.2cm} Suhas is a third-year graduate student at the University of Illinois, Urbana-Champaign. He is interested in condensed matter theory (specifically, unconventional superconductivity).
    
    \vspace{0.5cm}
    \item[\textbf{\large Sarthak Talukdar}]
    \item \vspace{0.2cm} Sarthak is a PhD student at King's College London, interested in M theory, Superconformal Field Theories, and conformal Bootstrap. 
\end{description}

%% file: FrontMatter/FM_Abstract.tex
\chapter*{Abstract} 

This book is a self-contained, tutorial introduction to topology, differential geometry, and general relativity for students and researchers in physics. Its distinguishing feature is a Mathematica-based approach that makes abstract constructions concrete through explicit, runnable notebooks and worked examples. A key contribution is a large collection of original Mathematica notebooks, written by the authors themselves, that let readers run, verify, and extend every demonstration. The first part covers point-set topology, topological spaces, continuity of maps, homotopy, and the fundamental group with applications that highlight the role of topology in modern physics. The second and largest part develops differential geometry from the ground up: manifolds, tangent and dual spaces, vector fields, pullbacks and pushforwards, Lie brackets and Lie algebras, local flows, and the Lie derivative. It then treats tensors, differential forms, the exterior derivative, volume forms, the metric tensor, and Hodge duality, emphasizing coordinate-free formulations and their computational realization. These tools are applied to Maxwell's equations in the language of forms and the generalized Stokes theorem, while Lie groups, fiber bundles, connections, and curvature bridge geometry and gauge-theoretic physics. The final part uses this framework to present general relativity, computing curvature tensors and field equations for standard spacetimes with dedicated Mathematica packages. Throughout, the book balances mathematical rigor with hands-on computation, enabling readers both to understand the theory and to reproduce every result themselves.

%% file: FrontMatter/FM_HowToRead.tex
\chapter*{How to Read This Book}
\chaptermark{How to Read This Book}
\addcontentsline{toc}{chapter}{How to Read This Book}

This book provides an introduction to the broad subjects of topology and differential geometry. It is intended primarily for undergraduate and graduate students in physics or applied physics who plan to pursue research in theoretical condensed-matter physics or high-energy physics. The text is conceived mainly as course material and as a structured guide to the relevant literature, rather than as an exhaustive textbook. Its principal objective is to highlight the role of Mathematica as a practical computational tool in modern theoretical physics and to enable students to apply it effectively in their own research. \par

Throughout the book, Mathematica is used to reproduce, illustrate, and explore results presented in several standard references. The exposition follows, in large part, the lecture notes of Amitabha Lahiri, one of the co-authors of this work. These lecture notes are freely available on his \href{https://bose.res.in/~amitabha/Diffgeom-lectures.html}{personal website}.

A comprehensive account of all applications of topology and differential geometry in physics is beyond the scope of this book, since these subjects arise throughout many areas of the discipline. We have therefore selected a modest collection of accessible but non-trivial examples that illustrate their physical relevance and are suitable for students encountering the subject for the first time. \par

Readers are encouraged to use this book alongside the accompanying GitHub repository, available at \href{https://github.com/diffgeo-topology-mathematica/DiffGeo_Topology_in_Mathematica}{https://github.com/diffgeo-topology-mathematica/DiffGeo\_Topology\_in\_Mathematica}. The notebooks are developed and tested with Mathematica 12.3 \footnote{Wolfram Research, Inc.,
\textit{Mathematica}, Version 12.3, Champaign, IL, 2021.}. Whenever a piece of code is discussed in the text, the corresponding notebook may be opened and executed in parallel. Readers should also experiment with the notebooks by modifying the examples and exploring related calculations, as this is an effective way to develop familiarity with the underlying concepts and computational methods. The material is also well suited to collaborative student projects aimed at developing further Mathematica-based applications and making the resulting notebooks available to the wider community. \par

%% file: FrontMatter/FM_Acknowledgements.tex
\chapter*{Acknowledgements}
\chaptermark{Acknowledgements}
\addcontentsline{toc}{chapter}{Acknowledgements}

We thank Frank Zizza for correspondence and for providing us the updated version of his package, which is now available in the GitHub repository. 

Souvik Bera has been supported by an appointment to the JRG Program at the APCTP through the Science and Technology Promotion Fund and Lottery Fund of the Korean Government and by the Korean Local Governments~--~Gyeongsangbuk-do Province and Pohang City.

Suhas Sheikh would like to thank Cianán Yuichi Conefrey-Shinozaki and Io Kovach for useful discussions regarding manifolds, homotopy groups and transition maps that formed part of the content for Chapter 4.

%% file: FrontMatter/FM_ListofNotebooks.tex
\chapter*{Mapping of Book Chapters to Mathematica Notebooks}

This document lists, chapter by chapter, the Mathematica notebooks that accompany the text.

\bigskip

\section*{Part I --- Topology}

\subsection*{Chapter 1: Introduction to Topological Spaces}
\begin{itemize}
    \item \texttt{Topology.nb}
\end{itemize}

\subsection*{Chapter 2: Maps and Continuity}
\begin{itemize}
    \item \texttt{Cover.nb}
    \item \texttt{Heine Borel.nb}
\end{itemize}

\subsection*{Chapter 3: Homotopy}
\begin{itemize}
  \item \texttt{Homotopy.nb} 
\end{itemize}

\subsection*{Chapter 4: Special Topics in Topology}
\begin{itemize}
  \item \texttt{Perron-Frobenius.nb} 
  \item \texttt{berrymagnetic.nb} 
  \item \texttt{berry\_curvature.nb}
  \item \texttt{Visualize\_Berry\_curvature\_z.nb}
  \item \texttt{hopf\_fibration\_visualization.nb}
\end{itemize}

\section*{Part II --- Differential Geometry}

\subsection*{Chapter 5: Introduction to Manifolds}
\begin{itemize}
  \item \texttt{Manifold\_examples.nb} 
  \item \texttt{Stereographic\_Projection.nb}
\end{itemize}

\subsection*{Chapter 6: Tangent Vectors and Tangent Spaces}
\begin{itemize}
  \item \texttt{Tangent\_vector.nb} 
  \item \texttt{Tangent\_space.nb} 
\end{itemize}

\subsection*{Chapter 8: Vector Fields}
\begin{itemize}
  \item \texttt{Vector\_fields.nb} 
  \item \texttt{torus-vector-field.nb}
  \item \texttt{Integral\_curve.nb}
\end{itemize}

\subsection*{Chapter 9: Pullback and pushforward}
\begin{itemize}
    \item \texttt{pullback-1.nb}
    \item \texttt{Push\_forward.nb}
\end{itemize}

\subsection*{Chapter 10: Lie brackets and Lie Algebra}
\begin{itemize}
    \item \texttt{Lie-bracket-matrix.nb}
\end{itemize}

\subsection*{Chapter 11: Local flows}
\begin{itemize}
    \item \texttt{localflow.nb}
\end{itemize}

\subsection*{Chapter 12: Lie derivative}
\begin{itemize}
    \item \texttt{LieDerivative.nb}
    \item \texttt{Lie\_Derivative\_Schutz.nb}
\end{itemize}

\subsection*{Chapter 13: Tensors}
\begin{itemize}
    \item \texttt{Tensor-Lie.nb}
\end{itemize}

\subsection*{Chapter 14: Differential forms}
\begin{itemize}
    \item \texttt{Forms.nb}
\end{itemize}

\subsection*{Chapter 15: Exterior derivative}
\begin{itemize}
    \item \texttt{Exterior-derivative.nb}
    \item \texttt{exactform.nb}
\end{itemize}

\subsection*{Chapter 17: Metric tensor}
\begin{itemize}
    \item \texttt{Metric3d.nb}
\end{itemize}

\subsection*{Chapter 18: Hodge duality}
\begin{itemize}
    \item \texttt{Hodge.nb}
    \item \texttt{mukhi.nb}
\end{itemize}

\subsection*{Chapter 19: Maxwell equations}
\begin{itemize}
    \item \texttt{Maxwell.nb}
\end{itemize}

\subsection*{Chapter 20: Stokes' theorem}
\begin{itemize}
    \item \texttt{Vector-calculus.nb}
    \item \texttt{SphericalCoordinates.nb}
    \item \texttt{Stokesexamples.nb}
\end{itemize}

\subsection*{Chapter 21: Lie groups}
\begin{itemize}
    \item \texttt{Introduction to SU2.nb}
    \item \texttt{SO(3).nb}
    \item \texttt{Introduction to SU3.nb}
\end{itemize}

\subsection*{Chapter 23: One parameter subgroups}
\begin{itemize}
    \item \texttt{rotations.nb}
\end{itemize}

\subsection*{Chapter 24: Fiber bundles, Connections \& Curvature}
\begin{itemize}
    \item \texttt{connection-curvature-test.nb}
\end{itemize}

\section*{Part III --- General Relativity}
\subsection*{Chapter 25: The Theory of Relativity}
\begin{itemize}
    \item \texttt{Geodesic\_example\_2D.nb}
    \item \texttt{PolarGeodesic2D.nb}
    \item \texttt{riemann-ricci-scalar.nb}
    \item \texttt{christoffel-curvature-geodesic-einstein.nb}
    \item \texttt{schwarzschildmetric.nb}
    \item \texttt{Schwarzschild\_tensors.nb}
    \item \texttt{SchwarzschildGeodesic.nb}
    \item \texttt{FRW.nb}
\end{itemize}


%% file: FrontMatter/FM_math-info.tex
\chapter*{\resizebox{\textwidth}{!}{\bfseries Introduction to Mathematica for Beginners}}


This will be a naive introduction to the syntax of the language Mathematica\footnote{This content also appears in~\cite{Ananthanarayan:2024ffu}}. 
It is a symbolic manipulation language, which is similar to programming languages like Python or C++ in that it can be used to manipulate numbers or arrays of numbers. However, it has greater strengths in terms of the way it allows one to use human-readable analytical expressions instead of a fully numerical setup like say, Python. This is a great reason to use this language in a book like this, which focuses on theoretical development on par with the advances in modern technology.

In the chapters following this, the reader will be assumed to have ample knowledge of the basic syntax used in Mathematica, since most of the demonstrations will be done with the help of this language. Feel free to skip this chapter if you are already familiar with the language.

Let us introduce you to some of the terms and definitions associated with the Mathematica language, in a completely informal manner. Throughout this introductory chapter, we assume the reader has basic familiarity with any programming language. 

\begin{enumerate}
    \item \textbf{Notebook:} A file in Mathematica is called a notebook, and users usually make a single notebook to solve a particular problem or to demonstrate a bunch of closely related ideas. For example, you may choose to create a notebook to solve a single problem.
    \item \textbf{Cell:} It is the fundamental unit of operation in a notebook. It is a block of code. A cell usually contains a few lines of code, which are meant to be run together as a unit.
    \item \textbf{Types of cell:} A cell can be a text cell, which is used to display information besides the code. There are input cells where the code is typed in, and output cells which display the output generated by an input cell.
    \item \textbf{Suppressing an output:} Sometimes the output from an input is long and lacks presentability. It is useful to put a semicolon (;) at the end of that line in the input cell, which suppresses or hides the output of that line when the code is run.
    \item \textbf{Function:} A function in Mathematica is very similar to that in other programming languages. It is denoted by a specific symbol. When \textit{called} with a set of arguments (or parameters), it processes the arguments to produce a result. In Mathematica, the syntax is to put the arguments in square brackets following the name of the function. For example, \texttt{\textbf {Sqrt}} is a function that returns the square root of its argument. So, for the following input:
    \begin{lstlisting}[style=in]
Sqrt[36]
    \end{lstlisting}
    You will get this output:
    \begin{lstlisting}[style=out]
6
    \end{lstlisting}

    \item \textbf{User-defined function:} Mathematica allows users to write their own code for a custom function. The way a function is defined is pretty similar to how it is done in ordinary mathematics.
    \begin{lstlisting}[style=in]
MyFunction[x_]:=x^2-1
    \end{lstlisting}
    The function we defined above is named \texttt{\textbf{MyFunction}}. It takes an argument \texttt{\textit{x}} and returns the expression on the right-hand side: \texttt{\textit{x$^2$-1}}.

    Notice the syntax used in the above definition. On the left-hand side, the argument is stated with an underscore (\_) following it. The definition uses a colon before the equal sign (:=). Also note that this cell does not produce any output, because we haven't called this function yet. When called with a specific value for the input, it returns an output.
    \begin{lstlisting}[style=in]
MyFunction[11]
    \end{lstlisting}
    \begin{lstlisting}[style=out]
120
    \end{lstlisting}

    In fact, the argument does not have to be a numerical value. You can even put any valid symbol as the argument and get a result:
    \begin{lstlisting}[style=in]
MyFunction[p]
    \end{lstlisting}
    \begin{lstlisting}[style=out]
$p^2-1$
    \end{lstlisting}

    \item \textbf{Lists:} A list is an array of numbers. A list is bound by curly brackets, such as
    \[\{1, 4, 2, 7, 9\}\]
    One can create a list whose $i$-th element is known in terms of $i$, using the built-in function \texttt{\textbf{Table}}, as follows:
    \begin{lstlisting}[style=in]
list1=Table[i^3-i^2, {i, 1, 4}]
    \end{lstlisting}
    \begin{lstlisting}[style=out]
{0, 6, 18, 48}
    \end{lstlisting}
    The first argument is the expression for the $i$-th element in terms of $i$. The second element is a tuple, which denotes that $i$ goes from 1 to $n$. So this returns a list of $i^3-i^2$ for $i=1,2,3,4$. We could have used alternative notations for this tuple, such as \texttt{\textbf{ \{i, 4\} }} or \texttt{\textbf{ \{i, 1, 4, 1\} }}. The former is used when it is understood that the counting starts from 1. The latter has an extra argument (the 1 at the 4th position) which denotes the steps by which $i$ increments or decrements, which is 1 by default. It could also be replaced by the tuple \texttt{\textbf{ \{i, list0\} }}, where \texttt{\textbf{list0}} is another list of numbers (or symbols). This would return the value of $i^3-i^2$ for every element $i$ in \texttt{\textbf{list0}}.

    \item \textbf{Matrices:} A matrix is a nested list, or a list of lists. Hence it has two indices, the first denoting the row and the second, the column. The function \code{MatrixForm} can be used to display a matrix in the rectangular format. For example:
    \begin{lstlisting}[style=in]
mat={{1,0}, {2, 1}};
MatrixForm[mat]
    \end{lstlisting}
    \begin{lstlisting}[style=out]
$\pmqty{1&0\\2&1}$
    \end{lstlisting}
    Often the function \code{MatrixForm} is used in the suffix format, as \code{mat//MatrixForm} which is equivalent to \code{MatrixForm[mat]}. A matrix whose $(i,j)$-th element is known in terms of $i$ and $j$ can be created using the \code{Table} function. For example:
    \begin{lstlisting}[style=in]
Table[2i+j, {i, 1, 2}, {j, 1, 3}]//MatrixForm
    \end{lstlisting}
    \begin{lstlisting}[style=out]
$\pmqty{3&4&5\\5&6&7}$
    \end{lstlisting}
Note that the syntax now contains two tuples instead of 1, to denote the two indices. Similarly one can build tensors of higher rank using the \code{Table} function.
\item \textbf{Module:} In Mathematica, a module is just like a function, in the sense that it takes arguments and returns a result, but has additional features compared to an ordinary function. It allows one to define local variables and execute multiple lines of code when the function is called, as opposed to a single line executed in a usual function. It is helpful in running a lengthy calculation at a function call. For example,
\begin{lstlisting}[style=in]
f[x_] := Module[{a = Log[x], b, c}, b = a x; c = b^2; x - c]
\end{lstlisting}
    Note the structure of the module. It contains two arguments separated by a comma. The first is a list of expressions or symbols where the variables local to the module are defined. These variables \code{a}, \code{b}, and \code{c} cannot be referred to outside the module. The second argument is a list of expressions separated by semicolons, where different steps of the calculation are performed. The last expression usually does not end with a semicolon, and the result of this line is returned as the value of the function.

    \item \textbf{Indices:} Indices are used to refer to a particular position in a list. For example, \code{list1[[2]]} refers to the 2nd item of the list \code{list1}. Note that indices in a list begin from 1 instead of the usual 0 in other programming languages. For the $(i,j)$-th element of a matrix, the syntax to use is \code{mat[[i, j]]}.

    \item \textbf{Non-alphanumeric symbols:} The basic arithmetic operators are very similar to those in other languages, with \code{+}, \code{-}, \code{*}, \code{/}, and \code{\^} denoting their usual actions. Please note that \code{//} is \textit{not} the modulo operator, rather the function \code{Mod[a,b]} is used for this. \code{\%} returns the output of the last executed cell, and \code{\%3} for example returns the output of the cell in the notebook that is numbered as \texttt{Out[3]}. \code{\#} and \code{\&} are often used for defining functions in shorthand. For example, 
    \begin{lstlisting}[style=in]
g = #^2 - # + 1 &;
g[9]
    \end{lstlisting}
    \begin{lstlisting}[style=out]
73
    \end{lstlisting}
    The first line defines the function \code{g} without using any arguments on the left-hand side, and using a simple \code{=} instead of \code{:=}. The right-hand side uses \code{\#} as a dummy variable (when there is more than one argument, they are denoted using \code{\#1}, \code{\#2} etc.) and the \code{\&} is used to mark the end of the definition.
\end{enumerate}

The above definitions are supposed to be a brief introduction for a beginner to the language. Please note that this chapter is not intended to be a complete tutorial on Mathematica. That purpose has to be served by specially designed professional tutorials that focus more on the structure, variations, errors, efficiency and other detailed aspects. What this chapter aims at is to help make the reader familiar with the code snippets that form an integral part of this text. For more details, readers are advised to learn Mathematica on their own from the official website\footnote{\url{https://www.wolfram.com/language/fast-introduction-for-math-students/en/}} and other tutorials available online. There are AI-based tools such as ChatGPT that too can help in understanding the syntax in case one gets stuck. We encourage the use of technology to get a better grasp of the concepts taught in the text.

The information about any expressions, variables or inbuilt/user defined functions can be obtained using the \code{Information} command.

    \begin{lstlisting}[style=in]
Information[Sin]
    \end{lstlisting}
    \begin{lstlisting}[style=out]
Sin[z] gives the sine of z
    \end{lstlisting}
Alternatively the information about a function can be yielded by a special input form \code{?} as below 
    \begin{lstlisting}[style=in]
?MyFunction
    \end{lstlisting}
\begin{lstlisting}[style=out]
Global`MyFunction

MyFunction[x_] := x^2 - 1
    \end{lstlisting}

    The first line in the output dictates the context of the function and the second line shows the definition.

\section*{Special Mathematica modules for Differential Geometry}
We use a bunch of specialized modules, courtesy of their respective owners and developers, in order to implement, simplify and visualize topics in differential geometry. Some of them are listed below:

\begin{itemize}
	\item \code{Ricci.m} --- \textbf{John M.\ Lee} (Univ.\ of Washington),
	\emph{Ricci: A Mathematica Package for Doing Tensor Calculations in
		Differential Geometry}, Version~1.63, 1992--2022.
	\url{https://sites.math.washington.edu/~lee/Ricci/}
	
	\item \code{DifferentialForms.m} --- \textbf{Frank Zizza} (Willamette Univ.;
	later Colorado State Univ.--Pueblo), \emph{Differential Forms},
	Version~3.2, 2014 (orig.\ 1991); Wolfram Library Archive, MathSource~\#482.
	\url{https://library.wolfram.com/infocenter/MathSource/482/}
	
	\item \code{Difform.m} --- \textbf{J\"org Enderlein} (Univ.\ of G\"ottingen),
	\emph{Difform: A Mathematica Package for Differential Form Algebra}, 2018.
	\url{https://www.joerg-enderlein.de/software}
	
	\item \code{EDCRGTC.m} --- \textbf{Sotirios Bonanos} (NCSR ``Demokritos''),
	\emph{EDC \& RGTC: Exterior Differential Calculus and Riemannian
		Geometry \& Tensor Calculus}; see S.\ Bonanos, ``Capabilities of the
	Mathematica Package `Riemannian Geometry and Tensor Calculus','' in
	\emph{Recent Developments in Gravity} (NEB~XI), World Scientific, 2003.
	\url{https://library.wolfram.com/infocenter/MathSource/4484/}
\end{itemize}

We will not assume the reader's familiarity with the specialized modules if and when we use them in the text. We have tried to give clear descriptions for the underlying syntax and coding that goes into using these modules. Readers are encouraged to play around with these modules, taking help from the notebooks in order to develop a better understanding and working knowledge of the packages.

%% file: DiffGeom/DG_topology.tex
\chapter{Introduction to Topological Spaces}\label{topo}

We start by defining a topological space.  Furthermore, we will present
some simple Mathematica commands to check basic properties in order to set
the stage for its use throughout this book.

\md
    A \db{topological space} is a set ${\cal S}$ together with a collection ${\scr O}$ of subsets called \db{open sets} such that the following are true:
\begin{list}{$\roman{enumii})$} {\usecounter{enumii}\topsep=5pt}
\item the empty set $\emptyset$ and ${\cal S}$ are open, and $\emptyset, {\cal S}\in {\scr O}$,
\item the intersection of a \textit{finite} number of open sets is open; if $U_1, U_2 \in \so$, then $U_1\cap U_2 \in \so $, and
\item the union of any number of open sets (in contrast to the case
of intersection) is open, if $U_i\in \so$, then  $\bigcup\limits_i U_i \in \so$  irrespective of the range of $i$. \hfill$\Box$.
\end{list}

It is the pair $\{\cs, \so \}$ which is, precisely speaking, a topological space, or alternatively  a \db{space with topology}. However, it is a common practice to refer to $\cs$ as a topological space which has been given a \db{topology} by specifying $\so$.  The
following are some well-known examples. 


\begin{example}~{\rm 
    $\cs = \mathbb{R}$, the real line, with the open sets being open intervals $\open{a}{b},$ i.e. the sets $\{x\in \mathbb{R}\, |\, a< x <b\}$ and their unions, plus $\emptyset$ and $\mathbb{R}$ itself. Then $(i)$ above is true by definition. }

For two such open sets $U_1 = \open{a_1}{b_1}$ and $U_2 = \open{a_2}{b_2},$ we can suppose $a_1 < a_2$. Then if $b_1 \leqslant a_2\,,$ the intersection $U_1\cap U_2 = \emptyset \in \so\,.$ Otherwise $U_1\cap U_2 = \open{a_2}{b_1}$ which is  an open interval and thus $U_1\cap U_2\in \so.$ So $(ii)$ is true. Furthermore, $(iii)$ is also true by definition. 

\end{example}
\begin{example}~
    
{\rm

$\mathbb{R}^n$ can be given a topology via open rectangles, i.e. via the sets $\{(x_1, \cdots, x_n)\in \mathbb{R}^n\, |\, a_i< x_i <b_i\}$. This is called the \db{standard} or usual topology of $\mathbb{R}^n$.
\hfill$\Box$
}
\end{example}

\begin{example}\label{ex:topospace}~
{\rm 
We will now look at one more example to determine whether a given set along with a collection of its subsets forms a topological space. This will introduce us to a simple application of Mathematica. We begin by defining a discrete set, which is quite easy to handle in Mathematica. The function \code{Subsets} gives a list of all subsets of the set.}
\begin{lstlisting}[style=in]
S = {1, 2, 3};
Subsets[S]
\end{lstlisting}
\begin{lstlisting}[style=out]
{{}, {1}, {2}, {3}, {1, 2}, {1, 3}, {2, 3}, {1, 2, 3}}
\end{lstlisting}
{\rm 
The function defined below checks the axioms in the definition of topological space stated above.}
\begin{lstlisting}[style=in]
IsTopologyQ[set_, subsets_] := Module[{test},
    If[SubsetQ[set, DeleteDuplicates[Flatten[subsets]]],
    test = 
    Sort[DeleteDuplicates[Union /@ Flatten /@ Subsets[subsets]]];
    Return[If[test === Sort[subsets], True, False]];
    , Return[False]];
  ]
\end{lstlisting}
{\rm 
In the function defined above, one has to input the set as the first variable, and a collection of sets (perhaps subsets of the first set) as the second variable. The function returns whether the given pair forms a topological space.

Now let us apply this function to a few such `candidate' topological spaces, in each of which we fix the first variable to be the set ${\cal S}=\{1,2,3\}$ that we defined before:}

\begin{enumerate}{\rm 
\item We begin with the following collection of sets:
\begin{lstlisting}[style=in]
IsTopologyQ[{1, 2, 3}, {{}, {1}, {2}, {3}, {4}}]
\end{lstlisting}
\begin{lstlisting}[style=out]
False
\end{lstlisting}
Clearly the given pair does not form a topological space by definition because $\{4\}$ is not even a subset of ${\cal S}=\{1,2,3\}$.

\item Next, we consider the following collection of subsets: $\{\emptyset, {\cal S}\}$
\begin{lstlisting}[style=in]
IsTopologyQ[{1, 2, 3}, {{}, {1, 2, 3}}]
\end{lstlisting}
\begin{lstlisting}[style=out]
True
\end{lstlisting}
Clearly, both of these are subsets of $\cal{S}$, and the collection satisfies the required criteria.

\item Our next candidate involves all possible subsets of the topology.
\begin{lstlisting}[style=in]
IsTopologyQ[{1, 2, 3},
{{}, {1}, {2}, {3}, {1, 2}, {1, 3}, {2, 3}, {1, 2, 3}}]
\end{lstlisting}
\begin{lstlisting}[style=out]
True
\end{lstlisting}

\item The final example includes all the subsets from the previous case, except for the set $\cal{S}$ itself, which, by definition, no longer meets the requirements to earn the status of a topology.
\begin{lstlisting}[style=in]
IsTopologyQ[{1, 2, 3}, 
{{}, {1}, {2}, {3}, {1, 2}, {1, 3}, {2, 3}}]
\end{lstlisting}
\begin{lstlisting}[style=out]
False
\end{lstlisting}
}
\end{enumerate}

\end{example}

Let us quickly run through a few more useful definitions:

\md
The \db{trivial} topology on $\cs$ consists of $\so = \{\emptyset, \cs\}.$ \hfill$\Box$

We verified that it is indeed a topology in \autoref{ex:topospace} for a specific example of a set, but 
it is easy to verify for any $\cs$.

The \db{discrete} topology on a set $\cs$ is defined by $\so = \{ A\,|\, A\subset\cs\},$ i.e., $\so$ consists of all subsets of $\cs.$ \hfill$\Box$.

We verified in \autoref{ex:topospace} for a particular case that this is indeed a topology.

\md
A set $A$ is \db{closed} if its complement in $\cs,$ also written $\cs\backslash A$ or as $A^{\complement}$, is open. \hfill$\Box$

Closed rectangles in $\mathbb{R}^n$ are closed sets, as are closed balls and single point sets. 

A set {\it can be} neither open nor closed, or {\it can be} both open and closed. In a discrete topology, every set $A\subset \cs$ is both open and closed, whereas in a trivial topology, any set $A\neq\emptyset$ or $\cs$ is neither open nor closed.

The collection $\mathscr{C}$ of closed sets in a topological space $\cs$ satisfies the following:
\begin{list}{$\roman{enumii})$} {\usecounter{enumii}\topsep=0pt}
\item the empty set $\emptyset$ and ${\cal S}$ are closed, $\emptyset, {\cal S}\in {\scr C}$,
\item the union of a \textit{finite} number of closed sets is closed; if $A_1, A_2 \in \scr{C}$, then $A_1\cup A_2 \in \scr{C} $, and 
\item the intersection of any number of closed sets is closed, if $A_i\in \scr{C}$, then  $\bigcap\limits_i A_i \in \scr{C}$  irrespective of the range of $i$.
\end{list}

Closed sets can also be used to define a topology. Given a set $\cs$ with a collection $\mathscr{C}$ of subsets satisfying the above three properties of closed sets, we can always define a topology, since the complements of closed sets are open. (Exercise!)

\md
An \db{open neighbourhood} of a point $P$ in a topological space $\cs$ is an open set containing $P$. A \db{neighbourhood} of $P$ is a set containing an open neighbourhood of $P$. Neighbourhoods can be defined for sets as well in a similar fashion. \hfill$\Box$

Let us see a few examples of the concept of a neighbourhood:
\begin{example}~{\rm 
For a point $x \in \mathbb{R},$ and for any $\epsilon > 0,$\\
$\open{x-\epsilon}{x + \epsilon}$ is an open neighbourhood of $x,$ \\
$\left[{x-\epsilon},{x + \epsilon}\right[$ is a neighbourhood of $x,$\\
$\left[ x, x+\epsilon \right[$ is not a neighbourhood of $x.$ \hfill$\Box$
}
\end{example}

A topological space is said to be \db{Hausdorff} if two distinct points have disjoint neighbourhoods. \hfill$\Box$ 

An example of a topological space of an $n-$element set is provided in the notebook (\texttt{topspacediag\_nelements}) in the repository.

%% file: DiffGeom/DG_maps.tex
\chapter{Maps and Continuity}
We next turn to maps and continuity.  In particular,
topology is useful to us in defining continuity of maps.
\md {\rm 
 A map $f: \cs_1 \to \cs_2$ is \db{continuous} if given any open set $U\subset \cs_2$ its inverse image (or pre-image, what it is an image of) $f\inv(U) \subset \cs_1$ is open
 (a counter-intuitive but necessary definition). }
 \hfill$\Box$   

Let us look at the case where the function maps from a general set to a Euclidean space ($\rn$). For the case of functions from a set to $\rn$, the general definition reduces to the following:
\md
 $f: \cs \to \rn$ is \db{continuous} at $s_0 \in \cs$ if given $\epsilon >0,$ we can always find an open neighbourhood $U$ of $s_0$ such that $|f(s) - f(s_0)| < \epsilon$ whenever $s \in U.$ \hfill$\Box$

Finally, when this definition is applied to functions from $\mathbb{R}^m$ to $\mathbb{R}^n,$ it reduces to the usual $\epsilon - \delta$ definition of continuity, which says the following:

\md
{\rm 
$f: \mathbb{R}^m \to \mathbb{R}^n$ is \db{continuous} at $\vec x_0$ if given $\epsilon > 0,$ we can always find a  $\delta > 0$ such that $|f(\vec x) - f(\vec{x}_0)| < \epsilon$ whenever $|\vec x - \vec{x}_0| < \delta$. \hfill$\Box$}
\md 
One of the most important ideas in topology is that of a \db{homeomorphism}, which pertains to the existence of a one-to-one correspondence between points in two spaces. Thus, we define a bijection between two topological spaces.
\md 
{\rm 
If a map $f: \cs_1 \to \cs_2$ is one-to-one and onto, i.e. a \db{bijection}, and both $f$ and $f\inv$ are continuous, $f$ is called a \db{homeomorphism} and we say that $\cs_1$ and $\cs_2$ are \db{homeomorphic}. \hfill$\Box$ }

    The composition of two continuous maps is a continuous map.
This may be proved as follows:  
if $f: \cs_1 \to \cs_2$ and $g:\cs_2 \to \cs_3$ are continuous maps, and $U$ is some open set in $\cs_3,$ then its pre-image $g\inv(U)$ is open in $\cs_2.$ So $f\inv(g\inv(U)),$ which is the pre-image of that, is open in $\cs_1.$ Thus $(g\circ f)\inv (U) = f\inv(g\inv(U))$ is open in $\cs_1.$ Thus $g\circ f$ is continuous.


Another central idea in topology is that of \db{connectedness}, which is, as intuitively suggested, related to the concept of whether a set consists of disconnected or disjoint sub-parts.

\md 
{\rm 
A topological space $(\cal S, \scr O)$ is \db{connected} if it cannot be expressed as the union of two disjoint open sets (both smaller than $\cal S$ and larger than $\emptyset$) in its topology. On the other hand, if $\exists U_1, U_2 \in \scr O$ such that $U_1 \cap U_2 = \emptyset$, $U_1 \cup U_2 = \cal S$, then the topological space is called \db{disconnected}. \hfill $\Box$.    } 

\begin{exercise}~{\rm 
Check that in a connected topological space $(\cal S, \scr O)$, the only sets which are both open and closed are $\cal S$ and $\emptyset$.}
\end{exercise}

\begin{example}~{\rm
Note that the connectivity depends not only on the set but also on the topology. Consider the set ${\cal S} = \{a, b, c\}$ and the topology ${\scr O} = \{\emptyset , {\cal S}, \{ a \}, \{ b, c \}\}$. Clearly, $\cal S$ is disconnected with respect to this topology. But ${\scr T} = \{\emptyset, {\cal S}, \{a, b\}, \{b, c\}, \{b\}\}$ is also a topology, and $\cal S$ is connected with respect to $\scr T$.}   
\end{example}

Another important idea in topology is that of \db{compactness}. Again, compactness in day-to-day life would suggest that the set is `bounded', but that is not the complete picture, as we will see. For example, in Euclidean space, \emph{closed} and \emph{bounded} subsets of $\R^n$ are called compact. They show many interesting properties. But to generalize this notion to arbitrary topological spaces, first, we need a new definition and a new theorem.

\md
{\rm 
A cover of a set $X$ is a family of sets $\{F_{\alpha}\}$ such that their union contains $X$, i.e., $X \in \cup_{\alpha} F_{\alpha}$. For a topological space $(S, \scr O)$, if $F_{\alpha} \in \scr O \s \s \forall \alpha$, then $\{F_{\alpha}\}$ is called an \db{open cover}. \hfill $\Box$}

In essence, a cover `covers' a set completely. Let us look at an example in Mathematica to check if a collection of sets is a cover under the given topology.
\begin{lstlisting}[style=in]
originalSet = {a, b, c, d, e};
collectionOfSets = {{a, b}, {c, d, e}, {b, c}};
unionOfSets = Union @@ collectionOfSets;
isCover = unionOfSets == originalSet;
Print[isCover]
\end{lstlisting}
\begin{lstlisting}[style=out]
True
\end{lstlisting}
In the above case, the collection of sets is a cover for the topology. It is easy to construct a collection that is not a cover as shown below.
\begin{lstlisting}[style=in]
originalSet = {a, b, c, d, e};
collectionOfSets = {{a, b}, {c, d}, {b, c}};
unionOfSets = Union @@ collectionOfSets;
isCover = unionOfSets == originalSet;
Print[isCover]
\end{lstlisting}
\begin{lstlisting}[style=out]
False
\end{lstlisting}
\md 
We note that a subcover is a subset of an existing cover of a set that still covers the original set, whereas
a finite subcover is defined as a finite collection of open sets from an open cover of a topological space that still covers the entire space.
We mention the statement of an important theorem that we will not prove here.  The Heine-Borel Theorem states that
every open cover of a closed bounded subset of $\R^n$ (in the usual topology) admits a finite subcover. \hfill $\Box$


For a demonstration of the theorem in Mathematica, the reader is advised to look up the corresponding notebook in the repository. 
\begin{exercise}
{\rm~  
Show that no finite subset of $F_n = \left(-1 , 1 - \frac{1}{n}\right)$ for $n = 2, 3, \dots$ covers the open set $(-1, 1)$  }
\end{exercise}

\md 
Inspired by this, we define \db{compactness} as follows: 
{\rm 
Given a topological space $(S, U)$, a set $X \in S$ is said to be \db{compact} if every open cover of $X$ admits a finite subcover \hfill $\Box$.}

%% file: Topology/TP_homotopy.tex
\chapter{Homotopy}\label{homotopy} \label{chap:homotopy}
The connectivity of topological spaces is largely understood by studying \emph{closed loops} and the possibility of deforming them into each other. Let us begin by establishing some terminology. 

\md    A \db{path} $\alpha(t)$ in a topological space $S$, from $x_0 \in \cal S$ to $x_1 \in \cal S$ is a continuous map $\alpha : [0, 1] \rightarrow \cal S$ such that $\alpha(0) = x_0$ and $\alpha(1) = x_1$. If $\exists$ a path between any two points in $\cal S$, $\cal S$ is called \emph{path connected}.\hfill $\Box$

    If $\cal S$ is path connected, then $\cal S$ is connected. \hfill $\Box$

\md {\rm 
    A closed path or \db{loop} in $S$ \emph{based at} $x_0$  is a path $\alpha(t)$ for which $\alpha(0) = \alpha(1)=x_0$. \hfill $\Box$}

  \md
    If $\alpha(t)$ and $\beta(t)$ are two loops based at $x_0$, the \db{product loop} $\gamma = \alpha * \beta$ is defined by 
\begin{equation}
    \gamma(t) = \begin{cases}
        \alpha(2t) & 0 \leq t \leq \frac{1}{2}\\
        \beta(2t-1) & \frac{1}{2} \leq t \leq 1
    \end{cases}
\end{equation}
If one were to think of $t$ as `time', $\gamma(t)$ follows $\alpha$ for the first half of the total time and follows $\beta$ in the second half. \hfill $\Box$

\md From the definition of the product loop, it is easy to define an 
\db{inverse loop} as follows: 
    The inverse of a loop $\alpha(t)$ is defined by $\alpha\inv (t) = \alpha(1 - t)$ for $0 \leq t \leq 1$. This is the same loop traced backwards, and it is easy to check that $\alpha * \alpha \inv$ just returns to the same point without closing a loop. \hfill $\Box$

 \md   The \db{constant loop} is the map $\alpha(t) = x_0$ for $0 \leq t \leq 1$. The image of this loop is a single point. \hfill $\Box$

The idea behind loops is that different loops based at the same point may be deformed into each other \emph{continuously}. This leads
to the notion of \db{homotopy}, which is an equivalence
relation. 

 \md 
    Two loops $\alpha(t)$ and $\beta(t)$ based at $x_0$ are \db{homotopic} to each other if $\exists$ a continuous map $H: [0, 1] \times [0, 1] \rightarrow \cal S$ such that $H(t, 0) = \alpha(t)$, $H(t, 1) = \beta(t)$, and $H(0, s) = H(1,s) = x_0$ for $0 \leq t, s \leq 1$. The new \emph{time} $s$ keeps track of the evolution of the loop. If two loops $\alpha(t)$ and $\beta(t)$ are homotopic, we write the equivalence as $\alpha \homotopy \beta$. \hfill $\Box$ 

\md Homotopy is an \emph{equivalence} relation, i.e.,
\begin{list}{$\roman{enumii})$} {\usecounter{enumii}\topsep=5pt}
\item $\alpha \homotopy \alpha$ (reflexivity),
\item $\beta \homotopy \alpha \Rightarrow \alpha \homotopy \beta$ (symmetry), and
\item $\alpha \homotopy \beta$, $\beta \homotopy \gamma \Rightarrow \alpha \homotopy \gamma$ (transitivity).
\end{list} \hfill $\Box$

Let us choose $H(t, s) = \alpha(t) \s \forall \s s$ to see the 
property of reflexivity. Given $H(t, s): \alpha \rightarrow \beta$, define the new homotopy $H(t, 1-s): \beta \rightarrow \alpha$, which proves the symmetry property. For transitivity, construct three homotopies  $H_1$, $H_2$, and $H_3$ corresponding to $\alpha \homotopy \beta$, $\beta \homotopy \gamma$, and $\alpha \homotopy \gamma$ respectively as follows 
\begin{equation*}
    H_3(t, s) = \begin{cases}
        H_1(t,s) & 0 \leq s \leq \frac{1}{2}, \\
        H_2(t,2s-1) & \frac{1}{2} \leq s \leq 1.
    \end{cases}
\end{equation*}
In a sense, it is similar to taking the product loop with respect to the \emph{new time} $s$. 

\md Other Properties of Homotopy:
\begin{itemize}
    \item $\alpha \homotopy \beta \Leftrightarrow \alpha \inv \homotopy \beta \inv$, and  
    \item $\alpha \homotopy \beta$, $\alpha' \homotopy \beta' \Rightarrow \alpha * \alpha' \homotopy \beta * \beta'$
\end{itemize} \hfill $\Box$

\md 
Equivalence relations partition a collection of objects into disjoint classes, called \emph{equivalence classes}. In the present context, the equivalence of loops lets us define $[\alpha]$, which is the class of all loops homotopic to $\alpha(t)$. Again, the collection of \emph{all distinct} homotopy classes of loops in $\cal S$ based at $x_0$ is $\pi_1({\cal S}, x_0)$, called the \textbf{fundamental group} or \textbf{first homotopy group} of $\cal S$ at $x_0$. Since the term \emph{group} exists in the definition, there must be a multiplication relation of homotopy classes, which is defined as 
\begin{equation}
    [\alpha] * [\beta] = [\alpha * \beta]
\end{equation}
\md
It remains to prove that $\pi_1({\cal S}, x_0)$ is a {group} under this composition. If $[\alpha] \in \pi_1$, $[\beta] \in \pi_1$, then quite trivially $[\alpha] * [\beta] = [\alpha* \beta] \in \pi_1$, which proves the \emph{closure}. 

Defining the identity of the group $[i]$ as the class of loops homotopic to the constant loop at $x_0$, $[\alpha] * [i] = [i] * [\alpha] = [\alpha]$. This identity element gives the existence of an inverse loop as $[\alpha^{-1}] = [\alpha]^{-1}$. For associativity, consider 
\begin{align}
    \alpha * (\beta * \gamma) &= \begin{cases}
        \alpha(2t) & 0 \leq t \leq \frac{1}{2}, \\
        (\beta * \gamma) (2t-1) & \frac{1}{2} \leq t \leq 1,
    \end{cases}\\
    &= \begin{cases}
        \alpha(2t) & 0 \leq t \leq \frac{1}{2}, \\
        \beta(4t-2) & \frac{1}{2} \leq t \leq \frac{3}{4},\\
        \gamma(4t-3) & \frac{3}{4} \leq t \leq 1.
    \end{cases}
\end{align}
Similarly, 
\begin{align}
    = \begin{cases}
        \alpha(4t) & 0 \leq t \leq \frac{1}{4}, \\
        \beta(4t-1) & \frac{1}{4} \leq t \leq \frac{1}{2},\\
        \gamma(2t-1) & \frac{1}{2} \leq t \leq 1.
    \end{cases}
\end{align}
For a pictorial depiction of this proof, the reader is referred to
the book~[\cite{nash2011topology}].

\begin{exercise}~{\rm 
    Construct an $H(t, s)$ such that $H(t, 0) = \alpha * (\beta * \gamma)$ and  $H(t, 1) = (\alpha * \beta) * \gamma$.}
\end{exercise}
Hence, $(\alpha * \beta) * \gamma \homotopy \alpha * (\beta * \gamma)  \implies ([\alpha] * [\beta]) * [\gamma] = [\alpha] * ([\beta] * [\gamma])$. \par
Now, we will show that the fundamental groups in $\cal S$, in some sense, are independent of the base point. For that, we will need two new definitions. First, we extend the definition of the product loop to open paths where the initial point of $\alpha$ and the final point of $\beta$ are not the same. 

\md 
{\rm 
    A map that preserves the group multiplication $\left(.\right)$, i.e., 
$\phi: g \inv \in G \implies \phi(g \inv) = (\phi(g))\inv \in H$, $\phi(g_1. g_2) \in G \implies \phi(g_1 . g_2) = \phi(g_1) . \phi(g_2) \in H$, and $\phi: i \in G \implies \phi(i) = i' \in H$ is called a \textbf{homomorphism}.   If the homomorphism is bijective, it is called an \textbf{isomorphism}. Now, we are ready to state the notion of independence of the fundamental group with respect to the base point. \hfill $\Box$}

\md 
{\rm 
     It may be proved that if a topological space $\cal S$ is path-connected, $\pi_1({\cal S}, x_0)$ and $\pi_1({\cal S}, x_1)$ are isomorphic as groups. \hfill $\Box$}

Consider the loop $\alpha(t)$ based at $x_0$ and its equivalence class $[\alpha] \in \pi_1({\cal S}, x_0)$. Consider $\gamma(t)$, an open path from $x_0$ to $x_1$. Now, define the map $\phi: \pi_1({\cal S}, x_0) \rightarrow \pi_1({\cal S}, x_1)$ to be $\phi([\alpha]) = [\gamma \inv * \alpha * \gamma]$. 

\begin{exercise}~{\rm 
    Check that $\phi$ satisfies the conditions for an isomorphism. }
\end{exercise}
Hence, for path-connected spaces, it is enough to write $\pi_1(\cal S)$ instead of $\pi_1({\cal S}, x_0)$. \hfill $\Box$

\md 
{\rm 
    Two topological spaces $\cal X$ and $\cal Y$ are of the same \db{homotopy type} if $\exists$ continuous maps $f: X \rightarrow Y$ and $g: Y \rightarrow X$ such that $f \circ g: X \rightarrow X$ and $g \circ f: Y \rightarrow Y$ are homotopic to the identity elements $\mathbf{I}_X$ and $\mathbf{I}_Y$. \hfill $\Box$}

{\rm
    If $\cal X$ and $\cal Y$ are two topological spaces of the same homotopy type, $\pi_1({\cal X}, x_0)$ is isomorphic to $\pi_1({\cal Y}, y_0)$ where $x_0 \in {\cal X}$ and $y_0 \in {\cal Y}$. \hfill $\Box$}

Since $\mathbf{I}_X$ and $\mathbf{I}_Y$ are homotopic to each other, two 
\db{homeomorphic spaces} are of the same homotopy type. Hence, the statement boils down to: 
\md 
{\rm A corollary is that
    if $\cal X$ and $\cal Y$ are two path-connected homeomorphic topological spaces, $\pi_1({\cal X}, x_0)$ is isomorphic to $\pi_1({\cal Y}, y_0)$ where $x_0 \in {\cal X}$ and $y_0 \in {\cal Y}$.\hfill $\Box$}

\md This corollary establishes the fundamental group as a \emph{topological invariant}. 

\md 
{\rm 
    A \db{retract} $\cal A$ of a topological space $\cal X$  is a subspace of $\cal X$  such that $\exists$ a continuous map $r:{\cal X} \rightarrow {\cal A}$ which preserves the position of all points in that subspace, i.e., $r(a) = a$ $\forall a \in {\cal A} \subset {\cal X}$. \hfill $\Box$}

\md 
{\rm 
    A subset $\cal A$ of a topological space $\cal X$ is a \db{deformation retract} if $\exists$ a retraction $r: {\cal X} \rightarrow {\cal A}$ and a homotopy $H: {\cal X} \times [0, 1] \rightarrow {\cal X}$ such that $H(x, 0) = x$, $H(x, 1) = r(x)$, and $H(a, t) = a$ $\forall a \in {\cal A}$ where $t \in [0, 1]$. \hfill $\Box$}
It is immediately clear from these definitions that not all retracts are deformation retracts. 

\md 
{\rm 
    If a point $a \in {\cal X}$ is a deformation retraction of $\cal X$, $\cal X$ is said to be contractible. The homotopy connecting $a$ and $\cal X$ is called a contraction. \hfill $\Box$}

\md 
In the context of fundamental groups, deformation retracts become useful because of the following theorem:
{\rm 
    If $\cal X$ is a path-connected topological space, and $\cal A$ is a deformation retract of $\cal X$, $\pi_1({\cal X}, a)$ is isomorphic to $\pi_1({\cal A}, a)$ for any $a \in \cal A$. \hfill $\Box$}
\newline

Hence, if we can show that a small space $\cal A$ is a deformation retract of a bigger space $\cal X$ and if we can calculate the fundamental group of that smaller space, we, in effect, know the fundamental group of the bigger space.

\begin{example}~
{\rm 
    
    \begin{itemize}
    \item Let $Y = {0}$ denote a deformation retract of $\rn$ that contains only the single point $\{0\}$. Define $H: \rn \times [0, 1] \rightarrow \rn$ in the following manner that establishes this claim: 
    \begin{equation}
        H(x, t) = (1-t) x, \s x \in \rn \s \text{and} \s t \in [0, 1]
    \end{equation}
    We also know that $\pi_1(\rn, 0)$ is isomorphic to $\pi_1(\{0\}, 0)$, which is the identity element. 

    \item A closed unit disc with origin removed $\mathbb{D}^n-\{0\}$ has a deformation retract $S^{n-1}$. Try to visualize it for $n= 2$, which is the isomorphism between the fundamental groups of the annulus and the circle: $\pi_1(\mathbb{D}^2-\{0\}) = \pi_1(S^1)$. These are isomorphic to the group of integers $\Z$ under addition because we are searching for homotopic maps that take $S^1 \rightarrow S^1$. Label two circles by $\theta$ and $\phi$. Define the homotopy map as follows: 
    \begin{equation}
        H(\theta, t) = \theta(1-t) + \phi(\theta) t
    \end{equation}
    and $\phi(0) =0$ and $\phi(2 \pi) = 2 n \pi$ are satisfied. Thus, $\phi(\theta) = n \theta$. Therefore, the equivalence classes are characterised by integers. $n$ is called the \emph{Winding Number}.
    \item Two circles with a point in common form a deformation retract of $\mathbb{D}^2 -\{p\} - \{q\}$. The group $\pi_1(\mathbb{D}^2 -\{p\} - \{q\})$ is \emph{not} abelian. 
    \item For a cylinder, $\pi_1(S^1 \times \R^1) = \Z$. This can be generalised to the statement, $\pi_1(X \times Y) = \pi_1(X) \oplus  \pi_1(Y)$. This allows us to write the fundamental group of the $n-$dimensional torus as $\pi_1(T^n) = \pi_1(\underbrace{S^1 \times \cdots \times S^1}_{n}) = \underbrace{\Z \oplus \cdots \oplus \Z}_{n}$. 
    \item $\pi_1(S^n) = 0$ for $n \geq 2$, i.e., a loop can always avoid the \emph{defect} and can therefore be shrunk to a point. 
\end{itemize}
}
\end{example}

\md 
The fundamental group classifies the homotopy classes of loops in a topological space $\cal X$. Still, we can go ahead and classify many other objects in $\cal X$, e.g., homotopy classes of spheres in $\cal X$ or tori in $\cal X$. This is the motivation behind higher homotopy groups. Intuitively, it can be immediately seen that a loop cannot detect a point-like defect in $\R^3$, but a 2-sphere can. Higher homotopy groups are obtained by suitably defining higher dimensional analogues of loops. \par

\md 
We start by defining an $n-$cube as $I^n=\{(s_1, s_2, \cdots, s_n)| 0 \leq s_i \leq 1, \forall i\}$, whose boundary is defined as $\partial I^n = \{(s_1, s_2, \cdots, s_n) \in I^n | s_i = 0 \s \text{or} \s s_i = 1,  \forall i\}$. Loops are defined such that their initial and final points are identified. Analogously, one considers continuous maps $\alpha: I^n \rightarrow \cal X$ with properties $\alpha(s) = x_0~ \forall s \in \partial I^n$, i.e., all points on the boundary of the $n-$cube are mapped to a single point in $\cal X$. One can now introduce the homotopy $F: I^n \times [0, 1] \rightarrow \cal X$ requiring 
\begin{align}
    F(s_1, s_2, \cdots, s_n, 0) &= \alpha(s_1, s_2, \cdots, s_n) \nonumber\\
    F(s_1, s_2, \cdots, s_n, 1) &= \beta(s_1, s_2, \cdots, s_n) \\
    F(s_1, s_2, \cdots, s_n, t) &= x_0, \s \forall (s_1, s_2, \cdots, s_n) \in \partial I^n \nonumber 
\end{align}
If such a homotopy map exists, $\alpha$ and $\beta$ are said to be homotopic. \hfill $\Box$

\md This is an equivalence relation between $n-$loops, which partitions the space of $n-$loops into disjoint classes. The set of equivalence classes for pathwise connected spaces is denoted by $\pi_n({\cal X}) = \{\alpha | \alpha: I^n \rightarrow {\cal X}, \s \alpha(s \in \partial I^n) = x_0\}$. We can impose an algebraic structure on this by introducing the composition of $\alpha$ and $\beta$ by connecting them along a common part of the boundary, say $s_1 = 1$: 
\begin{equation}
    \alpha \circ \beta (s_1, \cdots, s_n) = \begin{cases}
        \alpha(2s_1, s_2, \cdots, s_n) &\s 0 \leq s_1 \leq \frac{1}{2}\\
        \beta(2s_1-1, s_2, \cdots, s_n) &\s \frac{1}{2} \leq s_1 \leq 1
        \end{cases}
\end{equation}
\begin{align}
    \alpha\inv (s_1, \cdots, s_n) &= (1-s_1, s_2, \cdots, s_n) \\
    e(s_1, \cdots, s_n) &= x_0
\end{align} \hfill $\Box$

\md Some important results relevant to mapping between spheres are:
\begin{itemize}
    \item $\pi_n({\cal X})$ is abelian for $n > 1$.
    \item $\pi_n(S^n)$ is isomorphic to the group of integers under addition. 
    \item $\pi_m(S^n) = 0$ for $m<n$. 
    \item $\pi_m(S^n)$ for $m>n$ is non-trivial. A famous result is $\pi_3(S^2) \sim \Z$
\end{itemize} 

{\rm 

\md 
{\rm
    Let $\cal X$ and $\Tilde{\cal X}$ be connected topological spaces. The pair $(\Tilde{\cal X}, p)$ is called the \db{covering space} of $\cal X$ if $\exists$ a continuous map $p: \Tilde{\cal X} \to \cal X$ such that $p$ is surjective and $\forall x \in {\cal X}~ \exists$ connected open set ${\scr U} \subset \cal X$ containing $x$ such that $p \inv ({\scr U})$ is a disjoint union of open sets in $\Tilde{\cal X}$, each of which is mapped homeomorphically onto $\scr U$ by $p$. \hfill $\Box$}

\md 
{\rm 
    If $\Tilde{\cal X}$ is simply connected (path-connected and every path between two points can be continuously transformed into any other such path while preserving the two endpoints), $(\Tilde{\cal X}, p)$ is called the \db{Universal Covering Space} of $\cal X$. \hfill $\Box$}

\begin{example}~
{\rm 
    $\R$ is a universal covering space of $S^1$. Take the map $p: x \rightarrow e^{i 2 \pi x}$ that takes entries from $\R$ and sends them to $U(1)$. Clearly, $p$ is surjective, and if ${\scr U} = \{e^{i 2 \pi x}| x \in (x_0-\epsilon, x_0 + \epsilon)\}$, $p \inv ({\scr U}) = \bigcup\limits_{n \in \Z} (x_0 - \epsilon + n, x_0 + \epsilon + n)$, which is a disjoint union of open sets of $\R$. }
\end{example}

\md 
{\rm We now state a theorem: 
    let $(\Tilde{\cal X}, p)$ be a universal covering space of a connected topological space $\cal X$. If $x_0 \in \cal X$ and $\Tilde{x}_0 \in \Tilde{\cal X}$ are base points such that $p(\Tilde{x_0}) = x_0$, the induced homomorphism $p_* : \pi_n(\Tilde{\cal X}, \Tilde{x}_0) \rightarrow \pi_n ({\cal X}, x_0)$ is an isomorphism for $n \geq 2$. \hfill $\Box$}

\begin{example}~
{\rm 
     $\pi_n(\R) = 0$ since $\R$ is contractible. Hence, $\pi_n(S^1) \cong \pi_n(U(1)) = 0$ for $n \geq 2$. }
\end{example}
}
Demonstration of homotopy in Mathematica:
\begin{lstlisting}[style=in]
Manipulate[Graphics[{LightGray, Disk[{0, 0}, 1],
White, Disk[{0, 0}, (1 - t)*0.2],
Blue, Disk[{0, 0}, (1 - t)*1], White, Disk[{0, 0}, (1 - t)*0.2]},
PlotRange -> {{-1.2, 1.2}, {-1.2, 1.2}}, Axes -> True, 
AxesOrigin -> {0, 0}], {{t, 0}, 0, 0.9}]
\end{lstlisting}

\begin{figure}[ht]
    \centering
\begin{subfigure}{0.33\linewidth}
    \centering
    \includegraphics[width=0.95\linewidth]{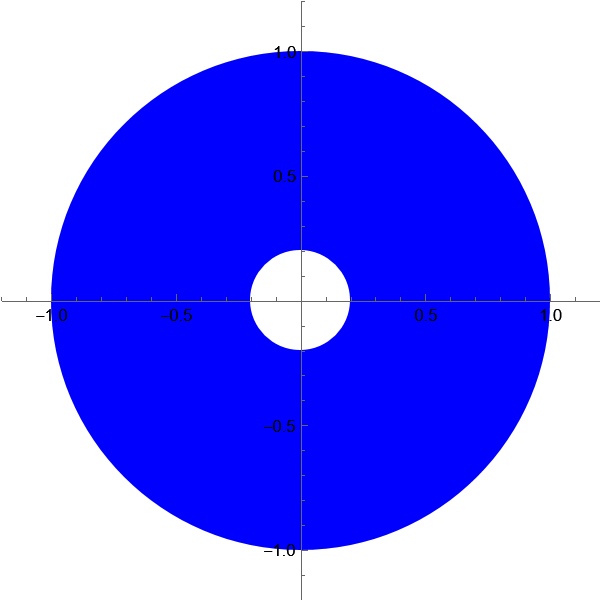}
    \subcaption{}
\end{subfigure}
\begin{subfigure}{0.33\linewidth}
    \centering
    \includegraphics[width=0.95\linewidth]{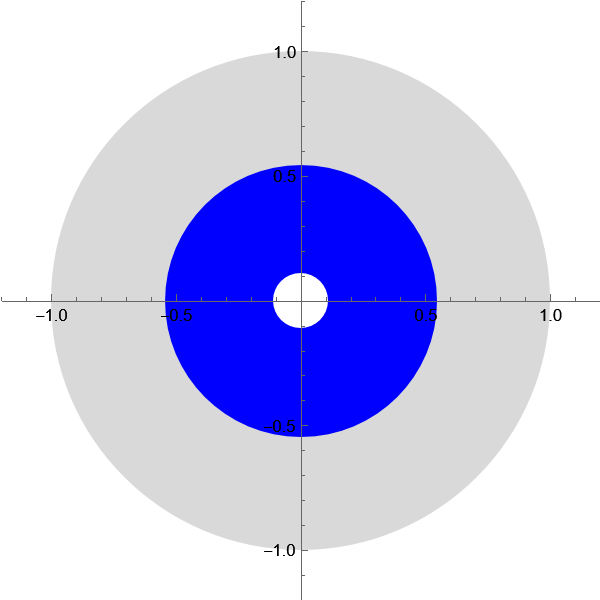}
    \subcaption{}
\end{subfigure}
\begin{subfigure}{0.33\linewidth}
    \centering
    \includegraphics[width=0.95\linewidth]{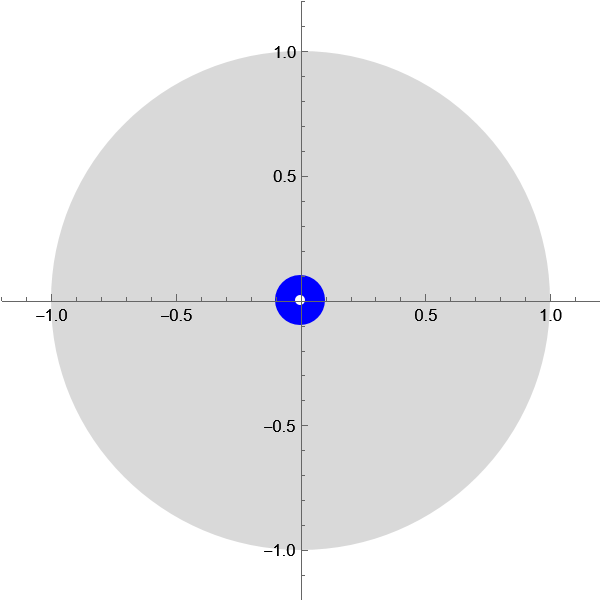}
    \subcaption{}
\end{subfigure}
\caption{Homotopy of an annular disk}
\label{fig:homotopy-disk-hole}
\end{figure}

\begin{lstlisting}[style=in]
Animate[Show[Graphics3D[Sphere[{0,0,0}]],
ParametricPlot3D[{Cos[u]Sin[1-v] ,Sin[u] Sin[1-v],Cos[1-v]},{u,0,2Pi},
PlotStyle->Directive[Red,Thick]]],{v,-1/2,1}]
\end{lstlisting}

\begin{figure}[ht]
    \centering
\begin{subfigure}{0.33\linewidth}
    \centering
    \includegraphics[width=0.95\linewidth]{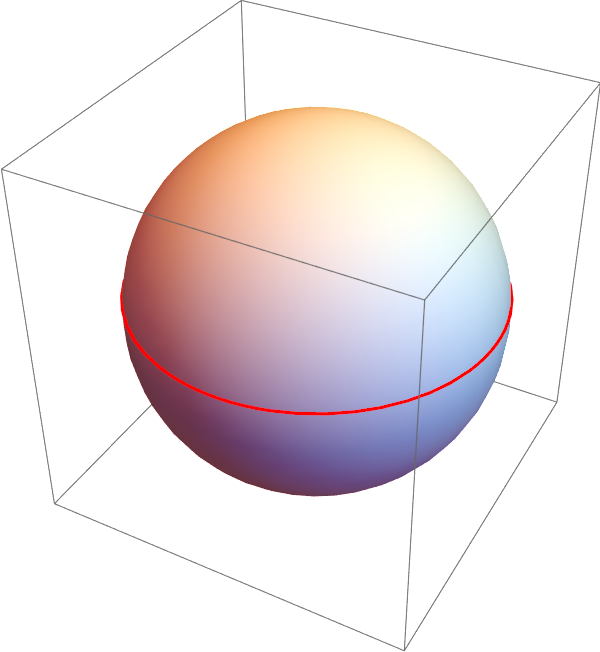}
    \subcaption{}
\end{subfigure}
\begin{subfigure}{0.33\linewidth}
    \centering
    \includegraphics[width=0.95\linewidth]{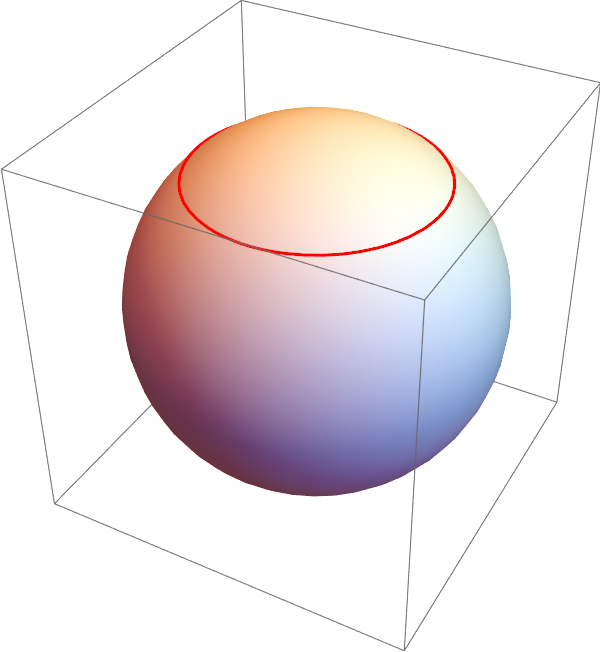}
    \subcaption{}
\end{subfigure}
\begin{subfigure}{0.33\linewidth}
    \centering
    \includegraphics[width=0.95\linewidth]{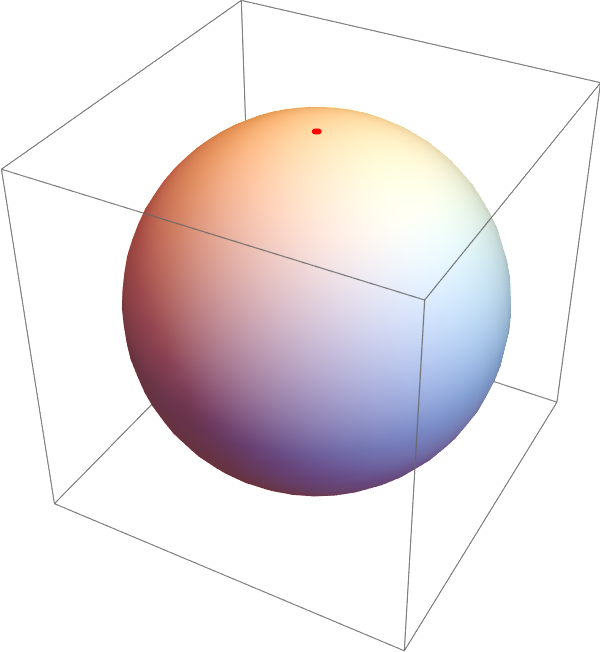}
    \subcaption{}
\end{subfigure}
\caption{Homotopy of a loop on a manifold without a hole}
\label{fig:homotopy-nohole}
\end{figure}

\begin{lstlisting}[style=in]
Animate[Show[{Graphics3D[{Sphere[{0, 0, 0}]}], 
   Graphics3D[{Black, 
     Polygon[Table[{0.12 Cos[\[Theta]], 0.12 Sin[\[Theta]], 
        1}, {\[Theta], 0, 2 Pi, Pi/40}]]}], 
   ParametricPlot3D[{Cos[u]*Sin[\[Theta]], Sin[u]*Sin[\[Theta]], 
      Cos[\[Theta]]} /. \[Theta] -> (Pi/2) - 
       v*((Pi/2) - ArcSin[0.12]), {u, 0, 2 Pi}, 
    PlotStyle -> {Red, Thick}]}, Boxed -> True, Axes -> False, 
  SphericalRegion -> True, 
  PlotRange -> {{-1.2, 1.2}, {-1.2, 1.2}, {-1.2, 1.2}}], {v, 0, 1}]
\end{lstlisting}

\begin{figure}[ht]
    \centering
\begin{subfigure}{0.33\linewidth}
    \centering
    \includegraphics[width=0.95\linewidth]{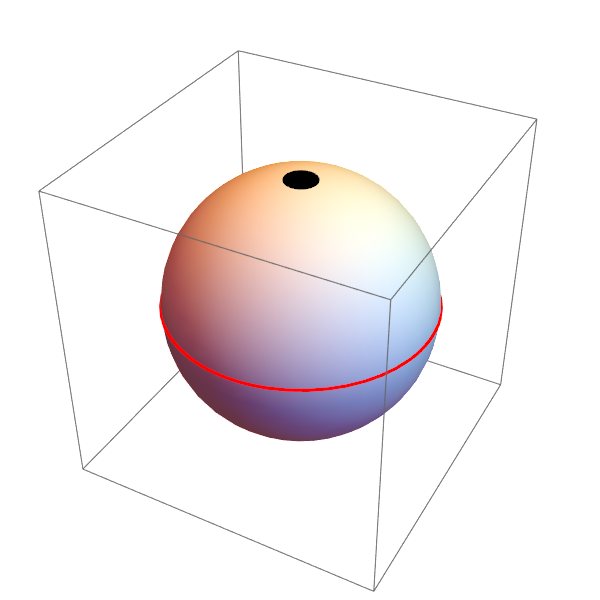}
    \subcaption{}
\end{subfigure}
\begin{subfigure}{0.33\linewidth}
    \centering
    \includegraphics[width=0.95\linewidth]{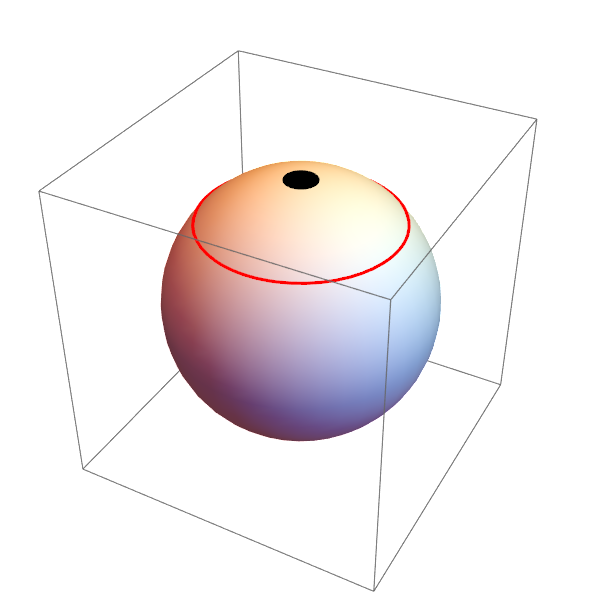}
    \subcaption{}
\end{subfigure}
\begin{subfigure}{0.33\linewidth}
    \centering
    \includegraphics[width=0.95\linewidth]{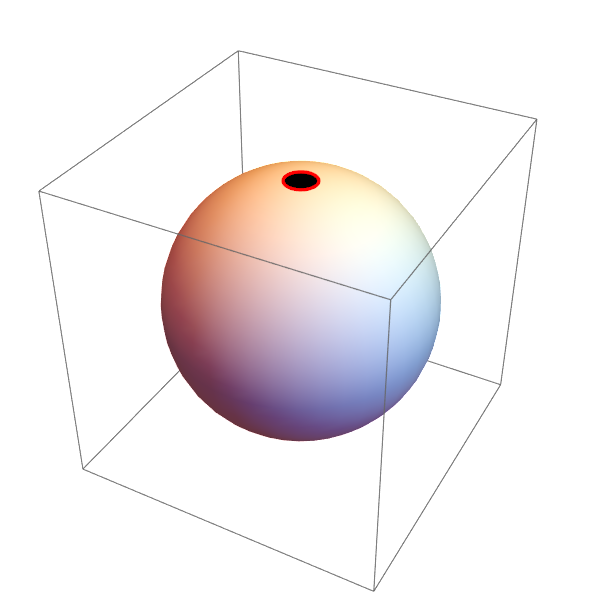}
    \subcaption{}
\end{subfigure}
\caption{Homotopy of a loop on a manifold with a hole}
\label{fig:homotopy-hole}
\end{figure}

\begin{lstlisting}[style=in]
Manipulate[
Graphics[{LightGray, Disk[{0, 0}, 1], Blue, Disk[{0, 0}, 1 - t]  },
PlotRange -> {{-1.2, 1.2}, {-1.2, 1.2}}, Axes -> True, 
AxesOrigin -> {0, 0}], {{t, 0}, 0, 0.95}]
\end{lstlisting}

\begin{figure}[hb]
    \centering
\begin{subfigure}{0.33\linewidth}
    \centering
    \includegraphics[width=0.95\linewidth]{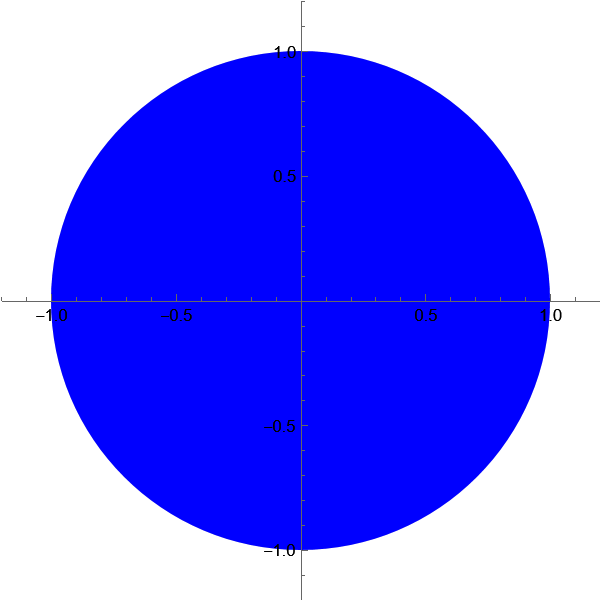}
    \subcaption{}
\end{subfigure}
\begin{subfigure}{0.33\linewidth}
    \centering
    \includegraphics[width=0.95\linewidth]{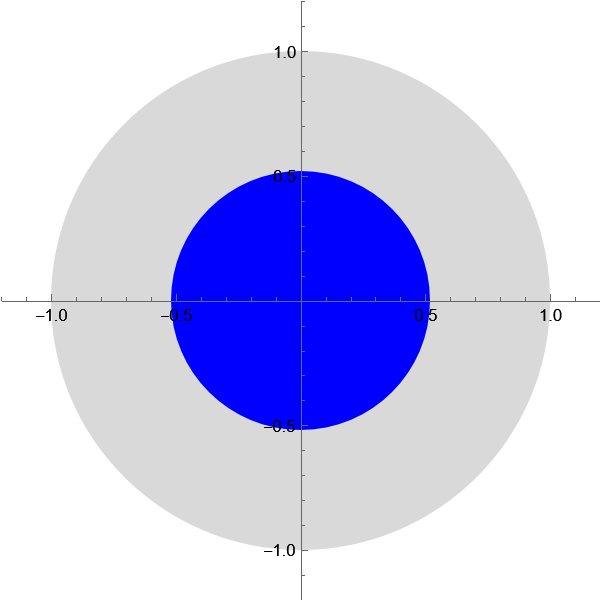}
    \subcaption{}
\end{subfigure}
\begin{subfigure}{0.33\linewidth}
    \centering
    \includegraphics[width=0.95\linewidth]{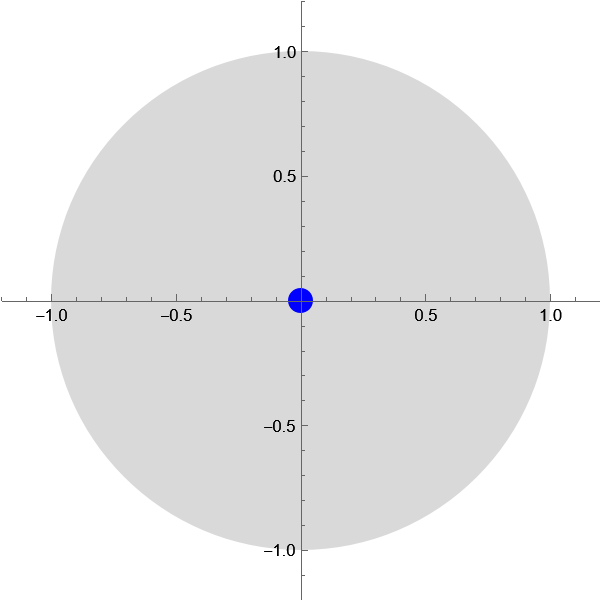}
    \subcaption{}
\end{subfigure}
\caption{Contractible manifold}
\label{fig:contractible-manifold}
\end{figure}

%% file: Topology/TP_topophysics.tex
\chapter{Topology: Some Special Topics}\label{topophysics}
Given the plethora of applications that incorporate topology (or topological concepts) in physics, we find it rather instructive to elucidate some examples in this chapter.
This chapter is in a sense a digression from the main pedagogical intent of this
manuscript, but brings the reader to the forefront of some important research
applications, and can be skipped on a first reading.

\section{Topology and the Integer Quantum Hall Effect}
An important example is the Integer Quantum Hall Effect (IQHE) and how it arises as a topological phenomenon.
Background information on the IQHE can be readily obtained in any modern standard graduate condensed matter text such as~[\cite{Altland:2006si}].
In the following, we briefly discuss its origin via a topological lens. 
Soon after the IQHE was experimentally discovered, Robert Laughlin~[\cite{PhysRevB.23.5632}] presented his well-known argument to explain the quantization of Hall conductance under certain general conditions. We provide a rough sketch of the argument here.

Due to its complete universality, the quantization can be assumed to be insensitive to continuous deformations of the sample geometry. Exercising this freedom, Laughlin proposed going from the regular ``Hall bar" geometry to an annular (ring-like) geometry with higher symmetry. Following through with this construction, the external bias voltage is replaced by the electromotive force generated by a weakly time-dependent flux through the annular region. 
It was then proposed that one could consider the response of the system to the application of an adiabatic  variation of the flux threading the annulus. 

We motivate this idea as follows:
For specific values of the annular flux, namely integer multiples of a flux quantum $\phi = 2\pi n, n=0,1,2,3...$, the Hamiltonian 
of the system is gauge equivalent to the Hamiltonian in the absence of flux. This is because an integer multiple of the flux quantum can be removed by performing the gauge transformation $\psi_{a} \rightarrow e^{in\theta} \psi_{a}$, on the wavefunctions of the system. (Note: For non-integral fluxes, this transformation changes the single-valuedness of the wavefunctions encoded in the boundary condition $\psi(r,2\pi) = \psi(r,0)$ and therefore alters the Hilbert space of the problem) 

As the flux is gradually increased from $\phi = 0$ to $\phi = 2\pi$ (we assume that the appropriate gauge transformation has been performed to redefine the flux dependence of the problem to a change in the azimuthal boundary conditions; the eigenstates for different values of the flux are then truly distinct), each state moves along a path ``perpendicular" to the collective set of eigenstates. Eventually, for $\phi = 2\pi$, we return to the original, initial $\phi = 0$ basis (This is merely because $\hat{H}_{\phi = 2\pi}$ can be mapped onto $\hat{H}_{\phi = 0}$ via a gauge transformation that leaves the boundary conditions intact). However, this does not imply that the individual basis states map onto themselves upon the traversal of the path $\phi = 0 \rightarrow \phi = 2\pi$. While the entire set of eigenstates is  reproduced in this picture, permutations of individual states are fully consistent with the problem's gauge invariance. The non-invariance of individual states upon completion of a round trip back to a gauge-equivalent Hamiltonian is called \textbf{spectral flow}. Upon the addition of a flux quantum through the ring, $n$ states radially centered at the inner perimeter are pushed above the Fermi energy $E_{F}$ ($n$ is the number of Landau levels below $E_{F}$). Simultaneously, $n$ states at the outer perimeter sink below $E_{F}$. To regain thermal equilibrium, the system responds by transferring $n$ electrons (the ones that get pushed above the Fermi energy $E_{F}$) from the inner to the outer perimeter. This process takes place on timescales that roughly equal the time $t_{0}$, time that it takes to adiabatically send the flux quantum through the system i.e., the transverse current $I_{2} = \frac{n}{t_{0}}$ (we work in units where $e = 1$). The electromotive force driving this process is merely $V_{1} = \dot{\phi} = \frac{2\pi}{t_{0}}$. The corresponding Hall conductance is then given by:

\begin{equation}
\sigma_{12} = \frac{I_{2}}{V_{1}} = \frac{n}{2\pi}
\end{equation}

To further elaborate on the concept of spectral flow, we direct readers to Chapter 9 of~[\cite{Altland:2006si}], which illustrates it succinctly using a set of examples that focus on the resolution of individual Landau levels (specifically, the lowest and first excited Landau levels; assumptions on disorder strength, that depends on the inverse elastic scattering time, relative to the separation between Landau levels, are made and delocalized bulk states are duly postulated). 

As it turns out (obtained by Avron and Seiler~[\cite{PhysRevLett.51.51}], and separately by Thouless, Kohmoto, Nightingale and den Nijs~[\cite{PhysRevLett.49.405}]), the quantum Hall conductance is indeed a topological invariant - the first Chern class of the $U(1)$ principal bundle over the two-dimensional torus, $\mathbb{S}^{1} \times \mathbb{S}^{1}$.

To develop an insight into the first Chern class, we now demonstrate a rudimentary calculation of the first Chern number for a very simple system: a spin-1/2 particle in a constant external magnetic field, $B$. The Hamiltonian describing the system is given by:

\begin{equation}
H = \mu \sigma \cdot \textbf{B}
\end{equation}
where $\mu$ is the magnetic moment, $\sigma$ denotes the Pauli matrices and $\textbf{B}$ is the magnetic field. In 3 spatial dimensions, the eigenstates of the Hamiltonian have energies $\pm \mu B$ and the corresponding eigenvectors are:

\begin{equation}
\ket{u_{-}} =\begin{bmatrix}
\sin{\frac{\theta}{2}} e^{-i \phi} \\
-\cos \frac{\theta}{2}
\end{bmatrix}
\end{equation}
and
\begin{equation}
\ket{u_{+}} =\begin{bmatrix}
\cos{\frac{\theta}{2}} e^{-i \phi} \\
\sin \frac{\theta}{2}
\end{bmatrix}
\end{equation}
where ($\pm$) denote the corresponding energy labels. \newline
Consider $\ket{u_{-}}$, whose Berry connection can readily be computed as:
\begin{equation}
\mathcal{A}_{\theta} = \bra{u_{-}} i \frac{1}{r} \partial_{\theta} \ket{u_{-}} = 0
\end{equation}
and
\begin{equation}
\mathcal{A}_{\phi} = \bra{u_{-}} i \frac{1}{r \sin \theta} \partial_{\phi} \ket{u_{-}} = \frac{\sin^{2} \frac{\theta}{2}}{r \sin \theta}
\end{equation}

The Berry curvature is then computed as:
\begin{equation}
\Omega_{r} = \frac{1}{r \sin \theta} [\partial_{\theta} (\mathcal{A}_{\phi} \sin \theta) - \partial_{\phi} \mathcal{A_{\theta}}] \hat{r} = \frac{1}{2r^{2}} \hat{r}
\end{equation}
Consequently, in momentum space, we find:
\begin{equation}
\Omega_{k} = \frac{1}{2\abs{k}^{2}} \hat{k}
\end{equation}
(We use $\hat{k}$ to mean the unit radial vector pointing radially outwards, on the sphere in momentum space).

Choosing a new gauge via the transformation(s) $\ket{u_{-}}^{\prime} = e^{i \alpha \theta} \ket{u_{-}}$ ($\alpha \in \mathbb{R}$) would only change the Berry connections (since they are gauge-dependent) while the Berry curvature would remain the same (since it is a gauge independent observable). 

The Berry curvature per solid angle is then given by:
\begin{equation}
\overline{\Omega}_{\theta \phi} = \frac{\Omega_{\theta \phi}}{\sin \theta} = \frac{1}{2}. 
\end{equation}
For this case, the Berry phase corresponding to any given path on the unit 3-sphere $\mathcal{S}^{2}$ in magnetic-field space is just exactly half the solid angle subtended by the path. The first Chern number for this system is then computed as (in spherical polar coordinates):

\begin{equation}
C = \frac{1}{2\pi} \oint d^{3}k \cdot  \Omega_{k} = \frac{2\pi}{2\pi} = 1
\end{equation}

Let us consider another approach: as the Berry curvature per solid angle is $\frac{1}{2}$, the Berry curvature when integrated (summed) over the entire $4\pi$ solid angle yields $\frac{1}{2} \cdot 4\pi = 2\pi$ and we have $C = 1$ as before. We thus establish a non-trivial Chern number, which indeed functions as a topological invariant for this simple two-level system. A good rule of thumb to keep in mind is the following: 

The Berry phase~[\cite{Berry1984}] is a geometric phase (one of many geometric phases obtained from regular adiabatic evolution in QM) which remains invariant (modulo $2\pi$) under an arbitrary gauge transformation. It only depends on the path taken in parameter space (which often happens to be reciprocal/momentum space, for systems with well defined energy dispersions i.e., band dispersions $\epsilon(k)$ in crystals; one can define the wavevector as a parameter and the Bloch waves serve as the bands/eigenstates). The Berry connection is the quantity that when integrated over a closed path yields the Berry phase~[\cite{Berry1984}]. The Berry curvature is an antisymmetric rank-$2$ tensor obtained from the Berry connection and can be viewed as a pseudovector in $3$-dimensional parameter space. The Chern theorem manifests itself in the Berry phase via the fact that the integral of the Berry curvature (which is related to the Berry phase) over a closed manifold is quantized in units of $2\pi$. This quantized number is the Chern number and serves as a very useful and ubiquitous tool for understanding and probing various quantization effects. It is a topological invariant (for two-dimensional topological insulators with non-degenerate bands) because no regular perturbation to the Hamiltonian can change the Chern number. A change in the Chern number is one of the signatures for identifying a topological phase transition.

Topological invariants can discriminate only between
Hamiltonians which are not homotopic, so the set of possible values for a topological invariant is equivalent to the
set of band structures up to homotopy.~[\cite{SRyu2009}] 

The Berry curvature can also be written as a sum over all other eigenstates and can thus be viewed as a result of the residual interaction of these projected-out eigenstates~[\cite{RevModPhys.82.1959}]

For a demonstration of Berry connection, curvature etc. in Mathematica, the reader is encouraged to look up the corresponding notebooks (\texttt{berrymagnetic, berry\_curvature, Visualize\_Berry\_curvature\_z}) in the repository. 

\begin{figure}[ht]
    \centering
    \includegraphics[width=0.5\linewidth]{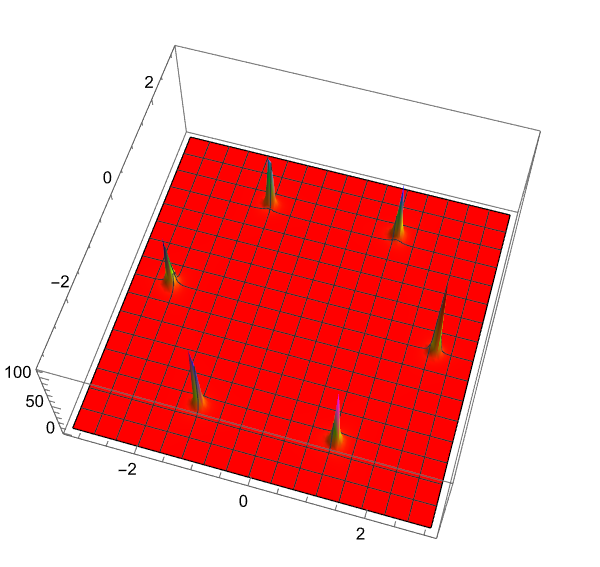}
    \caption{Visualisation of Berry curvature}
    \label{berry-curvature}
\end{figure}

\section{Hopf Fibration}
The next example we turn to is known as the Hopf Fibration~[\cite{BenjaminHSmith}].
The Hopf Fibration (also known as the Hopf bundle or the Hopf map) is a many-to-one continuous function from the 3-sphere onto the 2-sphere such that each distinct point of the 2-sphere is mapped from a distinct great circle of the 3-sphere. The 3-sphere is thought to be composed of fibers, with each fiber being a circle - one for each point of the 2-sphere. This fiber bundle structure is denoted as:
\begin{equation}
S^{1} \rightarrow S^{3} \xrightarrow{\text{p}} S^{2}
\end{equation}
such that the fiber space $S^{1}$ (a circle) is embedded in the total space $S^{3}$ (the 3-sphere), and $p:S^{3} \rightarrow S^{2}$ (the Hopf map) projects $S^{3}$ onto the base space $S^{2}$ (the usual 2-sphere). 

The Hopf fibration possesses the property that it is locally a product space. We now briefly discuss the definition and construction of this ``Hopf map".

For any $n \in \mathbb{N}$, an $n$-dimensional sphere ($n$-sphere) can be defined as the set of points in an $(n+1)$-dimensional space which are at a fixed constant distance from a central point. Without loss of generality, the central point can be taken to be the origin, and the distance of the points on the sphere from this origin can be assumed to be of unit length. With this convention, the $n$-sphere comprises the points $(x_{1},x_{2},\dots ,x_{n+1})$ in $\mathbb{R}^{n+1}$ with $\sum_{i = 1}^{i = n + 1} x_{i}^{2} = 1$.

The Hopf fibration $p:S^{3} \rightarrow S^{2}$ of the 3-sphere over the 2-sphere can be defined in the following manner~[\cite{BenjaminHSmith}]: 

Associate (``Identify") $\mathbb{R}^{4}$ with $\mathbb{C}^{2}$ and $\mathbb{R^{3}}$ with $\mathbb{C} \times \mathbb{R}$ ($\mathbb{C}$ here denotes the complex numbers) via the following:
\begin{equation}
(x_1,x_2,x_3,x_4) \leftrightarrow (z_0,z_1) = (x_1 + ix_2, x_3 + ix_4)
\end{equation}
and
\begin{equation}
(x_1,x_2,x_3) \leftrightarrow (z,x) = (x_1 + ix_2, x_3)
\end{equation}

$S^{3}$ is identified with the subset of all $(z_0,z_1)$ in $\mathbb{C}^{2}$ s.t. $\abs{z_0}^{2} + \abs{z_1}^{2} = 1$, and $S^{2}$ is identified with the subset of all $(z,x)$ in $\mathbb{C} \times \mathbb{R}$ s.t. $\abs{z}^{2} + x^{2} = 1$. The Hopf fibration $p$ is then defined by:
\begin{equation}
p(z_0,z_1) = (2z_0 z_1^{*}, \abs{z_{0}}^{2} - \abs{z_{1}}^{2})
\end{equation}
The first component is a complex number whereas the second component is a real number. Any point on the 3-sphere must preserve the property that $\abs{z_0}^{2} + \abs{z_1}^{2} = 1$. That then implies that $p(z_0,z_1)$ lies on the unit 2-sphere in $\mathbb{C} \times \mathbb{R}$ (which can be easily verified by adding the squares of the absolute values of the complex and the real components of $p$ and checking that they sum to $1$). Moreover, if two points on the 3-sphere map to the same point on the 2-sphere, i.e., if $p(z_0,z_1) = p(w_0,w_1)$ then $(w_0,w_1) = (\alpha z_0, \alpha z_1)$ for some $\alpha \in \mathbb{C}$ s.t. $\abs{\alpha}^{2} = 1$. As the set of complex numbers $\alpha$ satisfying $\abs{\alpha}^{2} = 1$ forms the unit circle in the complex plane, it immediately follows that for each point $m$ in $S^{2}$, the inverse image $p^{-1}(m)$ is a circle, i.e., $p^{-1}m \cong S^{1}$. Hence, the 3-sphere is realized as a disjoint union of these circular fibers.

A direct parametrization of the 3-sphere which employs the Hopf map is as follows~[\cite{BenjaminHSmith}]:
\begin{equation}
z_0 = e^{i \frac{\zeta_1 + \zeta_2}{2}} \sin \eta,
\end{equation}
\begin{equation}
z_1 = e^{i \frac{\zeta_2 - \zeta_1}{2}} \cos \eta
\end{equation}
or in Euclidean $\mathbb{R}^{4}$:
\begin{equation}
x_1 = \cos \left (\frac{\zeta_1 + \zeta_2}{2} \right) \sin \eta,
\end{equation}
\begin{equation}
x_2 = \sin \left (\frac{\zeta_1 + \zeta_2}{2} \right) \sin \eta,
\end{equation}
\begin{equation}
x_3 = \cos \left (\frac{\zeta_2 - \zeta_1}{2} \right) \cos \eta,
\end{equation}
\begin{equation}
x_4 = \sin \left (\frac{\zeta_2 - \zeta_1}{2} \right) \cos \eta,
\end{equation}
where $\eta \in [0, \frac{\pi}{2}]$, $\zeta_1 \in [0, 2\pi]$, $\zeta_2 \in [0,4\pi]$. Every value of $\eta$ excluding $0$ and $\frac{\pi}{2}$ (which specify circles), specifies a separate flat torus in the 3-sphere and one round trip ($0$ to $4\pi$) of either $\zeta_1$ or $\zeta_2$ results in the parametrization making one full circle of both limbs of the torus. A mapping of the aforementioned parametrization to the 2-sphere is given by:
\begin{equation}
z = \cos (2\eta),
\end{equation}
\begin{equation}
x = \sin(2\eta) \cos(\zeta_1),
\end{equation}
\begin{equation}
y = \sin(2\eta) \sin(\zeta_1)
\end{equation}
with points on the circles parametrized by $\zeta_2$. \par

A very nice visualisation of the Hopf Fibration is available online on the following site:
\begin{center}
    \href{https://samuelj.li/hopf-fibration/}{Visualization of Hopf Fibration}
\end{center}
This will take the reader to the Mathematica demonstration of Hopf Fibration.
We provide an explicit example below.
\begin{lstlisting}[style=in]
A = 1; R = 1;
Dnew = R^2 + x[t]^2 + y[t]^2 + z[t]^2;
ux = (2 * A/Dnew^2) * (x[t]*z[t] - R*y[t]);
uy = (2*A/Dnew^2)* (R*x[t] + y[t]*z[t]);
uz = (1/2)*(A/Dnew^2)*(R^2 - x[t]^2 - y[t]^2 + z[t]^2);
u = Sqrt[ux^2 + uy^2 + uz^2];
\end{lstlisting}
\begin{lstlisting}[style=in]
soln[x0_?NumericQ, y0_?NumericQ, z0_?NumericQ] := 
NDSolve[{x'[t] == ux/Abs[u], y'[t] == uy/Abs[u], z'[t] == uz/Abs[u],
   x[0] == x0, y[0] == y0, z[0] == z0}, {x, y, z}, {t, 0, 200}];
\end{lstlisting}
\begin{lstlisting}[style=in]
ParametricPlot3D[
Evaluate[{x[t], y[t], z[t]} /. soln[#, #, #] & /@ 
  Range[2, 5, .1]], {t, 0, 200}, PlotRange -> All, MaxRecursion -> 8, AxesLabel -> {"x", "y", "z"}]
\end{lstlisting}

\begin{figure}[ht]
    \centering
    \includegraphics[width=0.5\linewidth]{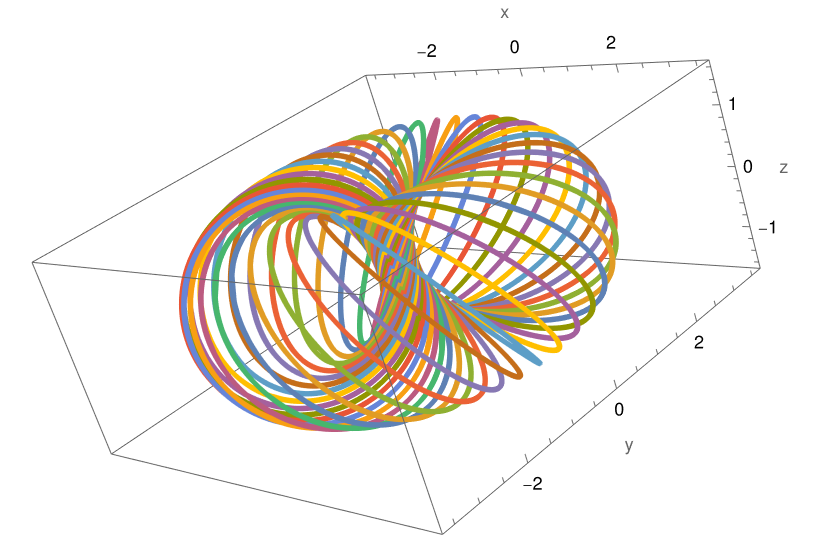}
    \caption{Visualisation of Hopf fibration}
    \label{hopf-fibr-vis}
\end{figure}

\section{A constructive example of a nontrivial bundle and its corresponding topological invariant}
Our next example is that of a nontrivial bundle and its corresponding
topological invariant.
We consider the
Weyl Hamiltonian, which is an intriguing 2-parameter family of Hamiltonians. Let us construct a nontrivial complex vector bundle over the 2-sphere, $S^{2}$.

The Hamiltonian is in general given by:
\begin{equation}
H = -\vec{d}\cdot \vec{\sigma}= - \begin{pmatrix}
d_z & d_x - id_y \\
d_x + id_y & -d_z \\
\end{pmatrix}
\end{equation}
Let $(d_x,d_y,d_z) \in S^{2} \subset \mathbb{R}^{3}$. The bundle can be defined by explicitly constructing the projection matrix $P(\hat{d}) = \ket{\psi_-}\bra{\psi_-}$, where $\ket{\psi_-}$ is the lower eigenvector of $H$, and then using the correspondence between projection matrix-valued functions over a manifold $\mathcal{M}$ and the complex vector bundles over $\mathcal{M}$. The bundle can be constructed more concretely by defining local trivializations and inspecting the transition maps. 

We represent $(d_x,d_y,d_z)$ by the spherical polar coordinate parametrization:$ (\sin \theta \cos \phi, \sin \theta \sin \phi, \cos \theta)$. For arbitrary $(\theta, \phi)$, the lower eigenvector is given by:
\begin{equation}
\ket{\psi(\theta,\phi)} = \begin{pmatrix}
\cos \frac{\theta}{2} \\
\sin \frac{\theta}{2} e^{i \phi} \\
\end{pmatrix}
\end{equation}
(Recall that we already looked at this eigenvector while looking at the example of computing the first Chern number for a simple two-level system in the presence of an external constant magnetic field).

We now run into a problem, which is that this eigenvector is not defined at the south pole of the sphere $\theta = \pi$. No choice of eigenvector at $\theta = \pi$ will make $\ket{\psi(\theta,\phi)}$ a smooth function on the entire sphere. This suggests that it is impossible to have a global non-vanishing section of this $\mathbb{C}$-bundle over $S^{2}$ i.e., this bundle is non-trivial. To put it another way, the wavefunctions on the sphere twist around in a sufficiently complicated way.

To better understand the structure of this complex line bundle, we describe it in terms of an open cover and transition functions. Points in the bundle are given by $\ket{\psi(\theta,\psi)}$. $(\theta,\psi)$ gives the point on the base manifold $S^{2}$ and the eigenvector $\ket{\psi(\theta,\psi)}$ is an element of the fiber (which is $\mathbb{C}$ in this case, since this is a complex line bundle). 

We now construct the local trivializations. 
$S^{2}$ is covered by two contractible open sets, called the Northern Hemisphere (parametrized by $\theta < \frac{\pi}{2} + \epsilon$) and the Southern Hemisphere (parametrized by $\theta > \frac{\pi}{2} - \epsilon$). The local trivializations are then given by:
\begin{equation}
\varphi_N^{-1} : (\hat{d}, \psi(\hat{d}) \rightarrow (\hat{d}, U_N^{\dagger}(\hat{d})\psi(\hat{d})),
\end{equation}
and
\begin{equation}
\varphi_S^{-1} : (\hat{d}, \psi(\hat{d}) \rightarrow (\hat{d}, U_S^{\dagger}(\hat{d})\psi(\hat{d})),
\end{equation}
where $U_N^{\dagger}(\hat{d})$ is the matrix that diagonalizes the effective $k \cdot p$ Hamiltonian $\hat{H}(\hat{d}) = -\hat{d} \cdot \vec{\sigma}$ (such that $U^{\dagger} H U$ is diagonal), so that $U_N^{\dagger}(\hat{d}) \psi(\hat{d})$ is always a vector with a vanishing lower component. Thus, the co-domain of this map is a trivial bundle $U \times \mathbb{C}$, where $U \subset S^{2}$ is the subset of points $\hat{d}$ where the map is well-defined. This is exactly what a local trivialization is supposed to be. (Had the bundle been trivial, we would have been able to find a global trivialization, but the fact is that neither $U_N(\hat{d})$ nor $U_{S}(\hat{d})$ are smooth functions on all of $S^{2}$) \newline
The definition of $U_{S}$ is valid only on the South Hemisphere, and the definition of $U_{N}$ is valid only on the North Hemisphere. We now construct $U_{N}$ and $U_{S}$ explicitly, given that we know the eigenvectors of $\hat{H}$:
\begin{equation}
U_N(\hat{d}) = \begin{pmatrix}
\cos \left (\frac{\theta}{2} \right) & \sin \left (\frac{\theta}{2} \right) e^{i\phi} \\
\sin \left (\frac{\theta}{2} \right) e^{i \phi} & -\cos \left (\frac{\theta}{2} \right) \\
\end{pmatrix}
\end{equation}
\begin{equation}
U_S(\hat{d}) = \begin{pmatrix}
\cos \left (\frac{\theta}{2} \right) e^{-i \phi} & \sin \left (\frac{\theta}{2} \right) \\
\sin \left (\frac{\theta}{2} \right) & -\cos \left (\frac{\theta}{2} \right) e^{-i \phi} \\
\end{pmatrix}
\end{equation}
The transition function is given by $t_{NS} :(\hat{d}, \psi(\hat{d})) \rightarrow (\hat{d}, U_S^{\dagger} U_N  \psi(\hat{d})$. Computing this explicitly, we find that the transition function on the "Equator" $\left (\frac{\pi}{2} - \epsilon < \theta < \frac{\pi}{2} + \epsilon\right)$ is given by:
\begin{equation}
t_{NS}: \psi(\theta,\phi) \rightarrow e^{i \phi} \psi(\theta, \phi)
\end{equation}

We observe that the phase of the transition map winds around the equator exactly once. We call this the winding number. 
Two bundles that are homotopic are isomorphic to each other (this can be seen in the following way: Let $F$ be a bundle over $M \times [0,1]$. By definition, the bundles $\tau^{*}E \vert_{M \times {0}}$ and $\tau^{*} E \vert_{M \times {1}}$ are homotopic, where $\tau$ is the restriction; two bundles are obtained by pulling back along the restriction at the endpoints). To discuss the isomorphism classes of bundles over $S^{2}$, we should discuss their homotopy classes, which are given by homotopy classes of the transition maps $t_{NS}: Equator \rightarrow U(1)$. (Notice here that the transition function of a line bundle is usually given by $GL(1,\mathbb{C})$ but this is homotopic to $U(1)$. Therefore, we work with a $U(1)$ structure group). \newline
If a bundle over $S^{2}$ is trivial, we should be able to deform $t_{NS}$ into the identity map. However, the winding number of $t_{NS}$ is homotopy invariant: we cannot deform a transition map from winding number $1$ to winding number 0. We then conclude that the bundle we described above is nontrivial: it is impossible to deform it into the trivial bundle $S^{2} \times \mathbb{C}$. There exists an immediate corollary which goes as follows:

There is an associated principal $U(1)$ bundle over $S^{2}$ with the same transition function $t_{NS}:g \rightarrow e^{i\phi} g$, which is nontrivial. (One cannot choose a global section of a nontrivial principal $U(1)$ bundle over $S^{2}$)\newline
Replacing $\vec{d}$ with $\vec{k}$ in the $\vec{k} \cdot \vec{p}$ Hamiltonian ($\vec{k}$ is a $3$-d wavevector in the Brillouin zone) yields an approximate Bloch Hamiltonian near a Weyl point~[\cite{PhysRevLett.121.106402}]. The homotopy class ("winding number") of $t_{NS}$ is referred to as the "topological charge" of the 'Weyl node'.~[\cite{PhysRevB.75.121306}]

We can generalize
the above as follows: transition functions $t_{ij}$ for a rank-$N$ vector bundle are elements of $GL(N,\mathbb{C})$. It is true that:
\begin{equation}
Vect_N(S^{n+1}) = \pi_{n}(GL(N,\mathbb{C})),
\end{equation}
and when $N > \frac{n}{2}$, 
\begin{equation}
\pi_{n}(GL(N,\mathbb{C})) = \mathbb{Z} \hspace{0.5mm}\forall n = 2k + 1, k \in \mathbb{Z},
\end{equation}
\begin{equation}
\pi_{n} = 0 \hspace{0.5mm} \forall n = 2k, k \in \mathbb{Z},
\end{equation}
where $Vect_{N}(S^{n+1})$ refers to the homotopy equivalence class (isomorphism class) of rank-$N$ vector bundles over $S^{n+1}$. 

\section{\texorpdfstring{$\mathbb{Z}_{2}$}{Z2} topological invariants}
Our next example pertains to $\mathbb{Z}_{2}$ topological invariants.
As discussed earlier, the Chern invariant is odd under time reversal, which means that topologically nontrivial Chern states can only arise when time reversal symmetry is broken either by an external magnetic field or by some form of magnetic order. A consequence of time reversal and inversion symmetry being preserved is that the Berry curvature has to vanish. In a material with only spin-orbit coupling present, the Chern numbers always vanish. In 2005, Kane and Mele~[\cite{PhysRevLett.95.146802}]
obtained a new topological invariant in  Time Reversal Invariant (TRI) fermionic systems. It is a parity 'odd' or 'even' invariant (hence the name $\mathbb{Z}_{2}$). Systems which fall in the 'odd' class are $2$D topological insulators. 

We briefly review the concept of time reversal symmetry. A physical system possesses time reversal symmetry if it remains invariant under the transformation $T:t \rightarrow -t$. In the Brillouin zone, $T$ (the time reversal symmetry operator) changes momentum $\vec{k}$ to $-\vec{k}$. It was shown by Kramers~[\cite{knaw_pu00014621}] that $T^{2} = (-1)^{2s}$ where $s$ is the total spin quantum number of a state. Thus, the time reversal symmetry operator takes the complex conjugate of the wavefunction and rotates the spin. Generically, the time reversal symmetry operator for spin-1/2 particles is given by the anti-unitary operator $\Theta = e^{i\pi \sigma_{y}}K$, where $K$ denotes complex conjugation. Quickly note that $T^{2} = -1$. Anti-unitary operators act on two generic states $\ket{\phi}$ and $\ket{\psi}$ as follows:
\begin{equation}
\bra{\Theta \phi}\Theta \psi \rangle = \langle \psi \ket{\phi} = \langle\phi \ket{\psi}^{*}.
\end{equation}

A consequence of time reversal symmetry is Kramers' theorem that says that all eigenstates of a time reversal invariant Hamiltonian are at least two-fold degenerate. In the absence of spin-orbit coupling interactions, Kramers degeneracy is essentially the degeneracy between up and down spins. The two partners of a Kramers pair live in the same fiber. With the additional structure given by the time reversal symmetry, the Bloch bundle modeling the band structure becomes a quaternionic vector bundle. If a Bloch Hamiltonian is $T$ invariant, it implies:
\begin{equation}
\Theta H(\vec{k})\Theta^{-1} = H(-\vec{k}).
\end{equation}
To obtain a $\mathbb{Z}_{2}$ invariant, we begin by defining a unitary matrix:
\begin{equation}
w_{mn}(\vec{k}) = \bra{u_{m}(\vec{k})}\Theta \ket{u_{n}(-\vec{k})},
\end{equation}
which is called the sewing matrix, where $\ket{u_{m}(\vec{k})}$ are occupied Bloch states. Using the fact that $\Theta$ is anti-unitary and that $\Theta^{2} = -1$, we have:
\begin{equation}
w^{T}(\vec{k}) = -w(-\vec{k}).
\end{equation}
This implies that the matrix $w$ is anti-symmetric at points of the Brillouin zone $T^{d}$ where $\vec{K} = -\vec{K}$. There are a total of $2^{d}$ such points in $T^{d}$. In $2$D, there are $4$ such special points $\vec{K_{a}}$ in the bulk of the Brillouin zone where this happens. At each of these TRI points, the filled band fiber of a $\mathbb{Z}_{2}$ topological insulator is equipped with a quaternionic structure. For an anti-symmetric matrix, the determinant is the square of its Pfaffian. We define:
\begin{equation}
\delta_{a} = \frac{Pf[w(\vec{K}_{a})]}{\sqrt{det[w(\vec{K}_{a})]}} = \pm 1.
\end{equation}
It is always possible to choose $\ket{u_{m}(\vec{k})}$ continuously throughout the Brillouin zone, so the branch of the square root can be specified globally and the $\mathbb{Z}_{2}$ invariant is:
\begin{equation}
(-1)^{\nu} = \prod_{i = 1}^{4}\delta_{i}.
\end{equation}
The above formulation can also be generalized to $3$D topological insulators. 
[\cite{PhysRevB.75.121306,PhysRevLett.98.106803,PhysRevLett.95.146802}].

The association $\vec{k} \rightarrow H(\vec{k})$ is a map $H:T^{d} \rightarrow H_{her}$, where $H_{her}$ is the space of $n \times n$ Hermitian matrices. Taking into consideration the action of $T$ (the time reversal symmetry operator) on $T^{d}$, to specify a map with time reversal symmetry it is sufficient to know the restriction of $H$ onto a fundamental domain of the $T$ action. Given that the time reversal symmetry operator maps the fibers at $\vec{k}$ and $-\vec{k}$, we realize that there is an apparent redundancy in the description of the system on the whole Brillouin torus. Such a choice of the fundamental domain is called the Effective Brillouin Zone (EBZ). The EBZ consists of half of the Brillouin torus, keeping only one member of each Kramers pair ($\vec{k},-\vec{k}$) except at the boundary. In the $2$D case, the EBZ is given by the region $[0,\pi] \times [-\pi,\pi]$ which can be thought of as a cylinder $C = S^{1} \times I$. The boundary of this EBZ (denoted by $\partial$EBZ) still has a nontrivial action on $T$, and we can consider a fundamental domain of the boundary of the EBZ, which we call $R^{d-1}$. In $2$D, we have $R^{1} = [0,\pi]$. 

Using this idea, Fu and Kane~[\cite{PhysRevB.76.045302}]
proposed another form of the $\mathbb{Z}_{2}$ invariant which is given by:
\begin{equation}
\nu = \frac{1}{2\pi} \left[\oint_{\partial EBZ}A - \int_{EBZ}F \right] ({\rm mod}\hspace{0.5mm}2),
\end{equation}
where $A$ is the total Berry connection (the sum of the Berry connections of all bands) constructed from Kramers pairs, and $F = dA$ is the Berry curvature. In the trivial case, $\nu = 0$ and $\nu = 1$ for the nontrivial case. They also showed the equivalence between this expression and the previous one. 

Unlike Chern insulators, $\mathbb{Z}_{2}$ topological insulators also exist in $3$ dimensions. In $3$-dimensional topological insulators, the topological invariants are parametrized by $4$ binary invariants $(\nu_{0},\nu_{1},\nu_{2},\nu_{3})$ in $(\mathbb{Z}_{2})^{2}$. The hallmark signature of $\mathbb{Z}_{2}$ topological order in $3$D is the existence of surface states with a linear dispersion which obey the Dirac equation. 

\section{Euler Characteristic}

Topological Invariants are extremely useful in Physics. One of the most celebrated topological invariants is the Euler Characteristic, which will be discussed in this section. It describes the shape of the polyhedron, regardless of the way it is bent. The Euler Characteristic 
\begin{equation}
    \chi = V - E + F
\end{equation}
where $V$, $E$, and $F$ are respectively the number of vertices, edges and faces of the given polyhedron. Any polyhedron with $s$ voids and $g$ tunnels satisfies $\chi = 2 (s-g)$. For example, $\chi = 2$ for any convex polyhedron, which again coincides with the 2-sphere. 

We note here that
the \href{https://en.wikipedia.org/wiki/Euler_characteristic}{Wikipedia Page} on the Euler Characteristic is a good place
to start. Interested readers may dig deep into this topic starting from this page. Here, we only demonstrate the use of certain Mathematica commands that calculate Euler characteristics of polyhedra. 

Turning now to 
the Mathematica command \href{https://reference.wolfram.com/language/ref/Polyhedron.html}{Polyhedron}, note that it creates a polyhedron given the vertices and surfaces of the polyhedron as arguments. For example, to define a unit cube with the origin as one of the vertices and edges coinciding with positive coordinates, we define it as follows:
\begin{lstlisting}[style=in]
P = 
Polyhedron[{{0, 0, 0}, {1, 0, 0}, {1, 1, 0}, {0, 1, 0}, {0, 1, 
   1}, {0, 0, 1}, {1, 0, 1}, {1, 1, 1}}, {{1, 2, 3, 4}, {5, 6, 7, 
   8}, {1, 2, 7, 6}, {5, 4, 3, 8}, {2, 7, 8, 3}, {1, 6, 5, 4}}];
   Graphics3D[P]
\end{lstlisting}
The entries inside the first curly braces are the vertices of the desired cube, and the entries inside the second one define the surfaces of the cube. The output will be:
\begin{figure}[ht]
    \centering
    \includegraphics[width=0.3\linewidth]{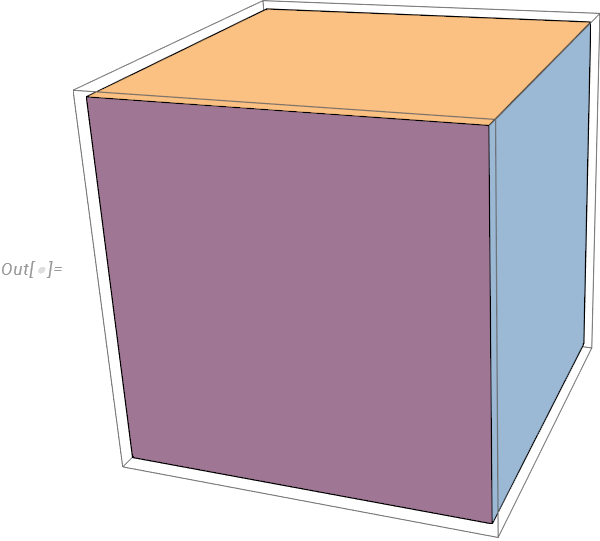}
    \caption{Unit cube}
    \label{fig:unitcube}
\end{figure}

The following command calculates the volume of the region encapsulated by the defined polyhedron, in this case, the unit cube. 
\begin{lstlisting}[style=in]
Volume[P]
\end{lstlisting}
\begin{lstlisting}[style=out]
1
\end{lstlisting}
Finally, the Euler characteristic can be calculated as follows:
\begin{lstlisting}[style=in]
EulerCharacteristic[P]
\end{lstlisting}
\begin{lstlisting}[style=out]
2
\end{lstlisting}
as expected for any convex polyhedron. We will demonstrate the tetrahedron as another example in the following:
\begin{lstlisting}[style=in]
P = Polyhedron[{{0., 0., 0.6}, {-0.3, -0.5, -0.2}, {-0.3, 
0.5, -0.2}, {0.6, 0., -0.2}}, {{2, 3, 4}, {3, 2, 1}, {4, 1, 2}, {1,
 4, 3}}];
 Graphics3D[P]
\end{lstlisting}
\begin{figure}[ht]
    \centering
    \includegraphics[width=0.3\linewidth]{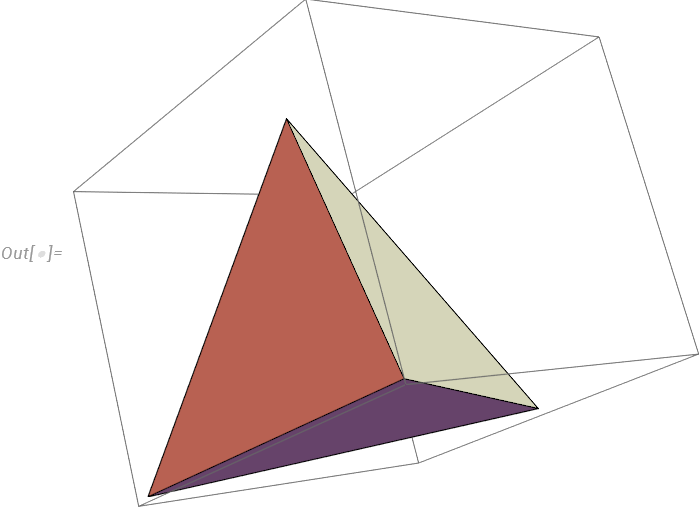}
    \caption{Tetrahedron}
    \label{fig:tetrahedron}
\end{figure}
\begin{lstlisting}[style=in]
Volume[P]
\end{lstlisting}
\begin{lstlisting}[style=out]
0.12
\end{lstlisting}
\begin{lstlisting}[style=in]
EulerCharacteristic[P]
\end{lstlisting}
\begin{lstlisting}[style=out]
2
\end{lstlisting}

A slightly non-trivial example is a cube inside another bigger cube. The syntax to define such a polyhedron is slightly different. Vertices are defined in the same way, but one just needs to put a `$\rightarrow$' between surfaces of the outer and inner polyhedra. 
\begin{lstlisting}[style=in]
P = Polyhedron[{{0, 0, 0}, {0, 3, 0}, {3, 3, 0}, {3, 0, 0}, {0, 0, 3}, {0,
 3, 3}, {3, 3, 3}, {3, 0, 3}, {1, 1, 1}, {1, 2, 1}, {2, 2, 1}, {2,
 1, 1}, {1, 1, 2}, {1, 2, 2}, {2, 2, 2}, {2, 1, 
2}}, {{{2, 3, 4, 1}, {1, 4, 8, 5}, {4, 3, 7, 8}, {3, 2, 6, 7}, {2,
   1, 5, 6}, {5, 8, 7, 
  6}} -> {{{10, 11, 12, 9}, {9, 12, 16, 13}, {12, 11, 15, 
   16}, {11, 10, 14, 15}, {10, 9, 13, 14}, {13, 16, 15, 14}}}}];
   Graphics3D[{Opacity[0.5], P}]
\end{lstlisting}
\begin{figure}[ht]
    \centering
    \includegraphics[width=0.3\linewidth]{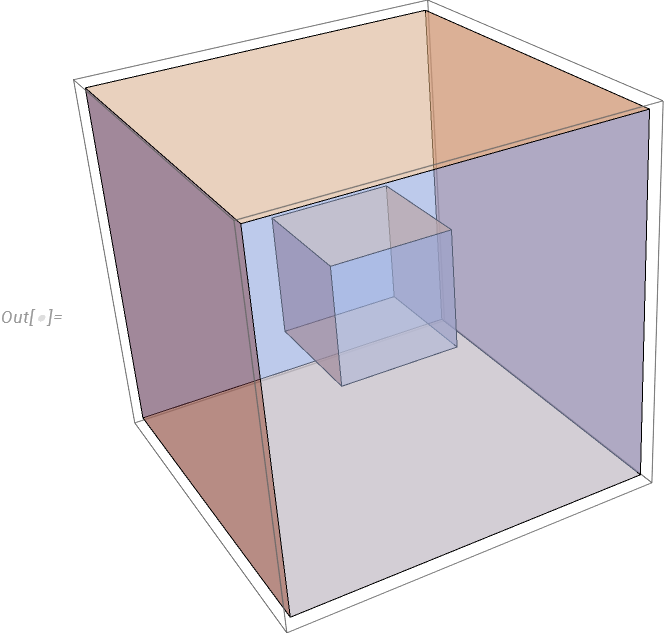}
    \caption{Unit Cube inside a Cube of length 3}
    \label{fig:cubeincube}
\end{figure}
\begin{lstlisting}[style = in]
Volume[P]
\end{lstlisting}
\begin{lstlisting}[style = out]
26
\end{lstlisting}
Mathematica also has inbuilt commands to extract information about the inner and outer polyhedra:
\begin{lstlisting}[style = in]
PolyhedronCoordinates[OuterPolyhedron[P]]
\end{lstlisting}
\begin{lstlisting}[style = out]
{{0, 3, 0}, {3, 3, 0}, {3, 0, 0}, {0, 0, 0}, {0, 0, 3}, {3, 0, 3}, {3, 3, 3}, {0, 3, 3}}
\end{lstlisting}
\begin{lstlisting}[style = in]
PolyhedronCoordinates[InnerPolyhedron[P]]
\end{lstlisting}
\begin{lstlisting}[style = out]
{{1, 1, 1}, {2, 1, 1}, {2, 2, 1}, {1, 2, 1}, {1, 1, 2}, {2, 1, 2}, {2, 2, 2}, {1, 2, 2}}
\end{lstlisting}

\section{Genus}
Our final example is
another popular topological invariant, the Genus, which is defined as the largest number of nonintersecting simple closed curves that can be drawn on the surface without separating it. Roughly speaking, it is the number of holes in a surface (for example, the genus of a torus is 1, but the genus of a double torus is 2). You will come across \textit{Orientability} of a surface later in this book, but for now, let us state a relation between orientability, genus and Euler characteristic of a surface. 
\begin{itemize}
    \item For an orientable surface such as Sphere (genus 0), Torus (genus 1), etc. 
    \begin{equation}
        \chi = 2 - 2 g
    \end{equation}
    \item For a non-orientable surface such as the Real Projective Plane (genus 1), the Klein bottle (genus 2), etc.
    \begin{equation}
        \chi = 2 - g
    \end{equation}
\end{itemize}
If the reader is interested in generating interesting surfaces in Mathematica, there is an extremely rich library of available notebooks on the Wolfram MathWorld website. 
\begin{center}
    \href{https://mathworld.wolfram.com/topics/SolidGeometry.html}{Solid Geometry in Wolfram MathWorld}
\end{center}
Advanced readers are also advised to look into different subsections of the Geometry section of Wolfram Demonstration Projects. There are a large number of notebooks, much more than enough to gain a visualization of the concepts introduced in the chapters of this book:
\begin{center}
    \href{https://demonstrations.wolfram.com/topics/mathematics}{Wolfram Demonstrations of Mathematics Topics}    
\end{center}


%% file: DiffGeom/DG_manifolds.tex
\chapter{Introduction to Manifolds}\label{mani}
Now that we have the notions of open sets and continuity, we are ready to define the fundamental object that we will study.

\md A \db{manifold} is a topological space which is locally like $\mathbb{R}^n.$ \hfill$\Box$ 

That is, every point of a manifold has an open neighbourhood with a one-to-one map onto some open set of $\rn.$ We will deal with smooth manifolds; hence, more precisely, we have the following definition for a smooth manifold.

\md A topological space $M$ is a \db{smooth} $n$-\db{dimensional manifold} if the following are true:

\begin{list}{$\roman{enumii})$} {\usecounter{enumii}\topsep=0pt}
\item We can \db{cover} the space with open sets $U_\alpha,$ i.e. every point of $M$ lies within some $U_\alpha$.
\item $\exists$ a map $\varphi_\alpha: U_\alpha \to \rn,$ where $\varphi_\alpha$ is one-to-one and onto some open set of $\rn.$ $\varphi_\alpha$ is continuous, $\varphi_\alpha\inv$ is continuous, i.e. $\varphi_\alpha \to V_\alpha \in \mathbb{R}^n$ is a homeomorphism for $V_\alpha.$

$(U_\alpha, \varphi_\alpha)$ is called a \db{chart} ($U_\alpha$ is called the \db{domain} of the chart). The collection of charts is called an \db{atlas}. 
\item In any intersection $U_\alpha\cap U_\beta,$ the maps $\varphi_\alpha\circ\varphi_\beta\inv,$ which are called \db{transition functions} and take open sets of $\rn$ to open sets of $\rn,$ i.e. $\varphi_\alpha\circ\varphi_\beta\inv: \varphi_\beta(U_\alpha\cap U_\beta) \to \varphi_\alpha(U_\alpha\cap U_\beta),$ are smooth maps. 
\end{list}

\md In the above definition, as one would guess, $n$ is called the \db{dimension} of $M.$ \hfill$\Box$

We have defined smooth manifolds. A more general definition is that of a $C^k$ manifold, in which the transition functions are $C^k$, i.e. $k$ times differentiable.
Smooth means $k$ is large enough for the purpose at hand. In practice, $k$ is taken to be as large as necessary, up to $C^\infty.$ We get \db{real analytic manifolds} when the transition functions are real analytic, i.e. have a Taylor expansion at each point, which converges. Smoothness of a manifold is useful because then we can say unambiguously if a function on the manifold is smooth as we will see below. A \db{complex analytic manifold} is defined similarly by replacing $\rn$ with ${\mathbb{C}}^n$ and assuming the transition functions  $\varphi_\alpha\circ\varphi_\beta\inv$ to be holomorphic (complex analytic). \hfill$\Box$

To proceed further, we must define a coordinate system, which is the most commonly used formalism when dealing with manifolds in day-to-day life.

\md 
Given a chart $(U_\alpha, \varphi_\alpha)$ for a neighbourhood of some point $P,$ the image $(x_1, \cdots, x_n) \in \rn$ of $P$ is called the \db{coordinates} of $P$ in the chart $(U_\alpha, \varphi_\alpha).$ A chart is also called a \db{local coordinate system}.\hfill$\Box$

In this language, a manifold is a space on which a local coordinate system can be defined, and coordinate transformations between different local coordinate systems are smooth. Often we will suppress $U$ and write only $\varphi$ for a chart around some point in a manifold. We will always mean a smooth manifold when we mention a manifold.

\begin{example}~{\rm 
$\rn$ (with the usual topology) is a manifold. \hfill$\Box$}
\end{example}

\begin{example}~
{\rm Here we consider a
sphere and stereographic projection.
Another typical example of a manifold is the sphere.
A Mathematica example of a $2$-sphere, i.e.  $2$-dimensional sphere, is given below.
\begin{lstlisting}[style=in]
ParametricPlot3D[{Cos[u] Sin[v], Sin[u] Sin[v], Cos[v]}, {v, 0, 
Pi}, {u, 0, 2 Pi}, Mesh -> 20, BoundaryStyle -> Black, 
PlotStyle -> FaceForm[Red, Yellow]]
\end{lstlisting}
This generates a unit sphere as shown below.
\begin{figure}[h]
    \centering
    \includegraphics[width=0.3\linewidth]{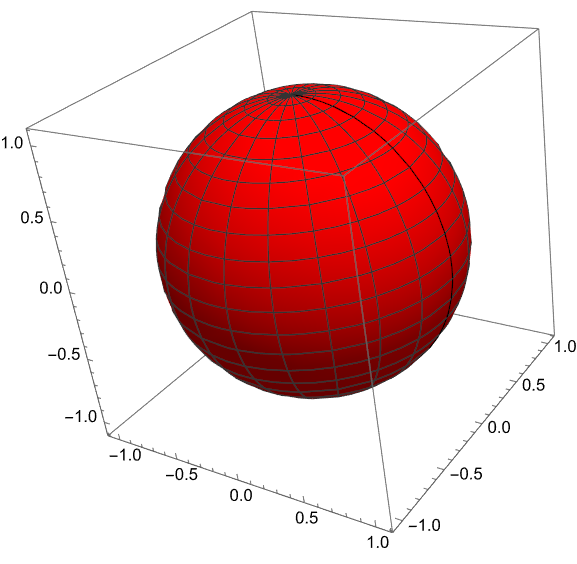}
    \caption{Unit sphere as $2$ dimensional manifold}
    \label{fig:unitsphere}

\end{figure}

Consider the sphere $S^n$ as a subset of ${\rn}^{+1}$:
\begin{eqnarray}
(x^1)^2 + \cdots + (x^{n + 1})^2 = 1
\end{eqnarray}
It is not possible to cover the sphere by a single chart, but it is possible to do so by two charts.\footnote{
The reason that it is not possible to cover the sphere with a single chart is that the sphere is a compact space, and the image of a compact space under a continuous map is compact. Since $\rn$ is non-compact, there cannot be a homeomorphism between $S^n$ and $\rn$.}

For the two charts, we will construct what is called the \db{stereographic projection}. It is most convenient to draw this for a circle in the plane, i.e. $S^1$ in ${\mathbb R}^2$, for which the equatorial `plane' is simply an infinite straight line. 

Of course the construction works for any $S^n.$ Consider the `equatorial plane' defined as $x^1 = 0$, i.e.  the set $\{(0, x^2, \cdots, x^{n+1})\}$, which is simply $\rn$ when we ignore the first zero. We will find homeomorphisms from open sets on $S^n$ to open sets on this $\rn$. Let us start with the north pole ${\cal N}$, defined as the point $(1, 0, \cdots, 0)$. 

We draw a straight line from ${\cal N}$ to any point on the sphere. If that point is in the upper hemisphere $(x^1 >0)$ the line is extended till it hits the equatorial plane. The point where it hits the plane is the image of the point on the sphere which the line has passed through. For points on the lower hemisphere, the line first passes through the equatorial plane (image point) before reaching the sphere (source point). Then using similarity of triangles we find (Exercise!) that the coordinates on the equatorial plane $\rn$ of the image of a point on $S^n\backslash\{{\cal N}\}$ are given by
\begin{eqnarray}
\varphi_N &:& \left(x^1, x^2, \cdots, x^{n+1}\right) \mapsto \left(\frac{x^2}{1-x^1}, \cdots, \frac{x^{n+1}}{1-x^1} \right).
\end{eqnarray}

A Mathematica code showing the stereographic projection of a unit circle is given below.
\begin{lstlisting}[style=in]
\[Delta] = 10^-3; Manipulate[ Show[ ContourPlot[{x^2 + y^2 == 1,
x  == 0, y == 0}, {x, -5, 5}, {y, -5, 5}, PlotPoints -> 100,    PlotRange -> {{-2, 2}, {-2, 2}}, PlotLegends -> "Expressions"],
Plot[1 + Tan[\[Theta]] x, {x, -3, 3}, PlotStyle -> Red],
Graphics[
Locator[{-1/Tan[\[Theta]], 0}, Style["X", FontSize -> 10]]],
Graphics[
Locator[#,  Style["O", FontSize -> 10]] & /@ ({x, y} /. (NSolve[y == 1 + Tan[\[Theta]] x && ({x, y} \[Element] Circle[]), {x, y}]))]], {\[Theta], 0 - \[Delta], Pi - \[Delta]}]
\end{lstlisting}
The output of the code is a dynamical plot that shows the stereographic projection of different points. We show the projection for one particular point below.
\begin{figure}[h]
    \centering
    \includegraphics[width=0.3\linewidth]{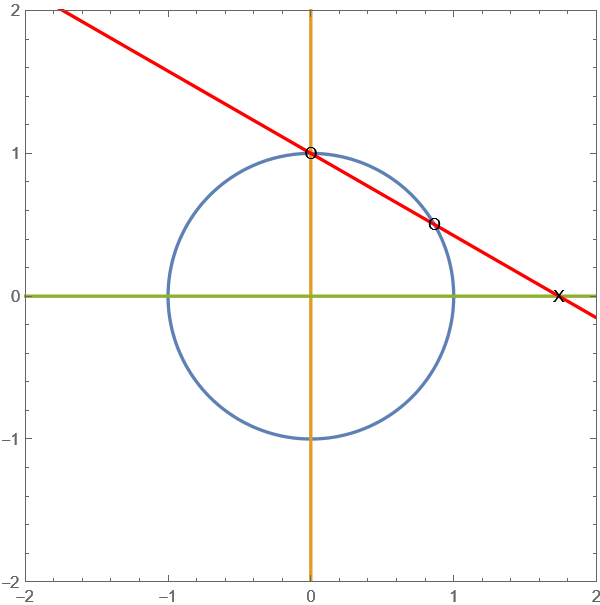}
    \caption{Stereographic projection of unit circle}
    \label{fig:stereographic-projection}
\end{figure}

Similarly, the stereographic projection from the south pole is 
\begin{eqnarray}
\varphi_S &:& S^n\backslash\{{\cal S}\} \to \rn, \nonumber \\
&&\left(x^1, x^2, \cdots, x^{n+1}\right) \mapsto \left(\frac{x^2}{1+x^1}, \cdots, \frac{x^{n+1}}{1+x^1} \right).
\end{eqnarray}
If we write 
\begin{eqnarray}
z = \left(\frac{x^2}{1-x^1}, \cdots, \frac{x^{n+1}}{1-x^1} \right),
\end{eqnarray}
we find that 
\begin{eqnarray}
|z|^2 \equiv \left(\frac{x^2}{1-x^1}\right)^2 +  \cdots + 
\left(\frac{x^{n+1}}{1-x^1} \right)^2 = \frac{1 - (x^1)^2}{(1 - x^1)^2} = \frac{1 + x^1}{1 - x^1}
\end{eqnarray}
The overlap between the two sets is the sphere without the poles. Then the transition function between the two projections is
\begin{eqnarray}
\varphi_S \circ \varphi_N :  \rn\backslash\{0\} \to \rn\backslash\{0\}, \qquad z \mapsto \frac{z}{|z|^2}.
\end{eqnarray}
These are differentiable functions of $z$ in $\rn\backslash\{0\}.$ This shows that the sphere is an $n$-dimensional differentiable manifold. \hfill$\Box$ }
\end{example}

\begin{example}~
    
{\rm 
The M\"obius strip is a 2-dimensional manifold. \hfill$\Box$

Mathematica code for generating a M\"obius strip:
\begin{lstlisting}[style=in]
ParametricPlot3D[{Cos[t] (3 + r Cos[t/2]), Sin[t] (3 + r Cos[t/2]), r Sin[t/2]}, {r, -1, 1}, {t, 0, 2 Pi}, Mesh -> {15, 20}, PlotStyle -> FaceForm[Red, Yellow]]
\end{lstlisting}

\begin{figure}[h]
    \centering
    \includegraphics[width=0.3\linewidth]{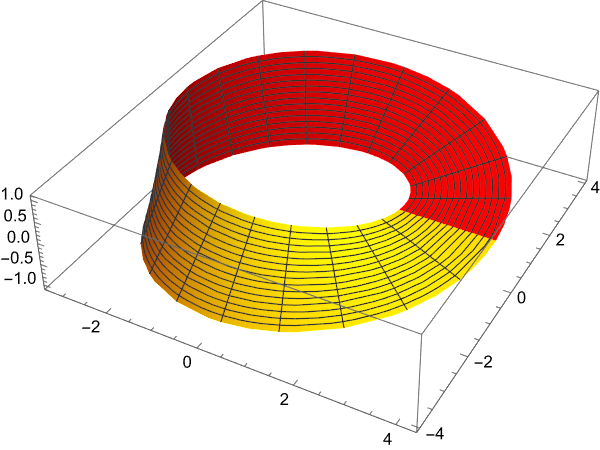}
    \caption{M\"obius strip}
    \label{fig:mobius}
\end{figure}
}
\end{example}

Finite dimensional vector spaces are manifolds. \hfill$\Box$

Infinite dimensional vector spaces with finite norm (e.g. Hilbert spaces) are manifolds. \hfill$\Box$

\md 
A \db{Lie group} is a group $G$ which is also a smooth (real analytic for the cases we will consider) manifold such that group composition written as a map $(x, y)\mapsto xy\inv$ is smooth. \hfill$\Box$

Another way of defining a Lie group is to start with an $n$-\db{parameter continuous group} $G$ which is a group that can be parametrized by $n$ (and only $n$) real continuous variables. $n$ is called the \db{dimension} of the group, $n=\dim G$. (This is a different definition of the dimension. The parameters are global, but do not in general form a global coordinate system.)

Then any element of the group can be written as $g(a)$ where $a = \left(a_1, \cdots, a_n\right).$ Since the composition of two elements of $G$ must be another element of $G,$ we can write $g(a)g(b) = g(\phi(a,b))$ where $\phi = \left(\phi_1, \cdots, \phi_n \right)$ are $n$ functions of $a$ and $b.$ Then for a \db{Lie group}, the functions $\phi$ are smooth (real analytic) functions of $a$ and $b$. 

These definitions of a Lie group are equivalent, i.e. define the same objects, if we are talking about finite dimensional Lie groups. Further, it is sufficient to define them as smooth manifolds if we are interested only in finite dimensions, because all such groups are also real analytic manifolds. Apparently there is another definition of a Lie group as a \db{topological group} (like an $n$-parameter continuous group, but without an a priori restriction on $n$, in which the composition map $(x,y) \mapsto xy\inv$ is continuous) in which it is always possible to find an open neighbourhood of the identity which does not contain a subgroup. 

Any of these definitions makes a Lie group a smooth manifold; an $n$-dimensional Lie group is an $n$-dimensional manifold. \hfill$\Box$

The phase space of $N$ particles is a $6N$-dimensional manifold, $3N$ coordinates and $3N$ momenta. \hfill$\Box$

The space of functions with some specified properties is often a manifold. For example, linear combinations of solutions of the Schr\"odinger equation which vanish outside some region form a manifold. \hfill$\Box$

\md A \db{connected} manifold cannot be written as the disjoint union of open sets. Alternatively, the only subsets of a connected manifold which are both open and closed are $\emptyset$ and the manifold itself. \hfill$\Box$

SO(3), the group of rotations in three dimensions, is a 3-dimensional connected manifold. O(3), the group of rotations plus reflections in three dimensions, is also a 3-dimensional manifold, but it is not connected since it can be written as the disjoint union SO(3)$\cup$PSO(3) where P is reflection. \hfill$\Box$

${\cal L}_+^{\uparrow},$ the group of proper (no space reflection) orthochronous (no time reflection) Lorentz transformations, is a 6-dimensional connected manifold. The full Lorentz group is a 6-dimensional manifold, not connected. \hfill$\Box$

\md Rotations in three dimensions can be represented by $3\times3$ real \db{orthogonal} matrices $R$ satisfying $R^\trans R = \mathbb{I}.$  The space of $3\times3$ real orthogonal matrices is a connected manifold. \hfill$\Box$

The space of all $n\times n$ real non-singular matrices is called GL$(n, R)$. This is an $n^2$-dimensional Lie group. \hfill$\Box$

%% file: DiffGeom/DG_TgtVec.tex
\chapter{Tangent Vectors and Tangent Space}\label{tan}

\section{Tangent Vectors}
Vectors on a manifold are to be thought of in terms of tangents to the manifold, which is a generalization of tangents to curves and surfaces, and will be defined shortly. But a tangent to a curve is like the velocity of a particle at that point, which of course comes from motion along the curve, which is its trajectory. And motion means comparing things at nearby points along the trajectory. And comparing functions at nearby points leads to differentiation. So in order to get to vectors, let us first start with the definitions of these things. 
\begin{definition}
    
\end{definition}
\md A function $f: {\cal M} \to \mathbb{R}$ is \db{differentiable} at a point $P \in M$ if in a chart $\varphi$ at $P,$ the function $f\circ\varphi\inv: \rn \to \mathbb{R}$ is differentiable at $\varphi(P).$ \hfill$\Box$

This definition does not depend on the chart. If $f\circ{\varphi_\alpha}\inv$ is differentiable at $\varphi_\alpha(P)$ in a chart $(U_\alpha, \varphi_\alpha)$ at $P$, then $f\circ\varphi_\beta\inv$ is differentiable at $\varphi_\beta(P)$ for any chart $(U_\beta, \varphi_\beta)$ because 
\begin{eqnarray}
f\circ\varphi_\beta\inv = (f\circ\varphi_\alpha\inv) \circ (\varphi_\alpha\circ\varphi_\beta\inv)
\end{eqnarray}
and the transition functions $(\varphi_\alpha\circ\varphi_\beta\inv)$ are differentiable. 

This should be thought of as a special case of functions from one manifold to another. Consider two manifolds ${\cal M}$ and ${\cal N}$ of dimensions $m$ and $n$, and a mapping $f: \cm \to {\cal N}\,, \, P\mapsto Q.$ Consider local charts $ (U, \varphi) $ around $ P $ and $ (W, \psi) $ around $Q$. Then $ \psi\circ f \circ \varphi\inv $ is a map from $ \mathbb{R}^m \to \rn $ and represents $f$ in these local charts.

\md $f$ is \db{differentiable} at $P$ if $ \psi\circ f \circ \varphi\inv $ is differentiable at $\varphi(P).$ In other words, $f$ is differentiable at $P$ if the coordinates $ y^i = f^i (x^\mu) $ of $Q$ are differentiable functions of the coordinates $x^\mu$ of $P$. \hfill$\Box$

\md If $f$ is a bijection (i.e.  one-to-one and onto) and $f$ and $f\inv$ are both differentiable, we say that $f$ is a \db{diffeomorphism} and that $\cm$ and ${\cal N}$ are \db{diffeomorphic}. \hfill$\Box$

In all of these definitions, differentiable can be replaced by $C^k$ or smooth. 

\md Two Lie groups are \db{isomorphic} if there is a diffeomorphism between them which is also a group homomorphism. \hfill$\Box$

\md A \db{curve} in a manifold $\cm$ is a map $\gamma$ of a closed interval $ \mathbb{R}$ to $\cm$. (This definition can be given also when $\cm$ is a topological space.) \hfill$\Box$

We will take this interval to be $I = [0, 1] \subset \mathbb{R}.$ Then a curve is a map $\gamma: I \to \cm.$ If $ \gamma(0) = P $ and $ \gamma(1) = P', $ for some $\gamma,$ we say that $\gamma$ joins $P$ and $P'.$

\md A manifold $\cm$ is \db{connected} (actually \db{arcwise connected})
if any two points in it can be joined by a continuous curve in $\cm$. \hfill$\Box$

As for any map, a curve $\gamma$ is called smooth iff its image in a chart is smooth in $\rn,$ i.e., iff $ \varphi \circ \gamma : I \to \rn$ is smooth in $\rn.$

Note that the definition of a curve implies that it is parametrized. So the same collection of points in $\cm$ can stand for two different curves if they have different parametrizations.

We are now ready to define tangent vectors and the tangent space to a manifold. There are different ways of defining tangent vectors.
\begin{list}{$\roman{enumii})$} {\usecounter{enumii}\topsep=0pt}
\item Coordinate approach: Vectors are defined to be objects satisfying certain transformation rules under a change of chart, i.e. coordinate transformation, $ (U_\alpha, \varphi_\alpha) \to (U_\beta, \varphi_\beta). $
\item Derivation approach: A vector is defined as a derivation of functions on the manifold. This is thinking of a vector as defining a ``directional derivative''.
\item Curves approach: A vector tangent to a manifold is tangent to a curve on the manifold.  
\end{list}
The approaches are equivalent in the sense that they end up defining the same objects and the same space. We will follow the third approach, or perhaps a mix of the second and the third approaches. Later we will briefly look at the derivation approach more carefully and compare it with the way we have defined tangent vectors.

Consider a smooth function $f: \cm \to {\mathbb R}.$ Given a curve $ \gamma: I \to \cm, $ the map $ f\circ \gamma : I \to \mathbb{R} $ is well-defined, with a well-defined derivative. The rate of change of $f$ along $\gamma$ is written as $ \displaystyle{ \frac{df}{dt} } $.

Suppose another curve $\mu(s)$ meets $\gamma(t)$ at some point $P$, where $s=s_0$ and $t = t_0,$ such that 
\begin{eqnarray}
\frac{d}{dt} (f \circ \gamma)\Big|_P = \frac{d}{ds} (f \circ \mu)\Big|_P \qquad \forall f \in C^\infty(\cm)
\label{tan.twocurves}
\end{eqnarray}
That is, we are considering a situation where two curves are tangent to each other both in geometric and parametric senses. Let us introduce  a convenient notation. In any chart $ \varphi $ containing the point $P,$ let us write $ \varphi(P) = ( x^1, \cdots, x^n). $ Then we can write $ f\circ \gamma = (f \circ \varphi\inv)\circ( \varphi\circ \gamma), $ so that the maps are 
\begin{eqnarray}
f \circ \varphi\inv &:& \rn \to {\mathbb R}, \qquad \vec{x} \mapsto f( \vec{x})\; \text{or}\; f(x^i) \\
\varphi\circ \gamma &:& I \to \rn, \qquad t \mapsto \{x^i(\gamma(t))\}\,,
\end{eqnarray}
the last being the coordinates of the curve in $ \rn. $ 

Using the chain rule for differentiation, we find
\begin{eqnarray}
\frac{d}{dt}(f\circ \gamma) = \frac{d}{dt}f(\vec{x}(\gamma(t))) = \frac{\partial f}{\partial x^i}\frac{dx^i(\gamma(t))}{dt}.
\end{eqnarray}
Similarly, for the curve $ \mu $  we find
\begin{eqnarray}
\frac{d}{ds}(f\circ \mu) = \frac{d}{ds}f(\vec{x}(\mu(s))) = \frac{\partial f}{ \partial x^i}\frac{dx^i(\mu(s))}{ds}. 
\end{eqnarray}
Since $f$ is arbitrary, we can say that two curves $\gamma, \mu$ have the same tangent vector at the point $P\in \cm$ (where $t=t_0$ and $s=s_0$) iff 
\begin{eqnarray}
\left.\frac{dx^i(\gamma(t))}{dt}\right|_{t = t_0} = 
\left.\frac{dx^i(\mu(s))}{ds}\right|_{s = s_0}\;.
\label{tan.same}
\end{eqnarray}

We can say that these numbers completely determine the rate of change of any function along the curve $\gamma$ or $\mu$ at $P.$ So we can define the tangent to the curve.

\md The \db{tangent vector} to a curve $\gamma$ at a point $P$ on it is defined as the map 
\begin{eqnarray}
\dot{\gamma}_{_P} : C^\infty(\cm) \to \mathbb{R}, \qquad f \mapsto \dot{\gamma}_{_P}(f) \equiv \frac{d}{dt}\left.(f \circ \gamma) \right|_P \;.
\end{eqnarray}

As we have already seen, in a chart with coordinates $\{x^i\}$ we can write using the chain rule
\begin{eqnarray}
\dot{\gamma}_{_P} (f) = \frac{dx^i(\gamma(t))}{dt} \left.\frac{\partial f}{\partial x^i}\right|_{ \varphi(P)}
\label{tan.components}
\end{eqnarray}
The numbers $ \displaystyle{\left.\frac{dx^i(\gamma(t))}{dt}\right|_{ \varphi(P)}} $ are thus the components of $\dot{\gamma}_{_P}$. We will often write a tangent vector at $P$ as $v_{_P}$ without referring to the curve it is tangent to.

We note here that there is another description of tangent vectors based on curves. Let us write $\gamma \sim \mu$ if $ \gamma $ and $\mu$ are tangent to each other at the point $P$. It is easy to see, using Eq.~(\ref{tan.same}) for example, that this relation $\sim$ is transitive, reflexive, and symmetric. In other words, $\sim$ is an equivalence relation, for which the equivalence class $[\gamma]$ contains all curves tangent to $\gamma$ (as well as to one another) at $P$.

\md A \db{tangent vector} at $P\in \cm$ is an equivalence class of curves under the above equivalence relation. \hfill$\Box$

The earlier definition is related to this by saying that if a vector $v_{_P}$ is tangent to some curve $\gamma$ at $P$, i.e. if $v_{_P} = \dot{\gamma}_{_P},$ we can write $v_{_P} = [\gamma].$

\begin{figure}
    \centering
\begin{subfigure}{0.45\linewidth}
    \includegraphics[width=1.0\linewidth]{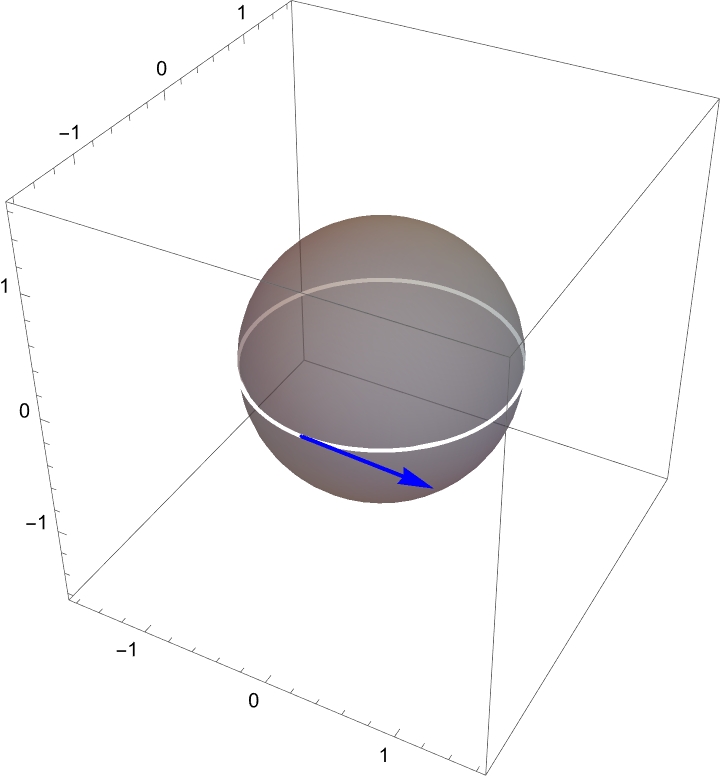}
    \subcaption{Tangent vectors on a torus}
    \label{fig:tangent_vector_fig1}
\end{subfigure}
\begin{subfigure}{0.45\linewidth}
    \includegraphics[width=1.0\linewidth]{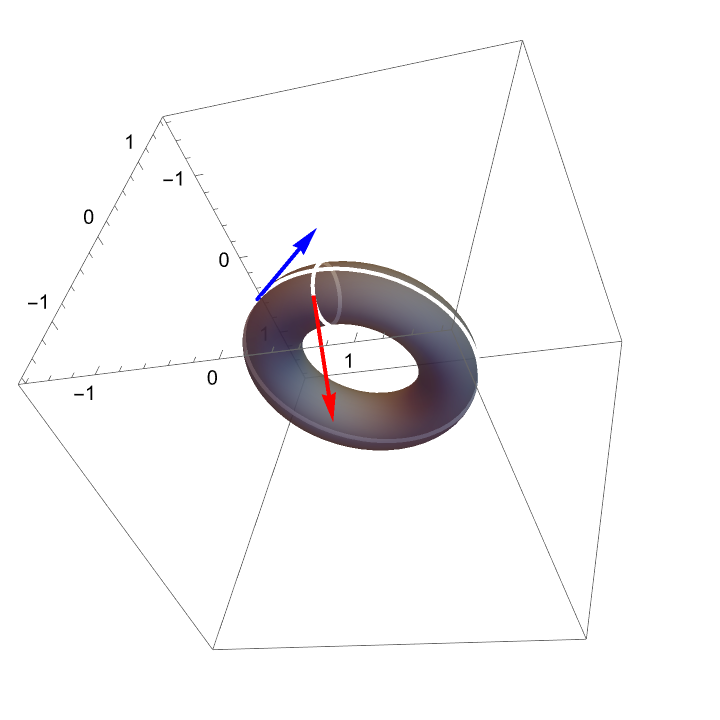}
    \subcaption{Tangent vectors on a torus}
    \label{fig:tangent_vector_fig4}
\end{subfigure}
\end{figure}

\begin{example}~
{\rm 
We now consider a tangent on a sphere. 
    Here we will give an example of tangent vectors along a curve in a manifold. Since manifolds of more than two dimensions are difficult to represent on paper,
we will try to visualize tangent vectors on a spherical surface using Mathematica:
\begin{lstlisting}[style=in]
Curve = {Cos[#], Sin[#], 0} &;
tangent = FrenetSerretSystem[Curve[t], t][[-1, 1]];
arrowtangent = Arrow[{Curve[t], Curve[t]+tangent}];
\end{lstlisting}

\begin{lstlisting}[style=in]
Manipulate[Show[ParametricPlot3D[Curve[s], {s, 0, 4 Pi}, PlotStyle -> {Thick, White}],
Graphics3D[{Thick, Blue, arrowtangent, Gray, Opacity[0.5], Sphere[]}], PlotRange -> 1.5] // Evaluate, {t, 0, 4Pi, Appearance -> {"Open", "Labelled"}}]
\end{lstlisting}

Fig.~\ref{fig:tangent_vector_fig1} shows a tangent vector on the sphere along an equatorial curve. The \code{Manipulate} function in Mathematica allows one to change the parameter along the curve to determine at which point the tangent vector is drawn.}
\end{example}

\begin{exercise}~
{\rm 
    Draw similar tangents on other curves (such as an 8-figure) on the surface of a sphere (See attached notebook for an example)}
\end{exercise}

\begin{example}~
{\rm We consider now a tangent on a torus. 
    One can similarly define a tangent vector on the surface of a torus. This example demonstrates two curves on a torus, belonging to different homotopy classes (See Chap.~\ref{chap:homotopy} for a recap).
\begin{lstlisting}[style=in]
Curve[t_] = {Cos[t], Sin[t], 0};
tangent = FrenetSerretSystem[Curve[t], t][[-1, 1]];
arrowtangent = Arrow[{Curve[t], Curve[t] + tangent}];
\end{lstlisting}

\begin{lstlisting}[style=in]
Curve2[t_] = {0, -3/4 + 1/4 Cos[t], 1/4 Sin[t]};
tangent2 = FrenetSerretSystem[Curve2[t], t][[-1, 1]];
arrowtangent2 = Arrow[{Curve2[t], Curve2[t] + tangent2}];
\end{lstlisting}

\begin{lstlisting}[style=in]
Manipulate[
 Show[ParametricPlot3D[Curve[t], {t, 0, 2 Pi}, 
    PlotStyle -> {Thick, White}],
   ParametricPlot3D[Curve2[t], {t, 0, 2 Pi}, 
    PlotStyle -> {Thick, White}],
    Graphics3D[{Thick, Blue, arrowtangent, Gray, Opacity[0.5], 
     Torus[]}],
   Graphics3D[{Thick, Red, arrowtangent2, Gray, Opacity[0.5], 
     Torus[]}], PlotRange -> 1.5] // Evaluate, {t, 0, 2 Pi, 
  Appearance -> {"Open", "Labelled"}}]
\end{lstlisting}

Fig.~\ref{fig:tangent_vector_fig4} shows the tangent vectors to the two curves on a torus.}
\end{example}

%% file: DiffGeom/DG_TgtSpc.tex
\section{Tangent Space}\label{tpm}

\md The \db{tangent space} $T_P\cm$ at a point $P\in \cm$ of a manifold $\cm$ is the set of all tangent vectors (to all curves) at $P.$

 \md   $T_P\cm$ is a vector space with the same dimensionality as the manifold $\cm$.

\begin{proof}
    Let us consider an $n$-dimensional manifold $\cm$. First we need to show that the tangent space $T_P\cm$ at a point $P$ is a vector space, i.e.  $\forall X_P,Y_P \in T_P\cm; $ and $ a \in \mathbb{R}$,
%
\begin{eqnarray}
X_P + Y_P \in T_P\cm\,,\qquad &&\hfill\\ a X_P \in T_P\cm\,. \qquad && \hfill \label{eqn_tangentspace2}
\end{eqnarray}
That is, given curves $\gamma, \mu$ passing through $P$ such that $X_P = \dot{\gamma}_{_P}, Y_P = \dot{\mu}_{_P},$ we need a curve $\lambda$ passing through $P$ such that $\dot\lambda_{_P}(f) = X_P(f) + Y_P(f) \forall f\in C^\infty(\cm).$ 

Let us define $ \bar{\lambda}: I \to \rn $ in some chart $ \varphi $ around $P$ by $ \bar{\lambda} = \varphi\circ \gamma + \varphi \circ \mu - \varphi(P). $ Then $\bar\lambda$ is a curve in $\rn,$ and 
\begin{eqnarray}
\lambda = \varphi^{-1} \circ \bar{ \lambda }\, :\, I \to M
\end{eqnarray}
is a curve with the desired property. \hfill$\Box$

Note that we cannot define $ \lambda = \gamma + \mu - P $ because addition does not make sense on the right hand side.

The proof of Eq.~(\ref{eqn_tangentspace2}) works similarly. 

Next, to show that $T_P\cm$ has $n$ basis vectors, we consider a chart $ \varphi $ with coordinates $x^i.$ Then take $n$ curves $ \lambda_k$ such that
\begin{eqnarray}
\varphi \circ \lambda_k(t) = \left( x^1(P), \cdots, x^k(P) + t, \cdots, x^n(P) \right)\,,
\end{eqnarray}
i.e., only the $k$-th coordinate varies along $t.$ So $ \lambda_k$ is like the axis of the $k$-th coordinate (but only in some open neighbourhood of $P$).

Let us now denote the tangent vector to $\lambda_k$ at $P$ by $ \displaystyle\left( \frac{\partial}{\partial x^k} \right)_P\,, $ i.e., 
\begin{eqnarray}
\left( \frac{\partial}{\partial x^k} \right)_P f = \dot{\lambda}_k(f)\Big|_P = \left. \frac{d}{dt}\left(f \circ \lambda_k\right)\right|_P\,.
\end{eqnarray}
This notation makes sense when we remember Eq.~(\ref{tan.components}). Using it we can write 
\begin{eqnarray}
\dot{\lambda}_k(f)\Big|_P = \left( \frac{\partial f}{\partial x^k} \right)_P \forall f \in C^\infty(\cm).
\end{eqnarray}

Note that $ \displaystyle{\left( \frac{\partial}{\partial x^k} \right)_P} $ is no more than a convenient notation. We should understand this as 
\begin{eqnarray}
\left(\frac{\partial}{\partial x^k} \right)_P f = \left.\frac{\partial}{\partial x^k}\left(f \circ \varphi\inv\right)\right|_{ \varphi(P)} \equiv \left.\frac{\partial f}{\partial x^k}\right|_{ \varphi(P)}
\label{tpm.basis}
\end{eqnarray} 
for every smooth $f$ in a chart around $P$. The $ \displaystyle{\left( \frac{\partial}{\partial x^k} \right)_P}  $ are defined only when this chart is given, but these are vectors on the manifold at $P$, not on $\rn$.

Let us now show that the tangent space at $P$ has $\dot\lambda_k|_{_P}$ as a basis. Take any vector $v_{_P} \in T_P\cm\,,$ which is the tangent vector to some curve $\gamma$ at $P$. (We may sometimes refer to $P$ as $\gamma(0)$ or as $t=0$.) Then
\begin{eqnarray}
v_{_P}(f) &=& \left.\frac{d}{dt}\left( f\circ \gamma \right) \right|_{t=0} \\
&=& \left.\left.\frac{d}{dt}( (f\circ \varphi\inv )\right|_{\varphi(P)} \circ ( \varphi \circ \gamma) ) \right|_{t=0}.
\end{eqnarray}
Note that $ \varphi\circ \gamma : I \to \rn,\, t \mapsto (x^1(\gamma(t)), \cdots, x^n(\gamma(t))) $ are the coordinates of the curve $ \gamma, $ so we can use the chain rule of differentiation to write 
\begin{eqnarray}
v_{_P}(f) &=& \left. \frac{\partial}{\partial x^i} (f\circ \varphi\inv )\right|_{\varphi(P)}  \left.
\frac{d}{dt}(x^i\circ\gamma)\right|_{t=0}\\
&=& \left. \frac{\partial}{\partial x^i} (f\circ \varphi\inv )\right|_{\varphi(P)}  v_{_P}(x^i)\,.
\end{eqnarray}
The first factor is exactly as shown in Eq.~(\ref{tpm.basis}), so we can write 
\begin{eqnarray}
v_{_P}(f) &=& \left(\frac{\partial}{\partial x^k} \right)_P f v_{_P}(x^i)\,  
\qquad \forall f \in C^\infty(\cm)
\end{eqnarray}
i.e., we can write 
\begin{eqnarray}
v_{_P} = v^i_{_P} \left(\frac{\partial}{\partial x^k} \right)_P\, \qquad \forall v_{_P} \in T_P\cm
\label{tpm.components}
\end{eqnarray}
where $ v^i_{_P} = v_{_P}(x^i) $. Thus the vectors $ {\left(\frac{\partial}{\partial x^k} \right)_P} $ span $T_P\cm\,.$ These are to be thought of as tangents to the coordinate curves in $\varphi.$ These can be shown to be linearly independent as well, so $ {\left(\frac{\partial}{\partial x^k} \right)_P} $ form a basis of $T_P\cm$ and $ v^i_{_P} $ are the components of $v_{_P}$ in that basis.

The $ {\left(\frac{\partial}{\partial x^k} \right)_P} $ are called \db{coordinate basis vectors} and the set $ \left\{\left(\frac{\partial}{\partial x^k} \right)_P\right\} $ is called the \db{coordinate basis}.

It can be shown quite easily that for any smooth  function $f\,$,  a vector $v_{_P}\,$ defines a derivation $f \mapsto v_{_P}(f)$\,, i.e., satisfies linearity and the Leibniz rule,
\begin{eqnarray}
v_{_P}(f + \alpha g) &=& v_{_P}(f) + \alpha v_{_P}(g) \\
v_{_P}(fg) &=& v_{_P}(f) g(P) + f(P) v_{_P}(g) \\
&& \qquad \forall f,g \in C^1(\cm)\, \text{and} \,\alpha\in \mathbb{R} \notag
\end{eqnarray}
\end{proof} 

%% file: DiffGeom/DG_dual.tex
\chapter{Dual space}\label{dual}

In this chapter, we will introduce the concepts of the dual space, covectors, and cotangent vectors. For the purposes of the discussions, we will restrict ourselves to the real vector spaces.

\md 
The \db{dual space} $T^*_P\cm$ of $T_P\cm$ is the space of linear mappings $\omega : T_P\cm \to \mathbb{R}.$ \hfill$\Box$
 
We will write the action of $\omega$ on $v_P\in  T_P\cm$ as $ \omega(v_P)$ or sometimes as $\braket{\omega}{v_P}\,.$

Linearity of the mapping $\omega$ means 
\begin{eqnarray}
\omega(u_P + a v_P) &=& \omega(u_P) + a \omega(v_P)\,, \\
&&\qquad\qquad\forall u_P, v_P \in T_P\cm\, \text{ and }\, a\in \mathbb{R}\,.\nonumber
\end{eqnarray}
The dual space is a vector space under the operations of vector addition and scalar multiplication defined by 
\begin{eqnarray}
a_1\omega_1 + a_2 \omega_2 : v_P \mapsto a_1\omega_1(v_P) + a_2 \omega_2(v_P)\,.
\end{eqnarray}
%

\md 
The elements of $T^*_P\cm$ are called \db{dual vectors}, \db{covectors},  \db{cotangent vectors} etc.

A dual space can be defined for any vector space $V$ as the space of linear mappings $V \to \mathbb{R}$ (or $V \to \mathbb{C}$ if $V$ is a complex vector space). 

\textbf{Example:} \begin{tabbing} 
\hspace*{1in} Vector \hspace*{1in} \=  Dual vector \hfill\\
\hspace{1in} column vectors \>   row vector \\
\hspace*{1in} kets $\ket{\psi}$ \>  bras $\bra{\phi}$ \\
\hspace*{1in} functions \> linear functionals, etc.\hfill$\Box$\\
\end{tabbing}
%

 \md 
     Given a function on a manifold $f: \cm \to \mathbb{R}\,,$ every vector at $P$ produces a number, $v_{_P}(f) \in \mathbb{R}\quad \forall v_{_P}\in T_P\cm\,.$ Thus $f$ defines a covector $\d f\,,$ given by $\d f(v_{_P}) = v_{_P}(f)$ called the \db{differential} or \db{gradient} of $f$.

Since the space $T_P\cm$ is linear, so is $\d f\,,$
\begin{eqnarray}
\d f(v_{_P} + a w_{_P}) 
&=& (v_{_P} + a w_{_P})(f) \nonumber \\
&=& v_{_P}(f) + a w_{_P}(f) \\
&& \qquad \forall v_{_P}, w_{_P} \in T_P\cm, a\in \mathbb{R}\,. \nonumber
\end{eqnarray}
Thus $\d f\, \in T_P^*\cm\,.$

 \md  We now have a proposition:
     $T^*_P\cm$ is also $n$-dimensional.
 This may be proved by considering
    a chart $ \varphi $ with coordinate functions $x^i\,.$ Each $x^i$ is a smooth function $x^i: \cm \to \mathbb{R}\,.$ Then the differentials $\d x^i$ satisfy 
\begin{eqnarray}
\d x^i \left(\frac{\partial}{\partial x^j}\right)_P = 
\left(\frac{\partial}{\partial x^j}\right)_P (x^i) 
= \left.\frac{\partial}{\partial x^j}\left(x^i \circ \varphi\inv \right)\right|_{\varphi(P)} = \delta^i_j\,.
\end{eqnarray}

The differentials $\d x^i$ are covectors, as we already know. So we have constructed $n$ covectors in $T_P^*\cm\,.$ Next consider a linear combination of these covectors, $\omega = \omega_i\d x^i.$ If this vanishes, it must vanish on every one of the basis vectors. In other words,
\begin{eqnarray}
\omega = 0 
&\implies& \omega \left(\frac{\partial}{\partial x^j}\right)_P =0 \nonumber \\
&\implies& \omega_i\d x^i\left(\frac{\partial}{\partial x^j}\right)_P =0 \nonumber \\
&\implies& \omega_i \delta^i_j =0 \quad i.e. \quad \omega_j = 0\,.
\end{eqnarray}
So the $\d x^i$ are linearly independent.

\md 
Finally, given any covector $\omega\,,$ consider the covector $\lambda = \omega - \omega\left(\frac{\partial}{\partial x^i}\right)_P  \d x^i\,.$
Then letting this act on a coordinate basis vector, we get
\begin{eqnarray}
\lambda \left(\frac{\partial}{\partial x^j}\right)_P \nonumber \\
&=& \omega\left(\frac{\partial}{\partial x^j}\right)_P 
- \omega\left(\frac{\partial}{\partial x^i}\right)_P \d x^i\left(\frac{\partial}{\partial x^j}\right)_P  \nonumber \\
&=& \omega \left(\frac{\partial}{\partial x^j}\right)_P - \omega \left(\frac{\partial}{\partial x^i}\right)_P \delta_j^i = 0, \quad \forall j\,.
\end{eqnarray}
So $\lambda$ vanishes on all vectors, since the $\displaystyle{\left(\frac{\partial}{\partial x^j}\right)_P }$ form a basis. Thus the $\d x^i$ span $T_P^*\cm\,,$ so $T_P^*\cm$ is $n$-dimensional.

Also, as we have just seen, any covector $\omega \in T_P^*\cm$ can be written as 
\begin{eqnarray}
\omega = \omega_i \d x^i \qquad \text{where} \quad \omega_i = \omega\left(\frac{\partial}{\partial x^i}\right)_P\,,
\end{eqnarray}
so in particular for $\omega = \d f\,,$ we get 
\begin{eqnarray}
\omega_i \equiv (\d f)_i = \d f \left(\frac{\partial}{\partial x^i}\right)_P = \left(\frac{\partial f}{\partial x^i}\right)_{\varphi(P)}\,.
\label{dual.df}
\end{eqnarray}
This justifies the name gradient.

It is straightforward to calculate the effect of switching to another overlapping chart, i.e. a coordinate transformation. In a new chart $\varphi'$ where the coordinates are $y^i$ (and the transition functions are thus $y^i(x)$) we can use Eq.~(\ref{dual.df}) to write the gradient of $y^i$ as
\begin{eqnarray}
\d y^i = \left(\frac{\partial y^i}{ \partial x^j }\right)_P\d x^j\,.
\end{eqnarray} 
This is the result of coordinate transformations on a basis of covectors.

Since $ \displaystyle{\left\{\left(\frac{\partial}{ \partial x^i }\right)_P\right\}} $ is the dual basis in $T_P\cm$ to $\{\d x^i\}$, in order for $\displaystyle{\left\{\left(\frac{\partial}{ \partial y^i }\right)_P\right\}} $ to be the dual basis to $\{\d y^i\}$ we must have 
\begin{eqnarray}
\left(\frac{\partial}{ \partial y^i }\right)_P = \left(\frac{\partial x^j}{ \partial y^i }\right)_P\left(\frac{\partial}{ \partial x^j }\right)_P\,.
\end{eqnarray}
These formulae can be generalized to arbitrary bases.

\md 
Given a vector $v$, it is not meaningful to talk about its dual, but given a basis $ \{e_a\}, $ we can define its dual basis $ \{\omega^a\} $ by $ \omega^a(e_b) = \delta^a_b\,. $

We can make a change of bases by a linear transformation,
\begin{eqnarray}
\omega^a \mapsto \omega'^a = A^a_{b} \omega^b\,,\qquad \qquad e_a \mapsto e'_a = (A\inv)^b_a e_b\,,
\end{eqnarray}
with $A$ a non-singular matrix, so that $\omega'^a(e'_b) = \delta_b^a\,.$

Given a 1-form $\lambda$ we can write it in both bases,
\begin{eqnarray}
\lambda = \lambda_a \omega^a = \lambda'_a \omega'^a = \lambda'_a A^a_b \omega^a\,,
\end{eqnarray}
from which it follows that $ \lambda'_a = (A\inv)^b_a \lambda_b\,.$

Similarly, if $v$ is a vector, we can write
\begin{eqnarray}
v = v^a e_a = v'^a e'_a = v'^a (A\inv)^b_a e_b\,,
\end{eqnarray}
and it follows that $v^a = A^a_b v^b\,.$ 

Quantities which transform like $\lambda_a$ are called \db{covariant}, while those transforming like $v^a$ are called \db{contravariant}.
    

A visual example in Mathematica showing covariant and contravariant vectors on simple $2D$ surfaces can be found here: \href{https://demonstrations.wolfram.com/TheDualBasisForTheTangentSpaceOfA2DSurface/}{The Dual Basis}

%% file: DiffGeom/DG_vfld.tex
\chapter{Vector fields}\label{vfld}
 \md 
    Consider the (disjoint) union of tangent spaces at all points,
\begin{eqnarray}
T\cm = \bigcup\limits_{P \in \cm} T_P\cm\,.
\end{eqnarray}
This is called the \db{tangent bundle} of $\cm$.

\md
    A \db{vector field} $v$ chooses an element of $T_P\cm$ for every $P$, i.e. $v: P\mapsto v(P) \equiv v_{_P} \in T_P\cm.$

We will often write $v(f)\vert_{_P} = v_{_P}(f).$

Given a chart, $v$ has components $v^i$ in the chart,
\begin{eqnarray}
v_{_P} = v^i \left(\frac{\partial}{\partial x^i}\right)_P\,, \qquad (v^i)_{_P} = v_{_P} (x^i)\,.
\end{eqnarray}
%

 \md 
    The vector field $v$ is said to be a \db{smooth vector field} if the functions $v^i = v(x^i)$ are smooth for any chart (and thus for all charts).

\md
    A rule that selects a covector from $T^*_P\cm$ for each $P$ is called a \db{one$-$form} (often written as a \db{1$-$form}).

 \md 
    Given a smooth vector field $v$ (actually $C^1$ is sufficient) we can define an \db{integral curve} of $v$, which is a curve $\gamma$ in $\cm$ such that $\dot\gamma(t)\vert_{_P} = v_{_P}$ at every $P\in \gamma.$ (One curve need not pass through all $P\in \cm.$)

Suppose $\gamma$ is an integral curve of a given vector field $v$, with $\gamma(0) = P.$ Then in a chart containing $P$, we can write 
\begin{eqnarray}
\dot{\gamma}(t) = v \Rightarrow \frac{d}{dt}x^i(\gamma(t)) = v^i(\vec{x}(t))\,,
\end{eqnarray}
with initial condition $x^i(0) = x^i\vert_{_P}$. This is a set of ordinary first order differential equations. If $v^i$ are smooth, the theory of differential equations guarantees, at least for small $t$ (i.e. locally), the existence of exactly one solution. The uniqueness of these solutions implies that the integral curves of a vector field do not cross. 

One use of integral curves is that they can be thought of as coordinate lines. Given a smooth vector field $v$ such that $\displaystyle{v|_{_P} \neq 0},$ it is possible to define a coordinate system $\{x^i\}$ in a neighbourhood around $P$ such that $\displaystyle{v = \frac{\partial}{\partial x^i}}.$ 

\begin{figure}[t]
    \centering
\begin{subfigure}[t]{0.45\linewidth}
     \centering
    \includegraphics[width=1.0\linewidth]{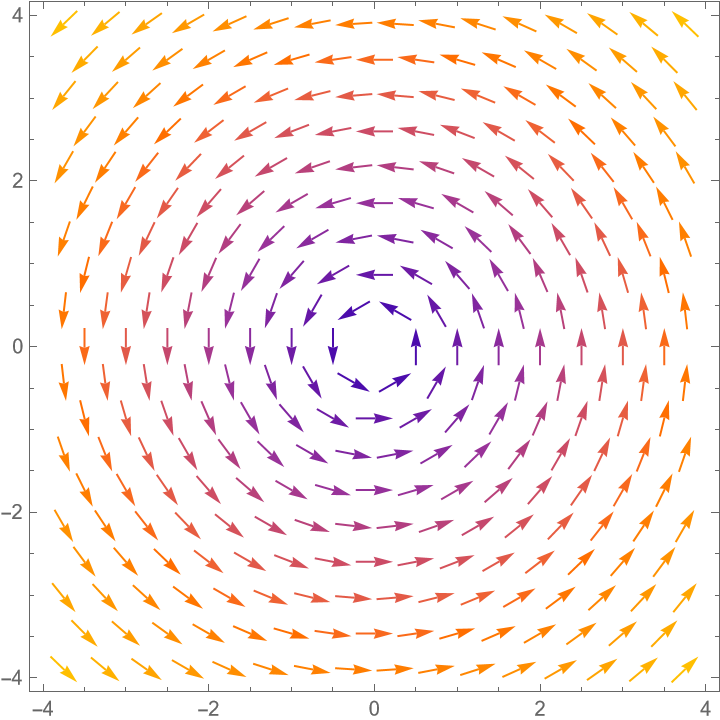}
    \subcaption{A vector field of Eq.~(\ref{eqn:ex:vector_field_curling}) curling about the origin}
    \label{fig:vector_fields_fig1}
\end{subfigure}
\hfill
\begin{subfigure}[t]{0.45\linewidth}
     \centering
    \includegraphics[width=1.0\linewidth]{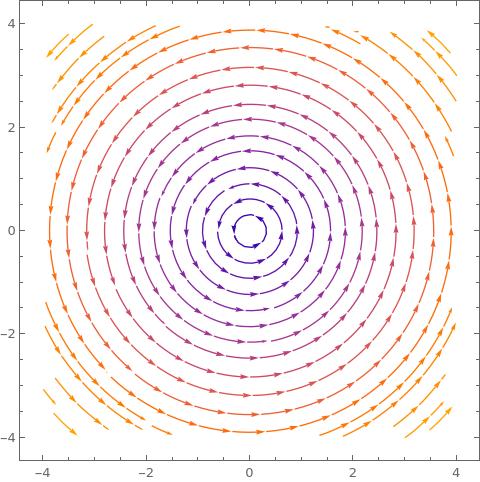}
    \subcaption{A few integral curves corresponding to the vector field}
    \label{fig:vector_fields_fig2}
\end{subfigure}
\end{figure}

\begin{example}\label{ex:vector_field_curling}
    {\rm 
    We will demonstrate the concept of vector fields on a manifold using a simple example on $\mathbb{R}^2$. Let us define the field as:
    \begin{eqnarray}\label{eqn:ex:vector_field_curling}
    v_P=\qty(-y\pdv{x}+x\pdv{y})_P    
    \end{eqnarray}
    Then one can visualize this on a plot as follows:
\begin{lstlisting}[style=in]
VectorPlot[{-y, x}, {x, -4, 4}, {y, -4, 4}]
\end{lstlisting}
    This describes a vector field which is always tangential to a circle about the origin (see Fig.~\ref{fig:vector_fields_fig1}). The integral curves of this field on the manifold are thus circles centered at the origin. We can use another line of code to plot the integral curves of this vector field:
\begin{lstlisting}[style=in]
StreamPlot[{-y, x}, {x, -4, 4}, {y, -4, 4}, StreamScale -> None]
\end{lstlisting}
}
    
\end{example}

\begin{example}~{\rm 

Let us consider another example of a vector field on a curved manifold, namely a torus. As before, we embed the torus in three-dimensional Euclidean space using the parametrization
\begin{align}
    \notag
    (x,y,z)=((R+r\cos v)\cos u,(R+r\cos v)\sin u,r\sin v)\,,
\end{align}
where ($0\leq u,v\leq 2\pi$), and ($R>r>0$) denote the major and minor radii of the torus, respectively. In Mathematica, we may define this parametrization as follows:
\begin{lstlisting}[style=in]
(* Torus parameters *)
R = 3;
r = 1;

(* Parametrization of torus *)
torus[u_, v_] := {
(R + r Cos[v]) Cos[u],
(R + r Cos[v]) Sin[u],
r Sin[v]
};
\end{lstlisting}

The coordinate basis vectors on the torus are obtained by differentiating the parametrization with respect to the coordinates ($u$) and ($v$). These are the tangent vectors $ \pdv{\vb*{r}}{u}$ and $\pdv{\vb*{r}}{v}$,
where ($\vb*{r}(u,v)$) denotes the embedding of the torus. In Mathematica, these tangent vectors are computed by
\begin{lstlisting}[style=in]
(* Tangent vector in u-direction *)
Vu[u_, v_] := D[torus[uu, vv], uu] /. {uu -> u, vv -> v};

(* Tangent vector in v-direction *)
Vv[u_, v_] := D[torus[uu, vv], vv] /. {uu -> u, vv -> v};
\end{lstlisting}

We now define a vector field on the torus by taking a linear combination of these basis vectors, $\frac{1}{2}\pdv{u}+\pdv{v}.$

At each point of the torus, this vector field points in a direction obtained by combining motion around the major circular direction with motion around the minor circular direction. For visualization purposes, the vectors are normalized so that all arrows have the same length. The corresponding Mathematica code is
\begin{lstlisting}[style=in]
L = 1;
arrows = Table[
Module[{p, vec},
p = torus[u, v];
vec = Normalize[0.5 Vu[u, v] + Vv[u, v]];
Arrow[{p - (L/2) vec, p + (L/2) vec}]
],
{u, 0, 2 Pi - 2 Pi/20, 2 Pi/20},
{v, 0, 2 Pi - 2 Pi/10, 2 Pi/10}
];
\end{lstlisting}

Finally, the torus together with the vector field can be displayed using
\begin{lstlisting}[style=in]
Show[
ParametricPlot3D[
torus[u, v],
{u, 0, 2 Pi}, {v, 0, 2 Pi},
Mesh -> None,
PlotStyle -> Directive[Opacity[0.6], LightBlue]
],
Graphics3D[{Red, Arrowheads[0.03], Flatten[arrows]}],
Boxed -> False,
Axes -> False
]
\end{lstlisting}

The resulting figure shows a smooth vector field tangent to the torus, with vectors spiraling around the surface due to the combined contributions from the ($\pdv{u}$) and ($\pdv{v}$) directions. The plot is shown in Fig.~\ref{fig:vector_fields_fig9}.


\begin{figure}
    \centering
    \includegraphics[width=0.5\linewidth]{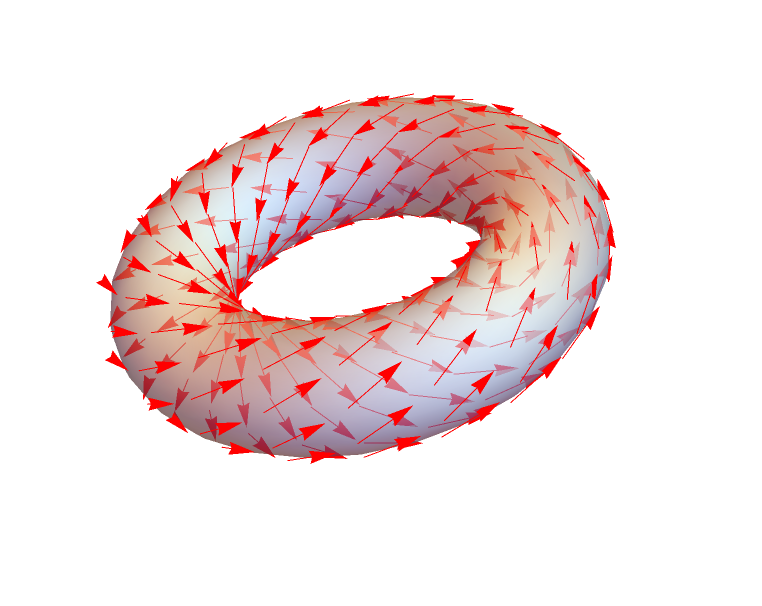}
    \caption{Vector field on a torus}
    \label{fig:vector_fields_fig9}
\end{figure}
}
\end{example}

\begin{exercise}~{\rm 
Construct a parametrization of a 2-sphere using colatitude $(\theta)$ and azimuth $(\phi)$ angles and plot the vector fields:
\begin{enumerate}
    \item $\pdv{\theta}+\sin{\theta}\pdv{\phi}$
    \item $\qty(\theta-\pi/2)^3\pdv{\theta}+\pdv{\phi}$
\end{enumerate}
Recall that these plots resemble different ways of `combing the hair' on a sphere, and observe how the combed field is not continuous at all points on the sphere.}
\end{exercise}

 \md 
    A vector field $v$ is said to be \db{complete} if for every point $P\in\cm$ the integral curve  $\gamma(t)$ of $v$ passing through $P$ can be extended to all $t\in \mathbb{R}\,.$

The tangent bundle $T\cm$ is a product manifold, i.e., a point in $T\cm$ is an ordered pair $(P, v)$ where $P\in\cm$ and $v \in T_P\cm.$ The topological structure and differential structure are given appropriately.

 \md
    The map $\pi: T\cm \to \cm, (P, v)\mapsto P$ (where $v\in T_P\cm$) is called the \db{canonical projection} (or simply \db{projection}).

 \md 
    For each $P\in \cm,$ the pre-image $\pi\inv(P)$ is $T_P\cm$. It is called the \db{fiber} over $P$. Then a vector field can be thought of as a \db{section} of the tangent bundle.

Given a smooth vector field $v,$ we can define an integral curve $ \gamma $ through any point $P$ by $ \dot{\gamma}(t) = v\,, $ i.e., 
\begin{eqnarray}
\frac{d}{dt}x^i(\gamma(t)) &=& v^i(\gamma(t)) \equiv v(x^i\left(\gamma(t))\right)\,, \\
\gamma(0) &=& P\,.
\end{eqnarray}
We could also choose $ \gamma(t_0) = P. $

Then in any neighbourhood $U$ of $P$ we also have $ \gamma_{_Q}\,, $ the integral curve through $Q$. So we can define a map $\phi: I\times U \to \cm$ given by $\phi(t, Q) = \gamma_{_Q}(t)$ where $\gamma_{_Q}(t)$ satisfies 
\begin{eqnarray}
\frac{d}{dt}x^i(\gamma_{_Q}(t)) &=& v(x^i(\gamma_{_Q}(t))\,, \\
\gamma_{_Q}(0) &=& Q\,.
\end{eqnarray}
%
 \md Given a vector field, one may consider an integral curve passing through each point of a sufficiently small neighbourhood. Equivalently, every point in that neighbourhood can be chosen as the initial point of an integral curve of the vector field. This $\phi$ defines a map $ \phi_t: U \to \cm $ at each $t$ by $ \phi_t(Q) = \phi(t, Q) = \gamma_{_Q}(t)\,, $ i.e. for given $t$, $ \phi_t $ takes a point by a parameter distance $t$ along the curve $ \gamma_{_Q}(t). $ This $ \phi_t $ is called the \db{local flow} of $v$.

The local flow has the following properties:
\begin{list}{$\roman{enumii})$} {\usecounter{enumii}\topsep=5pt}
\item $ \phi_0 $ is the identity map of $U$\,;
\item $\phi_s\circ\phi_t = \phi_{s+t}$ for all $s, t, s+t \in U$\,;
\item each flow is a diffeomorphism with $ \phi_t\inv = \phi_{-t}\,. $
\end{list}

The first property is obvious, while the second property follows from the uniqueness of integral curves, i.e. of solutions to first order differential equations. Then the integral curve passing through the point $ \gamma_{_Q}(s)$ is the same as the integral curve passing through $Q$, so that moving a parameter distance $t$ from $ \gamma_{_Q}(s) $ finds the same point on $\cm$ as by moving a parameter distance $s+t$ from $ \gamma_{_Q}(0) \equiv Q\,. $

A vector field can also be thought of as a map from the space of differentiable functions to itself $v: C^\infty(\cm) \to C^\infty(\cm),\, f \mapsto v(f)\,,$ with $v(f): \cm\to \mathbb{R},\, P \mapsto v_{_P}(f)\,.$ Often $v(f)$ is called the Lie derivative of $f$ along $v$ and denoted $\lie_v f\,.$ 

The map $v: f \mapsto v(f)$ has the following properties:
\begin{eqnarray}
v(f + \alpha g) &=& v(f) + \alpha v(g)\\
v(fg) &=& f v(g) + v(f) g \nonumber \\
&& \qquad\qquad \forall f,g \in C^\infty(\cm),\qquad  \alpha \in \mathbb{R} 
\end{eqnarray}
The set of all (real) vector fields $V(\cm)$ on a manifold $\cm$ has the structure of a (real) vector space under vector addition defined by
\begin{eqnarray}
(u + \alpha v)(f) = u(f) + \alpha v(f),\qquad \qquad u, v\in V(\cm), \qquad \alpha \in \mathbb{R}\,.
\end{eqnarray} 
It is possible to replace $\alpha$ by some function in $C^\infty(\cm)$. If $u, v$ are vector fields on $\cm$ and $\alpha$ is now a smooth function on $\cm$, define $u + \alpha v$ by 
\begin{eqnarray}
(u + \alpha v)_{_P}(f) = u_{_P}(f) + \alpha(P) v_{_P}(f)\qquad  \forall f\in C^\infty(\cm), \qquad P\in \cm\,.
\end{eqnarray}
This looks like a vector space but actually it is what is called a \db{module}. 

 \md 
    A \db{ring} $R$ is a set or space with addition and multiplication defined on it, satisfying $(xy)z = x(yz)\,, x(y+z) = xy + xz\,, (x+y)z = xz + yz\,,$ and two special elements 0 and 1, the additive and multiplicative identity elements, $0+x = x+0 = x\,, 1x=x1 = x\,.$
 \md 
    A \db{module} $X$ is an Abelian group under addition, with scalar multiplication by elements of a ring defined on it.

\md 
    A module becomes a \db{vector space} when this ring is a \db{commutative division ring}, i.e. when the ring multiplication is commutative, $xy = yx$, and an inverse exists for every element except 0. Given a smooth function $\alpha$, in general $\alpha\inv\notin C^\infty(\cm)$, so the space of vector fields on $\cm$ is in general a module, not a vector space.

Given a vector field $v$, in an open neighbourhood of some $P\in \cm$ and in a chart, and for any $f\in C^\infty(\cm)\,,$ we have
\begin{eqnarray}
v(f)\Big\vert_{_P} = v_{_P}(f) = v^i_{_P}\left(\frac{\partial f}{\partial x^i}\right)_{_P}\,, \qquad \text{where} \quad v^i_{_P} = v_{_P}(x^i)\,.
\end{eqnarray}
Thus we can write 
\begin{eqnarray}
v = v^i \frac{\partial}{\partial x^i}\qquad\text{with} \quad v^i = v(x^i)\,,
\end{eqnarray}
as an obvious generalization of vector space expansion to the module $V(\cm)$.

\md 
The $v^i$ are now the \db{components} of the vector field $v$, and $ \frac{\partial}{\partial x^i}  $ are now vector fields, which we will call the \db{coordinate vector fields}. Note that this is correct only in some open neighbourhood on which a chart can be defined. In particular, it may not be possible in general to define the coordinate vector fields globally, i.e. everywhere on $\cm$, and thus the components $v^i$ may not be defined globally either.

%% file: DiffGeom/DG_pbpf.tex
\chapter{Pullback and pushforward}\label{pbpf}
Two important concepts are those of pull back (or pull-back or pullback) and push forward (or push-forward or pushforward) of maps between manifolds.

\md 
    Given manifolds $\cm_1, \cm_2, \cm_3$ and maps $f: \cm_1 \to \cm_2\,, g: \cm_2 \to \cm_3\,,$ the \db{pullback} of $g$ under $f$ is the map $f^*g : \cm_1 \to \cm_3$ defined by   
\begin{eqnarray}
f^*g = g\circ f\,. 
\end{eqnarray}
Thus, in particular, if $\cm_1$ and $\cm_2$ are two manifolds with a map $f:\cm_1\to \cm_2$ and $g: \cm_2 \to \mathbb{R}$ is a function on $\cm_2\,,$ the pullback of $g$ under $f$ is a function on $\cm_1$\,,
\begin{eqnarray}
f^*g = g\circ f\,.
\end{eqnarray}
While this looks utterly trivial at this point, this concept will become increasingly useful later on.

Let $g: M_2 \to R$ be a smooth map and let $C \subset M_2$ be a contour of level $c$, i.e. $g(p) = c$ for all $p \in C$. Let $f :M_1 \to M_2$ be a smooth map. Then $f*g: M_1 \to R$ is the pullback of $g$ under $f$, and $C'$ is the preimage of $C$, i.e.$ g(f(p')) = c$ for all $p'$ in $C'$. To put the above in context, let us give a simple physical example.

\begin{example}~{\rm 
     Let $g: \cm_2 \to \mathbb{R}$ describe a set of contours for a function on the manifold $\cm_2$. Then the pullback $f^*g:\cm_1\to\mathbb{R}$ describes the corresponding set of contours on the manifold $\cm_1$, given the appropriate transformation between points on $\cm_1$ and those on $\cm_2$.}
\end{example}
Mathematica example:\\
\begin{lstlisting}[style=in]
Pullback[d[x, y, z], {x -> r Sin[theta] Cos[phi],y -> r Sin[theta] Sin[phi], z -> r Cos[theta]}] // Simplify
\end{lstlisting}
\begin{lstlisting}[style=out]
(r^2 Sin[theta]) dphi ^ dr ^ dtheta
\end{lstlisting}

 \md 
    Given two manifolds $\cm_1$ and $\cm_2$ with a smooth map $f: \cm_1\to \cm_2, P\mapsto Q$ the \db{pushforward} of a vector $v\in T_P\cm_1$ is a vector $f_*v \in T_Q\cm_2$ defined by 
\begin{eqnarray}
f_*v(g) = v(g\circ f) 
\end{eqnarray}
for all smooth functions $g: \cm_2 \to \mathbb{R}\,.$

Thus, we can write 
\begin{eqnarray}
f_*v(g) = v(f^*g)\,.
\end{eqnarray}

Physically, for a point $P$ and the corresponding tangent vector $v_P$ in the manifold $\cm_1$, we construct a corresponding vector $(f_* v)_Q$ at the point $Q=f(P)$ in the other manifold $\cm_2$. The criterion for selecting the pushed-forward vector $f_* v$ is that, for any set of contours near $Q$ in $\cm_2$, the Lie derivative (which will be discussed in a later chapter) along $f_* v$ of the function generating the contours must be equal to the Lie derivative along $v$ of the pulled-back function, near the point $P$ in $\cm_1$. If the idea of Lie derivative is not very clear at this moment, it might be helpful to consider it as being the rate of change of a function when one moves along the direction of a vector. In other words, push-forward generalizes the concept of transformation of vectors under a coordinate transformation to general manifolds.

The pushforward is linear,
\begin{eqnarray}
f_*(v_1 + v_2) &=& f_*v_1 + f_*v_2 \\
f_*(\lambda v) &=& \lambda f_*v\,.
\end{eqnarray}
And if $\cm_1, \cm_2, \cm_3$ are manifolds with maps $f: \cm_1 \to \cm_2\,, g: \cm_2 \to \cm_3$, it follows that 
\begin{eqnarray}
(g\circ f)_* &=& g_* f_*\,, \qquad i.e. \nonumber\\
(g\circ f)_*v &=& g_* f_* v \qquad \forall v\in T_P\cm_1\,.
\end{eqnarray}
Remember that we can think of a vector $v$ as an equivalence class of curves $[\gamma]$. The pushforward of an equivalence class of curves is 
\begin{eqnarray}
f_*v = f_*[\gamma] = [f\circ\gamma]
\end{eqnarray}

Note that for this pushforward to be defined, we do not need the original maps to be 1-1 or onto. In particular, the two manifolds may have different dimensions.

Suppose $\cm_1$ and $\cm_2$ are two manifolds with dimensions $m$ and $n$ respectively. So the respective tangent spaces $T_P\cm_1$ and $T_Q\cm_2$ are also of dimensions $m$ and $n$ respectively. So for a map $f: \cm_1 \to \cm_2, P\mapsto Q\,,$ the pushforward $f_*$ will not have an inverse if $m\neq n\,.$ 

Let us find the components of the pushforward $f_*v$ in terms of the components of $v$ for any vector $v$. Let us in fact consider, given charts $\varphi: P \mapsto (x^1, \cdots, x^m)\,, \psi: Q \mapsto (y^1, \cdots, y^n)$ the pushforward of the  basis vectors. 

For the basis vector $\left(\frac{\partial}{\partial x^\mu} \right)_P$, we want the pushforward $f_*\left(\frac{\partial}{\partial x^\mu} \right)_P\,,$ which is a vector in $T_Q\cm_2\,,$ so we can expand it in the basis $\left(\frac{\partial}{\partial y^\mu} \right)_Q\,,$ 
\begin{eqnarray}
f_*\left(\frac{\partial}{\partial x^i} \right)_P =\left(f_*\left(\frac{\partial}{\partial x^i} \right)_P\right )^\mu \left(\frac{\partial}{\partial y^\mu} \right)_Q
\end{eqnarray}
In any coordinate basis, the components of a vector are given by the action of the vector on the coordinates as in Chap.~\ref{tpm},
\begin{eqnarray}
v_{_P}^\mu = v_{_P}(y^\mu)
\end{eqnarray}
Thus we can write
\begin{eqnarray}
\left(f_*\left(\frac{\partial}{\partial x^i} \right)_P\right )^\mu &=& f_*\left(\frac{\partial}{\partial x^i} \right)_P \left(y^\mu \right)
\end{eqnarray}
But
\begin{eqnarray}
f_*v(g) &=& v(g\circ f)\,, 
\end{eqnarray}
so
\begin{eqnarray}
f_*\left(\frac{\partial}{\partial x^i} \right)_P \left(y^\mu \right)  &=& \left(\frac{\partial}{\partial x^i} \right)_P \left(y^\mu\circ f\right)\,.
\end{eqnarray}
But $y^\mu\circ f$ are the coordinate functions of the map $f\,,$ i.e., coordinates around the point $f(P) = Q\,.$ So we can write $y^\mu\circ f$ as $y^\mu(\vec{x})\,,$ which is what we understand by this. Thus 
\begin{equation}
\left(f_*\left(\frac{\partial}{\partial x^i} \right)_P\right )^\mu = \left(\frac{\partial}{\partial x^i} \right)_P\left(y^\mu\circ f\right) = \left.\frac{\partial y^\mu(\vec{x})}{\partial x^i}\right|_P\,.
\end{equation}
Because we are talking about derivatives of coordinates, these are actually done in charts around $P$ and $Q = f(P)\,,$ so the chart maps are hidden in this equation. 

\md The right hand side is called the \db{Jacobian matrix} (of $y^\mu(\vec{x})= y^\mu\circ f$ with respect to $x^i$). Note that since $m$ and $n$ may be unequal, this matrix need not be invertible and a determinant may not be defined for it. In such a case, there is no inverse of the pushforward. \hfill$\Box$

For the basis vectors, we can then write
\begin{eqnarray}
f_*\left(\frac{\partial}{\partial x^i} \right)_P = \left.\frac{\partial y^\mu(\vec{x})}{\partial x^i}\right|_P \, \left(\frac{\partial}{\partial y^\mu} \right)_{f(P)}
\end{eqnarray}
Since $f_*$ is linear, we can use this to find the components of $\left(f_*v\right)_Q$ for any vector $v_{_P}$,
\begin{eqnarray}
f_*v_{_P} &=& f_*\left[v^i_{_P}\left(\frac{\partial}{\partial x^i}\right)_P\right]\nonumber \\
&=& v^i_{_P} f_*\left(\frac{\partial}{\partial x^i}\right)_P \nonumber \\
&=& v^i_{_P} \left.\frac{\partial y^\mu(\vec{x})}{\partial x^i}\right|_P \, \left(\frac{\partial}{\partial y^\mu} \right)_{f(P)}\\
\Rightarrow \qquad \qquad \left(f_*v_{_P}\right)^\mu &=& v^i_{_P} \left.\frac{\partial y^\mu(\vec{x})}{\partial x^i}\right|_P\,.\label{eq:pf_coords}
\end{eqnarray}

Note that since $f_*$ is linear, we know that the components of $f_*v$ should be linear combinations of the components of $v\,,$ so we can already guess that $\left(f_*v_{_P}\right)^\mu = A^\mu_i v^i_{_P}$ for some matrix $A^\mu_i\,.$ The matrix is made of first derivatives because vectors are first derivatives.

\begin{example}~{\rm 
    We will further demonstrate the idea of pull-back and push-forward with an example in Mathematica. For simplicity, we will work with the 2-dimensional Euclidean space $\mathbb{R}^2$, and with a simple vector field:
\begin{equation} \label{eq:pushforward_example}
v=xy\pdv{x}+x^2\pdv{y}
\end{equation}
    
    Let us first plot this vector field on the manifold to visualize it:
    \begin{lstlisting}[style=in]
X = {x y , x^2}
    \end{lstlisting}
\begin{lstlisting}[style=in]
VectorPlot[#, {x, -5, 5}, {y, -5, 5}, VectorColorFunction -> None] & /@ {X, Y}
\end{lstlisting}
This plots the vector field as shown in Fig.~\ref{fig:pbpf_fig1}.

Let us now define a transformation $f$. We will keep the calculation simple and choose the co-domain manifold to also be $\mathbb{R}^2$. Consider the transformation:
\begin{equation}\label{eq:ex-transfm}
    (x,y)\to(x_1,y_1) = (x+y, x-y)
\end{equation}
which is equivalent to a rotation followed by a dilation about the origin.

Now, we choose a point $P$ on the manifold. For convenience, we choose the point $P=(1,1)$. 
We define the transformation:
\begin{lstlisting}[style=in]
f[{x_, y_}] = { x + y, x - y};
\end{lstlisting}

Let us see where the point $P$ is mapped to.
\begin{lstlisting}[style=in]
fp = f[p]
\end{lstlisting}
\begin{lstlisting}[style=out]
{2, 0}
\end{lstlisting}
So, $Q = (2,0)$.
Using Eq.~(\ref{eq:pf_coords}), we can easily construct the pushed forward vector at the point $Q=f(P)$, given the ingredients we have at hand.

\begin{lstlisting}[style=in]
p = {1, 1};
Xp = X /. MapThread[Rule, {{x, y}, p}]
\end{lstlisting}
\begin{lstlisting}[style=out]
{1, 1}
\end{lstlisting}
This step defines the point under consideration, and produces the vector at this point that we want to push forward. So,
\[v_{P} = \pdv{x}+\pdv{y}\]

\begin{figure}[t]
    \centering
\begin{subfigure}[t]{0.45\linewidth}
    \centering
    \includegraphics[width=0.95\linewidth]{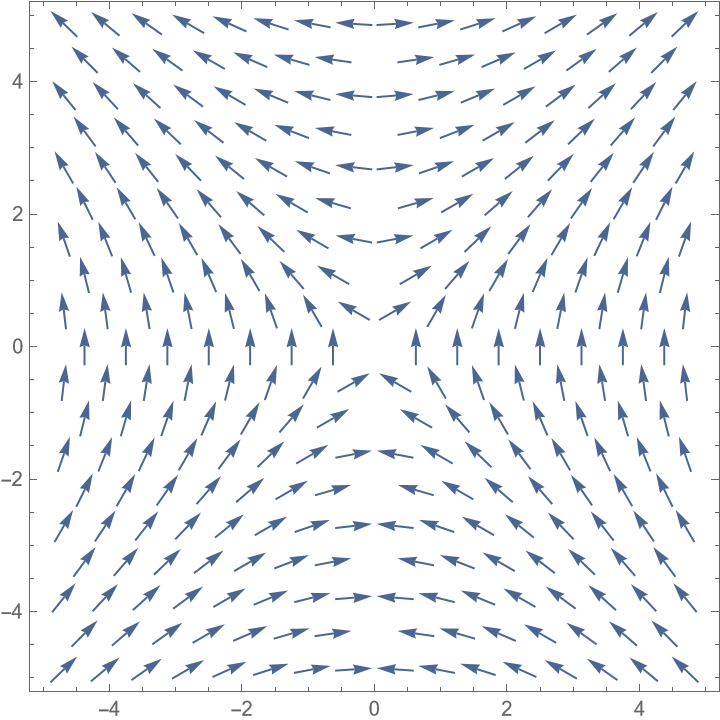}
    \subcaption{Vector field $v$ on $\mathbb{R}^2$}
    \label{fig:pbpf_fig1}
\end{subfigure}
\hfill
\begin{subfigure}[t]{0.45\linewidth}
    \centering
    \includegraphics[width=0.95\linewidth]{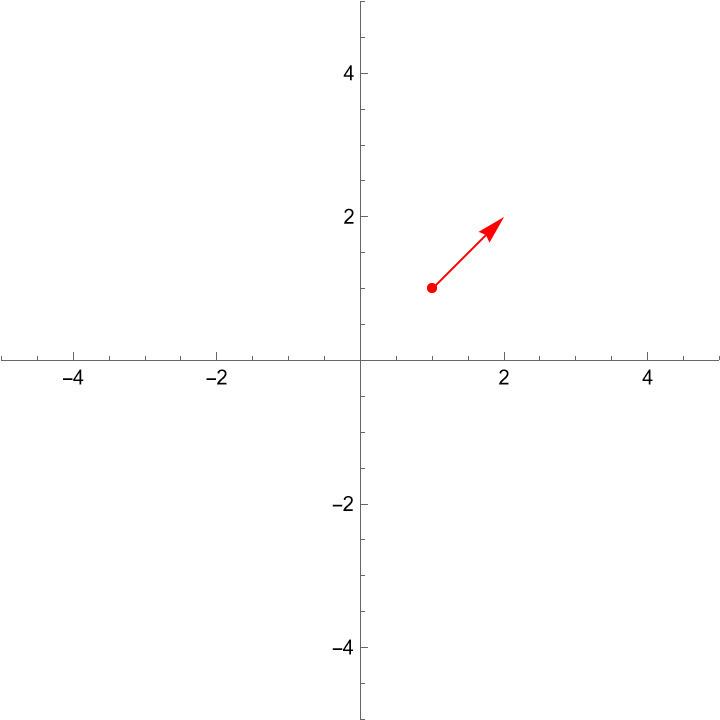}
    \subcaption{Vector $v_P$ at $P=(1,1)$}
    \label{fig:pbpf_fig3}
\end{subfigure}
\caption{}
\label{fig:pbpf_fig1-2}
\end{figure}

\begin{figure}[t]
\centering
\begin{subfigure}[t]{0.45\linewidth}
    \centering
\includegraphics[width=0.95\linewidth]{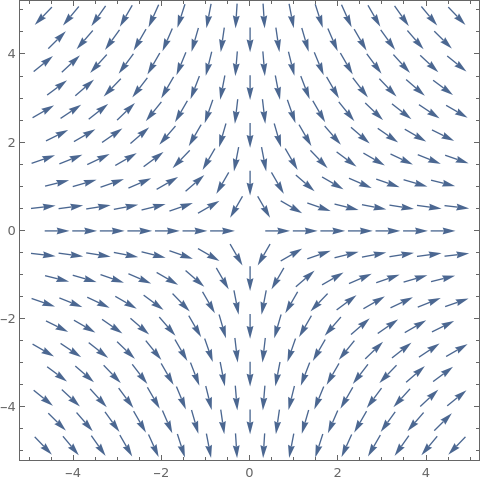}
\caption{The pushed-forward vector field $f_* v $ (given by Eq.~(\ref{eq:pushforwarded}))}
\label{fig:pbpf_fig5}
\end{subfigure}
\hfill
\begin{subfigure}[t]{0.45\linewidth}
    \centering
    \includegraphics[width=0.95\linewidth]{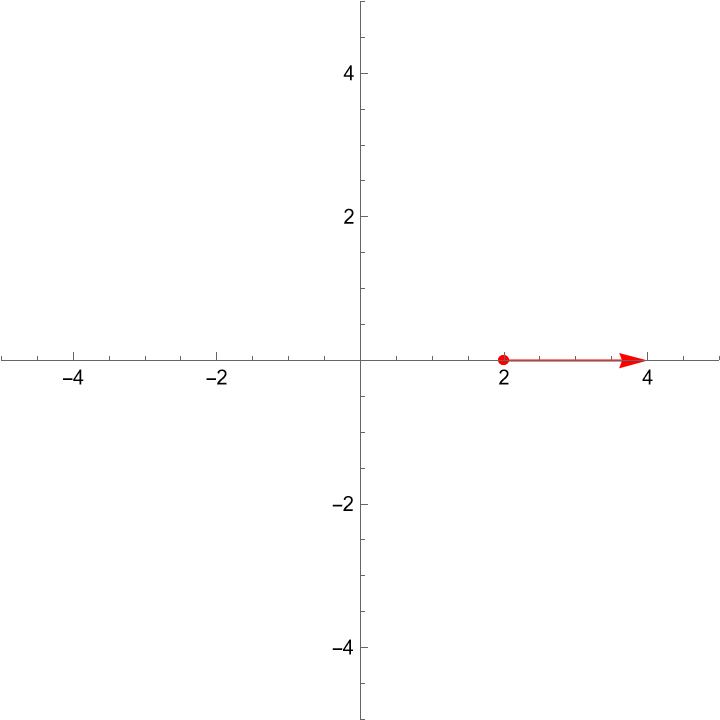}
    \subcaption{$f_* v_Q$ at $Q=(2,0)$}
    \label{fig:pbpf_fig4}
\end{subfigure}
\caption{}
\label{fig:pbpf_fig4-5}
\end{figure}

 Now we apply the pushforward operation to the vector at point $P$ using Eq.~(\ref{eq:pf_coords}). This is implemented using:
\begin{lstlisting}[style=in]
XXp = Xp*Table[D[i, j], {i, f[{x, y}]}, {j, {x, y}}] /.MapThread[Rule, {{x, y}, p}] // Total
\end{lstlisting}
where the matrix generated by \code{Table} is the Jacobian matrix defined before. This gives the following output:
\begin{lstlisting}[style=out]
{2,0}
\end{lstlisting}
Thus,
\[f_*(v_P) = 2 \pdv{x_1}\]
This is the vector obtained by pushing forward the vector at point $P$ under the transformation of Eq.~(\ref{eq:ex-transfm}). We plot the vector $v_P$ (Fig.~\ref{fig:pbpf_fig3}) and the pushed-forward vector $f_* v_Q$ (Fig.~\ref{fig:pbpf_fig4}) as follows:
\begin{lstlisting}[style=in]
Show[ListPlot[{p}, PlotRange -> {{-5, 5}, {-5, 5}}, PlotStyle -> Red], Graphics[{Red, Arrow[{p, p + Xp}]}, PlotRange -> {{-5, 5}, {-5, 5}}], AspectRatio -> 1]
\end{lstlisting}
\begin{lstlisting}[style=in]
Show[ListPlot[{fp}, PlotRange -> {{-5, 5}, {-5, 5}}, PlotStyle -> Red], Graphics[{Red, Arrow[{fp, fp + XXp}]}, PlotRange -> {{-5, 5}, {-5, 5}}], AspectRatio -> 1]
\end{lstlisting}

Notice that the pushed-forward vector at the point $Q$ is not the same as the vector (corresponding to the original vector field) at the point $Q$. This is expected since in general the manifolds $\cm_1$ and $\cm_2$ may not be identical at all. Only in a simple example such as this did we consider them to be identical.
}
\end{example}

\begin{exercise}~{\rm 
Find out the pushed-forward vector for the same vector field, but now under the transformation $f:\mathbb{R}^2\mapsto \mathbb{R}^3$ such that $f(x,y)=(x,y,x^2+y^2)$, given $P=(1,1)$
}
\end{exercise}

Another example of the pushforward map is the following. Remember that tangent vectors are derivatives along curves. Suppose $v_{_P}\in T_P\cm$ is the derivative along $\gamma\,.$ Since $\gamma: I\to \cm$ is a map, we can consider pushforwards under $\gamma\,,$ of derivatives on $I\,.$ Thus for $\gamma: I\to \cm\,, t\mapsto \gamma(t)=P\,,$ and for some $g:\cm \to \mathbb{R}\,,$
\begin{eqnarray}
\gamma_*\left(\frac{d}{dt}\right)_{t=0} g &=& \frac{d}{dt}\left.\left(g\circ\gamma\right)\right|_{t=0} \nonumber \\
&=& \left.\dot{\gamma}_{_P}(g)\right|_{t=0} = v_{_P}(g)\,,
\end{eqnarray}
so
\begin{equation}
\gamma_*\left(\frac{d}{dt}\right)_{t=0} = v_{_P}
\end{equation}

\md We can use this to give another definition of integral curves. Suppose we have a vector field $v$ on $\cm\,.$ Then the  integral curve of $v$ passing through $P\in \cm$ is a curve $\gamma: t\mapsto \gamma(t)$ such that $\gamma(0) = P\,$ and 
\begin{eqnarray}
\gamma_*\left(\frac{d}{dt}\right)_t = \left.v\right|_{\gamma(t)}
\end{eqnarray}
for all $t$ in some interval containing $P\,.$ \hfill$\Box$

Even though in order to define the pushforward of a vector $v$ under a map $f\,,$ we do not need $f$ to be invertible,
the pushforward of a vector field can be defined only if $f$ is both one-to-one and onto. 

If $f$ is not one-to-one, different points $P$ and $P'$ may have the same image, $f(P) = Q = f(P')\,.$ Then for the same vector field $v$ we must have 
\begin{eqnarray}
\left.f_*v\right|_Q = f_*(v_{_P}) = f_*(v_{_{P'}})\,,
\end{eqnarray}
which may not be true. And if $f:\cm \to \cn$ is not onto, $f_*v$ will be meaningless outside some region $f(\cm)\,,$ so $f_*v$ will not be a vector field on $\cn\,.$ 

If $f$ is one-to-one and onto, it is a diffeomorphism, in which case vector fields can be pushed forward, by the rule
\begin{equation}
\left(f_*v\right)_{f(P)} = f_*\left(v_{_P}\right)\,.
\end{equation}

The vector field defined in Eq.~(\ref{eq:pushforward_example}) can be pushed forward under the map in Eq.~(\ref{eq:ex-transfm}) as follows
\begin{lstlisting}[style=in]
newX = X*Table[D[i, j], {i, f[{x, y}]}, {j, {x, y}}] /. 
    Flatten[Solve[{x + y == x1, x - y == y1}, {x, y}]] // Total // 
  Simplify
\end{lstlisting}
\begin{lstlisting}[style=out]
{1/2 x1 (x1 + y1), -(1/2) y1 (x1 + y1)}
\end{lstlisting}
Thus,
\begin{equation}\label{eq:pushforwarded}
f_* v = \frac{1}{2} x_1 \left(x_1+y_1\right)\pdv{x_1} -\frac{1}{2} y_1 \left(x_1+y_1\right) \pdv{y_1}    
\end{equation}

This vector field is plotted in Fig.~\ref{fig:pbpf_fig5}.


%% file: DiffGeom/DG_lie.tex
\chapter{Lie brackets and Lie Algebra}\label{Lie}
A vector field $v$ is a linear map $C^\infty(\cm) \to C^\infty(\cm)$ since it is basically a derivation at each point, $v: f\mapsto v(f)\,.$ In other words, given a smooth function $f\,, v(f)$ is a smooth function on $\cm\,.$ Suppose we consider two vector fields $u\,, v\,.$  Then $u(v(f))$ is also a smooth function, linear in $f\,.$ But is $uv \equiv u\circ v$ a vector field? To find out, we consider
\begin{eqnarray}
u(v(fg)) &=& u(fv(g) + v(f)g) \nonumber \\
&=& u(f)v(g) + fu(v(g)) + u(v(f))g + v(f)u(g)\,.
\end{eqnarray}
We reorder the terms to write this as
\begin{equation}
uv(fg) = f uv(g) + uv(f) g + u(f) v(g) + v(f) u(g)\,,
\end{equation}
so the Leibniz rule is not satisfied by $uv\,.$ But if we also consider the combination $vu\,,$ we get
\begin{equation}
vu(fg) = f(vu(g) + vu(f) g + v(f)u(g) + u(f)v(g)\,.
\end{equation}
Thus 
\begin{equation}
(uv - vu)(fg) = f(uv - vu)(g) + (uv - vu)(f) g\,,
\end{equation}
which means that the combination 
\begin{equation}
\comm{u}{v} := uv - vu\,
\end{equation}
is a vector field on $\cm\,,$ with the product $uv$ signifying successive operation on any smooth function on $\cm\,.$ 

\md This combination is called the \db{commutator} or \db{Lie bracket} of the vector fields $u$ and $v$\,. \hfill$\Box$

In any chart around the point $P\in \cm\,,$ we can write a vector field in local coordinates
\begin{eqnarray}
v(f) = v^i \frac{\partial f}{\partial x^i}\,,
\end{eqnarray}
so that 
\begin{eqnarray}
u(v(f)) &=& u^j\frac{\partial}{\partial x^j}\left(v^i \frac{\partial f}{\partial x^i}\right)\, \nonumber \\
&=& u^j\frac{\partial v^i}{\partial x^j} \frac{\partial f}{\partial x^i} + u^j v^i \frac{\partial^2 f}{\partial x^j x^i}\,, \nonumber \\
v(u(f)) &=& v^j\frac{\partial u^i}{\partial x^j} \frac{\partial f}{\partial x^i} + u^j v^i \frac{\partial^2 f}{\partial x^j x^i}\,.
\end{eqnarray}
Subtracting, we get
\begin{equation}
u(v(f)) - v(u(f)) = u^j\frac{\partial v^i}{\partial x^j} \frac{\partial f}{\partial x^i} - v^j\frac{\partial u^i}{\partial x^j} \frac{\partial f}{\partial x^i} \,,
\end{equation}
from which we can read off the components of the commutator,
\begin{equation}
\comm{u}{v}^i = u^j\frac{\partial v^i}{\partial x^j} - v^j\frac{\partial u^i}{\partial x^j} 
\end{equation}
\md The commutator is antisymmetric, $\comm{u}{v} = - \comm{v}{u}\,,$ and satisfies the \db{Jacobi identity}
\begin{equation}
\comm{\comm{u}{v}}{w} + \comm{\comm{v}{w}}{u} + \comm{\comm{w}{u}}{v} = 0\,.
\end{equation}
%
%
%
%
%


Any set of $n$ linearly independent vector fields may be chosen as a basis, but they need not form a coordinate system. In a coordinate system,
\begin{equation}
\comm{\frac{\partial}{\partial x^i}}{\frac{\partial}{\partial x^j}} = 0\,,
\end{equation}
because partial derivatives commute. So $n$ vector fields will form a coordinate system only if they commute, i.e., have vanishing commutators with one another. Then the coordinate lines are the integral curves of the vector fields. For analytic manifolds, this condition is sufficient as well.

\begin{example}~{\rm 
A simple example is the polar coordinate system in $\mathbb{R}^2\,.$ The unit vectors are
\begin{eqnarray}
e_r &=& e_x \cos\theta + e_y \sin\theta \nonumber \\
e_\theta &=& - e_x \sin\theta + e_y \cos\theta\,,
\end{eqnarray}
with $e_x = \displaystyle{\frac{\partial}{\partial x}}$ and $e_y = \displaystyle{\frac{\partial}{\partial y}}$ being the Cartesian coordinate basis vectors, and
\begin{equation}
\cos\theta = \frac{x}{r}\,,\qquad \sin\theta = \frac{y}{r}\,, \qquad r = \sqrt{x^2 + y^2}
\end{equation}
Using these expressions, it is easy to show that $\comm{e_r}{e_\theta} \neq 0\,,$ so $\left\lbrace e_r, e_\theta \right\rbrace$ do not form a coordinate basis.

We will further demonstrate this example with the use of Mathematica. These examples are taken from the book~[\cite{schutz1980}]. First define the unit vector $e_\theta$ using the derivative of a function $f(x,y)$. The following calculation gives $e_\theta f$ in terms of the Cartesian coordinates $x,y$:
\begin{lstlisting}[style=in]
thetaf[x_, y_] = -Sin[$\theta$] D[f[x, y], x] + Cos[$\theta$] D[f[x, y], y] /. $\theta$ -> ArcTan[x, y] // Simplify
\end{lstlisting}
\begin{lstlisting}[style=out]
$\frac{x f^{(0,1)}(x,y)-y f^{(1,0)}(x,y)}{\sqrt{x^2+y^2}}$
\end{lstlisting}
The next step involves calculating $e_r e_\theta f$, which is obtained by applying the operation $e_r$ to the result of the previous calculation:
\begin{lstlisting}[style=in]
rthetaf[x_, y_] =  Cos[$\theta$] D[thetaf[x, y], x] + Sin[$\theta$] D[thetaf[x, y], y] /. $\theta$ -> ArcTan[x, y] // Simplify
\end{lstlisting}
\begin{lstlisting}[style=out]
$\frac{\left(x^2-y^2\right) f^{(1,1)}(x,y)+x y f^{(0,2)}(x,y)-x y f^{(2,0)}(x,y)}{x^2+y^2}$
\end{lstlisting}

On the other hand, we could apply the $e_r$ operation first to the function $f(x,y)$ to get $e_r f$ as follows:
\begin{lstlisting}[style=in]
rf[x_, y_] =  Cos[$\theta$] D[f[x, y], x] + Sin[$\theta$] D[f[x, y], y] /.  $\theta$ -> ArcTan[x, y] //  Simplify
\end{lstlisting}
\begin{lstlisting}[style=out]
$\frac{y f^{(0,1)}(x,y)+x f^{(1,0)}(x,y)}{\sqrt{x^2+y^2}}$
\end{lstlisting}
 Then use this to calculate $e_\theta e_r f$:
 \begin{lstlisting}[style=in]
thetarf[x_, y_] = -Sin[$\theta$] D[rf[x, y], x] + Cos[$\theta$] D[rf[x, y], y] /. $\theta$ -> ArcTan[x, y] // Simplify
\end{lstlisting}
\begin{lstlisting}[style=out]
$\frac{x^2 f^{(1,1)}(x,y)-y^2 f^{(1,1)}(x,y)+x f^{(0,1)}(x,y)+x y f^{(0,2)}(x,y)-x y f^{(2,0)}(x,y)-y f^{(1,0)}(x,y)}{x^2+y^2}$
\end{lstlisting}
Then we can calculate the result of the Lie bracket, $[e_r, e_\theta]$ on $f$:
 \begin{lstlisting}[style=in]
rthetaf[x, y] - thetarf[x, y] // FullSimplify
\end{lstlisting}
\begin{lstlisting}[style=out]
$\frac{y f^{(1,0)}(x,y)-x f^{(0,1)}(x,y)}{x^2+y^2}$
\end{lstlisting}
which is clearly non-zero, demonstrating that $e_r$ and $e_\theta$ do not form a coordinate basis.
 }   
\end{example}

 \md 
    A (real) \db{algebra} is a (real) vector space equipped with a bilinear operation (product) under which the algebra is closed, i.e., for an algebra ${\scr A}$
\begin{list}{$\roman{enumii})$} {\usecounter{enumii}\topsep=5pt}
\item $x \bull y \in {\scr A} \qquad \forall x, y \in {\scr A}$
\item $(\lambda x + \mu y) \bull z = \lambda x \bull z + \mu y \bull z\, \\
x \bull (\lambda y + \mu z) = \lambda x \bull y + \mu x \bull z\qquad \forall x, y, z\in {\scr A}, \quad \lambda, \mu \in \mathbb{R}\,.$
\end{list}
If $\lambda, \mu$ are complex numbers and ${\scr A}$ is a complex vector space, we get a \db{complex algebra}.
%
 \md 
    A \db{Lie algebra} is an algebra in which the operation is 
\begin{list}{$\roman{enumii})$} {\usecounter{enumii}\topsep=5pt}
\item antisymmetric, $x \bull y = - y \bull x\,,$ and
\item satisfies the \db{Jacobi identity}\,,
\begin{equation}
(x \bull y)\bull z + (y \bull z) \bull x + (z \bull x) \bull y = 0\,.
\end{equation}
\end{list} 
%


\begin{example}~{\rm 
    A simple example of a Lie algebra is the algebra of vector products (cross products) in 3 dimensions.}
\end{example}

\begin{example}~{\rm 
    The space ${\scr M}_n = \{\text{all\,} n\times n$ matrices\} under matrix multiplication, $A\bull B = AB\,.$ This is an \db{associative algebra} since matrix multiplication is associative, $(AB)C = A(BC)\, $. A Lie algebra is  not associative in general.}
\end{example}
\begin{example}~{\rm 
    The same space ${\scr M}_n$ of all $n\times n$ matrices as above, but now with matrix commutator as the product,
\begin{equation}
A\bull B = \comm{A}{B} = AB - BA\,.
\end{equation}
This product is antisymmetric and satisfies the Jacobi identity, so ${\scr M}_n$ with this product is a Lie algebra.
}
\end{example}
\begin{example}~{\rm 
The \db{angular momentum algebra} in quantum mechanics. If $L_i$ are the angular momentum operators with $\comm{L_i}{L_j} = \I\epsilon_{ijk}L_k\,,$ we can write the elements of this algebra as
\begin{equation}
\mathbb{L} = \left\lbrace a = \sum_i \zeta_i L_i | \zeta_ i \in \mathbb{C}\right\rbrace
\end{equation}
If $a = \sum a_i L_i$ and $b = \sum b_i L_i\,,$ their product is 
\begin{equation}
a\bull b \equiv \comm{a}{b} = \sum a_i b_j \comm{L_i}{L_j} = \I\sum \epsilon_{ijk} a_i b_j L_k\,.
\end{equation}
This is a Lie algebra because it is antisymmetric, $\comm{a}{a} = 0$ and the Jacobi identity is satisfied.
}
\end{example}
\begin{example}~{\rm 
    The \db{Poisson bracket algebra} of a classical dynamical system consists of functions on the phase space, with the product defined by the Poisson bracket, 
\begin{equation}
f\bull g = \comm{f}{g}_{P.B.}\,.
\end{equation}
This is a Lie algebra. As a vector space it is infinite-dimensional.
}
\end{example}
\begin{example}~{\rm 
    Vector fields on a manifold form a real Lie algebra under the commutator bracket, since the Jacobi identity is a genuine identity, i.e. automatically satisfied, as we have seen in the previous chapter. This algebra is infinite-dimensional. (It can be thought of as the Lie algebra of the group of diffeomorphisms, $\rm{Diff}(\cm))\,.$
    }
\end{example}

%% file: DiffGeom/DG_flow.tex
\chapter{Local flows}\label{flow}
We met local flows and integral curves in Chap.~\ref{vfld}. Given a vector field $v\,$, let us write its local flow as $\phi_t\,.$  

\md The collection $\phi_t$ for $t< \epsilon$ (for some $\epsilon > 0$, or alternatively for $t< 1$) is a \db{one$-$parameter group of local diffeomorphisms}\,. \hfill$\Box$

Consider the vector field in a neighbourhood $U$ of a point $Q\in \cm\,.$ Since $\phi_t: U \to \cm, Q \mapsto \gamma_Q(t)$ is a \db{local diffeomorphism}\,, i.e. a diffeomorphism for sufficiently small values of $t\,,$ we can use $\phi_t$ to push forward vector fields. At some point $P$ we have the curve $\phi_t(P)\,.$ We push forward a vector field at $t=\epsilon$ to $t=0$ and compare with the vector field at $t=0\,.$

We recall that for a map $\varphi:\cm_1 \to \cm_2$ the pullback of a function $f \in C^\infty(\cm_2)$ is defined as
\begin{equation}
\varphi^*f = f\circ\varphi : \cm_1 \to \mathbb{R}\,,
\end{equation}
and $\varphi^*f \in C^\infty(\cm_1)$ if $\varphi$ is $C^\infty\,.$

The push forward of a vector $v_{_P}$ is defined by 
\begin{eqnarray}
\varphi_* v_{_P}(f) = v_{_P}(f\circ\varphi) &=& v_{_P}(\varphi^* f)\, \\
v_{_P}\in T_P\cm_1 \,, && \varphi_* v_{_P} \in T_{\varphi(P)}\cm_2\,.
\end{eqnarray} 
If $\varphi$ is a diffeomorphism, we can define the push forward of a vector field $v$ by
\begin{eqnarray}
\left.\varphi_*v(f)\right|_{\varphi(P)} &=& v\left.\left(f\circ\varphi\right)\right|_{P} \nonumber \\
i.e. \qquad \qquad \left.\varphi_*v(f)\right|_Q &=& 
v\left.\left(f\circ\varphi\right)\right|_{\varphi^{-1}Q}  \nonumber \\
&=& v\left.\left(\varphi^*f\right)\right|_{\varphi^{-1}Q}\,.
\end{eqnarray}
We can rewrite this definition in several different ways,
\begin{eqnarray}
(\varphi_*v)(f) &=& v\left(f\circ\varphi\right)\circ \varphi^{-1}\nonumber \\
&=& (\varphi^{-1})^*\left(v\left(f\circ\varphi\right)\right) \nonumber \\
&=& (\varphi^{-1})^*\left(v\left(\varphi^*f\right)\right)\,.
\label{flow.pushpull}
\end{eqnarray}

\md We see from this that, if $\varphi:\cm_1\to\cm_2$ is not invertible, $\varphi_*v$ is not a vector field on $\cm_2\,.$ If $\varphi^{-1}$ exists but is not differentiable, $\varphi_*v$ is not differentiable. But there may be some $\varphi$ and some $v$ such that $\varphi_*v$ is a differentiable vector field, even if $\varphi$ is not invertible or $\varphi^{-1}$ is not differentiable. Then $v$ and $\varphi_*v$ are said to be \db{$\varphi-$related}.\hfill$\Box$

\textbf{Proposition:} Given a diffeomorphism $\varphi:\cm_1 \to \cm_2$ (say both $C^\infty$ manifolds) the push forward $\varphi_*$ is an isomorphism on the Lie algebra of vector fields, i.e. 
\begin{eqnarray}
\varphi_*\comm{u}{v} = \comm{\varphi_*u}{\varphi_*v}\,.
\end{eqnarray}

\textbf{Proof:} 
\begin{eqnarray}
\varphi_*\comm{u}{v}(f) &=& \comm{u}{v}\left(f\circ\varphi\right)\circ\varphi^{-1} \nonumber \\
&=& u\left(v\left(f\circ\varphi\right)\right)\circ\varphi^{-1} - u \leftrightarrow v \,, \\
\text{while}\qquad\qquad \comm{\varphi_*u}{\varphi_*v}(f) &=& \varphi_*u\left(\varphi_*v\left(f\right)\right) - u \leftrightarrow v  \,\qquad \nonumber \\
&=& u\left(\varphi_*v\left(f\right)\circ\varphi\right)\circ\varphi^{-1} - u \leftrightarrow v \nonumber\\
&=& u\left(\left(v\left(f\circ\varphi\right)\circ\varphi^{-1}\right)\circ\varphi\right)\circ\varphi^{-1} - u \leftrightarrow v \nonumber \\
&=& u\left(v\left(f\circ\varphi\right)\right)\circ\varphi^{-1} - u \leftrightarrow v \,. 
\end{eqnarray}
\hfill$\Box$

\md A vector field $v$ is said to be \db{invariant} under a diffeomorphism $\varphi:\cm\to\cm$ if $\varphi_*v = v\,,$ i.e. if $\varphi_*(v_{_P}) = v_{\varphi(P)}\,$ for all $P\in \cm\,.$\hfill$\Box$

We can write for any $f\in C^\infty(\cm)$
\begin{eqnarray}
\left(\varphi_*v\right)(f) &=& \left(\varphi^{-1}\right)^*\left(v\left(\varphi^*f\right)\right) \, \nonumber \\
\Rightarrow \qquad\qquad \varphi^*\left(\left(\varphi_*v\right)\left(f\right)\right) &=& v\left(\varphi^*f\right)\,, \nonumber \\
\Rightarrow \qquad\qquad\qquad \varphi^*\circ\varphi_* v &=& v \circ \varphi^*\,.
\end{eqnarray} 
So if $v$ is an invariant vector field, we can write 
\begin{equation}
\varphi^*\circ v = v\circ \varphi^*\,.
\end{equation}
This expresses invariance under $\varphi\,,$ and is satisfied by all differential operators invariant  under $\varphi\,.$

Consider a vector field $u\,,$ and the local flow (or one-parameter diffeomorphism group) $\phi_t$ corresponding to $u\,,$
\begin{equation}
\phi_t(Q) = \gamma_{_Q}(t)\,, \qquad \dot{\gamma}_{_Q}(t) = u(\gamma_{_Q}(t))\,.
\end{equation} 
But for any $f\in C^\infty(\cm)\,,$
\begin{eqnarray}
\dot{\gamma}_{_Q}(f) &=&
\frac{d}{dt}\left(f\circ\gamma_{_Q}(t)\right) \nonumber \\
&=& \frac{d}{dt}\left(f\circ \phi_t(Q)\right) \nonumber \\
&=& \frac{d}{dt}\left(\phi_t^*(f)\right) = u_{\gamma_{_Q}(t)}(f) \equiv u(f)\Big|_{\gamma_{_Q}(t)}
\end{eqnarray}
At $t = 0$ we get the equation
\begin{equation}
\frac{d}{dt}\left(\phi_t^*(f)\right)\Big|_{t=0} = u(f) \Big|_Q
\label{flow.t=0}
\end{equation}
We can also write
\begin{equation}
\frac{d}{dt}\left(\phi_t^* f\right)(Q) = u(f) \left(\phi_t(Q)\right) = \phi_t^*u(f)(Q)\,.
\end{equation}
This formula can be used to solve linear partial differential equations of the form
\begin{equation}
\frac{\partial}{\partial t}f\left(\vec{x}, t\right) = \sum\limits_{i=1}^{n} v^i(\vec{x})\frac{\partial}{\partial x^i}f(\vec{x},t)\,
\end{equation}
with initial condition $f(\vec{x}, 0) = g(\vec{x})$ and everything smooth. This is an equation on $\mathbb{R}^{n+1}\,,$ so it can be on a chart for a manifold as well.  

We can treat $v^i(\vec{x})$ as components of a vector field $v\,.$ Then a solution to this equation is
\begin{eqnarray}
f(\vec{x}, t) &=& \phi^*_t g(\vec{x}) \nonumber \\
&\equiv& g\left(\phi_t(\vec{x})\right) \equiv g\circ\phi_t(\vec{x})\,,
\end{eqnarray}
where $\phi_t$ is the flow of $v\,.$

\textbf{Proof:} 
\begin{equation}
\frac{\partial}{\partial t}f(\vec{x}, t) = \frac{d}{dt}\left(\phi_t^* g\right) = v(f) \equiv v^i\frac{\partial f}{\partial x^i}\,,
\end{equation}
using Eq.~(\ref{flow.t=0})\,. \hfill$\Box$

Thus, the partial differential equation can be solved by finding the integral curves of $v$ (the flow of $v$) and then by pushing (also called \db{dragging}) $g$ along those curves. It can be shown, using well-known theorems about the uniqueness of solutions to first order partial differential equations, that this solution is also unique.

\textbf{Example:} Consider the equation in 2+1 dimensions
\begin{equation}
\frac{\partial}{\partial t}f(\vec{x}, t) = (x - y) \left(\frac{\partial f}{\partial x} - \frac{\partial f}{\partial y}\right)
\end{equation}
with initial condition $f(\vec{x}, 0) = x^2 + y^2\,.$ The corresponding vector field is $v(\vec{x}) = (x-y, -x+y)\,.$ The integral curve passing through the point $P = (x_0, y_0)$ is given by the coordinates
\begin{equation}
\gamma(t) = (v_x(P)t + x_0, v_y(P)t + y_0)\,,
\end{equation}
so the integral curve passing through $(x, y)$ in our example is given by 
\begin{eqnarray}
\gamma(t) &=& \left((x - y)t + x, (-x + y)t + y\right)\, \\
&=& \Phi_t(x, y)\,, \nonumber
\end{eqnarray}
the flow of $v$. So the solution is 
\begin{eqnarray}
f(\vec{x}, t) &=& \Phi_t^*f(\vec{x}, 0) = f(\vec{x}, 0) \circ \Phi_t(x, y) \nonumber \\
&=& \left[(x-y)t + x\right]^2 + \left[(-x + y)t + y\right]^2 \nonumber \\
&=& (x - y)^2 t^2 + x^2 + 2(x-y)xt + (x-y)^2 t^2 + y^2 - 2(x-y)yt \nonumber \\
&=& 2 (x - y)^2 t^2 + (x^2 + y^2)(1 + 2t) - 4xyt\,.
\end{eqnarray}
%
The reader can use the following command to easily check the result as shown below.
\begin{lstlisting}[style=in]
Table[Coefficient[(((x - y) t + x)^2 + ((-x + y) t + y)^2 // Expand // Simplify), t, n], {n, 0, 2} ] // FullSimplify
\end{lstlisting}
\begin{lstlisting}[style=out]
{x$^2$ + y$^2$, 2 (x - y)$^2$, 2 (x - y)$^2$}
\end{lstlisting}

%% file: DiffGeom/DG_lieder.tex
\chapter{Lie derivative}\label{lieder}
Given some diffeomorphism $\varphi$, we have Eq.~(\ref{flow.pushpull}) for pushforwards and pullbacks,
\begin{equation}
(\varphi_*v) f = \left(\varphi^{-1}\right)^* v \left(\varphi^*(f)\right)\,.
\end{equation}
We will apply this to the flow $\phi_t$ of a vector field $u\,,$ defined by
\begin{equation}
\frac{d}{dt}\left(\phi_t^* f\right)\Big|_{t=0} = u(f)\Big|_{_Q}\,.
\end{equation}
Applying this at $-t\,,$ we get
\begin{eqnarray}
\phi_{-t*}v(f) &=& \left(\phi^{-1}_{-t}\right)^*v \left(\phi_{-t}^*(f)\right) \nonumber \\
&=& \phi_t^*v \left(\phi_{-t}^*(f)\right)\,,
\end{eqnarray}
where we have used the relation $\phi^{-1}_t = \phi_{-t}\,.$
Let us differentiate this equation with respect to $t\,,$
\begin{eqnarray}
\frac{d}{dt} (\phi_{-t*}v)(f)\Big|_{t = 0} &=& \frac{d}{dt} \phi_t^*v \left(\phi_{-t}^*(f)\right)\Big|_{t=0} 
\end{eqnarray}
On the right hand side, $\phi_t^*$ acts linearly on vectors and $v$ acts linearly on functions, so we can imagine $A_t = \phi_t^*v$ as a kind of linear operator acting on the function $f_t = \left(\phi_{-t}^*f\right)\,.$ Then the right hand side is of the form 
\begin{eqnarray}
\left.\frac{d}{dt} A_t f_t \right|_{t=0} &=& \left.\left(\frac{d}{dt}A_t\right)f_t \right|_{t=0} + A_t \left.\frac{d}{dt}f_t\right|_{t=0} \nonumber \\
&=& \left.\left(\frac{d}{dt}\phi_t^* v\right)f_t \right|_{t=0} + \left.A_t\left(\frac{d}{dt}\phi_{-t}^*(f)\right)\right|_{t=0}\nonumber \\
&=& u\left(v(f)\right)\Big|_{t=0} - v\left(u(f)\right)\Big|_{t=0} \nonumber \\
&=& \comm{u}{v}(f)\Big|_{t=0}\,.
\end{eqnarray}
We can also write this as 
\begin{equation}
\lim_{t\to 0}\frac{\phi_{t*}v_{_{\phi_t(P)}} - v_{_P}}{t} = \comm{u}{v}\,.
\end{equation}
The things in the numerator are numbers, so they can be compared at different points, unlike  vectors which may be compared only on the same space.
\md This has the look of a derivative, and it can be shown to have the properties of a derivation on the module of vector fields, appropriately defined. So the Lie bracket is also called the \db{Lie derivative}, and written as
\begin{equation}
\lie_u v = \comm{u}{v}\,. 
\end{equation}

The derivation on functions by a vector field $u:C^\infty(\cm) \to C^\infty(\cm)\,, f\mapsto u(f)\,,$ can be defined similarly as 
\begin{equation}
u(f) = \lim_{t \to 0}\frac{\phi_t^*f - f}{t}\,.
\label{lieder.fn}
\end{equation}
\md So this can also be called the \db{Lie derivative} of $f$ with respect to $u\,,$ and written as $\lie_u f\,.$ \hfill$\Box$

Then it is easy to see that 
\begin{eqnarray}
\lie_u(fg) &=& \left(\lie_uf\right)g + f\left(\lie_ug\right)\,, \nonumber \\
\text{and} \qquad\qquad \lie_u(f + ag) &=& \lie_uf + a \lie_ug\, \qquad \text{for } a \in {\mathbb R}.
\end{eqnarray}
So $\lie_u$ is a derivation on the space $C^\infty(\cm)\,.$ Also,
\begin{eqnarray}
\lie_u(v + aw) &=& \lie_u v + a\lie_u w \,, \nonumber \\
\text{and} \qquad\qquad \lie_u(f v) &=& \left(\lie_u f\right)v + f\lie_u v \qquad \forall f \in C^\infty(\cm)\,. 
\end{eqnarray}
So $\lie_u$ is a derivation on the module of vector fields. Also, using the Jacobi identity, we see that 
\begin{equation}
\lie_u(v\bull w) = \left(\lie_u v\right)\bull w + v\bull\left(\lie_u w\right)\,,
\end{equation}
where $v\bull w = \comm{v}{w}\,,$ so $\lie_u$ is a derivation on the Lie algebra of vector fields. 

Lie derivatives are useful in physics because they describe invariances. For functions, $\lie_u f = 0$ means $\phi^*_t f = f\,,$ so the function does not change along the flow of $u\,.$ So the flow of $u$ preserves $f\,,$ or leaves $f$ invariant.

If there are two vector fields $u$ and $v$ which leave $f$ invariant, $\lie_u f = 0 = \lie_v f\,.$ But we know from Eq.~(\ref{lieder.fn}), which defines the Lie derivative of a function, that
\begin{eqnarray}
\lie_{u + av}f = \lie_u f + a \lie_v &=& 0\, \qquad \forall a\in \mathbb{R}\nonumber \\
\text{and} \qquad\qquad \comm{\lie_u}{\lie_v} f = \lie_\comm{u}{v} f &=& 0\,.
\end{eqnarray}
So the vector fields which preserve $f$ form a Lie algebra.

Similarly, a vector field is invariant under a diffeomorphism $\varphi$ if $\varphi_*v = v\,,$ as mentioned earlier. Using the flow of $u\,,$ we find that a vector field $v$ is invariant under the flow of $u$ if 
\begin{eqnarray}
\phi_{-t}v &=& v\,\nonumber \\
\Rightarrow \qquad\qquad \lie_u v &=& v\,.
\end{eqnarray}
So if a vector field $w$ is invariant under the flows of $u$ and $v\,,$ i.e. if $\lie_u w = 0 = \lie_v w\,,$ we find that 
\begin{equation}
0 = \lie_u\lie_v w - \lie_v \lie_u w = \lie_\comm{u}{v}w \,.
\end{equation}
Thus again the vector fields leaving $w$ invariant form a Lie algebra.

\md Let us also define the corresponding operations for 1-forms. Just as we have discussed the tangent bundle in Chap.~\ref{vfld}, a \db{1$-$form} is similarly a section of the \db{cotangent bundle}
\begin{equation}
T^*\cm = \bigcup_P T^*_P\cm\,.
\end{equation}
Alternatively, a 1-form is a smooth linear map from the space of vector fields on $\cm$ to the space of smooth functions on $\cm\,,$
\begin{equation}
\omega: v \mapsto \omega(v) \in C^\infty(\cm), \qquad \omega(u + av) = \omega(U) + a\omega(v)\,.
\end{equation}
A 1-form is a rule that (smoothly) selects a cotangent vector at each point. \hfill$\Box$

\md Given a smooth map $\varphi:\cm_1 \to \cm_2\,$ (say a diffeomorphism, for convenience), the \db{pullback} $\varphi^*\omega$ is defined by
\begin{equation}
\left(\varphi^*\omega\right)(v) = \omega\left(\varphi_*v\right)\,.
\end{equation}

\md We have already seen the \db{gradient} 1-form for a function $f:\cm \to \mathbb{R}\,,$ which is a linear map from the space of vector fields to functions,
\begin{equation}
df(u + av) = u(f) + av(f) \,,
\end{equation}
and which can be written as 
\begin{equation}
df = \frac{\partial f}{\partial x^i}dx^i
\end{equation}
in some chart. \hfill$\Box$

For an arbitrary 1-form $\omega\,,$ we can write in a chart and for any vector field $v\,,$
\begin{equation}
\omega = \omega_i dx^i\,,\qquad v = v^i\frac{\partial}{\partial x^i}\,,\qquad \omega(v) = \omega_i v^i\,.
\end{equation}

All the components $\omega_i\,, v^i$ are smooth functions, so is $\omega_i v^i\,.$ The space of 1-forms is a module. Since the function $\omega(v)$ is chart-independent, we can find the components $\omega_{i'}$ of $\omega$ in a new chart by noting that 
\begin{equation}
\omega(v) = \omega_i v^i = \omega{_{i'}} v^{i'}\,.
\end{equation}
Note that the notation is somewhat ambiguous here -- $i'$ also runs from 1 to $n\,,$ and the prime actually distinguishes the chart, or the coordinate system, rather than the index $i\,.$

If the components of $v$ in the new chart are related to those in the old one by $v^{i'} = A^{i'}_j v^j\,,$ it follows that
\begin{equation}
\omega_{i'} A^{i'}_j v^j = \omega_j v^j \qquad \Rightarrow \qquad \omega_{i'} A^{i'}_j = \omega_j
\end{equation}
Since coordinate transformations are invertible, we can multiply both sides of the last equation by $A^{-1}$ and write
\begin{equation}
\omega_{i'} = \left(A^{-1}\right)^j_{i'}\omega_j\,.
\end{equation}
For coordinate transformations from a chart $\{x^i\}$ to a chart $\{x'^{i'}\}\,,$
\begin{eqnarray}
A^{i'}_j = \frac{\partial x'^{i'}}{\partial x^j}\,, \qquad && \qquad \left(A^{-1}\right)^j_{i'} = \frac{\partial x^j}{\partial x'^{i'}}\,\\
\text{so} \qquad \qquad v^{i'} = \frac{\partial x'^{i'}}{\partial x^j} v^j\,, \qquad && \qquad \omega_{i'} = \frac{\partial x^j}{\partial x'^{i'}} \omega_j\,.
\end{eqnarray}

We can define the Lie derivative of a 1-form very conveniently by going to a chart, and treating the components of 1-forms and vector fields as functions, 
\begin{eqnarray}
\lie_u \omega(v) = \lie_u \left(\omega_i v^i\right) &=& u^j \frac{\partial}{\partial x^j} \left(\omega_i v^i\right) \nonumber \\ &=& u^j \frac{\partial \omega_i}{\partial x^j} v^i + u^j \omega_i \frac{\partial v^i}{\partial x^j} \,.
\label{lieder.contract}
\end{eqnarray}
But we want to define things such that 
\begin{equation}
\lie_u \omega(v) = \left(\lie_u \omega\right)(v) + \omega\left(\lie_u v\right)\,.
\label{lieder.leibniz}
\end{equation}
We already know the left hand side of this equation from Eq.~(\ref{lieder.contract}), and the right hand side can be calculated in a chart as
\begin{eqnarray}
\left(\lie_u \omega\right)(v) + \omega\left(\lie_u v\right) 
&=& \left(\lie_u \omega\right)_i v^i + \omega_i\left(\lie_u v\right)^i \nonumber \\
&=& \left(\lie_u \omega\right)_i v^i + \omega_i\comm{u}{v}^i \nonumber \\
&=& \left(\lie_u \omega\right)_i v^i + \omega_i\left(u^j\frac{\partial v^i}{\partial x^j} - v^j  \frac{\partial u^i}{\partial x^j}\right) \,.\qquad
\end{eqnarray}
Equating the right hand side of this with the right hand side of Eq.~(\ref{lieder.contract}), we can write 
\begin{equation}
\left(\lie_u \omega\right)_i = u^j \frac{\partial \omega_i}{\partial x^j} + \omega_j\frac{\partial u^j}{\partial x^i}\,.
\label{lieder.formcomp}
\end{equation}
These are the components of $\lie_u \omega$ in a given chart $\{x^i\}\,.$

For the sake of convenience, let us write down the Lie derivatives of the coordinate basis vector fields and basis 1-forms.
The coordinate basis vector corresponding to the $i$-th coordinate is
\begin{equation}
v = \frac{\partial}{\partial x^i} \qquad \Rightarrow \qquad v^j = \delta^j_i\,.
\end{equation} 
Putting this into the formula for Lie derivatives, we get
\begin{eqnarray}
\lie_u \frac{\partial}{\partial x^i} &=& \comm{u}{v}^j \frac{\partial}{\partial x^j} \nonumber \\
&=& \left(u^k \frac{\partial v^j}{\partial x^k} - v^k \frac{\partial u^j}{\partial x^k} \right) \frac{\partial}{\partial x^j} \nonumber \\
&=& \left(0 - \delta^k_i \frac{\partial u^j}{\partial x^k}\right) \frac{\partial}{\partial x^j}\nonumber \\
&=& - \left(\frac{\partial u^j}{\partial x^i}\right) \frac{\partial}{\partial x^j}\,.
\end{eqnarray}
Similarly, the 1-form corresponding to the $i$-th basis coordinate is
\begin{equation}
dx^i = \delta^i_j dx^j\,, \qquad i.e. \qquad \left(dx^i\right)_j = \delta^i_j\,.
\end{equation}
Using this in the formula Eq.~(\ref{lieder.formcomp}) we get
\begin{equation}
\lie_u dx^i = \delta^i_k \frac{\partial u^k}{\partial x^j}dx^j = \frac{\partial u^i}{\partial x^j} dx^j\,.
\end{equation}

There is also a geometric description of the Lie derivative of 1-forms, 
\begin{eqnarray}
\left.\lie_u \omega \right|_{_P} &=& \lim_{t\to 0} \frac{1}{t}\left[\left.\phi_t^*\omega\right|_{_{\phi_t(P)}} - \omega_{_P}\right] \nonumber \\
&=& \left. \frac{d}{dt} \phi_t^*\omega\right|_{_P}\,.
\end{eqnarray}
We will not discuss this in detail, but only mention that it leads to the same Leibniz rule as in Eq.~(\ref{lieder.leibniz}), and the same description in terms of components as in Eq.~(\ref{lieder.formcomp}).

Demonstration of the Lie derivatives of functions and vector fields in Mathematica (these examples need the package \code{DifferentialForms}):

The Lie derivative of a function $f = x^2 y + \sin y$ with respect to the vector field $u = x \frac{\partial}{\partial x} + y \frac{\partial}{\partial y}$ can be found as follows:
\begin{lstlisting}[style=in]
f = x^2 y + Sin[y];
LieDerivative[ x X[x] + y X[y], f ] 
\end{lstlisting}
\begin{lstlisting}[style=out]
2 x$^2$ y + y (x$^2$ + Cos[y])
\end{lstlisting}
Similarly, the Lie derivative of $d = x^2+  y^2$ with respect to the vector field $u = -y \frac{\partial}{\partial x} + x \frac{\partial}{\partial y}$ can be obtained as
\begin{lstlisting}[style=in]
d = x^2 + y^2;
LieDerivative[-y X[x] + x X[y], d]
\end{lstlisting}
\begin{lstlisting}[style=out]
0
\end{lstlisting}
Let us now find the Lie derivative of two vector fields $u = x \frac{\partial}{\partial x} + y \frac{\partial}{\partial y}$ and $v = x^2 \frac{\partial}{\partial x} + y \frac{\partial}{\partial y}$
\begin{lstlisting}[style=in]
LieDerivative[x X[x] + y X[y], x^2 X[x] + y X[y]]
\end{lstlisting}
\begin{lstlisting}[style=out]
(x$^2$) X[x]
\end{lstlisting}
Similarly, the Lie derivative of $u = x \frac{\partial}{\partial x} + y \frac{\partial}{\partial y}$ and $v= -y \frac{\partial}{\partial x} + x \frac{\partial}{\partial y}$ is
\begin{lstlisting}[style=in]
LieDerivative[x X[x] + y X[y], -y X[x] + x X[y]]
\end{lstlisting}
\begin{lstlisting}[style=out]
0
\end{lstlisting}
The Lie derivative of one form $\alpha = x^2 dx + y^2 dy $ with respect to $u = x \frac{\partial}{\partial x} + y \frac{\partial}{\partial y}$ is
\begin{lstlisting}[style=in]
alpha = x^2  d[x] + y^2 d[y];
LieDerivative[x X[x] + y X[y], alpha]
\end{lstlisting}
\begin{lstlisting}[style=out]
(3 x^2) dx+(3 y^2) dy
\end{lstlisting}
 A notebook named \texttt{Lie\_Derivative\_Schutz} contains a demonstration of the chain rule using the example from the book by B. Schutz~[\cite{schutz1980}].

%% file: DiffGeom/DG_tensor.tex
\chapter{Tensors}\label{tensor}
We have so far defined tangent vectors, cotangent vectors, and also vector fields and 1-forms. We will now define tensors. We will do this by starting with the example of a specific type of tensor.

\md A (1, 2) \db{tensor} $A_{_P}$ at $P\in \cm$ is a map
\begin{equation}
A_{_P} : T_{_P}\cm \times T_{_P}\cm \times T^*_{_P}\cm \to \mathbb{R}
\end{equation}
which is linear in every argument. \hfill$\Box$

Thus, given two vectors $u_{_P}, v_{_P}$ and a covector $\omega_{_P}\,,$
\begin{equation}
A_{_P} : \left(u_{_P}, v_{_P}, \omega_{_P}\right) \mapsto A_{_P}\left(u_{_P}, v_{_P}; \omega_{_P}\right) \in \mathbb{R}\,.
\end{equation}

Suppose $\{e_a\}, \{\lambda^a\}$ are bases for $T_{_P}\cm, T^*_{_P}\cm\,.$ Write
\begin{equation}
A^c_{ab} = A_{_P}\left(e_a, e_b; \lambda^c\right) \,.
\end{equation}
Then for arbitrary vectors $u_{_P} = u^a e_a\,, v_{_P} = v^a e_a\,,$ and covector $\omega_{_P} = \omega_a \lambda^a$ we get using linearity of the tensor map, 
\begin{eqnarray}
A_{_P}\left(u_{_P}, v_{_P}; \omega_{_P}\right) &=& A_{_P}\left(u^a e_a, v^b e_b; \omega_c \lambda^c\right) \nonumber \\
&=& u^a v^b \omega_c A^c_{ab}\,.
\end{eqnarray}

Note that it is a matter of convention as to whether $A$ as written above should be called a (1, 2) tensor or a (2, 1) tensor, and the convention varies 
from one book to another.  As a result, it is best to specify the tensor by writing indices as there is no confusion about $A^c_{ab}\,.$

A tensor of type $(p, q)$ can be defined in the same way,
\begin{equation}
\buildrel{\displaystyle{A_{_P}^{p,q}:\,}\underbrace{T_P\cm\times\cdots\times T_P\cm}}\over{q\;\text{times}}\, \buildrel{\,\displaystyle{\times}\;\underbrace{T^*_P\cm\times\cdots\times T^*_P\cm}
\displaystyle{\quad\to\quad\mathbb{R}}}\over{p\;\text{times}}
\end{equation}
in such a way that the map is linear in every argument. 

\md Alternatively, $A_{_P}$ is an element of the \db{tensor product space}
\begin{equation}
\buildrel{\displaystyle{A_{_P}\in\;}\underbrace{T_P\cm\otimes\cdots\otimes T_P\cm}}\over{p\,\text{times}}\,
\buildrel{\,\displaystyle{\otimes}\;\underbrace{T^*_P\cm\otimes\cdots\otimes T^*_P\cm}}\over{q\,\text{times}}
\end{equation}
We can now define the components of this tensor in the same way that we did for the (1, 2) tensor. Then a $(p, q)$ tensor has components which can be written as
\[
A^{a_1\cdots a_p}_{b_1\cdots b_q}\,.
\]

\md Some special types of $(p, q)$ tensors have special names. A (1,~0) tensor is a linear map $A_{_P}: T^*_{_P}\cm \to \mathbb{R}\,,$ so it is a tangent vector. A (0,~1) tensor is a cotangent vector. A $(p,~0)$ tensor has components with $p$ upper indices. It is called a \db{contravariant $p-$tensor}. A $(0,~q)$ tensor has components with $q$ lower indices. It is called a \db{covariant $q-$tensor}. \hfill$\Box$

It is only possible to add tensors of the same type, but not of different types, viz.,
\begin{equation}
A^{a_1\cdots a_p}_{b_1\cdots b_q} + B^{a_1\cdots a_p}_{b_1\cdots b_q} = (A + B)^{a_1\cdots a_p}_{b_1\cdots b_q}\,.
\end{equation}

\md A \db{tensor field} is a rule giving a tensor at each point. \hfill$\Box$

We can now define the Lie derivative of a tensor field by using the Leibniz rule in a chart. Let us first consider the components of a tensor field in a chart. For a (1, 2) tensor field $A\,,$ the components in a chart are
\begin{equation}
A^k_{ij} = A(\frac{\partial}{\partial x^i}, \frac{\partial}{\partial x^j}; dx^k)\,.
\end{equation}
The components are functions of $x$ in a chart. Thus, we can write this tensor field as 
\begin{equation}
A = A^k_{ij} dx^i\otimes dx^j \otimes \frac{\partial}{\partial x^k}\,,
\end{equation}
where the $\otimes$ indicates a `product', in the sense that its action on two vectors and a 1-form is a product of the respective components,
\begin{equation}
\left(dx^i\otimes dx^j \otimes \frac{\partial}{\partial x^k}\right) (u, v; \omega) = u^i v^j \omega_k\,.
\end{equation}
Thus we find, in agreement with the earlier definition,
\begin{equation}
A(u, v; \omega) = A^k_{ij} u^i v^j \omega_k\,.
\end{equation}

Under a change of charts, i.e. coordinate system $x^i \to x'^{i'}\,,$ the components of the tensor field change according to 
\begin{equation}
A = A^k_{ij}\; dx^i\otimes dx^j \otimes \frac{\partial}{\partial x^k} = A^{k'}_{i'j'}\; dx'^{i'}\otimes dx'^{j'} \otimes \frac{\partial}{\partial x'^{k'}}\,.
\end{equation}
Since
\begin{equation}
dx'^{i'} = \frac{\partial x'^{i'}}{\partial x^i}dx^i\,,\;\qquad \frac{\partial}{\partial x'^{i'}} = \frac{\partial x^i}{\partial x'^{i'}}\frac{\partial}{\partial x^i}\,
\end{equation}
($i$ and $i'$ are not equal in general), we get 
\begin{eqnarray}
A^k_{ij}\; dx^i\otimes dx^j \otimes \frac{\partial}{\partial x^k} = A^{k'}_{i'j'}\; \frac{\partial x'^{i'}}{\partial x^i}dx^i \otimes \frac{\partial x'^{j'}}{\partial x^j}dx^j \otimes \frac{\partial x^k}{\partial x'^{k'}}\frac{\partial}{\partial x^k}\,. \nonumber \\
\end{eqnarray}
Equating components, we can write
\begin{eqnarray}
A^k_{ij} &=& A^{k'}_{i'j'}\; \frac{\partial x'^{i'}}{\partial x^i} \frac{\partial x'^{j'}}{\partial x^j} \frac{\partial x^k}{\partial x'^{k'}} \\
A^{k'}_{i'j'} &=& A^k_{ij}\; \frac{\partial x^i}{\partial x'^{i'}} \frac{\partial x^j}{\partial x'^{j'}} \frac{\partial x'^{k'}}{\partial x^k}\,.
\end{eqnarray}

From now on, we will use the notation $\partial_i$ for $\displaystyle{\frac{\partial}{\partial x^i}}$ and $\partial_i f$ for $\displaystyle\frac{\partial f}{\partial x^i}$ unless there is a possibility of confusion. This will save some space and make the formulae more readable. 

We can calculate the Lie derivative of a tensor field (with respect to a vector field $u$, say) by using the fact that $\lie_u$ is a derivative on the modules of vector fields and 1-forms, and by assuming the Leibniz rule for tensor products. Consider a tensor field 
\begin{equation}
T = T^{m\cdots n}_{a\cdots b}\; \partial_m\otimes\cdots\otimes\partial_n\otimes dx^a\otimes\cdots\otimes dx^b\,.
\end{equation}
Then 
\begin{eqnarray}
\lie_u T &=& \left(\lie_u T^{m\cdots n}_{a\cdots b}\right)\; \partial_m\otimes\cdots\otimes\partial_n\otimes dx^a\otimes\cdots\otimes dx^b\, \nonumber \\
&+&  T^{m\cdots n}_{a\cdots b} \left(\lie_u \partial_m\right)\otimes\cdots\otimes
\partial_n\otimes dx^a\otimes\cdots\otimes dx^b\, + \cdots \nonumber \\
&+&  T^{m\cdots n}_{a\cdots b}\;\partial_m\otimes\cdots\otimes\partial_n
\otimes\left(\lie_u dx^a\right)\otimes\cdots\otimes dx^b\, + \cdots \,,
\end{eqnarray}
where the dots stand for the terms involving all the remaining upper and lower indices. Since the components of a tensor field are functions on the manifold, we have 
\begin{equation}
\lie_u T^{m\cdots n}_{a\cdots b} = u^i\partial_i T^{m\cdots n}_{a\cdots b}\,,
\end{equation}
and we also know that 
\begin{equation}
\lie_u \partial_m = -\frac{\partial u^i}{\partial x^m}\partial_i\,, \qquad
\lie_u dx^a = \frac{\partial u^a}{\partial x^i} dx^i\,.
\end{equation}
Putting these into the expression for the Lie derivative for $T\,$ and relabeling the dummy indices, we find the components of the Lie derivative,
\begin{eqnarray}
\left(\lie_u T\right)^{m\cdots n}_{a\cdots b} &=& u^i\,\partial_i T^{m\cdots n}_{a\cdots b} \nonumber \\
&-& \, T^{i\cdots n}_{a\cdots b}\,\partial_i u^m -\cdots - T^{m\cdots i}_{a\cdots b}\,\partial_i u^n \nonumber \\
&+& \, T^{m\cdots n}_{i\cdots b}\,\partial_a u^i + \cdots + T^{m\cdots n}_{a\cdots i}\,\partial_b u^i\,.
\end{eqnarray}
A demonstration using the package \code{Ricci} is provided in the
notebook entitled \texttt{Tensor-Lie}, where we provide a demo of the Lie derivative of a fourth rank tensor with two covariant and two contravariant indices in the language of \code{Ricci}. This package appears to be the only Mathematica package to work with free indices and without an explicit metric. The reader may check that the result is consistent.

In the following snippets, some inputs are shown.
\begin{lstlisting}[style=in]
DefineTensor[v, 1, Variance -> Con];
DefineTensor[TT, 4, Variance -> {Con, Con, Co, Co}];
Lie[v, TT];
BasisExpand[%];
TensorSimplify[%];
OutputForm[%]
\end{lstlisting}
The output may be seen in the notebook. 

%% file: DiffGeom/DG_forms.tex
\chapter{Differential forms}\label{forms}
There is a special class of tensor fields, which is so useful as to have a separate treatment. They are called \db{differential $p-$forms} or \db{$p-$forms} for short. 

\md A \db{$p-$form} is a $(0, p)$ tensor which is completely antisymmetric, i.e., given vector fields $v_1\,, \cdots, v_p\,,$
\begin{equation}
\omega\left(v_1\,,\cdots,v_i\,,\cdots,v_j\,,\cdots, v_p\right) 
= - \omega\left(v_1\,,\cdots,v_j\,,\cdots,v_i\,,\cdots, v_p\right)
\end{equation}
for any pair $i, j\,.$ \hfill$\Box$

A 0-form is defined to be a function, i.e. an element of $C^\infty(\cm)\,,$ and a 1-form is as defined earlier. 

The antisymmetry of any $p$-form implies that it will give a non-zero result only when the $p$ vectors are linearly independent. On the other hand, no more than $n$ vectors can be linearly independent in an $n$-dimensional manifold. So $p\leqslant n\,.$

Consider a 2-form $A\,.$ Given any two vector fields $v_1\,,v_2\,,$ we have $A(v_1\,,v_2) = - A(v_2\,, v_1)\,.$ Then the components of $A$ in a chart are
\begin{equation}
A_{ij} = A\left(\partial_i\,, \partial_j\right) = - A_{ji}\,.
\end{equation}
Similarly, for a $p$-form $\omega\,,$ the components are $\omega_{i_1\cdots i_p}\,,$ and components are multiplied by $(-1)$ whenever any two indices are interchanged.

It follows that a $p$-form has $\dbinom{n}{p}$ independent components in $n$-dimensions.

Any 1-form produces a function when acting on a vector field. So given a pair of 1-forms $A, B,$ it is possible to construct a 2-form $\omega$ by defining
\begin{equation}
\omega(u, v) = A(u) B(v) - B(u) A(v), \qquad \hfill\forall u, v\,.
\end{equation}
\md This is usually written as $\omega = A\otimes B - B\otimes A\,,$ where $\otimes$ is called the \db{outer product}. \hfill$\Box$

\md Then the above construction defines a product written as 
\begin{equation}
\omega = A\wedge B = -B\wedge A\,,
\end{equation}
and called the \db{wedge product}\,. Clearly, $\omega$ is a 2-form.\hfill $\Box$

Let us work in a coordinate basis, but the results we find can be generalized to any basis. The coordinate bases for the vector fields, $\left\{\partial_i\right\},$ and 1-forms, $\{dx^i\},$ satisfy $dx^i(\partial_j) = \delta^i_j\,.$ A 1-form $A$ can be written as $A = A_i dx^i,$ and a vector field $v$ can be written as $v = v^i\partial_i\,,$ so that $A(v) = A_i v^i.$ Then for 
$\omega$ defined above, and for any pair of vector fields $u, v,$
\begin{eqnarray}
\omega(u, v) &=& A(u) B(v) - B(u) A(v) \nonumber \\
&=& A_i u^i B_j v^j - B_i u^i A_j v^j \nonumber \\
&=& \left(A_i B_j - B_i A_j\right) u^i v^j\,.
\end{eqnarray}

The components of $\omega$ are $\omega_{ij} = \omega(\partial_i\,, \partial_j)\,,$ so that 
\begin{equation}
\omega(u, v) = \omega(u^i \partial_i\,, v^j \partial_j) = \omega_{ij} u^i v^j\,.
\end{equation}
Then $\omega_{ij} = A_i B_j - B_i A_j\,$ for the 2-form defined above. We can now construct a basis for 2-forms, which we write as $dx^i\wedge dx^j\,,$
\begin{equation}
dx^i \wedge dx^j = dx^i \otimes dx^j - dx^j \otimes dx^i\,.
\end{equation}
Then a 2-form can be expanded in this basis as
\begin{equation}
\omega = \frac{1}{2!} \omega_{ij} dx^i\wedge dx^j\,,
\end{equation}
because then
\begin{eqnarray}
\omega(u, v) &=& \frac{1}{2!} \omega_{ij}\left(dx^i \otimes dx^j - dx^j \otimes dx^i\right)(u, v) \nonumber \\
&=& \frac{1}{2!} \omega_{ij} \left(u^i v^j - u^j v^i\right) = \omega_{ij} u^i v^j\,.
\end{eqnarray}

Similarly, a basis for $p-$forms is 
\begin{equation}
dx^{i_1}\wedge\cdots\wedge dx^{i_p} = dx^{[i_1}\otimes\cdots\otimes dx^{i_p]}\,,
\end{equation}
where the square brackets stand for total antisymmetrization: all even permutations of the indices are added and all the odd permutations are subtracted. (Caution: some books define the `square brackets' as antisymmetrization with a factor $1/p!\,.$) 

For example, for a 3-form, a basis is
\begin{eqnarray}
dx^i \wedge dx^j \wedge dx^k &=& dx^i \otimes dx^j \otimes dx^k - dx^j \otimes dx^i \otimes dx^k \nonumber \\
&+&  dx^j \otimes dx^k \otimes dx^i - dx^k \otimes dx^j \otimes dx^i \nonumber \\
&+&  dx^k \otimes dx^i \otimes dx^j - dx^i \otimes dx^k \otimes dx^j\,.  \qquad\qquad
\end{eqnarray}
Then an arbitrary 3-form $\Omega$ can be written as 
\begin{equation}
\Omega = \frac{1}{3!}\Omega_{ijk} dx^i \wedge dx^j \wedge dx^k\,.
\end{equation}
Note that there is a sum over indices, so that the factorial goes away if we write each basis 3-form up to permutations, i.e. treating different permutations as equivalent. Thus a $p-$form $\alpha$ can be written in terms of its components as
\begin{equation}
\alpha = \frac{1}{p!}\alpha_{i_1 \cdots i_p}\, dx^{i_1}\wedge\cdots\wedge dx^{i_p}\,.
\label{forms.component}
\end{equation}

\textbf{Examples:} A 2-form in two dimensions can be written as
\begin{eqnarray}
\omega &=& \frac{1}{2!}\omega_{ij}\,dx^i\wedge dx^j \nonumber \\
&=& \frac{1}{2!}\left(\omega_{12} dx^1\wedge dx^2 + \omega_{21} dx^2\wedge dx^1\right) \nonumber \\
&=& \frac{1}{2!} \left(\omega_{12} - \omega_{21} \right) dx^1\wedge dx^2 \nonumber \\
&=& \omega_{12}\, dx^1\wedge dx^2\,. 
\end{eqnarray}
\rightline\hfill$\Box$

A 2-form in three dimensions can be written as
\begin{eqnarray}
\omega &=& \frac{1}{2!}\omega_{ij}dx^i\wedge dx^j \nonumber \\
&=& \omega_{12}\,dx^1\wedge dx^2 + \omega_{23}\,dx^2\wedge dx^3 + \omega_{31}\,dx^3\wedge dx^1
\end{eqnarray}
\rightline\hfill$\Box$

In three dimensions, consider two 1-forms $\alpha= \alpha_i dx^i\,, \beta = \beta_i dx^i\,.$ Then 
\begin{eqnarray}
\alpha\wedge\beta &=& \left(\alpha_i \beta_j - \alpha_j \beta_i\right)\,\frac{1}{2!}\,dx^i\wedge dx^j \nonumber \\
&=& \alpha_i \beta_j dx^i\wedge dx^j \nonumber \\
&=& \left(\alpha_1 \beta_2 - \alpha_2 \beta_1\right) dx^1 \wedge dx^2 \nonumber \\
&+& \left(\alpha_2 \beta_3 - \alpha_3 \beta_2\right) dx^2 \wedge dx^3 \nonumber \\
&+& \left(\alpha_3 \beta_1 - \alpha_1 \beta_3\right) dx^3 \wedge dx^1 \,.  
\end{eqnarray}
The components are like the cross product of vectors in three dimensions. So we can think of the wedge product as a generalization of the cross product.

\md We can also define the \db{wedge product} of a $p-$form $\alpha$ and a $q-$form $\beta$ as a $(p+q)-$form satisfying, for any $p+q$ vector fields $v_1, \cdots, v_{p+q}\,,$
\begin{equation}
\alpha\wedge\beta\left(v_1, \cdots, v_{p+q}\right) = \frac{1}{p!q!}\sum_P(-1)^{\deg P} 
\alpha\otimes\beta\left(P\left(v_1, \cdots, v_{p+q}\right)\right)\,.
\end{equation}
Here $P$ stands for a permutation of the vector fields, and $\deg P$ is 0 or 1 for even and odd permutations, respectively. In the outer product on the right hand side, $\alpha$ acts on the first $p$ vector fields in a given permutation $P$, and $\beta$ acts on the remaining $q$ vector fields. \hfill$\Box$

The wedge product above can also be defined in terms of the components of $\alpha$ and $\beta$ in a chart as follows.
\begin{eqnarray}
\alpha &=& \frac{1}{p!}\,\alpha_{i_1\cdots i_p}\,dx^{i_1}\wedge\cdots \wedge dx^{i_p}\nonumber \\
\beta &=& \frac{1}{q!}\,\beta_{j_1\cdots j_q}\,dx^{j_1}\wedge\cdots \wedge dx^{j_q}\nonumber \\
\alpha \wedge \beta &=& \frac{1}{p!q!}\,\alpha_{i_1\cdots i_p}\,\beta_{j_1\cdots j_q}\,\left(dx^{i_1}\wedge\cdots \wedge dx^{i_p}\right)\wedge\left(dx^{j_1}\wedge\cdots \wedge dx^{j_q}\right)\,. \nonumber \\
\end{eqnarray}
Note that $\alpha\wedge\beta=0$ if $p+q>n\,,$ and that a term in which some $i$ is equal to some $j$ must vanish because of the antisymmetry of the wedge product. 

It can be shown by explicit calculation that wedge products are associative,
\begin{equation}
\alpha\wedge\left(\beta\wedge\gamma\right) = \left(\alpha\wedge\beta\right)\wedge\gamma\,.
\end{equation}
Cross-products are not associative, so there is a distinction between cross-products and wedge products. In fact, for 1-forms in three dimensions, the above equation is analogous to the identity for the triple product of vectors,
\begin{equation}
\vec{a}\cdot\left(\vec{b}\times\vec{c}\right) = \left(\vec{a}\times\vec{b}\right)\cdot\vec{c}\,.
\end{equation}

For a $p$-form $\alpha$ and $q$-form $\beta\,,$ we find
\begin{equation}
\alpha\wedge\beta = (-1)^{pq}\beta\wedge\alpha\,.
\end{equation}
\textbf{Proof:} Consider the wedge product written in terms of the components. We can ignore the parentheses separating the basis forms since the wedge product is associative. Then we exchange the basis 1-forms. One exchange gives a factor of $-1\,,$
\begin{equation}
dx^{i_p}\wedge dx^{j_1} = - dx^{j_1}\wedge dx^{i_p}\,.
\end{equation}
Repeating this process, we get
\begin{eqnarray}
&&dx^{i_1}\wedge\cdots \wedge dx^{i_p}\wedge dx^{j_1}\wedge\cdots \wedge dx^{j_q} \nonumber \\
&& \qquad\qquad = (-1)^p  dx^{j_1}\wedge dx^{i_1}\wedge\cdots \wedge dx^{i_p}\wedge dx^{j_2}\wedge\cdots \wedge dx^{j_q} \nonumber \\
&& \qquad\qquad = \cdots\nonumber \\
&& \qquad\qquad = (-1)^{pq} dx^{j_1}\wedge\cdots \wedge dx^{j_q} \wedge dx^{i_1}\wedge\cdots \wedge dx^{i_p}\,.
\end{eqnarray}
Putting back the components, we find 
\begin{equation}
\alpha\wedge\beta = (-1)^{pq}\beta\wedge\alpha\,
\end{equation}
as desired. \hfill $\Box$

\md The wedge product defines an algebra on the space of differential forms. It is called a \db{graded commutative algebra}\,. \hfill $\Box$

\md Given a vector field $v,$ we can define its \db{contraction} with a $p$-form by
\begin{equation}
\iota_v\omega = \omega(v, \cdots)
\end{equation}
with $p-1$ empty slots. This is a $(p-1)$-form. Note that the position of $v$ only affects the sign of the contracted form. \hfill $\Box$

\textbf{Example:} Consider a 2-form made of the wedge product of two 1-forms, $\omega = \lambda\wedge\mu\, = \lambda\otimes\mu - \mu\otimes\lambda\,.$ Then contraction by $v$ gives 
\begin{equation}
\iota_v\omega = \omega(v, \bull) = \lambda(v)\mu - \mu(v)\lambda = -\omega(\bull, v)\,.
\end{equation}

If we have a $p$-form $\omega = \frac{1}{p!}\omega_{i_1\cdots i_p}\,dx^{i_1}\wedge\cdots \wedge dx^{i_p}\,,$ its contraction  with a vector field $v = v^i \partial_i$ is 
\begin{equation}
\iota_v\omega = \frac{1}{(p-1)!}\, \omega_{ii_2\cdots i_p} v^i dx^{i_2}\wedge\cdots\wedge dx^{i_p}\,.
\end{equation}
Note the sum over indices. To see how the factor becomes $\frac{1}{(p-1)!}\,,$ we write the contraction as
\begin{equation}
\iota_v\omega = \frac{1}{p!}\,\omega_{i_1\cdots i_p} dx^{i_1}\wedge\cdots\wedge dx^{i_p}\left(v^i\partial_i\right)\,.
\end{equation}
Since the contraction is done in the first slot, we consider the action of each basis 1-form $dx^{i_k}$ on $\partial_i$ by carrying $dx^{i_k}$ to the first position and then writing a $\displaystyle{\delta^{i_k}_i}\,.$ This gives a factor of $(-1)$ for each exchange, but we get the same factor by rearranging the indices of $\omega\,,$ thus getting a $+1$ for each index. This leads to an overall factor of $p\,.$

\md Given a diffeomorphism $\varphi:\cm_1 \to \cm_2\,,$ the \db{pullback} of a 1-form $\lambda$ (on $\cm_2$) is $\varphi^*\lambda\,,$ defined by 
\begin{equation}
\varphi^*\lambda(v) = \lambda(\varphi_*v)
\end{equation}
for any vector field $v$ on $\cm_1\,.$ \hfill$\Box$

Then we can consider the pullback $\varphi^*dx^i$ of a basis 1-form $dx^i\,.$ For a general 1-form $\lambda=\lambda_i dx^i\,,$ we have $\varphi^*\lambda = \varphi^*(\lambda_i dx^i)\,.$ But 
\begin{equation}
\varphi^*\lambda(v) = \lambda(\varphi_* v) = \lambda_i\, dx^i(\varphi_* v) \,.
\end{equation} 
Now, $dx^i(\varphi_* v)= \varphi^*dx^i(v)$ and the object on the right hand side is a function on $\cm_1\,,$ so we can write this as 
\begin{equation}
\varphi^*\lambda(v) = (\varphi^*\lambda_i)\varphi^*dx^i(v)\,,
\end{equation}
where $\varphi^*\lambda_i$ are now functions on $\cm_1\,,$ i.e.
\begin{equation}
\left.\left(\varphi^*\lambda_i\right)\right|_{_P} = \left.\lambda_i\right|_{_{\varphi(P)}} \,.
\end{equation}
Thus we can write $\varphi^*\lambda = \left(\varphi^*\lambda_i\right)\varphi^*dx^i\,.$ For the wedge product of two 1-forms, 
\begin{eqnarray}
\varphi^*(\lambda\wedge\mu) (u, v) &=& (\lambda\wedge\mu) (\varphi_* u\,,\varphi_* v) \\
&=& \lambda\otimes\mu(\varphi_* u\,,\varphi_* v)  - \mu\otimes\lambda(\varphi_* u\,,\varphi_* v) \\
&=& \lambda(\varphi_* u)\mu(\varphi_* v) - \mu(\varphi_* u)\lambda(\varphi_* v) \\
&=& \varphi^*\lambda(u)\varphi^*\mu(v) - \varphi^*\mu(u)\varphi^*\lambda(v) \\
&=& (\varphi^*\lambda\wedge\varphi^*\mu)(u\,, v)\,.
\end{eqnarray}
Since $u, v$ are arbitrary vector fields it follows that 
\begin{eqnarray}
\varphi^*(\lambda\wedge\mu) &=& \varphi^*\lambda\wedge\varphi^*\mu \\
\varphi^*(dx^i\wedge dx^j) &=& \varphi^*dx^i \wedge \varphi dx^j\,.
\end{eqnarray}

Since the wedge product is associative, we can write (by assuming an obvious generalization of the above formula)
\begin{eqnarray}
\varphi^*\left(dx^i\wedge dx^j\wedge dx^k\right) &=&  \varphi^*\left(\left(dx^i\wedge dx^j\right)\wedge dx^k\right) \\ &=& \varphi^*\left(dx^i\wedge dx^j\right)\wedge \varphi^* dx^k \\
&=& \varphi^* dx^i \wedge \varphi^* dx^j \wedge \varphi^* dx^k\,,
\end{eqnarray}
and we can continue this for any number of basis 1-forms. So for any $p$-form $\omega\,,$ let us define the pullback $\varphi^*\omega$ by 
\begin{equation}
\varphi^*\omega(v_1\,, \cdots, v_p) = \omega\left(\varphi_*v_1\,,\cdots, \varphi_*v_p\right)\,,
\end{equation}
and in terms of components, by
\begin{equation}
\varphi^*\omega = \frac{1}{p!}\left(\varphi^*\omega_{i_1\cdots i_p}\right)\varphi^* dx^{i_1}\wedge\cdots\wedge \varphi^*dx^{i_p}\,.
\end{equation}

We assumed above that the pullback of the wedge product of a 2-form and a 1-form is the wedge product of the pullbacks of the respective forms, but it is not necessary to make that assumption -- it can be shown explicitly by taking three vector fields and following the arguments used earlier for the wedge product of two 1-forms.

Then for any $p$-form $\alpha$ and $q$-form $\beta$ we can calculate from this that
\begin{equation}
\varphi^*(\alpha\wedge\beta) = \varphi^*\alpha\wedge\varphi^*\beta\,.
\end{equation}
Thus pullbacks commute with (are distributive over) wedge products.

Demonstrations in Mathematica:

All of the following use the \code{DifferentialForms} package.

A 1-form may be defined as follows.
\begin{lstlisting}[style=in]
$\omega$1=x*d[x]+y^2*d[y]+z*d[z];
Print["$\omega_1$ = ",$\omega$1] 
\end{lstlisting}
\begin{lstlisting}[style=out]
$\omega_1$ = (x) dx+(y^2) dy+(z) dz
\end{lstlisting}

A 2-form from the wedge product of two 1-forms can be seen as
\begin{lstlisting}[style=in]
$\alpha$1 = x*d[x];
$\alpha$2 = y*d[y];
$\omega$2 = $\alpha$1 $\wedge$ $\alpha$2;
Print["$\omega_2$ = $\alpha_1 \wedge \alpha_2$ = ", $\omega$2] 
\end{lstlisting}
\begin{lstlisting}[style=out]
$\omega_2$ = $\alpha_1 \wedge \alpha_2$ = (x y) dx ^ dy
\end{lstlisting}

Pullback of a 2-form under a map:
\begin{lstlisting}[style=in]
$\varphi$ = {x -> u, y -> v, z -> u*v};
pb$\omega$2 = Pullback[$\omega$2, $\varphi$];
Print["Pullback($\omega_2$) = ", pb$\omega$2]
\end{lstlisting}
\begin{lstlisting}[style=out]
Pullback($\omega_2$) = (u v) du ^ dv
\end{lstlisting}

%% file: DiffGeom/DG_extder.tex
\chapter{Exterior derivative}\label{extder}
The \db{exterior derivative} is a generalization of the gradient of a function. It is a map from $p$-forms to $(p+1)$-forms. This should be a derivation, so it should be linear,
\begin{equation}
d(\alpha + \omega) = d\alpha + d\omega \qquad \qquad \forall p\text{-forms\,} \alpha\,,\omega\,.
\end{equation}
This should also satisfy the Leibniz rule, but the algebra of $p$-forms is not a commutative algebra but a \db{graded commutator} algebra, i.e., involves a factor of $(-1)^{pq}$ for exchanges. So we need
\begin{equation}
d(\alpha\wedge\beta) = d\alpha\wedge\beta + (-1)^{pq} d\beta\wedge \alpha\,,
\end{equation}
or alternatively,
\begin{equation}
d(\alpha\wedge\beta) = d\alpha\wedge\beta + (-1)^p \alpha\wedge d\beta\,.
\end{equation}
This will be the Leibniz rule for wedge products. Note that it gives the correct result when one or both of $\alpha, \beta$ are 0-forms, i.e., functions. The two formulas are identical by virtue of the fact that $d\beta$ is a $(q+1)$-form, so that 
\begin{equation}
\alpha\wedge d\beta = (-1)^{p(q+1)} d\beta\wedge \alpha\,.
\end{equation}
We will try to define the exterior derivative in a way such that it has these properties.

Let us define the exterior derivative of a $p$-form $\omega$ in a chart as
\begin{equation}
d\omega = \frac{1}{p!}\,\partial_i\omega_{i_1\cdots i_p} dx^i \wedge dx^{i_1}\wedge \cdots \wedge dx^{i_p}
\end{equation}
This clearly has the first property of linearity. To check the (graded) Leibniz rule, let us write $\alpha\wedge\beta$ in components. Then 
\begin{eqnarray}
d(\alpha\wedge\beta) &=& \frac{1}{p!q!} \partial_i\left(\alpha_{i_1\cdots i_p}\beta_{j_1\cdots j_q}\right)\, dx^i\wedge dx^{i_1}\wedge\cdots \wedge dx^{j_q} \next
&=& \frac{1}{p!q!}\, \left[\left(\partial_i \alpha_{i_1\cdots i_p}\right) \beta_{j_1\cdots j_q} + \alpha_{i_1\cdots i_p} \left(\partial_i \beta_{j_1\cdots j_q}\right)\right]\,dx^i\wedge dx^{i_1}\wedge\cdots \wedge dx^{j_q} \next
&=& \frac{1}{p!q!}\, \left(\partial_i \alpha_{i_1\cdots i_p}\right) \beta_{j_1\cdots j_q} \,dx^i\wedge dx^{i_1}\wedge\cdots \wedge dx^{i_p}\wedge dx^{j_1}\wedge \cdots dx^{j_q} \nonumber \next
&& \quad + \frac{1}{p!q!}\,(-1)^p\,\alpha_{i_1\cdots i_p} \left(\partial_i \beta_{j_1\cdots j_q}\right)\,dx^{i_1}\wedge\cdots \wedge dx^{i_p}\wedge dx^i\wedge dx^{j_1}\wedge \cdots dx^{j_q} \next
&=& d\alpha\wedge\beta + (-1)^p \alpha\wedge d\beta\,.
\end{eqnarray}

A third property of the exterior derivative immediately follows from here,
\begin{equation}
d^2 = 0\,.
\end{equation}
To see this, we write
\begin{eqnarray}
d(d\omega) &=& \frac{1}{p!}\,d\left(\partial_i\omega_{i_1\cdots i_p}dx^i\wedge dx^{i_1}\wedge\cdots dx^{i_p}\right) \next
&=& \frac{1}{p!}\,\partial_j\partial_i\omega_{i_1\cdots i_p}dx^j\wedge dx^i\wedge dx^{i_1}\wedge\cdots dx^{i_p}\,.
\end{eqnarray}
But the wedge product is antisymmetric, $dx^j\wedge dx^i = -dx^i\wedge dx^j\,,$ and the indices are summed over, so the above object must be antisymmetric in $\partial_j\,,\partial_i\,.$ But that vanishes. So $d^2 = 0$ on all forms. 

Note that we can also write 
\begin{equation}
d\omega = \frac{1}{p!}\,\left(d\omega_{i_1\cdots i_p}\right)\wedge dx^{i_1}\wedge\cdots dx^{i_p}\,,
\end{equation}
where the object in parentheses is a gradient 1-form corresponding to the gradient of the component. 

Consider a 1-form $A = A_\mu dx^\mu$ where $A_\mu$ are smooth functions on $\cm\,.$ Then using this definition we can write 
\begin{eqnarray}
dA &=& (dA_\nu) \wedge dx^\nu \next
&=& \partial_\mu A_\nu dx^\mu \wedge dx^\nu \next
&=& \half \left(\partial_\mu A_\nu - \partial_\nu A_\mu\right) dx^\mu \wedge dx^\nu \next
\Rightarrow \qquad (dA)_{\mu\nu} &=& \partial_\mu A_\nu - \partial_\nu A_\mu \,.			
\end{eqnarray}
We can generalize this result to write for a $p$-form,
\begin{eqnarray}
\alpha &=& \frac{1}{p!} \alpha_{\mu_1\cdots\mu_p} dx^{\mu_1}\wedge\cdots\wedge dx^{\mu_p} \\
d\alpha &=& \frac{1}{p!} \left(\partial_\mu\alpha_{\mu_1\cdots\mu_p}\right) dx^{\mu_1}\wedge\cdots\wedge dx^{\mu_p} \next
&=& \frac{1}{(p+1)!} \partial_{[\mu}\alpha_{\mu_1\cdots\mu_p]} dx^\mu\wedge dx^{\mu_1}\wedge\cdots\wedge dx^{\mu_p} \next
\Rightarrow \qquad (d\alpha)_{\mu\mu_1\cdots\mu_p} &=& \partial_{[\mu}\alpha_{\mu_1\cdots\mu_p]}
\end{eqnarray}

\textbf{Example:} For $p=1$ i.e. for a 1-form $A\,$ we get from this formula $\left(dA\right)_{\mu\nu} = \partial_\mu A_\nu - \partial_\nu A_\mu\,,$ in agreement with our previous calculation.

For $p=2$ we have a 2-form, call it $\alpha$. Then using this formula we get
\begin{eqnarray}
\left(d\alpha\right)_{\mu\nu\lambda} &=& \partial_{[\mu}\alpha_{\nu\lambda]} \next
&=& \partial_\mu\alpha_{\nu\lambda} - \partial_\nu\alpha_{\mu\lambda} + \partial_\nu\alpha_{\lambda\mu} -\partial_\lambda\alpha_{\nu\mu} + \partial_\lambda\alpha_{\mu\nu} -\partial_\mu\alpha_{\lambda\nu}\,.
\end{eqnarray}
Note that $d$ is not defined on  arbitrary tensors, but only on forms. \hfill$\Box$

By definition, $d^2 = 0$ on any $p$-form. So if $\alpha = d\beta\,,$ it follows that $d\alpha = 0\,.$ But given a $p$-form $\alpha$ for which $d\alpha = 0\,,$ can we say that there must be some $(p-1)$-form $\beta$ such that $\alpha = d\beta$\,?

\md This is a good place to introduce some terminology. Any form $\omega$ such that $d\omega = 0$ is called \db{closed}, whereas any form $\alpha$ such that $\alpha = d\beta$ is called \db{exact}.\hfill$\Box$

Thus every exact form is closed. Is every closed form exact? The answer is yes, in a sufficiently small neighbourhood. We say that every closed form is {\bf locally exact}.  Note that if a $p$-form $\alpha = d\beta\,,$ we cannot uniquely specify the $(p-1)$-form $\beta$ since for any $(p-2)$-form $\gamma\,,$ we can always write $\alpha = d\beta'\,,$  where $\beta' = \beta + d\gamma\,.$

Thus a more precise statement is that given any $p$-form $\alpha$ such that $d\alpha = 0$ in a neighbourhood of some point $P$, there is some neighbourhood of this point and some $(p-1)$-form $\beta$ such that $\alpha = d\beta$ in that neighbourhood. But this may not be true globally. This statement is known as the \db{Poincar\'e lemma}. 
\hfill$\Box$

\textbf{Example:} In $\mathbb{R}^2$ remove the origin. Consider the 1-form 
\begin{equation}
\alpha = \frac{xdy - ydx}{x^2 + y^2}\,.
\end{equation}
Then
\begin{eqnarray}
d\alpha &=& \left(\frac{1}{x^2 + y^2} - \frac{2x^2}{\left(x^2 + y^2\right)^2}\right) dx\wedge dy - \left(\frac{1}{x^2 + y^2} - \frac{2y^2}{\left(x^2 + y^2\right)^2}\right) dy\wedge dx \next
&=& \frac{2}{x^2 + y^2}\, dx\wedge dy - 2 \frac{x^2 + y^2}{\left(x^2 + y^2\right)^2}\, dx\wedge dy = 0\,.
\end{eqnarray}
(It is not difficult to demonstrate the above on Mathematica and
our implementation is discussed a little later in this chapter.)

Introduce polar coordinates $r, \theta\,$ with $x = r\cos\theta\,, y = r\sin\theta\,.$ Then
\begin{eqnarray}
dx &=& dr\cos\theta - r\sin\theta d\theta\,, \qquad dy = dr\sin\theta + r\cos\theta d\theta \next \next
\alpha &=& \frac{r\cos\theta\left(\sin\theta dr + r\cos\theta d\theta\right)}{r^2} - \frac{r\sin\theta\left(\cos\theta dr - r\sin\theta d\theta\right)}{r^2} \next
&=& \frac{r^2\left(\cos^2\theta + \sin^2\theta\right)d\theta}{r^2} = d\theta\,.
\end{eqnarray}

Thus $\alpha$ is exact, but $\theta$ is multivalued so there is no function $f$ such that $\alpha = df$ everywhere. In other words, $\alpha = d\theta$ is exact only in a neighbourhood small enough that $\theta$ remains single-valued.

Examples in Mathematica (the following examples use the package \code{DifferentialForms}):
\begin{lstlisting}[style=in]
f = x^2*y*z;
ExteriorDerivative[f]   
\end{lstlisting}
\begin{lstlisting}[style=out]
(x$^2$ y) dz+(x$^2$ z) dy+(2 x y z) dx
\end{lstlisting}
\begin{lstlisting}[style=in]
$\omega$1 = x*d[x] + y^2*d[y] + z*d[z];
\end{lstlisting}
\begin{lstlisting}[style=out]
0
\end{lstlisting}
\begin{lstlisting}[style=in]
$\alpha$1=z*d[x];
$\alpha$2=d[y];
$\omega$2=$\alpha$1 $\wedge$ $\alpha$2;
ExteriorDerivative[$\omega$2] 
\end{lstlisting}
\begin{lstlisting}[style=out]
(1) dx ^ dy ^ dz
\end{lstlisting}

Given a closed form, we are trying to find the form whose exterior derivative is the required closed form at hand. The method of finding the form is trial and error~[\cite{nash2011topology}].
\begin{lstlisting}[style=in]
omega=-y/(x^2+y^2) d[x] + x/(y^2+x^2) d[y]
\end{lstlisting}
\begin{lstlisting}[style=out]
$\frac{\text{x}}{\text{x}^2+\text{y}^2} \text{dy}+(-\frac{\text{y}}{\text{x}^2+\text{y}^2}) \text{dx}$
\end{lstlisting}
\begin{lstlisting}[style=in]
d[omega] // Simplify 
\end{lstlisting}
\begin{lstlisting}[style=out]
0
\end{lstlisting}
\begin{lstlisting}[style=in]
eta=ArcTan[y/x]; 
\end{lstlisting}

\begin{lstlisting}[style=in]
D[eta,x] // Simplify
\end{lstlisting}
\begin{lstlisting}[style=out]
$-\frac{\text{y}}{\text{x}^2+\text{y}^2}$
\end{lstlisting}
\begin{lstlisting}[style=in]
D[eta,y] // Simplify
\end{lstlisting}
\begin{lstlisting}[style=out]
$\frac{\text{x}}{\text{x}^2+\text{y}^2}$
\end{lstlisting}
\begin{lstlisting}[style=in]
d[eta] // Simplify
\end{lstlisting}
\begin{lstlisting}[style=out]
$\frac{(\text{x})\text{dy} - (\text{y})\text{dx}}{\text{x}^2+\text{y}^2}$ 
\end{lstlisting}
We have shown a closed one form and the corresponding zero form here. A non-trivial example is the 3-D case.
It corresponds to the magnetic monopole one-form singular
along the negative $z$-axis (Dirac string). 
Replacing $r+z$ by $r-z$ has the effect of having the Dirac
string along the positive $z$-axis.
The solutions are not unique and are related by cyclic permutations of $x,~y,~z$, with the string now along the appropriate axis.  In the following snippets part of the implementation is
displayed.

\begin{lstlisting}[style=in]
eta2 = (-x d[y] + y d[x])/(r (r + z))
\end{lstlisting}
\begin{lstlisting}[style=out]
((-x) dy+(y) dx)/(Sqrt[x^2+y^2+z^2] (z+Sqrt[x^2+y^2+z^2]))
\end{lstlisting}
\begin{lstlisting}[style=in]
d[eta2] // Simplify
\end{lstlisting}
\begin{lstlisting}[style=out]
(-(x/(x^2+y^2+z^2)^(3/2))) dy ^ dz+(y/(x^2+y^2+z^2)^(3/2)) dx ^ dz+(-(z/(x^2+y^2+z^2)^(3/2))) dx ^ dy
\end{lstlisting}
\begin{lstlisting}[style=in]
d[%] // Simplify
\end{lstlisting}
\begin{lstlisting}[style=out]
0
\end{lstlisting}
It can be seen in the corresponding notebook \texttt{exactform.nb}.

%% file: DiffGeom/DG_vol.tex
\chapter{Volume form}\label{vol}
The space of $p$-forms in $n$ dimensions is $\dbinom{n}{p}$ dimensional. Hence the space of $n$-forms in $n$ dimensions is 1-dimensional, i.e., there is only one independent component, and all $n$-forms are scalar multiples of one another. 

Choose an $n$-form field. Call it $\omega$. Suppose $\omega \neq 0$ at some point $P$. Then given any basis $\left\{e_\mu\right\rbrace$ of $T_P\cm\,,$ we have $\omega(e_1, \cdots e_n) \neq 0$ since $\omega \neq 0\,.$ Thus all vector bases at $P$ fall into two classes, one for which $\omega(e_1, \cdots e_n) > 0$ and the other for which it is $< 0\,.$ 

Once we have identified these two classes, they are independent of $\omega\,.$ That is, if $\omega'$ is another $n$-form which is non-zero at $P\,,$ there must be some function $f\neq 0$ such that $\omega' = f\omega\,.$ Two bases which gave positive numbers under $\omega$ will give the same sign --- both positive or both negative --- under $\omega'$ and therefore will be in the same class. 

\md As a result of the above, every basis (set of $n$ linearly independent vectors) is a member of one of the two classes. These are called \db{right-handed} and \db{left-handed}. \hfill$\Box$

\md A manifold is called \db{orientable} if it is possible to define a continuous $n$-form field $\omega$ which is non-zero everywhere on the manifold. Then it is possible to choose a basis with the same handedness everywhere on the manifold continuously. \hfill$\Box$

It may be noted that whereas Euclidean space is orientable, the M\"obius band is not.

\md An orientable manifold is called \db{oriented} once an \db{orientation} has been chosen, i.e., once we have decided to choose basis vectors with the same handedness everywhere on the manifold. \hfill$\Box$

\md It is necessary to choose an oriented manifold when we discuss the integration of forms. On an $n$-dimensional manifold, a set of $n$ linearly independent vectors defines an $n$-dimensional parallelepiped. If we define an $n$-form $\omega \neq 0$  we can think of the value of these vectors as the volume of this parallelepiped. This $\omega$ is called a \db{volume form}. \hfill $\Box$

Once a volume form has been chosen, any set of $n$ linearly independent vectors will define a positive or negative volume. 

The integral of a function $f$ on $\rn$ is the sum of the values of $f\,,$ multiplied by infinitesimal volumes of coordinate elements. Similarly, we define the integral of a function $f$ on an oriented manifold as the sum of the values of $f\,,$ multiplied by infinitesimal volumes. The way to do that is the following:

Given a function $f,$ define an $n$-form in a chart by $\omega = f dx^1\wedge\cdots\wedge dx^n.$ To integrate over an open set $U\,,$ divide it up into infinitesimal `cells', spanned by vectors  
\[\displaystyle\left\{\Delta x^1\frac{\partial}{\partial x^1}, \Delta x^2 \frac{\partial}{\partial x^2}, \cdots, \Delta x^n \frac{\partial}{\partial x^n}  \right\}\,, \]
where the $\Delta x^i$ are small numbers. 

Then the integral of $f$ over one such cell is approximately 
\begin{eqnarray}
f \Delta x^1 \Delta x^2 \cdots \Delta x^n &=& f dx^1 \wedge \cdots \wedge dx^n \left(\Delta x^1\partial_1, \cdots, \Delta x^n \partial_n\right) \next
&=& \omega(cell)\,.
\end{eqnarray}
Adding up the contributions from all cells and taking the limit of cell size going to zero, we find 
\begin{equation}
\int\limits_U \omega = \int\limits_{\varphi(U)} f d^nx\,.
\end{equation}
The right hand side is the usual integration in calculus of $n$ variables, and the left hand side is our newly defined object, thereby establishing a notation. 

The right hand side can be seen to be independent of the choice of coordinate system. If we choose a different coordinate system, we get a Jacobian, but also a redefinition of the region $\varphi(U)\,.$ Let us check that the left hand side is also invariant under the choice of the coordinates. We will do this in two dimensions with $\omega = f dx^1\wedge dx^2\,.$ In another coordinate system $(y^1, y^2)\,$ corresponding to $\varphi'(U)$
\begin{eqnarray}
dx^1 &=& \frac{\partial x^1}{\partial y^1}\, dy^1 + \frac{\partial x^1}{\partial y^2}\, dy^2 \next
dx^2 &=& \frac{\partial x^2}{\partial y^1}\, dy^1 + \frac{\partial x^2}{\partial y^2}\, dy^2 \next
\Rightarrow \qquad dx^1 \wedge dx^2 &=& \left(\frac{\partial x^1}{\partial y^1}\frac{\partial x^2}{\partial y^1} - \frac{\partial x^1}{\partial y^2} \frac{\partial x^2}{\partial y^2} \right) dy^1 \wedge dy^2\next
&=& J dy^1 \wedge dy^2 \,,
\end{eqnarray}
and $J$ is the Jacobian. 

We now have
\begin{eqnarray}
\int\limits_U \omega &=& \int\limits_U f(x^1, x^2) dx^1 \wedge dx^2 \next
&=& \int\limits_U f(y^1, y^2) J dy^1 \wedge dy^2 \next
&=& \int\limits_{\varphi'(U)} f(y^1, y^2) J d^2y\,,
\end{eqnarray}
so we get the same result either way. 

Given the same $f,$ if we choose a basis with the opposite orientation, the integral of $\omega$ will have the opposite sign. This is why the choice of orientation has to be made {\bf before} integration. 

Manifolds become even more interesting if we define a metric, as we will see.

%% file: DiffGeom/DG_metric.tex
\chapter{Metric tensor}\label{metric}
\md A \db{metric} on a vector space $V$ is a function $g: V\times V \to \mathbb{R}$ which is 

\begin{list}{$\roman{enumii})$} {\usecounter{enumii}\topsep=0pt}
\item bilinear: 
\begin{eqnarray}
g(av_1 + v_2, w) &=& ag(v_1, w) + g(v_2, w)\next
g(v, w_1 + aw_2) &=& g(v, w_1) + a g(v, w_2)\,,
\label{metric.bilinear}
\end{eqnarray}
i.e., $g$ is a (0,2) tensor;
\item symmetric:
\begin{equation}
g(v, w) = g(w, v);
\label{metric.symmetric}
\end{equation}
\item non-degenerate: 
\begin{equation}
g(v, w) = 0 \qquad \forall w \qquad \Rightarrow v = 0\,.
\label{metric.nondegenerate}
\end{equation}
\end{list}
\hfill$\Box$

\md If for some $v, w \neq 0\,,$ we find that $g(v, w) = 0\,,$ we say that $v, w$ are  \db{orthogonal}.\hfill$\Box$

\md Given a metric $g$ on $V\,,$ we can always find an \db{orthonormal basis} $\{e_\mu\}$ such that $g(e_\mu, e_\nu) = 0$ if $\mu \neq \nu$ and $\pm 1$ if $\mu = \nu\,.$ \hfill$\Box$

\md If the number of $(+1)$'s is $p$ and the number of $(-1)$'s is $q\,,$ we say that the metric has \db{signature} $(p, q)\,.$ 

We have defined a metric for a vector space. We can generalize this definition to a manifold $\cm$ by the following.

\md A \db{metric} $g$ on a manifold $\cm$ is a (0, 2) tensor field such that if $(v, w)$ are smooth vector fields, $g(v, w)$ is a smooth function on $\cm\,,$ and has the properties Eq.~(\ref{metric.bilinear}), Eq.~(\ref{metric.symmetric}) and Eq.~(\ref{metric.nondegenerate}) mentioned earlier. \hfill$\Box$

It is possible to show that smoothness implies that the signature is constant on any connected component of $\cm\,$; we will assume that it is constant on all of $\cm\,.$

A vector space becomes related to its dual space by the metric. Given a vector space $V$ with metric $g\,,$ a vector $v$ defines a linear map $g(v, \cdot): V \to \mathbb{R}\,, w\mapsto g(v, w) \in \mathbb{R}\,.$ Thus $g(v,\cdot) \in V^*$ where $V^*$ is the dual space of $V\,.$ But $g(v,\cdot)$ is itself linear in $v\,,$ so the map $V\to V^*$ defined by $g(v,\cdot)$ is linear. Since $g$ is non-degenerate, this map is an isomorphism. It then follows that on a manifold we can use the metric to define a linear isomorphism between vectors and 1-forms.

In a basis, the components of the metric are $g_{\mu\nu} = g(e_\mu\,, e_\nu)\,.$ This is an $n\times n$ matrix in an $n$-dimensional manifold. We can thus write $g(v, w) = g_{\mu\nu} v^\mu w^\nu$ in terms of the components. Non-degeneracy implies that this matrix is invertible. Let $g^{\mu\nu}$ denote the inverse matrix. Then, by definition of an inverse matrix, we have
\begin{equation}
g_{\mu\nu} g^{\nu\lambda} = \delta^\lambda_\mu = g^{\lambda\nu} g_{\mu\nu}\,.
\end{equation}
Then the linear isomorphism takes the following form.

\begin{list}{$\roman{enumii})$} {\usecounter{enumii}\topsep=0pt}
\item If $v = v^\mu e_\mu$ is a vector field in a chart, and $\{\lambda^\mu\}$ is the dual basis to $\{e_\mu\}\,,$
\begin{equation}
g(v, \cdot) = v_\mu \lambda^\mu\,,
\end{equation}
where $v_\mu = g_{\mu\nu} v^\nu\,.$
\item If $A = A_\mu\lambda^\mu$ is a 1-form written in a basis $\{\lambda^\mu\}\,,$ the corresponding vector field is $A^\mu e_\mu\,,$ where $A^\mu = g^{\mu\nu}A_\nu\,.$
\end{list}

This is the isomorphism between vector fields and 1-forms. (We could of course define a similar isomorphism between vectors and covectors without referring to a manifold.) A similar isomorphism holds for tensors, e.g. in terms of components,
\begin{eqnarray}
&& T^{\mu\nu} \longleftrightarrow T^\mu_{\phantom{\mu}\nu} \longleftrightarrow T_\mu^{\phantom{\mu}\nu} \longleftrightarrow T_{\mu\nu} \\
&& T^{\mu\nu\rho\,\cdots} \longleftrightarrow T^{\mu\nu}{}_{\rho}^{\phantom{\rho}\,\cdots} \longleftrightarrow 
T^{\mu\nu}{}_{\rho\,\cdots} \longleftrightarrow T_{\mu\nu\rho}{}^{\,\cdots} 
\longleftrightarrow \cdots 
\end{eqnarray}
These correspondences are not equalities --- the components are not equal. What it means is that, if we know one set of components, say $T^{\mu\nu\rho\,\cdots}\,,$ and the metric, we also know every other set of components.

\md Using the fact that a non-degenerate metric defines a 1-1 linear map between vectors and 1-forms, we can define an \db{inner product of 1$-$forms}, by
\begin{equation}
\inprod{A}{B} = g^{\mu\nu} A_\mu B_\nu
\end{equation}
for 1-forms $A, B\,.$ This result is independent of the choice of basis, i.e. independent of the coordinate system, just like the \db{inner product of vector fields},
\begin{equation}
\inprod{v}{w} = g(v, w) = g_{\mu\nu} v^\mu w^\nu\,.
\end{equation}
\hfill$\Box$

Given a manifold with metric, there is a canonical volume form $dV\,$ (sometimes written as $vol$)\,, which in a coordinate chart reads
\begin{eqnarray}
dV = \sqrt{|\det g_{\mu\nu}|} dx^1\wedge\cdots\wedge dx^n\,.
\end{eqnarray}
Note that despite the notation, this is not a 1-form, nor the gradient of some function $V\,.$ This is clearly a volume form because it is an $n$-form which is non-zero everywhere, as $g_{\mu\nu}$ is non-degenerate.

We need to show that this definition is independent of the chart. Take an overlapping chart. Then in the new chart, the corresponding volume form is
\begin{equation}
dV' = \sqrt{|\det g'_{\mu\nu}|} dx'^1\wedge\cdots\wedge dx'^n\,.
\end{equation}
We wish to show that $dV' = dV\,.$ In the overlap,
\begin{equation}
dx'^\mu = \frac{\partial x'^\mu}{\partial x^\nu} dx^\nu = A^\mu_\nu dx^\nu\, \mathrm{(say)}
\end{equation}
Then $dx'^1\wedge\cdots\wedge dx'^n = (\det A) dx^1\wedge\cdots\wedge dx^n\,.$

On the other hand, if we look at the components of the metric tensor in the new chart,
\begin{eqnarray}
g'_{\mu\nu} &=& g(\partial'_\mu, \partial'_\nu) \next
&=& \left(\frac{\partial x^\alpha}{\partial x'^\mu}\,\partial_\alpha\,, \frac{\partial x^\beta}{\partial x'^\nu}\,\partial_\beta\right) \next
&=& g\left(\left(A\inv\right)^\alpha_\mu \partial_\alpha\,, \left(A\inv\right)^\beta_\nu \partial_\beta\right) \next
&=& \left(A\inv\right)^\alpha_\mu \left(A\inv\right)^\beta_\nu  g_{\alpha\beta}\,.
\end{eqnarray}
Taking determinants, we find
\begin{equation}
\det g'_{\mu\nu} = \left(\det A\right)^{-2} \left(\det g_{\mu\nu}\right)\,.
\end{equation}
Thus 
\begin{equation}
\sqrt{|\det g'_{\mu\nu}|} = \left|\det A\right|\inv\sqrt{|\det g_{\mu\nu}|}\,,
\end{equation}
and so $dV' = dV\,.$ 

\md This is called the \db{metric volume form} and written as 
\begin{equation}
dV = \sqrt{|g|} dx^1\wedge\cdots\wedge dx^n
\end{equation}
in a chart. \hfill$\Box$

When we write $dV\,,$ sometimes we mean the $n$-form as defined above, and sometimes we mean $\sqrt{|g|} d^nx\,,$ the measure for the usual integral. Another way of writing the volume form in a chart is in terms of its components, 
\begin{equation}
dV = \frac{\sqrt{|g|}}{n!}\, \epsilon_{\mu_1\cdots\mu_n} dx^{\mu_1}\wedge\cdots\wedge dx^{\mu_n}
\end{equation}
where $\epsilon$ is the totally antisymmetric Levi-Civita symbol, with $\epsilon_{12\cdots n} = +1\,.$ Thus $\sqrt{|g|}\, \epsilon_{\mu_1\cdots\mu_n}$ are the components of the volume form.

\section{Demonstration in Mathematica}
The following demonstrations use the Mathematica package \code{DifferentialForms} by Frank Zizza.

To define the Cartesian metric in $3d$ Euclidean space,
\begin{lstlisting}[style=in]
CartMetric = t[x, x] + t[y, y] + t[z, z]
\end{lstlisting}

\begin{lstlisting}[style=out]
(1) d[x] o d[x]+(1) d[y] o d[y]+(1) d[z] o d[z]
\end{lstlisting}

Spherical-Polar metric:
\begin{lstlisting}[style=in]
$\text{PolMetric}=t(r,r) + \theta + r^2 ~t(\theta ,\theta ) + r^2 \text{Sin}[\theta]^2 ~t(\phi ,\phi )$
\end{lstlisting}

\begin{lstlisting}[style=out]
$(1\text{) }\text{d[}r]\text{ o }\text{d[}r + (r^2\text{) }\text{d[}\theta ]\text{ o }\text{d[}\theta ]+(r^2 \text{Sin}^2 \theta \text{) }\text{d[}\phi ]\text{ o }\text{d[}\phi ]]$
\end{lstlisting}

Minkowski metric:
\begin{lstlisting}[style=in]
LorMetric = t[ct, ct] - t[x, x] - t[y, y]
\end{lstlisting}

\begin{lstlisting}[style=out]
(-1) d[x] o d[x]+(-1) d[y] o d[y]+(1) d[ct] o d[ct]
\end{lstlisting}

The inverse metric can be found as follows.
\begin{lstlisting}[style=in]
InvertMetric[CartMetric]
\end{lstlisting}

\begin{lstlisting}[style=out]
(1) X[x] o X[x]+(1) X[y] o X[y]+(1) X[z] o X[z]
\end{lstlisting}

\begin{lstlisting}[style=in]
InvertMetric[PolMetric]
\end{lstlisting}

\begin{lstlisting}[style=out]
$(1\text{) }\text{X[}r]\text{ o }\text{X[}r] + (\frac{1}{r^2}\text{) }\text{X[}\theta ]\text{ o }\text{X[}\theta ] + (\frac{1}{r^2 \text{Sin}^2 \theta }\text{) }\text{X[}\phi ]\text{ o }\text{X[}\phi ]$
\end{lstlisting}

\begin{lstlisting}[style=in]
InvertMetric[LorMetric]
\end{lstlisting}

\begin{lstlisting}[style=out]
(-1) X[x] o X[x]+(-1) X[y] o X[y]+(1) X[ct] o X[ct]
\end{lstlisting}

To find the inner product of two vectors, the following command may be used.
\begin{lstlisting}[style=in]
v1 = a X[x] + b X[y] + c X[z];
v2 = d X[x] + e X[y] + f X[z];

InnerProduct[v1, v2, CartMetric]
\end{lstlisting}

\begin{lstlisting}[style=out]
a d + b e + c f
\end{lstlisting}

For finding the inner product of two one forms you can use the following command.
\begin{lstlisting}[style=in]
$\text{u1}=a t(r)+b t(\theta )+c t(\phi );$
$\text{u2}=d t(r)+d t(\phi )+e t(\theta );$

$\text{InnerProduct}[\text{u1},~\text{u2},~\text{PolMetric}]$ 
\end{lstlisting}

\begin{lstlisting}[style=out]
$a d+\frac{b e}{r^2}+\frac{c d}{r^2 \text{Sin}^2\theta}$
\end{lstlisting}

Here we have shown only two examples of inner products; various other examples can be found in the repository.

%% file: DiffGeom/DG_hodge.tex
\chapter{Hodge duality}\label{hodge}
We will next define the Hodge star operator. We will define it in a chart rather than abstractly. 

\md The \db{Hodge star operator}, denoted $\star$ in an $n$-dimensional manifold, is a map from $p$-forms to $(n-p)$-forms given by
\begin{eqnarray}
\left(\star\omega\right)_{\mu_1\cdots\mu_{n-p}} 
&\equiv& \frac{\sqrt{|g|}}{p!}\, \epsilon_{\mu_1\cdots\mu_n} \, g^{\mu_{n-p+1}\nu_1} \,\cdots g^{\mu_n \nu_p}\, \omega_{\nu_1\cdots\nu_p}\,,
\end{eqnarray}
where $\omega$ is a $p$-form. \hfill$\Box$

The $\star$ operator acts on forms, not on components.

\textbf{Example:} Consider $\mathbb{R}^3$ with metric +++, i.e. $g_{\mu\nu} = \rm{diag}(1,1,1)\,.$ Then $|g| \equiv g = 1\,, g^{\mu\nu} = \rm{diag}(1,1,1)\,.$ Write the coordinate basis 1-forms as $dx, dy, dz\,.$ Their components are clearly 
\begin{equation}
(dx)_i = \delta^1_i\,, \, (dy)_i = \delta^2_i\,, \, (dz)_i = \delta^3_i\,,
\end{equation}
the $\delta$'s on the right hand sides are Kroenecker deltas.
So
\begin{eqnarray}
(\star dx)_{ij} &=& \epsilon_{ijk} g^{kl} (dx)_l = \epsilon_{ijk} g^{kl} \delta^1_l = \epsilon_{ijk} g^{k1} \next
\Rightarrow \qquad \star dx &=& \frac{1}{2!} (\star dx)_{ij}  dx^i\wedge dx^j = \frac{1}{2!}\epsilon_{ijk} g^{k1} dx^i\wedge dx^j \next
&& \qquad \qquad g^{k1} = 1\; \rm{for}\; k=1\,, 0 \;\rm{otherwise} \nonumber \next
\Rightarrow \qquad \star dx &=& \frac{1}{2!} \left(dx^2\wedge dx^3 - dx^3\wedge dx^2\right) = dx^2\wedge dx^3 = dy\wedge dz\,.
\end{eqnarray}
Similarly, $\star dy = dz\wedge dx \,, \qquad \star dz = dx\wedge dy\,.$ \hfill $\Box$

\textbf{Example:} Consider $p=0$ (scalar), i.e. a 0-form $\omega$ in $n$ dimensions. 
\begin{eqnarray}
(\star\omega)_{\mu_1\cdots\mu_n} &=& \sqrt{|g|} \epsilon_{\mu_1\cdots\mu_n} \omega \next
\Rightarrow \qquad (\star 1)_{\mu_1\cdots\mu_n} &=& \sqrt{|g|} \epsilon_{\mu_1\cdots\mu_n} \next
\Rightarrow \qquad (\star 1) &=& \frac{\sqrt{|g|}}{n!} \epsilon_{\mu_1\cdots\mu_n} dx^{\mu_1}\wedge\cdots\wedge dx^{\mu_n} \next &=& dV
\end{eqnarray}
\hfill$\Box$

\textbf{Example:} $p = n\,.$ Then
\begin{equation}
(\star \omega) = \frac{\sqrt{|g|}}{n!} \epsilon_{\mu_1\cdots\mu_n} g^{\mu_1\nu_1}\cdots g^{\mu_n\nu_n} \omega_{\nu_1\cdots \nu_n}\,.
\end{equation}
For the volume form,
\begin{eqnarray}
dV &=& \frac{\sqrt{|g|}}{n!} \epsilon_{\mu_1\cdots\mu_n} dx^{\mu_1}\wedge\cdots\wedge dx^{\mu_n} \next
(dV)_{\nu_1\cdots\nu_n} &=& \sqrt{|g|} \epsilon_{\nu_1\cdots\nu_n}\next
(\star dV) &=& \frac{|g|}{n!}\epsilon_{\mu_1\cdots\mu_n}g^{\mu_1\nu_1}\cdots g^{\mu_n\nu_n} \epsilon_{\nu_1\cdots \nu_n} \next 
&=& \frac{|g|}{n!}n!(\det g)\inv = \frac{|g|}{n!}\frac{n!}{g} = sign(g) = (-1)^s\,,
\end{eqnarray}
where $s$ is the number of $(-1)$ in $g_{\mu\nu}\,.$ \hfill $\Box$

So we find that 
\begin{equation}
\star(\star 1) = \star dV = (-1)^s\,,
\end{equation}
and 
\begin{equation}
\star(\star dV) = (-1)^s (\star 1) = (-1)^s dV\,,
\end{equation}
i.e., $(\star)^2 = (-1)^s$ on 0-forms and $n$-forms. 

In general, on a $p$-form in an $n$-dimensional manifold with signature $(s, n-s)\,,$ it can be shown in the same way that %
\begin{equation}
(\star)^2 = (-1)^{p(n-p)+s}\,.
\end{equation}
In particular, in four dimensional Minkowski space, $s=1, n=4\,,$ so 
\begin{equation}
(\star)^2 = (-1)^{p(4-p) + 1}\,.
\label{hodge.star2}
\end{equation}

It is useful to work out the Hodge dual of basis $p$-forms. Suppose we have a basis $p$-form $dx^{I_1}\wedge \cdots \wedge dx^{I_p}\,,$ where the indices are arranged in increasing order $I_p>\cdots >I_1\,.$ Then its components are $\displaystyle{p! \delta^{I_1}_{\mu_1}\cdots \delta^{I_p}_{\mu_p}}\,.$ So
\begin{eqnarray}
 \star\left(dx^{I_1}\wedge \cdots \wedge dx^{I_p}\right)_{\nu_1\cdots \nu_{n-p}}  &=& \frac{\sqrt{|g|}}{p!}\, \epsilon_{\nu_1\cdots\nu_{n-p}\mu_1\cdots\mu_p}g^{\mu_1\mu'_1}\cdots g^{\mu_p\mu'_p}\,p!\, \delta^{I_1}_{\mu'_1}\cdots \delta^{I_p}_{\mu'_p} \next
&=& \sqrt{|g|}\, \epsilon_{\nu_1\cdots\nu_{n-p}\mu_1\cdots\mu_p} g^{\mu_1 I_1}\cdots g^{\mu_p I_p}\,.
\label{hodge.dualbasis}
\end{eqnarray}
%
%
We will use this to calculate $\star\omega\wedge\omega\,.$

For a $p$-form $\omega\,,$ we have
\begin{eqnarray}
\omega &=& \frac{1}{p!} \omega_{\mu_1\cdots \mu_p} dx^{\mu_1}\wedge \cdots \wedge dx^{\mu_p}\next
&=& \sum\limits_{I} \omega_{I_1\cdots I_p} dx^{I_1}\wedge \cdots \wedge dx^{I_p}
\end{eqnarray}
where the sum over $I$ means a sum over all possible index sets $I = I_1<\cdots <I_p\,,$
but there is no sum over the indices $\{I_1, \cdots, I_p\}\,$ themselves; in a given index set the $I_k$ are fixed. 
Using the dual of basis $p-$forms, and Eq.~(\ref{forms.component}), we get 
\begin{eqnarray}
\star\omega &=& \sum\limits_{I}\omega_{I_1\cdots I_p} \star(dx^{I_1}\wedge \cdots \wedge dx^{I_p})
\next 
&=& \sum\limits_{I}\frac{\sqrt{|g|}}{(n-p)!} \epsilon_{\nu_1\cdots\nu_{n-p}\mu_1\cdots\mu_p} g^{\mu_1 I_1}\cdots g^{\mu_p I_p}\,\omega_{I_1\cdots I_p} dx^{\nu_1}\wedge \cdots \wedge dx^{\nu_{n-p}}\,.
\next
\end{eqnarray}

The sum over $I$ is a sum over different index sets as before, and the Greek indices are summed over as usual.
Thus we calculate
\begin{eqnarray}
\star\omega\wedge\omega &=& \frac{\sqrt{|g|}}{(n-p)!}\sum\limits_{I,J}\epsilon_{\nu_1\cdots\nu_{n-p}\mu_1\cdots\mu_p}g^{\mu_1 I_1}\cdots g^{\mu_p I_p}\omega_{I_1\cdots I_p}\times\nonumber\next 
&& \qquad\qquad dx^{\nu_1}\wedge \cdots \wedge dx^{\nu_{n-p}}\wedge\left(\omega_{J_1\cdots J_p} dx^{J_1}\wedge \cdots \wedge dx^{J_p} \right)  \next
&=& \frac{\sqrt{|g|}}{(n-p)!}\sum\limits_{I,J}\epsilon_{\nu_1\cdots\nu_{n-p}\mu_1\cdots\mu_p}g^{\mu_1 I_1}\cdots g^{\mu_p I_p}\omega_{I_1\cdots I_p}\omega_{J_1\cdots J_p}\times\nonumber \next
&& \qquad\qquad dx^{\nu_1}\wedge \cdots \wedge dx^{\nu_{n-p}}\wedge dx^{J_1}\wedge \cdots \wedge dx^{J_p}
\end{eqnarray}
We see that the set $\{\nu_1, \cdots , \nu_{n-p}\}$ cannot have any overlap with the set $J = \{J_1, \cdots, J_p\}$, because of the wedge product. 
On the other hand,  $\{\nu_1, \cdots , \nu_{n-p}\}$ cannot have any overlap with  $\{\mu_1, \cdots , \mu_{p}\}$ because $\epsilon$ is totally antisymmetric in its indices. So the set $\{\mu_1, \cdots , \mu_{p}\}$ must have the same elements as the set $J = \{J_1, \cdots, J_p\}\,,$ but they may not be in the same order. 

Now consider the case where the basis is orthogonal, i.e. $g^{\mu\nu}$ is diagonal. Then $g^{\mu_k I_k} = g^{I_k I_k}$ etc. and we can write 
\begin{eqnarray}
\star\omega\wedge\omega &=& \frac{\sqrt{|g|}}{(n-p)!}\sum\limits_{I,J}\epsilon_{\nu_1\cdots\nu_{n-p}I_1\cdots I_p}g^{I_1 I_1}\cdots g^{I_p I_p}\omega_{I_1\cdots I_p}\omega_{J_1\cdots J_p}\times\nonumber\next 
&& \qquad\qquad dx^{\nu_1}\wedge \cdots \wedge dx^{\nu_{n-p}}\wedge dx^{J_1}\wedge \cdots \wedge dx^{J_p} \,. \qquad
\end{eqnarray}
We see that in each term of the sum, the indices $\{I_1\cdots I_p\}$ must be the same as $\{J_1\cdots J_p\}$ because both sets are totally antisymmetrized with the indices $\{\nu_1\cdots\nu_{n-p}\}$. 

Since both sets are ordered, it follows that we can replace $J$ by $I$,
\begin{eqnarray}
\star\omega\wedge\omega &=& \frac{\sqrt{|g|}}{(n-p)!}\sum\limits_{I}\epsilon_{\nu_1\cdots\nu_{n-p}I_1\cdots I_p}g^{I_1 I_1}\cdots g^{I_p I_p}\omega_{I_1\cdots I_p}\omega_{I_1\cdots I_p}\times \nonumber\next 
&& \qquad\qquad dx^{\nu_1}\wedge \cdots \wedge dx^{\nu_{n-p}}\wedge dx^{I_1}\wedge \cdots \wedge dx^{I_p} \, \next
 &=& \frac{\sqrt{|g|}}{(n-p)!}\sum\limits_{I}\epsilon_{\nu_1\cdots\nu_{n-p}I_1\cdots I_p}\omega^{I_1\cdots I_p}\omega_{I_1\cdots I_p}\times\nonumber\next
 && \qquad\qquad dx^{\nu_1}\wedge \cdots \wedge dx^{\nu_{n-p}}\wedge dx^{I_1}\wedge \cdots \wedge dx^{I_p} \,.
\end{eqnarray}
In each term of this sum, the indices $\{\nu_1\cdots\nu_{n-p}\}$ are completely determined, so we can replace them by the corresponding ordered set $K = K_1 <\cdots <K_{n-p}$\,, which is completely determined by the set $I$\,, so that
\begin{eqnarray}
\star\omega\wedge\omega &=& \sqrt{|g|}\sum\limits_{I}\epsilon_{K_1\cdots K_{n-p}I_1\cdots I_p}\omega^{I_1\cdots I_p}\omega_{I_1\cdots I_p}\times \nonumber \next 
&& \qquad\qquad dx^{K_1}\wedge \cdots \wedge dx^{K_{n-p}}\wedge dx^{I_1}\wedge \cdots \wedge dx^{I_p} \,.
\end{eqnarray}

The indices on this $\epsilon$ are a permutation of $\{1, \cdots, n\}$\,, so $\epsilon$ is $\pm1$.
But this sign is the same as that for the permutation to bring the basis to the order $dx^{1}\wedge\cdots\wedge dx^{n}\,,$ so the overall sign to get both to the standard order is positive. Thus we get
\begin{eqnarray}
\star\omega\wedge\omega &=& \sqrt{|g|}\sum\limits_{I}\omega^{I_1\cdots I_p}\omega_{I_1\cdots I_p}\epsilon_{1\cdots n}\,dx^{1}\wedge \cdots\wedge dx^{n} \,\next
&=&  \sqrt{|g|}\,\frac{1}{p!}\omega^{\mu_1\cdots \mu_p}\omega_{\mu_1\cdots \mu_p}\, dx^{1}\wedge \cdots\wedge dx^{n}\next
&=& \frac{1}{p!}\omega^{\mu_1\cdots \mu_p}\omega_{\mu_1\cdots \mu_p}\,(vol)
\label{hodge.contraction}
\end{eqnarray}

If we are in a basis where the metric is not diagonal, it is still symmetric. So we can diagonalize it locally by going to an appropriate basis, or set of coordinates, at each point.  In this basis, the components of $\omega$ may be $\omega'_{\mu_1\cdots\mu_p}$, so we can write 
\begin{eqnarray}
\star\omega\wedge\omega &=& \left(\frac{1}{p!}\omega^{\mu'_1\cdots \mu'_p}\omega_{\mu'_1\cdots \mu'_p}\right)\,(vol')
\end{eqnarray}
But both factors are invariant under a change of basis. So we can now change back to our earlier basis, and find Eq.~(\ref{hodge.contraction}) even when the metric is not diagonal. Note that the metric may not be diagonalizable globally or even in an extended region. 

One may use the inbuilt command in Mathematica to evaluate the following:

\begin{lstlisting}[style=in]
MatrixForm[HodgeDual[{x,y,z}]]
\end{lstlisting}

\begin{lstlisting}[style=out]
$\begin{pmatrix}
0 & z & y\\
-z & 0 & x\\
y & -x & 0
\end{pmatrix}$
\end{lstlisting}

The Hodge dual of the Hodge dual should return the original vector as is shown below.
\begin{lstlisting}[style=in]
MatrixForm[HodgeDual[HodgeDual[{x,y,z}]]]
\end{lstlisting}

\begin{lstlisting}[style=out]
$\begin{pmatrix}
x\\
y\\
z
\end{pmatrix}$
\end{lstlisting}

Example with constant vector:
\begin{lstlisting}[style=in]
MatrixForm[HodgeDual[1,2]]
\end{lstlisting}

\begin{lstlisting}[style=out]
$\begin{pmatrix}
0 & 1\\
-1 & 0
\end{pmatrix}$
\end{lstlisting}

In order to work with the differential forms package, the \emph{HodgeStar} command should be used.

The following demonstrations also use the Mathematica package \code{DifferentialForms}.
\begin{lstlisting}[style=in]
HodgeStar[d[x,y],t[x,x] + t[y,y] + t[z,z]]
\end{lstlisting}

\begin{lstlisting}[style=out]
(1) dz
\end{lstlisting}

\begin{lstlisting}[style=in]
$\text{HodgeStar}\left(d(r),r^2 t(\theta ,\theta )+t(r,r)\right)$
\end{lstlisting}

\begin{lstlisting}[style=out]
$(\frac{1}{\sqrt{\frac{1}{r^2}}})\text{d}\theta $
\end{lstlisting}

An interesting result can be seen in $3$ dimensions~[\cite{mukunda2010lectures}]. The Hodge dual of the wedge product of two vectors can be expressed as a vector and is unique to $3$ dimensions. In fact, its components are precisely equal to those of the cross product of two vectors as shown below.

\begin{lstlisting}[style=in]
HodgeStar[v1 w2 d[x1,x2]+ v2 w1 d[x2,x1]+v1 w3 d[x1,x3]+ v3 w1 d[x3,x1]+ v2 w3 d[x2,x3]+v3 w2 d[x3,x2],{t[x1,x1],t[x2,x2],t[x3,x3]}]
\end{lstlisting}
\begin{lstlisting}[style=out]
(-v2 w1+v1 w2) dx3+(v3 w1-v1 w3) dx2+(-v3 w2+v2 w3) dx1
\end{lstlisting}

%% file: DiffGeom/DG_maxwell.tex
\chapter{Maxwell equations}\label{maxwell}
We will now consider a particular example in physics where differential forms are useful. The Maxwell equations of electrodynamics are, with $c = 1\,,$
\begin{eqnarray}
\vec{\nabla}\cdot\vec{E} &=& \rho \\
\vec{\nabla}\times\vec{B} - \frac{\partial\vec{E}}{\partial t} &=& \vec{j}\\
\vec{\nabla}\cdot\vec{B} &=& 0\\
\vec{\nabla}\times\vec{E} + \frac{\partial\vec{B}}{\partial t} &=& 0\,.
\end{eqnarray}
The electric and magnetic fields are all vectors in three dimensions, but these equations are Lorentz-invariant. We will write these equations in terms of differential forms.

Consider $\mathbb{R}^4$ with Minkowski metric $g_{\mu\nu} = \rm{diag}(-1, 1, 1, 1)\,.$ For the magnetic field define a 2-form 
\begin{equation}
B = B_x dy\wedge dz + B_y dz\wedge dx + B_z dx\wedge dy\,.
\end{equation}
For the electric field define a 1-form 
\begin{equation}
E = E_x dx + E_y dy + E_z dz\,.
\end{equation}
Combine these two into a 2-form $F = B + E\wedge dt\,.$ Let us calculate $dF = d(B + E\wedge dt) = dB + dE\wedge dt\,.$ As usual, we will write $1, 2, 3$ for the component labels $x, y, z\,.$
\begin{eqnarray}
dB &=& d(B_1 dy\wedge dz + B_2 dz\wedge dx + B_3 dx\wedge dy) \next
&=& \partial_t B_1 dt\wedge dy\wedge dz + \partial_1 B_1 dx\wedge dy\wedge dz \nonumber \next
&& \quad + \partial_t B_2 dt\wedge dz\wedge dx + \partial_2 B_2 dy\wedge dz \wedge dx \nonumber \next
&& \quad + \partial_t B_3 dt\wedge dx\wedge dy + \partial_3 B_3 dz\wedge dx\wedge dy\,.
\end{eqnarray}
And
\begin{eqnarray}
d(E\wedge dt) &=& d(E_1 dx\wedge dt + E_2 dy\wedge dt + E_3 dz\wedge dt)\next
&=& \partial_2 E_1 dy\wedge dx\wedge dt + \partial_3 E_1 dz\wedge dx\wedge dt \nonumber \next
&& \quad + \partial_1 E_2 dx\wedge dy\wedge dt + \partial_3 E_2 dz\wedge dy\wedge dt \nonumber \next
&& \quad + \partial_1 E_3 dx\wedge dz \wedge dt + \partial_2 E_3 dy\wedge dz\wedge dt\,.
\end{eqnarray}
Thus, recalling that the wedge product changes sign under each exchange, we can combine these two to get
\begin{eqnarray}
dF &=& \left(\partial_t B_1 + \partial_2 E_3 - \partial_3 E_2 \right)\, dt\wedge dy\wedge dz \nonumber \next
&& + \left(\partial_t B_2 + \partial_1 E_3 - \partial_3 E_1 \right)\, dt\wedge dz\wedge dx \nonumber \next
&& + \left(\partial_t B_3 + + \partial_1 E_2 - \partial_2 E_1 \right)\, dt\wedge dx\wedge dy \nonumber\next
&& + \left(\partial_1 B_1 + \partial_2 B_2 + \partial_3 B_3 \right)\, dx\wedge dy\wedge dz\, \next
&=& \left(\partial_t B_1 + ({\vec\nabla}\times{\vec E})_1\right) \, dt\wedge dy\wedge dz \nonumber \next
&& + \left(\partial_t B_2 + ({\vec\nabla}\times{\vec E})_2 \right)\, dt\wedge dz\wedge dx \nonumber \next
&& + \left(\partial_t B_3 + ({\vec\nabla}\times{\vec E})_3 \right)\, dt\wedge dx\wedge dy \nonumber \next
&& + \left({\vec \nabla}\cdot{\vec B} \right)\, dx\wedge dy\wedge dz\,.
\end{eqnarray}
Thus, two of Maxwell's equations are equivalent to $dF = 0\,.$

For the other two equations we need $\star F\,.$ Using the formula Eq.~(\ref{hodge.dualbasis}) for dual basis forms, it is easy to calculate that 
\begin{eqnarray}
\star(dx\wedge dy) &=& dt\wedge dz\,, \quad \star(dy\wedge dz) = dt\wedge dx\,, \quad \star(dz\wedge dx) = dt\wedge dy\,, \next
\star(dx\wedge dt) &=& dy\wedge dz\,, \quad \star(dy\wedge dt) = dz\wedge dx\,, \quad \star(dz\wedge dt) = dx\wedge dy\,.
\end{eqnarray}
We use these to calculate 
\begin{eqnarray}
{\star F} &=& \star(B + E \wedge dt) \next
&=& B_1 dt\wedge dx + B_2 dt\wedge dy + B_3 dt\wedge dz \nonumber \next 
&& + E_1 dy\wedge dz + E_2 dz\wedge dx + E_3 dx\wedge dy \,.
\end{eqnarray}
Then in the same way as for the previous calculation, we find
\begin{eqnarray}
d{\star F} &=& ({\vec \nabla}\cdot{\vec E}) \, dx\wedge dy\wedge dz \nonumber\next
&& + \left(\partial_t E_1 - ({\vec\nabla}\times{\vec B})_1\right) \, dt\wedge dy\wedge dz\, \nonumber \next 
&& + \left(\partial_t E_2 - ({\vec\nabla}\times{\vec B})_2 \right)\, dt\wedge dz\wedge dx  \nonumber \next
&& + \left(\partial_t E_3 + ({\vec\nabla}\times{\vec B})_3 \right)\, dt\wedge dx\wedge dy \,.
\label{maxwell.dstarF}
\end{eqnarray}
We need to relate this to the charge-current. 

Define the current four-vector as 
\begin{eqnarray}
j^\mu \partial_\mu = \rho\partial_t + j^1\partial_1 + j^2\partial_2 + j^3\partial_3 \,.
\end{eqnarray}
Then there is a corresponding one-form $j_\mu dx^\mu$ with $\displaystyle{j_\mu = g_{\mu\nu}j^\nu}\,.$ So in terms of components, 
\begin{eqnarray}
j_\mu dx^\mu = -\rho dt + j_1 dx^1 + j_2 dx^2 + j_3 dx^3\,.
\end{eqnarray}
Then using Eq.~(\ref{hodge.dualbasis}) it is easy to calculate that
\begin{eqnarray}
\star j &=& -\rho \, dx\wedge dy\wedge dz + j_1\, dt\wedge dy\wedge dz\, \nonumber \next 
&& + j_2 \, dt\wedge dz\wedge dx + j_3 \, dt\wedge dx\wedge dy\,.
\end{eqnarray}
Comparing this equation with Eq.~(\ref{maxwell.dstarF}) we find that the other two Maxwell equations can be written as 
\begin{eqnarray}
d{\star F} = -{\star j}\,.
\end{eqnarray}
Finally, using Eq.~(\ref{hodge.contraction}), we see that the action of electromagnetism can be written as
\begin{eqnarray}
-\frac12 \int F\wedge{\star F} 
\end{eqnarray}
This expression holds in both flat and curved spacetimes. For the latter, with local coordinates $(t,x,y,z)$ we find
\begin{eqnarray}
F\wedge{\star F} = ({\vec B}^2 - {\vec E}^2) \sqrt{-g}\, dt\wedge dx\wedge dy\wedge dz\,.
\end{eqnarray}
Mathematica demonstration: A demo notebook, using the \texttt{DifferentialForms.m}, is provided where the results appear in a self-explanatory notation consistent with Mathematica's syntax. It is easy to see in the notebook that the electromagnetic Lagrangian can be found as shown below (the notebook gives a detailed derivation of all the expressions in this chapter and can be very instructive).
\begin{lstlisting}[style=in]
Bvec = Bx[x, y, z, t] d[y, z] + By[x, y, z, t] d[z, x] + 
Bz[x, y, z, t] d[x, y]
\end{lstlisting}
\begin{lstlisting}[style=out]
(Bx[x,y,z,t]) dy ^ dz+(-By[x,y,z,t]) dx ^ dz+(Bz[x,y,z,t]) dx ^ dy
\end{lstlisting}

\begin{lstlisting}[style=in]
Evec = Ex[x, y, z, t] d[x] + Ey[x, y, z, t] d[y] + Ez[x, y, z, t] d[z]
\end{lstlisting}
\begin{lstlisting}[style=out]
(Ex[x,y,z,t]) dx+(Ey[x,y,z,t]) dy+(Ez[x,y,z,t]) dz
\end{lstlisting}

\begin{lstlisting}[style=in]
Fvec=Bvec + ExteriorProduct[Evec, d[t]] // Simplify
\end{lstlisting}
\begin{lstlisting}[style=out]
(Bx[x,y,z,t]) dy ^ dz+(-By[x,y,z,t]) dx ^ dz+(Bz[x,y,z,t]) dx ^ dy+(-Ex[x,y,z,t]) dt ^ dx+(-Ey[x,y,z,t]) dt ^ dy+(-Ez[x,y,z,t]) dt ^ dz
\end{lstlisting}

\begin{lstlisting}[style=in]
StarFvec=Bx[x,y,z,t] d[t,x]+By[x,y,z,t] d[t,y]+Bz[x,y,z,t] d[t,z] +Ex[x,y,z,t] d[y,z]+ Ey[x,y,z,t] d[z,x]+Ez[x,y,z,t] d[x,y]
\end{lstlisting}
\begin{lstlisting}[style=out]
(Bx[x,y,z,t]) dt ^ dx+(By[x,y,z,t]) dt ^ dy+(Bz[x,y,z,t]) dt ^ dz+(Ex[x,y,z,t]) dy ^ dz+(-Ey[x,y,z,t]) dx ^ dz+(Ez[x,y,z,t]) dx ^ dy
\end{lstlisting}

\begin{lstlisting}[style=in]
ExteriorProduct[Fvec, StarFvec] // Simplify  
\end{lstlisting}
\begin{lstlisting}[style=out]
(Bx[x,y,z,t]$^2$+By[x,y,z,t]$^2$+Bz[x,y,z,t]$^2$-Ex[x,y,z,t]$^2$-Ey[x,y,z,t]$^2$-Ez[x,y,z,t]$^2$) dt ^ dx ^ dy ^ dz
\end{lstlisting}

The relevant notebook is named \texttt{Maxwell}.  We note here that although this is a popular example, we have not found any Mathematica demonstration of what we consider to be a really beautiful exposition of the power and beauty of differential forms.  The student is encouraged to study this notebook thoroughly and compare with the definitions given in the discussion above. 

%% file: DiffGeom/DG_stokes.tex
\chapter{Stokes' theorem}\label{stokes}
We will next discuss a very beautiful result called Stokes' formula. This is actually a theorem, but we will not prove it, only state the result and discuss its applications. So for us although it is only a formula, it is still deep and beautiful.

\md A \db{submanifold} ${\cal S}$ is a subset of points in $\cm$ such that any point in ${\cal S}$ has an open neighbourhood in $\cm$ for which there is some chart where $(n - m)$ coordinates vanish. ${\cal S}$ is then $m$-dimensional. \hfill$\Box$

\md Suppose $\cu$ is a region of an oriented manifold $\cm\,.$  The \db{boundary} $\partial\cu\,$ of $\cu$ is a submanifold of dimension $n - 1$ which divides $\cm$ in such a way that any curve joining a point in $\cu$ with a point in $\cu^c$ must contain a point in $\partial\cu\,.$

Now suppose $\cu$ has an oriented smooth boundary $\partial\cu\,.$ Then $\partial\cu\,$ is automatically an oriented manifold, by considering the restrictions of the charts on $\cu$ to $\partial\cu\,.$ 

\md Consider a smooth $(n - 1)$ form in $\cm\,.$ \db{Stokes}' \db{formula} says that 
\begin{eqnarray}
\int\limits_{\cu} d\omega = \int\limits_{\partial\cu} \omega\,.
\end{eqnarray}
If $\cm$ is a compact manifold with boundary $\partial\cm\,,$ this formula can be applied to all of $\cm\,.$ If $\omega$ vanishes outside some compact region we can again set $\cu = \cm\,.$ Also, $\cu$ can be a submanifold in another manifold, like a 2-surface in a 3-manifold. \hfill$\Box$

\textbf{Example:} Let $\cu = [0, 1]\,.$ Then a function $f: \cm \to \mathbb{R}\,$ is a 0-form, and $df = f'(x)dx$ is a 1-form. Take the orientation of $\cm$ to be from 0 to 1. Then $\partial\cm$ consists of the points $x=0$ and $x=1\,,$ and Stokes' formula says that 
\begin{eqnarray}
\int\limits_\cm df &=& \int\limits_{\partial\cm}f \,\next
i.e. \qquad \int\limits_0^1 f'(x) dx &=& f(1) - f(0)\,.
\end{eqnarray}

\textbf{Example:} Consider a 2-d disk $D$ in $\mathbb{R}^2\,,$ with boundary $\partial D\,.$ Take a 1-form $A\,.$ Then Stokes' formula says 
\begin{eqnarray}
\int\limits_{\partial D} A = \int\limits_D dA\,.
\end{eqnarray}
Let us see this equation in a chart. We can write 
\begin{eqnarray}
A &=& A_i\, dx^i\,\next
dA &=& \partial_i A_j\, dx^i \wedge dx^j 
\end{eqnarray}
$A$ evaluated on $\partial D$ can be written as $\displaystyle{A\left(\frac{d}{dt}\right)}$ where $\displaystyle{\frac{d}{dt}}$ is tangent to $\partial D\,.$ So we can write $\displaystyle{A\left(\frac{d}{dt}\right) = A_i \frac{dx^i}{dt} dt}\,, $ and 
\begin{eqnarray}
\int\limits_{\partial D}A_i dx^i &=& \int\limits_D  \partial_i A_j\, dx^i \wedge dx^j \,\next
&=&  \int\limits_D \left( \partial_1 A_2 - \partial_2 A_1\right)\, dx^1 \wedge dx^2 \,\next
&=& \int\limits_{\varphi(D)} \left( \partial_1 A_2 - \partial_2 A_1\right)\, d^2x\,.
\end{eqnarray}
Similarly for higher forms  on higher dimensional manifolds. 

\md \db{Gauss}' \db{divergence theorem} is a special case of Stokes' theorem. Before getting to Gauss' theorem, we need to make a new definition. Consider an $n$-form $\omega\neq 0$ on an $n$-dimensional manifold. We can write this in a chart as 
\begin{eqnarray}
\omega &=& f dx^1\wedge\cdots\wedge dx^n \,\next
&=& \frac{1}{n!} f \epsilon_{\mu_1 \cdots \mu_n}\, dx^{\mu_1}\wedge\cdots\wedge dx^{\mu_n}\,.
\end{eqnarray}
Given a vector field $v\,,$ its contraction with $\omega$ is 
\begin{eqnarray}
\iota_v \omega = \omega(v, \cdots) &=& \frac{1}{(n-1)!} \omega_{\mu_1\mu_2\cdots\mu_n}\, v^{\mu_1}dx^{\mu_2}\wedge\cdots\wedge dx^{\mu_n} \,\next
&=& fv^1 dx^2\wedge\cdots\wedge dx^n - fv^2 dx^1\wedge dx^3\wedge\cdots\wedge dx^n + \cdots\,\next
\end{eqnarray}
Then we can calculate
\begin{eqnarray}
d(\iota_v\omega) = d\omega(v, \cdots) &=& \partial_1(fv^1)\, dx^1\wedge dx^2\wedge\cdots\wedge dx^n\, \nonumber\next
&& + \partial_2(fv^2)\, dx^1\wedge dx^2\wedge\cdots\wedge dx^n\, \nonumber \next
&& +\cdots + \partial_n(fv^n)\, dx^1\wedge dx^2\wedge\cdots\wedge dx^n\,\next
&=& \partial_\mu(fv^\mu)\, dx^1\wedge dx^2\wedge\cdots\wedge dx^n\,\next
&=& \frac{1}{f}  \partial_\mu(fv^\mu)\, \omega\,.
\end{eqnarray}
In particular, if $\omega$ is the volume form, we can write
\begin{eqnarray}
\omega &=& \frac{\sqrt{|g|}}{n!} \epsilon_{\mu_1\cdots \mu_n} dx^{\mu_1}\wedge\cdots\wedge dx^{\mu_n}\,,\next
d(\iota_v(vol)) &=& \frac{1}{\sqrt{|g|}}\partial_\mu(v^\mu\sqrt{|g|})(vol)\,.
\end{eqnarray}
\md This is called the \db{divergence} of the vector field $v\,.$ 

There is another expression for the divergence. Recall that given a vector field $v\,,$ we can define a one-form, also called $v\,,$ with components defined with the help of the metric,
\begin{eqnarray}
v_\mu = g_{\mu\mu'}v^{\mu'}
\end{eqnarray}
Consider $\star v\,,$ which has components
\begin{eqnarray}
(\star v)_{\mu_1\cdots\mu_{n-1}} &=& \sqrt{|g|} \epsilon_{\mu_1\cdots\mu_{n-1}\mu} g^{\mu\mu'}v_{\mu'}\,\next
&=& \sqrt{|g|} \epsilon_{\mu_1\cdots\mu_{n-1}\mu} v^\mu\,.\\
\Rightarrow\qquad \star v &=& \frac{\sqrt{|g|}}{(n-1)!} \epsilon_{\mu_1\cdots\mu_{n-1}\mu} v^\mu dx^{\mu_1}\wedge\cdots\wedge dx^{\mu_{n-1}}\,\next
d{\star v} &=& \partial_{\mu_n}\left(\frac{\sqrt{|g|}}{(n-1)!} \epsilon_{\mu_1\cdots\mu_{n-1}\mu} v^\mu\right)\, dx^{\mu_n}\wedge dx^{\mu_1}\wedge\cdots\wedge dx^{\mu_{n-1}}\, \nonumber \next
&=& \frac{(-1)^{n-1}}{(n-1)!}\epsilon_{\mu_1\cdots\mu_{n-1}\mu}\left(\partial_{\mu_n}\left(\sqrt{|g|} v^\mu\right)\right)dx^{\mu_1}\wedge\cdots\wedge dx^{\mu_n}\,.
\end{eqnarray}
Both $\mu$ and $\mu_n$ must be different from $(\mu_1, \cdots, \mu_{n-1})\,,$ so $\mu = \mu_n\,.$ Thus in each term of the sum, the choice of $(\mu_1, \cdots, \mu_{n-1})\,$ automatically selects $\mu_n (=\mu)\,,$ so a sum over $(\mu_1, \cdots, \mu_n)\,$ overcounts $n$ times. So we can write
\begin{eqnarray}
d{\star v} &=& \frac{(-1)^{n-1}}{(n-1)!}\epsilon_{\mu_1\cdots\mu_n}\left(\partial_\mu\left(\sqrt{|g|} v^\mu\right)\right)dx^{\mu_1}\wedge\cdots\wedge dx^{\mu_n}\,\nonumber \next
&=& (-1)^{n-1}\frac{1}{\sqrt{|g|}}\partial_\mu(\sqrt{|g|}v^\mu) (vol)\,.
\end{eqnarray}

Since this is an $n$-form in $n$ dimensions, we can calculate from here that
\begin{eqnarray}
\star{d{\star v}} &=& \frac{(-1)^{n+s-1}}{\sqrt{|g|}} \partial_\mu(\sqrt{|g|}v^\mu) \,,
\end{eqnarray}
where as before $s$ is the signature of the manifold, i.e. the number of negative entries in the metric in a locally diagonal form. 

Let us now go back to Stokes' formula. Take a region $\cu$ of $\cm$ which is covered by a single chart and has an orientable boundary $\partial\cu$ as before. Then we find
\begin{eqnarray}
\int\limits_\cu \frac{1}{\sqrt{|g|}}\partial_\mu(\sqrt{|g|}v^\mu) (vol)\, \nonumber &=& \int\limits_\cu d(\iota_v(vol)) \,\next
&=& \int\limits_{\partial\cu}\iota_v(vol)\,.
\end{eqnarray}
Now suppose $b$ is a 1-form normal to $\partial\cu\,,$ i.e. $b\displaystyle{\left(\frac{d}{dt}\right)} = 0$ for any vector $\displaystyle{\frac{d}{dt}}$ tangent to $\partial\cu\,,$ and $\alpha$ is an $(n-1)$-form such that $b\wedge\alpha = (vol)\,.$ Since all $n$-forms are proportional, $\alpha$ always exists. For the same reason, if $b\neq 0$ on 
$\partial\cu\,,$ it is unique up to a factor. And $b\neq 0$ on $\partial\cu$ because $\partial\cu$ is defined as the submanifold where one coordinate is constant, usually set to zero, so that one component of $\displaystyle{\frac{d}{dt}}$ vanishes at any point on $\partial\cu\,,$ and therefore the corresponding component of $b$ can be chosen to be non-zero.

Thus $b$ is unique up to a rescaling $b\to b' = fb$ for some nonvanishing function $f\,.$ But we can always scale $\alpha \to \alpha' = f^{\inv} \alpha\,$ so that $b'\wedge\alpha' = b\wedge\alpha\,.$ Further, if we restrict $\alpha$ to $\partial\cu\,,$ i.e. if $\alpha$ acts only on tangent vectors to $\partial\cu\,,$ we find that $\alpha$ is an $(n-1)$-form on an $(n-1)$-dimensional manifold, so it is unique up to scaling. Therefore, $\alpha$ is unique once $b$ is given. Finally, for any vector $v\,,$
\begin{eqnarray}
\iota_v(vol)\Big|_{\partial\cu} = \iota_v(b\wedge\alpha)\Big|_{\partial\cu}\,
\end{eqnarray}
is an $(n-1)$-form on $\partial\cu$ which acts only on vectors tangent to $\partial\cu\,.$
Then 
\begin{eqnarray}
 \iota_v(b\wedge\alpha)\Big|_{\partial\cu} = b(v)\alpha\Big|_{\partial\cu}
\end{eqnarray}
because all terms of the form $b\wedge\iota_v\alpha$ give zero for any choice of $(n-1)$ vectors on $\partial\cu\,.$ 

Then we have 
\begin{eqnarray}
\int\limits_\cu \frac{1}{\sqrt{|g|}}\partial_\mu(\sqrt{|g|}v^\mu) (vol)\, &=&  \int\limits_{\partial\cu} b(v)\alpha \,.
\end{eqnarray}
Often $b$ is taken to have norm 1. Then $\alpha$ is the volume form on $\partial\cu\,,$ and we can write
\begin{eqnarray}
\int\limits_\cu \frac{1}{\sqrt{|g|}}\partial_\mu(\sqrt{|g|}v^\mu) (vol)\, &=&    \int\limits_{\partial\cu}\left(n_\mu v^\mu\right) \sqrt{|g_{(\partial\cu)}|} d^{n-1}x\,.
\end{eqnarray}

Mathematica has in-built functions of vector calculus for calculating gradient, divergence, curl, Laplacian etc. We show some examples here.

The gradient of a scalar field is calculated with the command \texttt{Grad}.
\begin{lstlisting}[style=in]
Grad[x^2 + y^2 + z^2, {x, y, z}]
\end{lstlisting}

\begin{lstlisting}[style=out]
{2 x, 2 y, 2 z}
\end{lstlisting}

The divergence of a vector field is calculated with the \texttt{Div} command. Let us find the divergence of the vector field in the above output.

\begin{lstlisting}[style=in]
Div[{2 x, 2 y, 2 z}, {x, y, z}]
\end{lstlisting}

\begin{lstlisting}[style=out]
6
\end{lstlisting}

The Laplacian of a scalar field is calculated as \texttt{Laplacian}.

\begin{lstlisting}[style=in]
Laplacian[x^2 + y^2 + z^2, {x, y, z}]
\end{lstlisting}

\begin{lstlisting}[style=out]
6
\end{lstlisting}

The curl of a vector field can be found using the \texttt{Curl} command.
\begin{lstlisting}[style=in]
Curl[{y, z, x}, {x, y, z}]
\end{lstlisting}

\begin{lstlisting}[style=out]
{-1, -1, -1}
\end{lstlisting}

The coordinate system, by default, is Cartesian. In order to change to a different coordinate system, spherical for example, the following command may be used.
\begin{lstlisting}[style=in]
SetCoordinates[Spherical[r, $\theta$, $\phi$]]
\end{lstlisting}

Verification of Stokes' theorem in Mathematica: The following code uses the \code{DifferentialForms} package.
\begin{lstlisting}[style=in]
SetAttributes[R, Constant]
\end{lstlisting}
\begin{lstlisting}[style=in]
ball=Chain[-1,{x-> r Cos[theta] Sin[phi],y-> r Sin[theta] Sin[phi], z-> r Cos[phi]},{r,0,R},{theta,0,2 Pi},{phi,0,Pi}] 
\end{lstlisting}
\begin{lstlisting}[style=out]
Chain[-1,{x->r Cos[theta] Sin[phi],y->r Sin[phi] Sin[theta],z->r Cos[phi]},{r,0,R},{theta,0,2 $\pi$},{phi,0,$\pi$}]
\end{lstlisting}
\begin{lstlisting}[style=in]
hull=Boundary[ball]
\end{lstlisting}
\begin{lstlisting}[style=out]
Chain[-1, {x -> 0, y -> 0, z -> -r}, {r, 0, R}, {theta, 0, 2 $\pi$}] +
Chain[1, {x -> 0, y -> 0, z -> r}, {r, 0, R}, {theta, 0, 2 $\pi$}] +
Chain[1, {x -> R Cos[theta] Sin[phi], y -> R Sin[phi] Sin[theta], 
z -> R Cos[phi]}, {phi, 0, $\pi$}, {theta, 0, 2 $\pi$}]
\end{lstlisting}
\begin{lstlisting}[style=in]
Integral[d[x, y, z], ball]
\end{lstlisting}
\begin{lstlisting}[style=out]
$\frac{4 \pi R^3}{3}$
\end{lstlisting}
\begin{lstlisting}[style=in]
Integral[x d[y, z], hull]
\end{lstlisting}
\begin{lstlisting}[style=out]
$\frac{4 \pi R^3}{3}$
\end{lstlisting}
We have shown only one example here; the reader is encouraged to check various other examples given in the corresponding notebook,
including \texttt{Stokesexamples}. 

%% file: DiffGeom/DG_liegroup.tex
\chapter{Lie groups}\label{liegroup}
We start a brief discussion on Lie groups, mainly with an eye to their structure as manifolds and also their application to the theory of fiber bundles.

\md A \db{Lie group} is a group which is also an analytic manifold. ($\Box$
We did not define a Lie group in this way in Chap.~\ref{mani}\,, but said that a Lie group was a manifold in which the group product is analytic in the group parameters, or alternatively the group product and group inverse are both $C^\infty\,.$) 

The definition above comes from a theorem that given a continuous group $G$ in which the group product and group inverse are $C^\infty$ functions of the group parameters, it is always possible to find a set of coordinate charts covering $G$ such that the overlap functions are real analytic, i.e. are $C^\infty$ and their Taylor series at any point converge to their respective values.

\md A \db{Lie subgroup} of $G$ is a subset $H$ of $G$ which is a subgroup of $G\,,$ a submanifold of $G\,,$ and is a topological group, i.e., a topological space in which the group product and group inverse are continuous maps. \hfill$\Box$

\md Sometimes this is expressed in terms of another definition. ${\cal P}$ is an \db{immersed submanifold} of $\cm$ if the inclusion map $j: {\cal P} \hookrightarrow\cm$ is smooth and at each point $p\in {\cal P}$ its differential $dj_p$ is one to one, with $dj_p$ being defined by $dj_p: T_p{\cal P} \to T_{j(p)}\cm\,$ such that  $dj_p(v)(g) = v(g\cdot j_p)\,.$ \hfill$\Box$ 

We have mentioned some specific examples of Lie groups earlier. Let us mention some more  examples. 

{\bf Example:} ${\mathbb R}^n$ is a Lie group under addition. So is ${\mathbb C}^n\,.$

{\bf Example:} ${\mathbb R}^n\backslash\{0\}$ is a Lie group under multiplication. So is ${\mathbb C}^n\backslash\{0\}\,.$

{\bf Example:} The \db{direct product} of two Lie groups is itself a Lie group, with multiplication $(g_1, h_1)(g_2, h_2) = (g_1 g_2, h_1 h_2)\,. \hfill \Box$

{\bf Example:} The set of all $n\times n$ real invertible matrices forms a group under matrix multiplication, called the \db{General Linear group} $GL(n, {\mathbb R})\,.$ This is also the space of all invertible linear maps of ${\mathbb R}^n$ to itself. We can similarly define $GL(n, {\mathbb C})\,. \hfill \Box$

{\bf Example:} A Lie subgroup is an immersed submanifold of the Lie group.

The next few examples are Lie subgroups of $GL(n, {\mathbb R})\,.$

{\bf Example:} The \db{Special Linear group} $SL(n, {\mathbb R})$ is the subset of $GL(n, {\mathbb R})$ for which all the matrices have determinant $+1\,,$ i.e., $SL(n, {\mathbb R}) = \{ A \in GL(n, {\mathbb R}) | \det A = 1\}\,.$ One can define $SL(n, {\mathbb C})$ in a similar manner. \hfill$\Box$

{\bf Example:} The \db{Orthogonal group} $O(n) = \{ R \in GL(n, {\mathbb R})\,| $ $R^\trans R = {\mathbb I} \}\,.$ \hfill $\Box$

{\bf Example:} The \db{Unitary group} $U(n) = \{ U \in GL(n, {\mathbb C})\,| $ $U^\dagger U = {\mathbb I} \}\,.$ \hfill $\Box$

{\bf Example:} The \db{Symplectic group} $Sp(n)\,,$ defined as the subgroup of $U(2n)$ given by $A^\trans {\mathbb J} A = {\mathbb J}\,,$ where 
\begin{eqnarray}
{\mathbb J} = 
\left(
\begin{array}{cc}
0 & - {\mathbb I}_{n\times n}\\
{\mathbb I}_{n\times n} & 0
\end{array}
\right)
\nonumber
\end{eqnarray}
\hfill$\Box$

{\bf Example:} $O(p, q) =  \{ R \in GL(p+q, {\mathbb R})\,| $ $R^\trans \eta_{p,q} R = \eta_{p,q} \}\,,$ where 
\begin{eqnarray}
\eta_{p,q} = 
\left(
\begin{array}{cc}
{\mathbb I}_{p\times p} & 0\\
0 & -{\mathbb I}_{q\times q} 
\end{array}
\right)
\nonumber
\end{eqnarray}
\hfill$\Box$

{\bf Example:} $U(p, q) =  \{ U \in GL(p+q, {\mathbb C})\,| $ $U^\dagger \eta_{p,q} U = \eta_{p,q} \}\,.$ 

{\bf Example} The \db{Special Orthogonal group} $SO(n)$ is the subgroup of $O(n)$ for which the determinant is +1. Similarly, the \db{Special unitary group} $SU(n)$ is the subgroup of $U(n)$ with determinant +1. Similarly for $SO(p.q)$ and $SU(p,q)\,.$

The group $U(1)$ is the group of phases $U(1) = \{ e^{i\phi} | \phi\in {\mathbb R} \}\,.$ As a manifold, this is isomorphic to a circle $S^1\,.$ 

The group $SU(2)$ is isomorphic as a manifold to a three-sphere $S^3\,.$ These are the only two spheres (other than the point $S^0$) which admit a Lie group structure. 

An important property of a Lie group is that the tangent space at any point is isomorphic to the tangent space at the identity by an appropriate group operation. Of course, the tangent space at any point of a manifold is isomorphic to the tangent space  at any other point. For Lie groups, the isomorphism between the tangent spaces is induced by group operations, so is in some sense natural. 

For any Lie group $G\,,$ we can define diffeomorphisms of $G$ labelled by elements $g\in G\,,$ called 

\md \db{Left translation} $l_g: G\to G\qquad g' \mapsto gg'\,$; \hfill $\Box$

\md \db{Right translation} $r_g: G\to G \qquad g' \mapsto g'g\,.$ \hfill $\Box$

These can be defined for any group, but are diffeomorphisms for Lie groups. We see that 
\begin{eqnarray}
l_{g\inv}l_g(g') = l_{g\inv}(gg') = g\inv g g' = g' \quad & \Rightarrow & \quad (l_g)\inv = l_{g\inv} \next
r_{g\inv}r_g(g') = r_{g\inv}(g'g) = g' g g\inv = g' \quad & \Rightarrow & \quad (r_g)\inv = r_{g\inv} \,.\qquad 
\end{eqnarray}
It is easy to check that 
\begin{eqnarray}
l_{g_1}l_{g_2} = l_{g_1 g_2} \qquad r_{g_1} r_{g_2} = r_{g_2 g_1}
\end{eqnarray}
Further, $l_{g\inv}(g) = e$ and $r_{g\inv}(g) = e\,,$ so any element of $G$ can be moved to the identity by a diffeomorphism. The tangent space at the identity forms a Lie algebra, as we shall see. The left and right translations lead to diffeomorphisms which relate the tangent space at any point to this Lie algebra, as we shall see now.

\section{Demonstration in Mathematica of Lie algebra properties}
In ref.~[\cite{Ananthanarayan:2024ffu}], detailed discussions have been presented.
For the purposes of completeness, some of the discussions are being reproduced below.
\subsection{\texorpdfstring{$SU(2)$}{SU(2)}}
$SU(2)$ consists of $2\times2$ unitary matrices with determinant equal to $1$.  Let us demonstrate the $SU(2)$ group in two-dimensional matrix representation in terms of Pauli matrices. First, we define the basis and the Lie bracket (commutator).

\begin{lstlisting}[style=in]
ElementName = {Subscript[t, x], Subscript[t, y], Subscript[t, z]};
Elements = {1/2 {{0, 1}, {1, 0}}, 1/2 {{0, -I}, {I, 0}}, 
1/2 {{1, 0}, {0, -1}}};
NameToElem = Thread[ElementName -> Elements];
MatRep[x_] := x /. NameToElem;
ElemToName = Thread[Elements -> ElementName];
NameElem[x_] := x /. ElemToName;
Commutator[x_, y_] := (MatRep[x] . MatRep[y] - MatRep[y] . MatRep[x]);
\end{lstlisting}

Let us check the commutator relations $[t_i,t_j]=i\epsilon_{ijk} t_k$.
\begin{lstlisting}[style=in]
NameElem[Commutator[Subscript[t, x], Subscript[t, y]]/I]
\end{lstlisting}

\begin{lstlisting}[style=out]
$t_z$
\end{lstlisting}

\begin{lstlisting}[style=in]
NameElem[Commutator[Subscript[t, y], Subscript[t, z]]/I]
\end{lstlisting}

\begin{lstlisting}[style=out]
$t_x$
\end{lstlisting}

\begin{lstlisting}[style=in]
NameElem[Commutator[Subscript[t, z], Subscript[t, x]]/I]
\end{lstlisting}

\begin{lstlisting}[style=out]
$t_y$
\end{lstlisting}

It is easy to check the Jacobi identity $[[t_x,t_y],t_z]+[[t_y,t_z],t_x]+[[t_z,t_x],t_y] = 0$  as follows.

\begin{lstlisting}[style=in]
Commutator[Commutator[Subscript[t, x], Subscript[t, y]], Subscript[t, 
 z]] + Commutator[Commutator[Subscript[t, y], Subscript[t, z]], 
 Subscript[t, x]] + 
Commutator[Commutator[Subscript[t, z], Subscript[t, x]], Subscript[t,
  y]]
\end{lstlisting}

\begin{lstlisting}[style=out]
$\{\{0, 0\}, \{0, 0\}\}$
\end{lstlisting}

\subsection{\texorpdfstring{$SO(3)$}{SO(3)}}
The Lie group $SO(3)$ consists of $3\times3$ antisymmetric matrices with unit determinant. The generators are often labeled as $L_x,L_y,L_z$ with commutator relations
\begin{align}
    [L_i,L_j] = \epsilon_{ijk} L_k\,.
\end{align}
One can write Mathematica code using $3x3$ matrices, but it will not be given here.




\subsection{\texorpdfstring{$SU(3)$}{SU(3)}}
$SU(3)$ consists of $3\times3$ unitary matrices with determinant equal to $1$. Elements are obtained by exponentiating $iM$ with $M$ being a traceless $3\times3$ Hermitian matrix. The generators (Gell-Mann matrices) and commutator bracket may be defined as follows in Mathematica.  We suppress the Mathematica code here and provide the
output for the commutation relations of the generators in standard notation.

The output is generated as
\[
\begin{array}{|c|c|c|c|c|c|c|c|c|}
\hline
 \text{} & T_+ & T_- & T_z & U_+ & U_- & V_+ & V_- & \text{Y} \\
 \hline
 T_+ & 0 & 2 T_z & -T_+ & V_+ & 0 & 0 & -U_- & 0 \\
 \hline
 T_- & -2 T_z & 0 & T_- & 0 & -V_- & U_+ & 0 & 0 \\
 \hline
 T_z & T_+ & -T_- & 0 & -\frac{U_+}{2} & \frac{U_-}{2} & \frac{V_+}{2} & -\frac{V_-}{2} & 0 \\
 \hline
 U_+ & -V_+ & 0 & \frac{U_+}{2} & 0 & \frac{3 Y}{2}-T_z & 0 & T_- & -U_+ \\
 \hline
 U_- & 0 & V_- & -\frac{U_-}{2} & T_z-\frac{3 Y}{2} & 0 & -T_+ & 0 & U_- \\
 \hline
 V_+ & 0 & -U_+ & -\frac{V_+}{2} & 0 & T_+ & 0 & T_z+\frac{3 Y}{2} & -V_+ \\
 \hline
 V_- & U_- & 0 & \frac{V_-}{2} & -T_- & 0 & -T_z-\frac{3 Y}{2} & 0 & V_- \\
 \hline
 \text{Y} & 0 & 0 & 0 & U_+ & -U_- & V_+ & -V_- & 0 \\
 \hline
\end{array}
\]

%% file: DiffGeom/DG_identity.tex
\chapter{Tangent space  at the identity}\label{identity}
A point on the Lie group is a group  element. So a vector field on the Lie group selects a vector at each $g\in G\,.$ Since left and right translations are diffeomorphisms, we can consider the pushforwards due to them. 

\md A \db{left}-\db{invariant vector field} $X$ is invariant under left translations, i.e., 
\begin{eqnarray}
X = l_{g*} (X) \qquad\qquad \forall g\in G\,.
\end{eqnarray}
In other words, the vector (field) at $g'$ is pushed forward by $l_g$ to the same vector (field) at $l_g(g')$:
\begin{eqnarray}
l_{g*}(X_{g'}) = X_{gg'}\, \qquad\qquad \forall g, g'\in G\,.
\end{eqnarray}

\md Similarly, a \db{right}-\db{invariant vector field} $X$ is defined by 
\begin{eqnarray}
X &=& r_{g*} (X) \qquad \qquad \forall g\in G\,,\next
{\mathrm i.e.}\qquad r_{g*}(X_{g'}) &=& X_{g'g}\, \qquad\qquad \forall g, g' \in G\,.
\end{eqnarray}

A left or right invariant vector field has the important property that it is completely determined by its value at the identity element $e$ of the Lie group, since 
\begin{eqnarray}
l_{g*}(X_e) = X_g \qquad \qquad \forall g\in G\,,
\end{eqnarray}
and similarly for right-invariant vector fields. 

Write the set of all left-invariant vector fields on $G$ as $L(G)\,.$ Since the push-forward is linear, we get
\begin{eqnarray}
l_{g*}(aX + Y)  = al_{g*}X + l_{g*}Y\,,
\end{eqnarray}
so that if both $X$ and $Y$ are left-invariant, 
\begin{eqnarray}
l_{g*}(aX + Y) = aX + Y\,,
\end{eqnarray}
so the set of left-invariant vector fields forms a real vector space. 

We also know that push-forwards leave the Lie algebra invariant, i.e., for $l_{g*}\,,$ 
\begin{eqnarray}
\comm{l_{g*}X}{l_{g*}Y} = l_{g*}\comm{X}{Y}\,.
\end{eqnarray}
Thus if $X, Y \in L(G)\,,$
\begin{eqnarray}
 l_{g*}\comm{X}{Y} = \comm{l_{g*}X}{l_{g*}Y} = \comm{X}{Y}\,,
\end{eqnarray}
so $\comm{X}{Y} \in L(G)\,.$ Thus the set of all left-invariant vector fields on $G$ forms a Lie algebra. 

\md This $L(G)$ is called the \db{Lie algebra of} $G\,.$ \hfill$\Box$

The dimension of this Lie algebra is the same as that of $G$ because of the 

{\bf Theorem:} $L(G)$ as a real vector space is isomorphic to the tangent space $T_eG$ to $G$ at the identity of $G\,.$ 

{\bf Proof:} We will show that left translation leads to an isomorphism. 

For $X \in T_e G\,,$ define the vector field $L^X$ on $G$ by 
\begin{eqnarray}
L^X\big|_g \equiv L^X_g := l_{g*}X \qquad \qquad \forall g\in G
\end{eqnarray}
Then for all $g, g' \in G\,,$ 
\begin{eqnarray}
{l_{g'*}}(L^X_g) = l_{g'*} (l_{g*} X) = l_{g'g*}X = L^X_{g'g}\,.
\end{eqnarray}
Note that for two diffeomorphisms $\varphi_1, \varphi_2\,,$ we can write
\begin{eqnarray}
(\varphi_{1*}(\varphi_{2*}v))(f) &=& (\varphi_{2*}v)(f\circ\varphi_1) \,\next
&=& v(f\circ\varphi_1\circ\varphi_2) \,\next
&=& ((\varphi_1\circ\varphi_2)_*v)(f) \, \next 
\Rightarrow \qquad \varphi_{1*}(\varphi_{2*}v) &=& (\varphi_1\circ\varphi_2)_*v
\end{eqnarray}
Since left translation is a diffeomorphism, 
\begin{eqnarray}
l_{g'*} (l_{g*} X) = (l_{g'}\circ l_g)_* X  = (l_{g'g*})X
\end{eqnarray}
Thus it follows that $L^X$ is a left-invariant vector field, and we have a map $T_e G\,\to L(G)\,.$ Since the pushforward is a linear map, so is the map $X\to L^X\,.$ We need to prove that this map is 1-1 and onto.  

If $L^X = L^Y\,,$ we have
\begin{eqnarray}
L_g^X = L^Y_g\qquad \forall g\in G\,,
\end{eqnarray}
so 
\begin{eqnarray}
l_{g\inv*}L_g^X = l_{g\inv*}L_g^Y \qquad \Rightarrow \qquad X=Y \quad (\in T_e G)\,.
\end{eqnarray}
So the map $X \to L^X$ is 1-1. 

Now given $L^X\,,$ define $X_e\in T_eG$ by 
\begin{eqnarray}
X_e = l_{g\inv*}L_g^X \qquad \mathrm{for\, any\,} g\in G\,.
\end{eqnarray}
We can also write 
\begin{eqnarray}
X_e = L^X_e\,.
\end{eqnarray}
Then 
\begin{eqnarray}
l_{g*}X_e = l_{g*}l_{g\inv*}L^X_g = L^X_g\,.
\end{eqnarray}
Hence, the map $X \mapsto L^X$ is onto. 

Then we can define a Lie bracket on $T_eG$ by 
\begin{eqnarray}
\comm{u}{v} = \comm{L^u}{L^v}\big|_e\,.
\end{eqnarray}
The Lie algebra of vectors in $T_eG$ based on this bracket is thus the Lie algebra of the group $G\,.$ It follows that 
\begin{eqnarray}
\dim L(G) = \dim T_eG = \dim G\,.
\end{eqnarray}
Note that since commutators are defined for vector fields and not vectors, the Lie bracket on $T_eG$ has to be defined using the commutator of left-invariant vector fields on $G$ and the isomorphism $T_eG \leftrightarrow L(G)\,.$ 

\md If for an $n$-dimensional Lie group $G\,,\, \{t_1\,, \cdots\,, t_n\}$ is a set of basis vectors on $T_eG \simeq L(G)\,,$ the Lie bracket of any pair of these vectors must be a linear combination of them, so 
\begin{eqnarray}
\comm{t_i}{t_j} = \sum\limits_k C_{ij}^k\, t_k
\end{eqnarray}
for some set of real numbers $C_{ij}^k\,.$ These numbers are known as the \db{structure constants} of the Lie group or algebra. \hfill$\Box$

Since $L(G)$ is a Lie algebra, with the Lie bracket as the product, the Lie bracket is antisymmetric,
\begin{eqnarray}
\comm{t_i}{t_j} &=&  - \comm{t_j}{t_i} \nonumber \next
\Rightarrow \qquad \sum\limits_k C_{ij}^k\, t_k &=&  - \sum\limits_k C_{ji}^k t_k \nonumber\next
\Rightarrow \qquad C_{ij}^k &=&  - C_{ji}^k\,,
\end{eqnarray}
and the structure constants satisfy the Jacobi identity
\begin{eqnarray}
\comm{t_i}{\comm{t_j}{t_k}} + \comm{t_j}{\comm{t_k}{t_i}} + \comm{t_k}{\comm{t_i}{t_j}} &=& 0\,\nonumber \next
\Rightarrow \qquad C_{ij}^l\, C_{kl}^m + C_{jk}^l\, C_{il}^m + C_{ki}^l\, C_{jl}^m &=& 0\,.
\end{eqnarray}

A similar construction can be done using a set of right-invariant vector fields defined by 
\begin{eqnarray}
R_g^X := r_{g*} X\, \qquad \mathrm{for}\, X \in T_eG\,
\end{eqnarray}
and its `inverse' $X_e = r_{g\inv*}R^X_g\,.$

%% file: DiffGeom/DG_1ps.tex
\chapter{One parameter subgroups}\label{1ps}
There is another characterization of $T_eG$ for a Lie group $G$ as the set of its one parameter subgroups, which we will now define. This is also called the ``infinitesimal'' description of a Lie group, and what Lie called an infinitesimal group. 

\md A \db{one parameter subgroup}  of a Lie group $G$ is a smooth homomorphism from the additive group of real numbers to $G\,, \gamma:(\mathbb{R}, +) \to G\,.$ Then $\gamma:\mathbb{R}\to G$ is a curve such that $\gamma(s+t) = \gamma(s)\gamma(t)\,, \gamma(0) = e\,,$ and $\gamma(-t) = \gamma(t)\inv\,.\hfill \Box$

Also, since this is a homomorphism, the one parameter subgroup is Abelian.

{\bf Example:} For $G = (\mathbb{R}\backslash\{0\}, \times)$ the multiplicative group of non-zero real numbers, $\gamma(t) = e^t$ is a 1-p subgroup. 

{\bf Example:} $G= U(1)\,, \qquad \gamma(t) = e^{it}\,.$

{\bf Example:} $G = SU(2)\,, \qquad \gamma(t) = \left(
\begin{array}{rr}
\cos t & \sin t \\
-\sin t & \cos t
\end{array}
\right)\,.
$

{\bf Example:} $ G = GL(3, \mathbb{R})\,, \qquad \gamma(t) = \left(
\begin{array}{rrr}
\cos t & \sin t & 0\\
-\sin t & \cos t & 0 \\
0 & 0 & e^t
\end{array}
\right)\,.
$

The relation between 1-p subgroups and $T_eG$ is given by the 

{\bf Theorem:} The map $\gamma \mapsto \dot\gamma(0) = \dot\gamma\big|_e\,$ defines a 1-1 correspondence between 1-p subgroups of $G$ on the one hand, and $T_eG$ on the other.

{\bf Proof:} For any $X\in T_eG\,$ define $L^X = l_{g*}X\,$ as the corresponding left-invariant vector field. We need to find a smooth homomorphism from $\mathbb{R}$ to $G$ using $L^X\,.$ This homomorphism is provided by the flow or integral curve of $L^X\,,$ but let us work this out in more detail.

Denote the integral curve of $L^X$ by $\gamma^X(t)\,,$ i.e., 
\begin{eqnarray}
&&\gamma^X(0) = L^X_e = X \next
\mathrm{and}\qquad && \gamma^X(t) = L^X_{\gamma(t)} = l_{\gamma(t)*}X\,.
\end{eqnarray}
Since $L^X$ is left-invariant, $l_{g'*}L^X_g = L^X_{g'g}\,.$ Consider the equation 
\begin{eqnarray}
\frac{d}{dt}\gamma(t) = L^X_{\gamma(t)} = l_{\gamma(t)*}X \equiv \gamma_*\left.\left(\frac{d}{dt}\right)\right|_t\,.
\label{1ps.curve}
\end{eqnarray}
Given some $\tau\,,$ replace $\gamma(t)$ by $\gamma(\tau + t)$ to get
\begin{eqnarray}
\gamma(\tau + t) = l_{\gamma(\tau + t)*} X\,.
\end{eqnarray}
Recall that $\gamma(t)$ is an element of the group for each $t\,.$ Now replace $\gamma(t)$ in Eq.~(\ref{1ps.curve}) by $\gamma(\tau)\gamma(t)$ to get 
\begin{eqnarray}
\big(\gamma(\tau)\gamma(t)\big)_*\left(\frac{d}{dt}\right) &=& L^X_{\gamma(\tau)\gamma(t)} 
\end{eqnarray}
We see that $\gamma(t+\tau)$ and $\gamma(\tau)\gamma(t)$ are both integral curves of $L^X$\,, i.e. both satisfy the equation of the integral curve of $L^X\,,$ and at $t=0$ both curves are at the point $\gamma(\tau)\,.$ Thus by uniqueness these two are the same curve, 
\begin{eqnarray}
\gamma(\tau + t) = \gamma(\tau)\gamma(t)\,,
\end{eqnarray}
and $t\mapsto\gamma(t)$ is the homomorphism $\mathbb{R} \to G$ that we are looking for.

Thus for each $X\in T_eG$ we find a 1-p subgroup $\gamma(t)$ given by the integral curve of $L^X\,,$ 
\begin{eqnarray}
\dot\gamma(t) = L^X_{\gamma(t)} = l_{\gamma(t)*}X \,,
\end{eqnarray}
where, as mentioned earlier, $(\dot\gamma(0) = X)\,.$ \hfill$\Box$

In a compact connected Lie group $G\,,$ every element lies on some 1-p subgroup. This is not true in a non-compact $G\,,$ i.e. there are elements in $G$ which do not lie on a 1-p subgroup. However, an Abelian non-compact group will always have a 1-p subgroup, so this remark applies only to non-Abelian non-compact groups. 

For matrix groups, every 1-p subgroup is of the form 
\begin{eqnarray}
\gamma(t) = \left\{e^{tM}\, \Big| \, M \,\mathrm{fixed}, \, t\in \mathbb{R}\right\}\,.
\end{eqnarray}
Let us see why. Suppose $\{\gamma(t)\}$ is a 1-p subgroup of the matrix group. Then $\gamma(t)$ is a matrix for each $t\,,$ and
\begin{eqnarray}
\gamma(s)\gamma(t) = \gamma(s+t)\,.
\end{eqnarray}
Differentiate with respect to $s$ and set $s=0\,.$ Then 
\begin{eqnarray}
\dot\gamma(0) \gamma(t) = \dot\gamma(t) \,.
\end{eqnarray}
Write $\dot\gamma(0) = M\,.$ Since $G$ is a matrix group, $M$ is a matrix. Then the unique solution for $\gamma$ is 
\begin{eqnarray}
\gamma(t) = e^{tM}\,.
\end{eqnarray}
The properties of $M$ are determined by the properties of the group and vice versa; not every matrix $M$ will generate any group. 

The allowed matrices $\{M\}$ for a given group $G$ are the $\{\dot\gamma(0)\}$ for all the 1-p subgroups $\gamma(t)\,,$ so these are in fact the tangent vectors at the identity. The allowed matrices $\{M\}$ for a given matrix group $G$ thus form a Lie algebra with the Lie bracket being given by the matrix commutator. This Lie algebra is isomorphic to the Lie algebra of the group $G\,.$ (We will not give a proof of this here.)

We can find the Lie algebra of a matrix group by considering elements of the form $\gamma(t) = e^{tM}$ for small $t\,,$ i.e., 
\begin{eqnarray}
\gamma(t) = \mathbb{I} + tM 
\end{eqnarray}
for small $t\,.$ Conversely, once we are given, or have found, a Lie algebra with basis $\{t_i\}\,,$ we can exponentiate the Lie algebra to find the set of 1-p subgroups
\begin{eqnarray}
\left\{\gamma(a) = \exp a^i t_i \right\}
\end{eqnarray}
\md This is the \db{infinitesimal group}. For compact connected groups this is identical to the Lie group itself. So in such cases, the entire group can be generated by exponentiating the Lie algebra. Non-compact groups cannot be written as the exponential of the Lie algebra in general. This follows from the statement that there are elements in a non-compact group which do not lie on a 1-p subgroup. \hfill $\Box$

{\bf Example:} Consider $SO(N)\,,$ the group of $N\times N$ real orthogonal  matrices $R$ with $R^\trans R = \mathbb{I}\,, \det{R} = 1\,.$ Write $R = \mathbb{I} + A\,,$ then $A^\trans = -A\,,$ i.e. the Lie algebra is spanned by $N\times N$ real antisymmetric matrices. Let us construct a basis for this algebra. 

An $N\times N$ antisymmetric matrix has $N(N-1)/2$ independent elements. So we define $N(N-1)/2$ independent antisymmetric matrices, labelled by $\mu, \nu = 1, \cdots, N\,,$
\begin{eqnarray}
M_{\mu\nu} &=& - M_{\nu\mu}\, \qquad \mu\,,\nu\, \mathrm{are\, not\, matrix\, indices}\, \next
\left(M_{\mu\nu}\right)_{\rho\sigma} &=& -\left(M_{\mu\nu}\right)_{\sigma\rho}\,, \qquad \rho\,, \sigma\, \mathrm{are\, matrix\, indices}\,.
\end{eqnarray}
A convenient choice for the basis is given by
\begin{eqnarray}
\left(M_{\mu\nu}\right)_{\rho\sigma} = \delta_{\mu\rho}\delta_{\nu\sigma} - \delta_{\mu\sigma}\delta_{\nu\rho}\,.
\end{eqnarray}
Then the commutators are calculated to be
\begin{eqnarray}
\comm{M_{\mu\nu}}{M_{\alpha\beta}} = \delta_{\nu\alpha}M_{\mu\beta} - \delta_{\mu\alpha}M_{\nu\beta} + \delta_{\mu\beta}M_{\nu\alpha} - \delta_{\nu\beta}M_{\mu\alpha}\,.
\end{eqnarray}
This defines the Lie algebra. 

{\bf Example:} For $SU(N)\,,$ the group of $N\times N$ unitary matrices $U$ with $U^\dagger U = \mathbb{I}\,,\, \det{U}= 1\,,$ the 1-p subgroups are given by $\gamma(t) = e^{tM}$ with $M^\dagger + M = 0$ in the same way as above, and $\det(\mathbb{I} + tM) = 1 \Rightarrow \Tr M = 0\,.$ So the $SU(N)$ Lie algebra consists of traceless antihermitian matrices. Often the basis is multiplied by $i$ to write $\gamma(a) = \exp(i a_j t_j)\,,$ where $t_j$ are now Hermitian matrices, with 
\begin{eqnarray}
\comm{t_i}{t_j} = i f_{abc} t_c\,.
\end{eqnarray}
Examples of exponentiation in Mathematica:
\begin{lstlisting}[style=in]
M = {{0, -1}, {1, 0}}
\end{lstlisting}
\begin{lstlisting}[style=out]
{{0,-1},{1,0}}
\end{lstlisting}
This can be thought of as a basis matrix in the $sl(2,\mathbb{R})$ algebra, which consists of traceless real $2\times 2$ matrices.
\begin{lstlisting}[style=in]
Sum[ MatrixPower[M, n] t^n/n!, {n, 0, Infinity}] // FullSimplify
\end{lstlisting}
\begin{lstlisting}[style=out]
{{Cos[t], -Sin[t]}, {Sin[t], Cos[t]}}
\end{lstlisting}
There are other examples of exponentiation in the corresponding notebook named \texttt{rotations}.

An example of an element of $SL(2,\mathbb{R})$ which does not lie on a 1-p subgroup is the matrix 
\begin{align*}
\begin{pmatrix}
-2 & 0 \\
0 & -\frac{1}{2}
\end{pmatrix}\,.
\end{align*}
This cannot be written as the exponential (with a real parameter) of a matrix in the $sl(2,\mathbb{R})$ algebra.

%% file: DiffGeom/DG_FB_Conn_Curv.tex
\chapter{Fiber bundles, Connections \& Curvature}\label{FB_Conn_Curv}

\input{DiffGeom/DG_fiber}
\input{DiffGeom/DG_connection}

\input{DiffGeom/DG_curvature}

%% file: DiffGeom/DG_fiber.tex

\section{Fiber bundles}\label{fiber}

Consider a manifold $\cm$ with the tangent bundle $T\cm =  \bigcup\limits_{P \in \cm} T_P\cm\,.$ Let us look at this more closely. $T\cm$ can be thought of as the original manifold $\cm$ with a tangent space stuck at each point $P\in \cm\,.$ Thus there is a projection map $\pi: T\cm \to\cm\,, \quad T_P\cm \mapsto P\,,$ which associates the point $P\in \cm$ with $T_P\cm\,.$ 

Then we can say that $T\cm$ consists of points $P\in \cm$ and vectors $v\in T_P\cm$ as an ordered pair $(P, v_{_P})\,.$ Then in the neighbourhood of any point $P\,,$ we can think of $T\cm$ as a \db{product manifold}, i.e. as the set of ordered pairs $(P, v_{_P})\,.$

This is generalized to the definition of a \db{fiber bundle}. Locally a fiber bundle is a product manifold $E = B\times F$ with the following properties.

\md $B$ is a manifold called the \db{base manifold}, and $F$ is another manifold called the \db{typical fiber} or the \db{standard fiber}.

\md There is a projection map $\pi: E \to B\,,$ and if $P\in B\,,$ the pre-image $\pi\inv(P)$ is homeomorphic, i.e. bicontinuously isomorphic, to the standard fiber. \hfill$\Box$

$E$ is called the \db{total space}, but usually it is also called the bundle, even though the bundle is actually the triple $(E, \pi, B)\,.$ 

\md $E$ is locally a product space. We express this in the following way. Given an open set $U_i$ of $B\,,$  the pre-image $\pi\inv(U_i)$ is homeomorphic to $U_i\times F\,,$ or in other words there is a bicontinuous isomorphism $\varphi_i: \pi\inv(U_i) \to U_i\times F\,.$ The set $\left\{U_i\,, \varphi_i\right\}$ is called a \db{local trivialization} of the bundle.  \hfill$\Box$

\md If $E$ can be written globally as a product space, i.e. $E = B \times F\,,$ it is called a \db{trivial bundle}.\hfill$\Box$

\md This description includes a homeomorphism $\pi\inv(P) \to F$ for each $P \in U_i\,.$ Let us denote this map by $h_i(P)\,.$ Then at $P$ in some overlap $U_i\cap U_j$, the fiber $\pi^{-1}(P)$ has  homeomorphisms $h_i(P)$ and $h_j(P)$ onto $F\,.$ It follows that $h_j(P)\cdot h_i(P)\inv$ is a homeomorphism $F\to F\,.$ These are called \db{transition functions}. The transition functions $F\to F$ form a group, called the \db{structure group} of $F\,.\, \hfill \Box$

Let us consider an example. Suppose $B= S^1\,.$ Then the tangent bundle $E = TS^1$ has $F = \mathbb{R}$ and $\pi(P, v) \mapsto P\,,$ where $P \in S^1\,, v\in TS^1\,.$ Consider a covering of $S^1$ by open sets $U_i\,,$ and let the coordinates of $U_i \subset S^1$ be denoted by $\lambda_i\,.$ Then any vector in $T_P S^1$ can be written as $\displaystyle{v = a_i \frac{d}{d\lambda_i}}$ (no sum) for $P \in U_i\,.$ 

So we can define a homeomorphism $h_i(P): T_P S^1 \to \mathbb{R}, v \mapsto a_i$ (fixed $i$). If $P\in U_i \cap U_j$ there are two such homeomorphisms $TS^1 \to \mathbb{R}\,,$ and since $\lambda_i$ and $\lambda_j$ are independent, $a_i$ and $a_j$ are also independent. 

Then $h_i(P)\cdot h_j(P)\inv : F \to F$ (or $\mathbb{R}\to \mathbb{R}$) maps $a_j$ to $a_i\,.$ The homeomorphism, which in this case relates the component of the vector in two coordinate systems, is simply multiplication by the number $r_{ij} = \dfrac{a_i}{a_j} \in \mathbb{R}\backslash\{0\}\,.$ So the structure group is $\mathbb{R}\backslash\{0\}$ with multiplication.

For an $n$-dimensional  manifold $\cm\,,$ the structure group of $T\cm$ is $GL(n, \mathbb{R})\,.$ 

\md A fiber bundle where the standard fiber is a vector space is called a \db{vector bundle}\,.$\hfill\Box$

A cylinder can be made by glueing two opposite edges of a flat strip of paper. This is then a Cartesian product of a circle $S^1$ with a line segment $I\,.$ So $B=S^1\,, F = I$ and this is a trivial bundle, i.e. globally a product space. On the other hand, a M\"obius strip is obtained by twisting the strip and then glueing. Locally for some open set $U\subsetneq S^1$ we can still write a segment of the M\"obius strip as $U\times I\,,$ but the total space is no longer a product space. As a bundle, the M\"obius strip is non-trivial.


\md Given two bundles $(E_1, \pi_1, B_1)$ and $(E_2, \pi_2, B_2)\,,$ the relevant or useful maps between these are those which preserve the bundle structure locally, i.e. those which map fibers into fibers. They are called \db{bundle morphisms}. \hfill $\Box$

A bundle morphism is a pair of maps $(F, f)\,, F : E_1 \to E_2\,, f: B_1 \to B_2\,,$ such that $\pi_2\circ F = f\circ \pi_1\,.$ 

Not all systems of coordinates are appropriate for a bundle. But it is possible to define a set of \db{fiber coordinates} in the following way. Given a differentiable fiber bundle with $n$-dimensional base manifold $B$ and $p$-dimensional fiber $F\,,$ the coordinates of the bundle are given by bundle morphisms onto open sets of $\rn\times \mathbb{R}^p\,.$ \hfill$\Box$

\md Given a manifold $\cm$ with tangent space $T_P\cm\,,$ consider $A_P = (e_1, \cdots, e_n)\,,$ a set of $n$ linearly independent vectors at $P\,.$ $A_P$ is a basis in $T_P\cm\,.$ The typical fiber in the \db{frame bundle} is the set of all bases, $F = \left\{A_P\right\}\,. \hfill \Box$

Given a particular basis $\bar A_P = (\bar e_1, \cdots, \bar e_n)\,,$ any basis $A_P$ may be expressed as 
\begin{eqnarray}
e_i = a^j_i \bar e_j\,.
\end{eqnarray}
The numbers $a^j_i$ can be thought of as the components of a matrix, which must be invertible so that we can recover the original basis from the new one. Thus, starting from any one basis, any other basis can be reached by an $n\times n$ invertible matrix, and any $n\times n$ invertible matrix produces a new basis. So there is a bijection between the set of all frames in $T_P\cm$ and $GL(n\,, \mathbb{R})\,.$ 

Clearly the structure group of the typical fiber of the frame bundle is also $GL(n\,, \mathbb{R})\,.$

\md  A fiber bundle in which the typical fiber $F$ is identical (or homeomorphic) to the structure group $G$, and $G$ acts on $F$ by left translation is called a \db{principal fiber bundle}. \hfill $\Box$

{\bf Example:} 

\begin{itemize}
    \item  Typical fiber = $S^1$\,, structure group $U(1)$\,.
    \item  Typical fiber = $S^3$\,, structure group $SU(2)$\,.
    \item 
    Starting from any frame we can reach any other frame by a $GL(n\,, \mathbb{R})\,$ transformation, so the fiber is isomorphic (homeomorphic) to $GL(n\,, \mathbb{R})\,,$ which is also the structure group. Thus the frame bundle is a principal $GL(n\,, \mathbb{R})\,$ bundle. But an initial frame needs to be specified.
\end{itemize}

\md A \db{section} of a fiber bundle $(E, \pi, B)$ is a mapping $s: B \to E\,, p\mapsto s(p)\,,$ where $p\in B\,, s(p) \in \pi\inv(p)\,.$ So we can also say $\pi \circ s$ = identity. \hfill $\Box$

{\bf Example:} A vector field is a section of the tangent bundle,  $v: P\mapsto v_{_P}$\,. 

{\bf Example:} A function on $\cm$ is a section of the bundle which locally looks like $\cm\times\mathbb{R}$ (or $\cm\times\mathbb{C}$ if we are talking about complex functions). 

\md Starting from the tangent bundle we can define the \db{cotangent bundle}, in which the typical fiber is the dual space of the tangent space. This is written as $T^*\cm\,.$ As we have seen before, a section of $T^*\cm$ is a 1-form field on $\cm$\,. \hfill$\Box$

\md Recall that a \db{vector bundle} $F \to E \buildrel{\pi}\over{\rightarrow} B$ is a bundle in which the typical fiber $F$ is a vector space. \hfill$\Box$

\md A vector bundle $(E, \widetilde\pi, B, F, G)$ with typical fiber $F$ and structure group $G$ is said to be \db{associated} to the principal bundle $(P, \pi, B, G)$ by the representation $\left\{D(g)\right\}$ of $G$ on $F$ if its transition functions are the images under $D$ of the transition functions of $P$\,. \hfill$\Box$

Stated differently, suppose that we have a covering $\left\{U_i\right\}$ of $B$\,, and the local trivialization of $P$ with respect to this covering is $\Phi_i: \pi\inv(U_i) \to U_i\times G\,,$ which is essentially the same as writing $\Phi_{i,x}: \pi\inv(x) \to G\,, \; x\in U_i$\,. Then the transition functions of $P$ are of the form 
\begin{eqnarray}
g_{ij} = \Phi_i\circ \Phi_j\inv : U_i\cap U_j \to G\,.
\end{eqnarray}
The transition functions of $E$ corresponding to the same covering of $B$ are given by $\phi_i : \widetilde\pi\inv(U_i) \to U_i\times F$ with $\phi_i\circ \phi_j\inv = D(g_{ij})\,.$ That is, if $v_i$ and $v_j$ are images of the same vector $v_x\in F_x$ under overlapping trivializations $\phi_i$ and $\phi_j\,,$ we must have
\begin{eqnarray}
v_i = D\left(g_{ij}(x)\right) v_j\,.
\end{eqnarray}
A more physical way of saying this is that if two observers look at the same vector at the same point, their observations are related by a group transformation $(p, v) \simeq \left(p, D(g_{ij}v\right)$\,.

\md These relations  are called \db{gauge transformations} in physics, and $G$ is called the \db{gauge group}. Usually $G$ is a Lie group for reasons of continuity. \hfill$\Box$

Fields appearing in various physical theories are sections of vector bundles, which in some trivialization look like $U_\alpha\times V$ where $U_\alpha$ is some open neighborhood of the point we are interested in, and $V$ is a vector space. $V$ carries a representation of some group $G$\,, usually a Lie group, which characterizes the theory.

To discuss this a little more concretely, let us consider an associated vector bundle $(E, \widetilde\pi, B, F, G)$ of a principal bundle $(P, \pi, B, G)\,.$ Then the transition functions are in some representation of the group $G\,.$ Because the fiber carries a representation $\left\{D(g)\right\}$ of $G$\,, there are always linear transformations $T_x: E_x \to E_x$ which are members of the representation $\left\{D(g)\right\}$\,. Let us write the space of all sections of this bundle as $\Gamma(E)$\,. An element of $\Gamma(E)$ is a map from the base space to the bundle. Such a map assigns an element of $V$ to each point of the base space. 

\md We say that a linear map $T: \Gamma(E) \to \Gamma(E)$ is a \db{gauge transformation} if at each point $x$ of the base space, $T_x \in \left\{D(g)\right\}$ for some $g$\,, i.e. if
\begin{eqnarray}
T_x: (x, v)_\alpha \mapsto (x, D(g)v)_\alpha\,,
\end{eqnarray}
for some $g\in G$ and for $(x, v)_\alpha \in U_\alpha\times F\,.$ In other words, a gauge transformation is a representation-valued linear transformation of the sections at each point of the base space.  The right hand side is often written as $(x, gv)_\alpha$\,. \hfill$\Box$

This definition is independent of the choice of $U_\alpha\,.$ To see this, consider a point $x\in U_\alpha\cap U_\beta\,.$ Then 
\begin{eqnarray}
(x, v)_\alpha = (x, g_{\beta\alpha} v)_\beta\,.
\label{fiber.gv}
\end{eqnarray}
 In the other notation we have been using, $v_\alpha$ and $v_\beta$ are images of the same vector $v_x\in V_x$\,, and $v_\beta = D(g_{\beta\alpha})v_\alpha\,.$ A gauge transformation $T$ acts as 
\begin{eqnarray}
T_x: (x, v)_\alpha \mapsto (x, gv)_\alpha\,.
\end{eqnarray}
But we also have
\begin{eqnarray}
(x, gv)_\alpha = (x, g_{\beta\alpha}gv)_\beta
\label{fiber.gv2}
\end{eqnarray}
using Eq.~(\ref{fiber.gv})\,. So it is also true that 
\begin{eqnarray}
T_x: (x, g_{\beta\alpha}v)_\beta \mapsto (x, g_{\beta\alpha}gv)_\beta\,.
\label{fiber.Tx}
\end{eqnarray}

Since $F$ carries a representation of $G\,,$ we can think of $gv$ as a change of variables, i.e. define $v' = g_{\beta\alpha}v$\,. Then  Eq.~(\ref{fiber.Tx}) can be written also as 
\begin{eqnarray}
T_x: (x, v')_\beta \mapsto (x, g'v')_\beta \,,
\end{eqnarray}
where now $g' = g_{\beta\alpha} g g_{\beta\alpha}\inv\,.$ So $T$ is a gauge transformation in $U_\beta$ as well. The definition of a gauge transformation is independent of the choice of $U_\alpha\,,$ but $T$ itself is not. The set of all gauge transformations ${\mathscr G}$ is a group, with 
\begin{eqnarray}
(gh)(x) = g(x) h(x)\,, \quad (g\inv)(x) = \left(g(x)\right)\inv\,.
\end{eqnarray}

\md The groups $G$ and ${\mathscr G}$ are both called the \db{gauge group} in different treatments. \hfill$\Box$

%% file: DiffGeom/DG_connection.tex
\section{Connections}\label{conn}
There is no canonical way to differentiate sections of a fiber bundle. That is to say, no unique derivative arises from the definition of a bundle. Let us see why. The usual derivative of a function on $\mathbb{R}$ is of the form 
\begin{eqnarray}
f'(x) = \lim\limits_{\epsilon\to 0} \frac{f(x + \epsilon) - f(x)}{\epsilon}\,.
\end{eqnarray}
But a section of a bundle assigns an element of the fiber at any point $P \in \cm$ to the base point $P$. Each fiber is isomorphic to the standard fiber but the isomorphism is not canonical or unique. So there is no unique way of adding or subtracting points on different fibers. Thus there are many ways of differentiating sections of fiber bundles. 

\md Each way of taking a derivative, i.e. of comparing, is called a \db{connection}. Let us consider a bundle $\pi: E \to B\,,$ where $\Gamma(E)$ is the space of all sections. Then \db{a connection} $D$ on $B$ assigns to every vector field $v$ on $B$ a map $D_v: \Gamma(E) \to \Gamma(E)$ satisfying 
\begin{eqnarray}\label{eqn_conn_prop}
D_v(s + \alpha t) &=& D_vs +\alpha\, D_v t \nonumber \\
D_v(fs) &=& v(f)s + f D_v s \nonumber \\
D_{v+ fw} (s) &=& D_v s + f D_w s\,,
\end{eqnarray}
where $s, t $ are sections of the bundle, $s, t \in \Gamma(E)$\,, $v, w$ are vector fields on $B$\,, $f \in C^\infty(B)$ and $\alpha$ is a real number (or complex, depending on what the manifold is). \hfill $\Box$

Note that this is \db{a connection}, not some unique connection. In other words, we need to choose a connection before we can talk about it. In what follows, whenever we refer to a connection $D$\,, we mean that we have chosen a connection $D$ and that is what we are discussing.

\md We call $D_v s$ the \db{covariant derivative} of $s$\,. \hfill $\Box$

To be specific, let us consider the bundle to be a vector bundle on a manifold $\cm$\,, and try to understand the meaning of $D$ by going to a chart. Consider coordinates $x^\mu$ in an open set $U \subset \cm\,,$ with $\partial_\mu$ the coordinate basis vector fields. Write $D_\mu = D_{\partial_\mu}\,.$ Also choose a basis for sections, which is like a basis for the fiber (vector space) at each point of $\cm$\,, i.e. like a set of basis vector fields, but the vectors are not along $\cm$\,, but along the fiber at $\cm$\,.

Call this basis $\left\{e_i\right\}$\,; then $\left\{e_i(x)\right\}$ is a basis for the fiber at $P \in \cm\,,$ with $\{x\}$ being the set of coordinates at $P\,.$ Any element of $V \simeq F_x$ can be written uniquely as a linear combination of $e_i(x)$\,. But then $D_\mu e_j$ can be expressed uniquely as a linear combination of the $e_i$\,,
\begin{eqnarray} \label{eqn_conn_coeff}
D_\mu e_j = A_{\mu j}^{\phantom{\mu j}i} e_i\,.
\end{eqnarray}
\md These $A_{\mu j}^i $ are called \db{components} of the \db{vector potential} or the \db{connection one}-\db{form}. \hfill $\Box$

Given a section $s = s^i e_i$ with $s^i \in C^\infty(\cm)\,,$ we can write 
\begin{eqnarray}
D_v s = D_{v^\mu\partial_\mu} s = v^\mu D_\mu s\,.
\end{eqnarray}
Also, 
\begin{eqnarray}
D_\mu s = D_\mu(s^i e_i) &=&\left (\partial_\mu s^i\right) e_i + s^i D_\mu e_i \nonumber \\
&=& \left(\partial_\mu s^i\right) e_i + s^i A_{\mu i}^{\phantom{\mu i}j}e_j \nonumber \\
&=& \left(\partial_\mu s^i + A_{\mu j}^{\phantom{\mu j}i} s^j \right)e_i\,,
\end{eqnarray}
so writing $D_\mu s = \left(D_\mu s\right)^i e_i$\,, we can say
\begin{eqnarray}
\left(D_\mu s\right)^i = \partial_\mu s^i +  A_{\mu j}^{\phantom{\mu j}i} s^j\,.
\end{eqnarray}
We have considered connections on an associated vector bundle, which may be a principal fiber bundle such as a frame bundle. So we should be able to talk about gauge transformations. 

Remember that a gauge transformation is a linear map $T: E \to E\,, (x, v) \mapsto (x, gv)$ for some $g\in G$ and for all $v \in V \simeq F_x\,.$ Let us apply this idea to the section $s$\,. We claim that given a connection $D$\,, there is a connection $D'$ on $E$ such that 
\begin{eqnarray}
D' (g\phi) = g D_v \phi\,,
\end{eqnarray}
where $v$ is a vector field on $\cm$ and $\phi \in \Gamma(E)$ (i.e. $\phi = s$). 

Let us first check if the definition makes sense. Since $g(x) \in G$ for all $x \in \cm\,,$ we know that $g\inv(x)$ exists for all $x$. So 
\begin{eqnarray}
D'_v(\phi) &=& D'_v \left(g g\inv \phi\right) \nonumber \\
&=& g D_v \left( g\inv \phi\right)\,,
\end{eqnarray}
and thus $D'$ is defined on all $\phi$ for which $D_v\phi$ is defined, i.e. $D'$ exists because $D$ does. We have of course assumed that $g(x)$ is differentiable as a function of $x$.

Let us now check that $D'$ is a connection according to our definitions. $D'$ is linear since
\begin{eqnarray}
D'_v\left(\phi_1 + \alpha\phi_2\right) &=& g D_v\left( g\inv \left(\phi_1 + \alpha\phi_2\right)\right) \nonumber \\
&=& g D_v\left( g\inv \phi_1 \right) + \alpha g D_v\left(g\inv \phi_2\right) \nonumber \\
&=& D'_v\phi_1 + \alpha D'_v\phi_2\,.
\end{eqnarray}
And it satisfies the Leibniz rule because 
\begin{eqnarray}
D'_v (f\phi) &=& gD_v \left(g\inv f\phi\right) \nonumber \\
&=& g D_v\left(f\left(g\inv \phi\right)\right) \nonumber \\
&=& g v(f) g\inv\phi + gf D_v\left(g\inv\phi\right) \nonumber \\
&=& v(f)\phi + f g D_v\left(g\inv\phi\right) \nonumber \\
&=& v(f) \phi + f D'_v(\phi)\,.
\end{eqnarray}
Similarly,
\begin{eqnarray}
D'_{v + \alpha w} \phi &=& g D_{v + \alpha w}\left(g\inv \phi \right) \nonumber \\
&=& g\left( D_v\left(g\inv\phi\right) + \alpha D_w\left(g\inv\phi\right)\right) \nonumber \\
&=& g D_v(g\inv\phi) + \alpha g D_w\left(g\inv\phi\right) \nonumber \\
&=& D'_v \phi + \alpha D'_w\phi\,.
\end{eqnarray}
So $D'$ is a connection, i.e. there is a connection $D'$ satisfying $D'_v(g\phi) = g\left(D_v\phi\right)\,.$ 

Since $\phi$ is a section, i.e. $\phi \in \Gamma(E)$, so is $g\inv \phi$, and thus $D_v\left(g\inv\phi\right) \in \Gamma(E)$ and also $gD_v\left(g\inv\phi\right) \in \Gamma(E)\,.$ Therefore, $D'_v$ maps sections to sections, $D'_v: \Gamma(E) \to \Gamma(E)\,.$ This completes the definition of the gauge transformation of the connection. We can now write 
\begin{eqnarray}
D'_\mu\phi = \left(\partial_\mu\phi^i + A^{\prime \phantom{i}i}_{\mu\phantom{i}j}\,\phi^j\right) e_i\,.
\end{eqnarray}

Using the dual space, let us write
\begin{eqnarray}
A_\mu &=&  A^{\phantom{\mu}i}_{\mu\phantom{i}j}\, e_i \otimes \theta^j\,, \nonumber \\
A'_\mu &=& A^{\prime \phantom{i}i}_{\mu\phantom{i}j}\, e_i \otimes \theta^j\,,
\end{eqnarray}
where $\left\{\theta^i\right\}$ is the dual basis to $\left\{e_i\right\}$\,. The gauge transformation is then given by
\begin{eqnarray}
D'_v\phi &=& g D_v\left(g\inv \phi\right) \nonumber \\
\Rightarrow \qquad \qquad D'_\mu\phi &=& g D_\mu\left(g\inv\phi\right) \nonumber \\
\Rightarrow \qquad \left(\partial_\mu\phi^i + A^{\prime \phantom{i}i}_{\mu\phantom{i}j}\right) e_i
&=& g \left[\partial_\mu\left(g\inv\phi\right)^i + A^{\phantom{\mu}i}_{\mu\phantom{i}j}\left(g\inv\phi\right)^j \right] e_i \,, \nonumber \\
\end{eqnarray}
where, as always, the $g$'s are in some appropriate representation of $G$. 
Then we can write the right hand side as 
\begin{eqnarray}
\left[\partial_\mu\phi^i +\left(g\partial_\mu g\inv\right)^i_j \phi^j + \left(g A_\mu g\inv\right)^i_j \phi^j\right] e_i \nonumber \\
 \qquad \qquad = \left[ \partial_\mu\phi^i + \left(g\partial_\mu g\inv + g A_\mu g\inv\right)^i_j  \phi^j\right] e_i\,.
\end{eqnarray}
From this we can read off
\begin{eqnarray}
A'_\mu = gA_\mu g\inv + g\partial_\mu g\inv\,.
\end{eqnarray}
\md A connection which transforms like this is also called a $G$-\db{connection}. \hfill $\Box$

{\bf Example:} Consider $G = U(1)$\,. Suppose $E$ is a trivial complex line bundle over $\cm\,,$ i.e. $E = \cm \times {\mathbb C}$\,, so that the fiber over any point $p\in \cm $ is ${\mathbb C}$\,. A connection $D$ on $E$ may be written as $D_\mu = \partial_\mu + A_\mu\,.$ We can make $E$ into a $U(1)$ bundle by thinking of the fiber ${\mathbb C}$ as the fundamental representation space of $U(1)$\,. Then sections are complex functions, and a gauge transformation is multiplication by a phase. \hfill$\Box$

%% file: DiffGeom/DG_curvature.tex

\section{Introduction to Curvature}
We start with a connection $D\,,$ two vector fields $v, w$ on $B\,,$ and a section $s\,,$ all on some associated vector bundle of some principal $G$-bundle $E\,.$ Then $D_v\,, D_w$ are both maps $\Gamma(E) \to \Gamma(E)\,.$

We will define the curvature of this connection $D$ as a rule $F$ which, given two vector fields $v\,, w\,,$ produces a linear map $F(v, w): \Gamma(E) \to \Gamma(E)$ by 
\begin{equation}  \label{eqn_curvature}
F(v, w) s = D_v D_w s - D_w D_vs - D_{[v, w]} s\,.
\end{equation}
Recall that 
\begin{align}
D_v s &= D_{v^\mu\partial_\mu} s = v^\mu D_\mu s \nonumber \\
&= v^\mu \left( \partial_\mu s^i + A^{\phantom{\mu}i}_{\mu\phantom{i}j}\,s^j \right) e_i \nonumber \\
& = v(s^i) e_i + v^\mu A^{\phantom{\mu}i}_{\mu\phantom{i}j}\, s^j e_i \,.
\end{align}
$D_\mu s$ is again a section. As a result, we can act with $D$ on it and write 
\begin{align}
D_v D_w s 
&= D_v \left[ w(s^j) e_j + \left(w^\nu A^{\phantom{\nu}j}_{\nu\phantom{j}k}\, s^k \right) e_j\right] \nonumber \\
&= v\left(w\left(s^j\right)\right) e_j + v\left(w^\nu A^{\phantom{\nu}j}_{\nu\phantom{j}k}\, s^k \right) e_j \nonumber \\
&\qquad + \left(w\left(s^j\right) + w^\nu A^{\phantom{\nu}j}_{\nu\phantom{j}k}\, s^k \right) v^\mu A^{\phantom{\mu}i}_{\mu\phantom{i}j}\, s^j e_i \,.
\end{align}
Since the connection components $A^{\phantom{\nu}j}_{\nu\phantom{j}k}\,$ are functions, we can then write 
\begin{equation}
 v\left(w^\nu A^{\phantom{\nu}j}_{\nu\phantom{j}k}\, s^k \right) =  v(w^\nu) A^{\phantom{\nu}j}_{\nu\phantom{j}k}\, s^k +  w^\nu v(A^{\phantom{\nu}j}_{\nu\phantom{j}k}) s^k +  w^\nu A^{\phantom{\nu}j}_{\nu\phantom{j}k} v(s^k) \,. 
\end{equation}
Inserting this into the previous equation and writing $D_w D_v s $ similarly, we find
\begin{align}
D_v D_w s  - D_w D_v s &= \comm{v}{w}(s^i)e_i + \comm{v}{w}^\mu A^{\phantom{\mu}i}_{\mu\phantom{i}j}\,s^j e_i 
\nonumber \\
&\qquad + \left(w^\nu v\left(A^{\phantom{\nu}i}_{\nu\phantom{i}j}\right) - v^\mu w\left(A^{\phantom{\mu}i}_{\mu\phantom{i}j} \right)\right)s^j e_i \nonumber \\
& \qquad + v^\mu w^\nu \left(  A^{\phantom{\mu}i}_{\mu\phantom{i}j} A^{\phantom{\nu}j}_{\nu\phantom{j}k}
- A^{\phantom{\nu}i}_{\nu\phantom{i}j} A^{\phantom{\mu}j}_{\mu\phantom{j}k}
 \right) s^k e_i\,.
\end{align}
Also,
\begin{equation}
D_{\comm{v}{w}}s = \comm{v}{w}(s^i)e_i + \comm{v}{w}^\mu A^{\phantom{\mu}i}_{\mu\phantom{i}j}\,s^j e_i\,,
\end{equation}
so that
\begin{equation}
F(v, w) s = v^\mu w^\nu \left(\partial_\mu A^{\phantom{\nu}i}_{\nu\phantom{i}j} 
- \partial_\nu A^{\phantom{\mu}i}_{\mu\phantom{i}j} 
+ A^{\phantom{\mu}i}_{\mu\phantom{i}k} A^{\phantom{\nu}k}_{\nu\phantom{k}j}
- A^{\phantom{\nu}i}_{\nu\phantom{i}k} A^{\phantom{\mu}k}_{\mu\phantom{k}j} \right) s^j e_i\,.
\end{equation}
Thus we can define $F_{\mu\nu}$ by 
\begin{equation}
F(\partial_\mu, \partial_\nu) s = F_{\mu\nu}s = \left(F_{\mu\nu}s\right)^i e_i = \left(F_{\mu\nu}\right)^i_{\phantom{i}j} s^j e_i\,,
\end{equation}
so that 
\begin{equation}
\left(F_{\mu\nu}\right)^i_{\phantom{i}j} = \partial_\mu A^{\phantom{\nu}i}_{\nu\phantom{i}j} 
- \partial_\nu A^{\phantom{\mu}i}_{\mu\phantom{i}j} 
+ A^{\phantom{\mu}i}_{\mu\phantom{i}k} A^{\phantom{\nu}k}_{\nu\phantom{k}j}
- A^{\phantom{\nu}i}_{\nu\phantom{i}k} A^{\phantom{\mu}k}_{\mu\phantom{k}j}\,.
\end{equation}
We are providing a notebook titled \texttt{connection-curvature-test} that demonstrates the use of the commands of Ricci.m for the definition of fibre bundle, connection, and curvature; and confirms this equation. Attention must be paid to the factor of $\frac12$, which is explained in the manual of Ricci.m, and to the fact that $a,~b$ are the indices for the fibre and $i,~j$ for the tangent bundles in the grammar of Ricci.m. This includes a knowledge of the syntax of the tangent bundle, fibre bundles and the index structure of the tensors; in particular, the latter is useful for tensor manipulations which are otherwise very difficult. A couple of snippets are provided below.
\begin{lstlisting}[style=in]
O1 = Conn[L[a], U[b]];
O2 = Extd[O1] - Conn[L[a], U[c]]~Wedge~Conn[L[c], U[b]] // Simplify;
O2 /. SecondStructureRule;
O3 = TensorSimplify[%];
BasisExpand[%];
Result[i_, j_] = O3[L[i], L[j]];
OutputForm[%]
\end{lstlisting}

Note that since coordinate basis vector fields commute, $\comm{\partial_\mu}{\partial_\nu} = 0\,,$
\begin{equation}
F_{\mu\nu}s = F(\partial_\mu, \partial_\nu) s  = D_\mu D_\nu s - D_\nu D_\mu s = \comm{D_\mu}{D_\nu} s\,. 
\end{equation}

\md It is not very difficult to work out that the curvature acts linearly on the module of sections,
\begin{equation}
F(u, v) (s_1 + fs_2) = F(u, v)s_1 + f F(u, v) s_2\,,
\end{equation}
where $f \in C^\infty(B)\,.$ Also, 
\begin{equation}
F(u, v+ f w) \, s = F(u, v) s + f F(u, w) s\,.
\end{equation}

\md For coordinate basis vector fields $\comm{\partial_\mu}{\partial_\nu} = 0\,,$ so 
\begin{equation}
F_{\mu\nu}s = F(\partial_\mu, \partial_\nu) s = D_\mu D_\nu s - D_\nu D_\mu s = \comm{D_\mu}{D_\nu} s\,. 
\end{equation}
Since $F(\partial_\mu, \partial_\nu) s$ is a section, so is
\begin{equation}
D_\lambda \left(F_{\mu\nu}s \right) = D_\lambda \comm{D_\mu}{D_\nu} s\,.
\end{equation}
Similarly, since $D_\lambda s$ is a section, so is
\begin{equation}
F_{\mu\nu} D_\lambda s = \comm{D_\mu}{D_\nu}  D_\lambda s\,.
\end{equation}
Thus
\begin{equation}
D_\lambda \left(F_{\mu\nu}s \right) - F_{\mu\nu} D_\lambda s = \comm{D_\lambda}{\comm{D_\mu}{D_\nu}} s\,.
\end{equation}
Considering $C^\infty$ sections, and noting that maps are associative under map composition, we find that 
\begin{equation}
\comm{D_\lambda}{\comm{D_\mu}{D_\nu}} s + \rm{cyclic} = 0\,.
\end{equation}
On the other hand, 
\begin{equation}
F_{\mu\nu}s = \left(F_{\mu\nu}s \right)^i e_i = \left(F_{\mu\nu\phantom{i}j}^{\phantom{\mu\nu}i}s^j \right) e_i\,,
\end{equation}
where $F_{\mu\nu\phantom{i}j}^{\phantom{\mu\nu}i}$ and $s^i$ are in $C^\infty(B)\,.$ So we can write 
\begin{align}
D_\lambda \left(F_{\mu\nu}s \right)  &= \partial_\lambda \left(F_{\mu\nu\phantom{i}j}^{\phantom{\mu\nu}i}s^j \right) + \left(F_{\mu\nu\phantom{i}j}^{\phantom{\mu\nu}i}s^j \right) D_\lambda e_i \nonumber \\
&= \left(\partial_\lambda F_{\mu\nu\phantom{i}j}^{\phantom{\mu\nu}i} \right) s^je_i + F_{\mu\nu\phantom{i}j}^{\phantom{\mu\nu}i} \left(\partial_\lambda s^j\right) e_i 
+ F_{\mu\nu\phantom{i}j}^{\phantom{\mu\nu}i}s^j A_{\lambda\phantom{k}i}^{\phantom{\lambda}k} e_k
\nonumber \\
& = \left(\partial_\lambda F_{\mu\nu\phantom{i}j}^{\phantom{\mu\nu}i} +  F_{\mu\nu\phantom{i}j}^{\phantom{\mu\nu}k} A_{\lambda\phantom{i}k}^{\phantom{\lambda}i} \right) s^je_i + F_{\mu\nu\phantom{i}j}^{\phantom{\mu\nu}i} \left(\partial_\lambda s^j\right) e_i \nonumber \\
&\qquad\qquad - F_{\mu\nu\phantom{i}k}^{\phantom{\mu\nu}i} A_{\lambda\phantom{k}j}^{\phantom{\lambda}k} s^j e_i
+ F_{\mu\nu\phantom{i}j}^{\phantom{\mu\nu}i} A_{\lambda\phantom{j}k}^{\phantom{\lambda}j} s^k e_i \nonumber \\
& = \left(D_\lambda F_{\mu\nu}\right)^i_{\phantom{i}j} s^j e_i + F_{\mu\nu\phantom{i}j}^{\phantom{\mu\nu}i} \left(D_\lambda s\right)^j e_i\,,
\end{align}
where we have defined $\left(D_\lambda F_{\mu\nu}\right)$ by $\left(D_\lambda F_{\mu\nu}\right)^i_{\phantom{i}j}\,$
in this. Then this is a Leibniz rule, 
\begin{equation}
D_\lambda \left(F_{\mu\nu}s \right) = \left(D_\lambda F_{\mu\nu}\right) s + F_{\mu\nu}\left(D_\lambda s\right)\,.
\end{equation}
Then we can write
\begin{align}
& D_\lambda \left(F_{\mu\nu}s \right)  - F_{\mu\nu}\left(D_\lambda s\right) + \rm{cyclic} = 0\nonumber \\
\Rightarrow \qquad &\left(D_\lambda F_{\mu\nu}\right) s + \rm{cyclic} = 0 \qquad \forall s \nonumber \\
\Rightarrow \qquad & D_\lambda F_{\mu\nu} + \rm{cyclic}  = 0\,.
\end{align}
This is known as the \db{Bianchi identity}\,.\hfill $\Box$

Given $D$ and $g$ such that $g(x) \in G\,,$ we have $D'$ given by 
\begin{equation}
D'_v \phi = g D_v\left(g^{-1}\phi\right)\,.
\end{equation}
Then 
\begin{equation}
D'_u D'_v \phi = D'_u\left(g D_v\left(g\inv \phi\right)\right) = g D_u D_v \left(g\inv \phi\right)\,,
\end{equation}
and thus
\begin{align}
F'\left(u,v\right)\phi &\equiv \left(D'_u D'_v - D'_v D'_u -D'_{\comm{u}{v}}\right)\phi \nonumber \\
& = g D_u D_v \left(g\inv\phi\right) - D_v D_u \left(g\inv\phi\right) - g D_{\comm{u}{v}}\left(g\inv\phi\right) \nonumber \\
&= g F(u, v)g\inv \phi \nonumber \\
\Rightarrow \qquad\qquad  F'_{\mu\nu} & = g\circ F_{\mu\nu}\circ g\inv\,.
\end{align}
As before, $g$ is in some representation of $G\,,$ and $D$ (and thus $F$) acts on the same representation. This is the meaning of the statement that the curvature is \db{gauge covariant}\,. \hfill$\Box$

%% file: DiffGeom/GR.tex
\chapter{General Theory of Relativity}
\section{Introduction}

One of the greatest achievements of modern science is the General Theory of Relativity (GR), constructed by Einstein using common sense, ``pure thought'', and giant leaps of intuition. He realised that the gravitational field was a manifestation of the curvature of spacetime treated as a differentiable manifold. In previous Chapters we have discussed the geometrical structures which are required to set up GR. Now we will arrange them together to describe the physical theory, although we will not discuss the physics behind the equations. For a more complete discussion, we refer the reader to any of several comprehensive textbooks by~[\cite{weinberg2008cosmology,Weinberg:1972kfs}],~[\cite{Wald:1984rg}], and~[\cite{Misner:1973prb}].




We will consider GR on a four-dimensional manifold, which is the appropriate description for the spacetime we live in -- for three space dimensions and one time dimension. The manifold comes equipped with a metric which is a symmetric matrix when evaluated in an orthonormal basis of vector fields at any point. We will assume that this matrix, when diagonalised, has one negative and three positive entries. Such a manifold is called a $3+1$-dimensional \db{Lorentzian manifold}, with \db{signature} $(- \, +\,+\, + )$. 
Needless to say, GR can be built on other ($d+1$) dimensional spacetimes, as well as on \db{Euclidean} spacetimes in which the metric has signature $(+\,+\,+\,+)$.
While many interesting theories of physics are constructed on such spacetimes, we will not consider them, but only mention that most of our methods and results can be generalised to such spacetimes.



For this chapter we will always work in some chart, i.e., use a local coordinate system. The spacetime coordinates will be denoted by $x^\mu, \mu= 0, 1, 2, 3.$ These are related to Physics by $(x^0, x^1, x^2, x^3) \equiv (ct, x, y, z)$, where $c$ is the speed of light in vacuum. Usually we set $c=1$.

\section{Connection}
Let us recall from Chap.~\ref{metric} that a metric is a function  $g: V\times V \to \mathbb{R}$. In our manifold,
\begin{align}
    g: T_P\cm \times T_P\cm \rightarrow \mathbb{R}\,.
\end{align}
Let $(U, \phi)$ be a chart in the manifold $\cm$ and $\{x^\mu\}$ be the coordinates.  In terms of coordinates, we can write
\begin{align}
   g = g_{\mu\nu} d x^\mu dx^\nu\,.
\end{align}
Note that the metric transforms as a $(0,2)$ tensor. Under a change of coordinates, the metric transforms as
\begin{align}
    g_{\mu\nu}(x) \rightarrow \tilde{g}_{\mu\nu}(\tilde{x}) = \frac{\partial x^\sigma }{\partial \tilde{x}^\mu} \frac{\partial x^\lambda }{\partial \tilde{x}^\nu}  g_{\mu\nu}(x)\,.
\end{align}
It is possible to find a suitable coordinate system such that, at any given point, the eigenvalues of $g_{\mu\nu}$ are either $+1$ or $-1$. If the number of $(+1)$'s is $p$ and the number of $(-1)$'s is $q\,,$ we say that the metric has \db{signature} $(p, q)\,.$  In GR, we work with a manifold with signature $(3,1)$.



Recall from Ch.~\ref{FB_Conn_Curv} that we need a connection to find the derivative of a vector field.
Let $\{e_\mu\} = \{\partial_\mu\}$ be the coordinate in $T_P\cm$, and the affine connection be $\nabla$. Then we have, similarly to Eq.~(\ref{eqn_conn_coeff}),
\begin{align}
    \nabla_\nu e_\mu = \nabla_{e_\nu} e_\mu = \Gamma^\lambda_{\nu \mu} e_\lambda\,.
\end{align}
\md The $\Gamma^\lambda_{\nu \mu}$ are called the \db{connection coefficients}.

Note that we have used $\nabla_\mu$ for the connection instead of $D_\mu$. The former is the usual notation used in GR. The connection satisfies all the properties of Eq.~(\ref{eqn_conn_prop}). Thus the action of $\nabla$ on a vector $V \in T_P\cm$ is
\begin{align}
    \nabla_\nu V  = (\nabla_{\nu} V^\mu) e_\mu + V^\mu \nabla_{\nu} e_\mu =  \partial_\nu V^\mu e_\mu+  \Gamma^\lambda_{\nu \mu} V^\mu e_\lambda =  (\partial_\nu V^\mu +\Gamma^\mu_{\nu \lambda} V^\lambda)e_\mu\,.
\end{align}
The quantity $\nabla_\nu V$ is known as the \db{covariant derivative} of $V$. In terms of coordinates,
\begin{align}
    (\nabla_\nu V)^\mu \equiv  \nabla_\nu V^\mu = \partial_\nu V^\mu +\Gamma^\mu_{\nu \lambda} V^\lambda\,.
\end{align}
The action of $\nabla$ on a function  on the manifold $f\in C^\infty(\cm)$ is assumed to be  trivial, i.e., 
\begin{align}
    \nabla_\mu f =  \partial_\mu f\,.
\end{align}
Then we also assume that the covariant derivative is a derivative on the module of vector and tensor fields of all ranks. Then the covariant derivative is linear and obeys the Leibniz rule. 
It follows that for a $1-$form $\omega$ and a vector $V$,
\begin{align}
    \nabla_\mu(\omega_\nu V^\nu) =& \partial_\mu(\omega_\nu V^\nu) \,.    
\end{align}
We can apply the Leibniz rule to this product to find the action of $\nabla$ on a $1-$form as
\begin{align}
(\nabla_\mu\omega_\nu) V^\nu + \omega_\nu (\nabla_\mu V^\nu) =&\, (\partial_\mu\omega_\nu)V^\nu + \omega_\nu(\partial_\mu V^\nu)\,, \nonumber \\
\Rightarrow \qquad
   (\nabla_\nu \omega)_\mu \equiv 
    \nabla_\nu \omega_\mu =&\, \partial_\nu \omega_\mu -   \Gamma^\lambda_{\nu \mu} \omega_\lambda \,. 
\end{align}
The action of the covariant derivative on tensor fields can be found by generalizing this process.

Under a change of coordinates, the covariant derivative transforms covariantly,
\begin{align}
    \nabla_\mu V^\nu \rightarrow \tilde{\nabla}_{\tilde{\mu} } \tilde{V}^{\tilde{\nu}} = \frac{\partial x^\sigma }{\partial \tilde{x}^\mu} \frac{\partial \tilde{x}^\nu }{\partial x^\lambda}   \nabla_\sigma V^\lambda \,.
\end{align}

\noindent{\bf Exercise:} Using the transformation rule and the definition of the covariant derivative, find the transformation rule of the Christoffel symbols:
\begin{align}
    \Gamma^\lambda_{\mu \nu}\rightarrow  \tilde{\Gamma}^{\tilde{\lambda}}_{\tilde{\mu} \tilde{\nu}} = \frac{\partial \tilde{x}^\lambda }{\partial x^\sigma} \frac{\partial x^\alpha }{\partial \tilde{x}^\mu}   \frac{\partial x^\beta }{\partial \tilde{x}^\nu}  \Gamma^\sigma_{\alpha \beta} - \frac{\partial^2 \tilde{x}^\lambda}{\partial \tilde{x}^\mu \partial \tilde{x}^\nu} \,.
\end{align}


\md
The covariant derivative also helps us to define the parallel transport of a vector along a curve. Let $\gamma(t):(0,1)\rightarrow \cm$ be a curve on $\cm$ and $w$ be the tangent vector to the curve. Any vector $U$ is said to be \db{parallel transported}
along $\gamma$ if it satisfies
\begin{align}
    \nabla_w U = 0\,.
\end{align}
In terms of coordinates, $w^\mu = \dfrac{dx^\mu}{dt}$ and
\begin{align} 
    \nabla_w U^\mu = \frac{dx^\nu}{dt}\nabla_\nu U^\mu =
    \frac{dx^\nu}{dt}\left(\partial_\nu U^\mu +\Gamma^\mu_{\nu \lambda} U^\lambda\right) = \frac{dU^\mu}{dt}  + \Gamma^\mu_{\nu \lambda} \frac{dx^\nu}{dt} U^\lambda =0\,.
\end{align}
If the vector $U$ is the tangent vector along the curve, then we have
\begin{align}
   \frac{d^2 x^\sigma}{dt^2} + \Gamma^\sigma_{\mu\nu} \frac{dx^\mu}{dt} \frac{dx^\nu}{dt }  = 0\,.
\end{align}
The curve $\gamma(t)$ is called the \db{geodesic} and the above equation is called the geodesic equation.

\md A connection is called \db{metric compatible} if $\nabla_\sigma g_{\mu\nu} = 0$.

\md The \db{torsion tensor} is defined as $T^\lambda_{\mu\nu} = \Gamma^\sigma_{[\mu\nu]} = \Gamma^\sigma_{\mu\nu} -\Gamma^\sigma_{\nu\mu}$.

\noindent{\bf Exercise:} Show that $T^\lambda_{\mu\nu}$ transforms as a tensor.

\md The \db{Levi}-\db{Civita connection} is a unique connection that is metric compatible and torsion free, i.e., $\nabla_\sigma g_{\mu\nu} = 0$ and $T^\lambda_{\mu\nu} = 0$.

\noindent{\bf Exercise:} Show that the components of the Levi-Civita connection take the following form
\begin{align}
    \Gamma^\sigma_{\mu\nu} = \frac{1}{2} g^{\sigma\rho}\left(\partial_\mu g_{\rho\nu} + \partial_\nu g_{\rho\mu} - \partial_\rho g_{\mu\nu} \right)\,.
\end{align}

\section{Geodesic as shortest route}
We now present an alternative approach to deriving the geodesic equation from a physical perspective, namely, from the action of a particle moving on the spacetime manifold. 
We first define the proper time 
\begin{align}
    d\tau^2 = - g_{\mu\nu} d x^\mu dx^\nu\,.
\end{align}
The action of a particle moving in the spacetime is defined as
\begin{align}
    S = -m \int d\tau\,,
\end{align}
where $m$ is the mass of the particle and $\tau$ is the proper time along the path of the particle in the spacetime. Let us parametrize the path in the spacetime as $x^\mu(\lambda)$ with $\lambda\in(0,1)$; then we can write the action as
\begin{align}
    S = -m \int d\tau = - m \int_0^1 d\lambda \sqrt{ - g_{\mu\nu} \frac{d x^\mu}{d\lambda} \frac{dx^\nu}{d\lambda}}\,.
    \label{action.particle}
\end{align}
A \db{geodesic} is a curve for which the action is an extremum for fixed initial and final points. It can be shown that the geodesic must satisfy (exercise)
\begin{align}
    \frac{d^2 x^\sigma}{d\tau^2} +\Gamma^\sigma_{\mu\nu} \frac{dx^\mu}{d \tau} \frac{dx^\nu}{d \tau} = 0\,,
    \label{geodesic.extremum}
\end{align}
where $\Gamma^\sigma_{\mu\nu}$ are the \db{Christoffel symbols} defined earlier,  
\begin{align}
    \Gamma^\sigma_{\mu\nu} = \frac{1}{2} g^{\sigma\rho}\left(\partial_\mu g_{\rho\nu} + \partial_\nu g_{\rho\mu} - \partial_\rho g_{\mu\nu} \right)\,.
\end{align}
Since $d\tau$ can be thought of as the proper time required to traverse an infinitesimal distance by the particle, the geodesic curve is the `shortest' possible route between two given points in the manifold.

\noindent{\bf Exercise:} Show that extremising the action of Eq.~(\ref{action.particle}) produces the geodesic equation, Eq.~(\ref{geodesic.extremum}).

\section{Riemann and Ricci tensors and Ricci scalar}

Recall from Eq.~(\ref{eqn_curvature}) that given two vector fields $(v,w)$, the curvature is a map of the vector bundle to itself. In GR, we work with the tangent bundle. 

\md Given three vector fields $v, w, s$, the \db{Riemann tensor} is defined as 
\begin{equation}
R(v, w) s = \nabla_v \nabla_w s - \nabla_w \nabla_v s - \nabla_{\left[v, w\right]} s\,.
\end{equation}
Here $\left[v,w\right]$ is the commutator of two vector fields. We have used the notation $R$ for the \db{Riemann tensor}, commonly used in GR. The Riemann tensor is a measure of the curvature of the manifold. In terms of components, the Riemann tensor is expressed as
\begin{align}
    R^\lambda_{\phantom{\lambda}\mu\nu\sigma} V^\mu W^\nu S^\sigma=  V^\mu \nabla_\mu(W^\nu \nabla_\nu S^\lambda)-  W^\nu \nabla_\nu(V^\mu \nabla_\mu S^\lambda) -  (V^\mu \partial_\mu W^\rho -  W^\nu \partial_\nu V^\rho)\nabla_\rho S^\lambda\,.
\end{align}

\noindent{\bf Exercise:} Using the above expression and the definition of the covariant derivative, show that the Riemann tensor can be written as
\begin{align}
R^\lambda_{\phantom{\lambda}\mu\nu\sigma} = \partial_\nu \Gamma^\lambda_{\mu \sigma}- \partial_\sigma \Gamma^\lambda_{\mu \nu} + \Gamma^\lambda_{\nu\rho}\Gamma^\rho_{\mu \sigma} - \Gamma^\lambda_{\sigma\rho}\Gamma^\rho_{\mu \nu}\,.
\end{align}

It may be noted that not all components of the Riemann tensor are independent. It is possible to work out the symmetry property of the Riemann tensor and the number of independent components.

\noindent{\bf Exercise:} In the above text, the torsion tensor is defined as $T^\lambda_{\phantom{\lambda}\mu\nu} = \Gamma^\lambda_{\phantom{\lambda}\mu\nu} -\Gamma^\lambda_{\phantom{\lambda}\nu\mu} $. It should be noted that $\Gamma^\lambda_{\phantom{\lambda}\mu\nu}$ denotes the general affine connection and not the Christoffel symbols in this context. Show that this definition is consistent with the following definition which says that for two vector fields $u, w$,
\begin{align}
    T(u,w) = \nabla_u w - \nabla_w u - \left[u,w\right]\,.
\end{align}

\noindent{\bf Exercise:} Show that the Riemann tensor satisfies the \db{Bianchi identity}:
\begin{align}
\nabla_\lambda R_{\mu \nu \rho \sigma} + \nabla_\mu R_{\nu  \rho \lambda \sigma} + \nabla_\nu R_{  \rho \lambda \mu \sigma} = 0\,.
\end{align}

\md The \db{Ricci tensor} is defined as
\begin{align}
    R_{\mu \nu} = R^\lambda_{\phantom{\lambda}\mu\lambda\sigma}\,.
\end{align}
\md The \db{Ricci scalar} is the trace of the Ricci tensor
\begin{align}
    R = R^\mu_{\phantom{\mu}\mu}\,.
\end{align}
A demonstration notebook \texttt{riemann-ricci-scalar} using package \code{Ricci} displays the use of some of the expressions above, and also verifies the Bianchi identity mentioned above. We list below a couple of useful input commands after defining the tangent bundle.
\begin{lstlisting}[style=in]
Rm[U[i], L[j], L[k], L[l]] + 
Rm[U[i], L[l], L[j], L[k]] /. FirstBianchiRule;
CovD[Rm[L[i], L[j], L[k], L[l]], {L[m]}] + 
CovD[ Rm[L[i], L[j], L[l], L[m]], {L[k]}] /. SecondBianchiRule;
Tensor[Rm, {U[i], L[j], L[i], L[l]}, {}];
Tensor[Rc, {U[l], L[l]}, {}];
Conn[L[i], U[j]]
\end{lstlisting}

\section{The Einstein field equations}
\md Spacetime is Minkowskian if there is no matter. However, in the presence of matter, spacetime curves according to \db{The Einstein field equations}:
\begin{align}
    G_{\mu \nu} = 8 \pi G T_{\mu \nu}\,.
\end{align}
where $T_{\mu \nu}$ is the stress-energy tensor, $G$ is Newton's gravitational constant and $G_{\mu \nu}$ is the \db{Einstein tensor}, defined as
\begin{align}
    G_{\mu \nu} =  R_{\mu \nu} - \frac{1}{2} g_{\mu \nu} R\,.
\end{align}


\section{The Einstein-Hilbert Action}
\md The Einstein field equation can also be derived from the \db{Einstein}-\db{Hilbert action}. In the absence of matter, the action takes the form 
\begin{align}
    S = \int d^4x \sqrt{-g} R  \,.
\end{align}

It can be shown that the variation of the above action with respect to the metric tensor yields the vacuum Einstein field equation.


\section{Examples}

\noindent{\bf Example:} 
Consider the following metric
\begin{align}
    ds^2 = dr^2 + \sin^2\theta d\theta^2 \,.
\end{align}
There is only one non-vanishing Christoffel symbol:
\begin{align}
\Gamma^\theta_{\theta\theta} = \cot\theta \,.
\end{align}
Thus, the geodesic equations are:
\begin{align}
    \frac{d^2r}{d\lambda^2} &= 0\,,\\
    \frac{d^2\theta}{d\lambda^2} + \cot\theta \frac{d\theta}{d\lambda}&= 0\,.
\end{align}

We have obtained the same geodesic in the notebook titled \texttt{Geodesic\_example\_2D} using the \code{RGTC} package.

\noindent{\bf Example:} \db{Schwarzschild spacetime}

Let us consider the Schwarzschild metric
\begin{align}
    ds^2 = -\left(1- \frac{2GM}{r}\right) dt^2 + \left(1- \frac{2GM}{r}\right)^{-1}dr^2 + r^2 (d\theta^2 + \sin^2 \theta d\phi^2)\,.
\end{align}
 
Let us determine the Christoffel symbols, the Riemann tensor, the Ricci tensor, and the Ricci scalar for the Schwarzschild metric. It is easy to define these quantities in Mathematica. We have provided a notebook titled \texttt{Schwarzschild\_tensors} containing the commands to calculate these quantities.  After setting up the coordinates and the metric, we find
\begin{lstlisting}[style=in]
coord = {t, r, \[Theta], \[Phi]};

metric = {{-(1 - 2 GM/r), 0, 0, 0}, {0, 1/(1 - 2 GM/r), 0, 0}, {0, 0,  r^2, 0}, {0, 0, 0, r^2 Sin[\[Theta]]^2}};
\end{lstlisting}
The Christoffel symbols can be calculated with the following command:
\begin{lstlisting}[style=in]
conn = Simplify[Table[(1/2)*
Sum[(Inverse[metric][[i, s]])*(D[metric[[s, j]], coord[[k]]] + D[metric[[s, k]], coord[[j]]] - 
D[metric[[j, k]], coord[[s]]]), {s, 1, n}], {i, 1, n}, {j, 1, n}, {k, 1, n}]]
\end{lstlisting}
where \texttt{n} is the \texttt{Length[coord]}.
We find that the Christoffel symbols are:
\begin{align}
&\Gamma^r_{tt} = \frac{G M (r-2 G M)}{r^3}\,,\quad 
\Gamma^r_{rr} = \frac{G M}{2 G M r-r^2}\,, \quad
\Gamma^r_{\theta\theta} = 2 G M-r\,, \quad
\Gamma^r_{\phi\phi} =  (2 G M-r)\sin ^2\theta\,,\nonumber\\
&\Gamma^t_{t r} = -\frac{G M}{2 G M r-r^2}\,, \quad 
\Gamma^\theta_{r \theta} = \frac{1}{r}\,, \quad
\Gamma^\theta_{\phi\phi} = -\sin \theta \cos \theta \,, \quad
\Gamma^\phi_{r\phi} = \frac{1}{r}\, , \quad 
\Gamma^\phi_{\theta\phi} = \cot\theta\,. \qquad
\end{align}
Similarly, the Riemann tensor can be obtained using the following command:

\begin{lstlisting}[style=in]
Riemann = 
Simplify[Table[
D[conn[[i, j, l]], coord[[k]]] - D[conn[[i, j, k]], coord[[l]]] + 
Sum[conn[[s, j, l]] conn[[i, k, s]] - 
conn[[s, j, k]] conn[[i, l, s]], {s, 1, n}], {i, 1, n}, {j, 1, n}, {k, 1, n}, {l, 1, n}]]
\end{lstlisting}
We also present a few non-vanishing components of the Riemann tensor:
\begin{align}
&R^t_{\phantom{t}rtr} = -\frac{2 G M}{r^2 (2 G M-r)}\,, \hspace{.5cm} 
R^t_{\phantom{t}\theta t \theta} = -\frac{G M}{r}\,,\hspace{.5cm}
R^t_{\phantom{t}\phi t \phi} =  -\frac{G M \sin ^2\theta }{r}\,, \hspace{.5cm}
R^r_{\phantom{r}t t r} = \frac{2 G M (r-2 G M)}{r^4} \,, \nonumber\\
&R^\theta_{\phantom{\theta} t t \theta} = \frac{G M (2 G M-r)}{r^4}\,, \hspace{.5cm} 
R^\theta_{\phantom{\theta} \phi \theta \phi} = \frac{2 G M \sin ^2(\theta )}{r}\, \dots
\end{align}

These tensors can also be determined using the commands of the \code{RGTC} package, which is shown in the \texttt{SchwarzschildGeodesic} notebook. We also show how to construct the geodesic equation in the Schwarzschild spacetime.

\noindent{\bf Example:}  \db{Friedmann$-$Robertson$-$Walker $($FRW$)$ spacetime}

We now look into the geodesics on the FRW metric:
\begin{align}
    ds^2 =  -dt^2 + a(t)^2\left( \frac{dr^2}{1- k r^2} + r^2 \left(d\theta^2 + \sin^2\theta d \phi^2\right)\right)\,.
\end{align}

We use the \code{RGTC} package to find the Christoffel symbols and other tensors. After calling the package, we define the coordinates and the metric for the FRW spacetime. We refer the reader to the \texttt{FRW}  notebook for details.
\begin{lstlisting}[style=in]
coordinates={t,r,\[theta],\[phi]};

Metric = DiagonalMatrix[{-1, a[t]^2/(1 - k r^2), a[t]^2 r^2, a[t]^2 r^2 Sin[\[Theta]]^2}];
\end{lstlisting}

Now we use the following command to find the relevant quantities:
\begin{lstlisting}[style=in]
RGtensors[Metric, coordinates];
\end{lstlisting}

The Christoffel symbols $\Gamma^\lambda_{\mu \nu}\,$, Riemann tensor $R^\lambda_{\phantom{\lambda}\mu \nu \sigma}\,$, Ricci tensor $R_{\mu \nu}$ and Ricci scalar $R$ can be obtained by the following variables:
\begin{lstlisting}[style=in]
GUdd (*Christoffel symbols*)
RUddd (*Riemann tensor*)
Rdd (*Ricci tensor*)
R (*Ricci scalar*)
\end{lstlisting}

We do not reproduce the output here as it is too long. 

To obtain the Friedmann equations, we first define the stress-energy tensor.
\begin{lstlisting}[style=in]
Tdd = (\[Rho] + p) DiagonalMatrix[{1, 0, 0, 0}] + p gdd;
\end{lstlisting}
Now we try to solve the Einstein equation:
\begin{align}
    R_{\mu\nu}-\frac{1}{2}R g_{\mu\nu}+\Lambda g_{\mu\nu}-\kappa \
T_{\mu\nu}=0\,.
\end{align}

Note that we have added a term with the cosmological constant $\Lambda$. Let us store the expression as:
\begin{lstlisting}[style=in]
Zerodd = Rdd - 1/2 R gdd + \[CapitalLambda] gdd - \[Kappa] Tdd;
\end{lstlisting}

Equating \texttt{Zerodd} to zero, we find: 
\begin{align}
    \left(\frac{a'}{a}\right)^2 +\frac{k}{a^2} &=   \frac{\Lambda}{3}  
    +  \frac{\kappa \rho}{3}\,,\\
    2 \frac{a'}{a} +  \left(\frac{a'}{a}\right)^2 + \frac{k}{a^2}  &= \Lambda - \kappa p\,.
\end{align}

where $p$ is the pressure, $\rho$ is the density and $a' =  da/dt$. These are the famous Friedmann equations, which play an important role in the expansion of the Universe.

In this chapter, we have discussed the fundamental geometric ideas that form the basis of General Relativity. Beginning with connections, covariant derivatives, and geodesics, we introduced the concept of spacetime curvature through the Riemann tensor and its contractions, ultimately leading to Einstein's field equations. These equations reveal the profound relationship between geometry and gravitation, showing how matter and energy shape the structure of spacetime. The examples discussed provide a glimpse of the rich physical consequences of the theory and lay the groundwork for further studies of gravitation, black holes, and the evolution of the Universe.

%% file: data.tex
\chapter*{Data Declaration}
\chaptermark{Data Declaration }
\addcontentsline{toc}{chapter}{Data Declaration}

\textbf{Data Availability}\\
Data sharing is not applicable to this article as no datasets were generated or analyzed during the
current study.

\textbf{Conflict of interest}\\
We have no conflict of interest to disclose.

%% file: book_index.tex
\begin{theindex}
\item Atlas, 55
\indexspace
\item Bundle, 6, 42, 43, 45--47, 71, 74, 95, 149--151, 153, 154, 156, 157, 163
\item Bundles, 7, 45--47, 135, 149--151, 153
\indexspace
\item Chart, 55, 56, 61--63, 66, 68, 71, 75, 79, 83, 91, 95, 96, 100, 103, 109, 113, 116, 117, 119, 129, 130, 153, 161, 174
\item Christoffel, 20, 161--166, 173
\item Christoffel Symbols, 20, 161--163, 165, 166
\item Connection, 42, 48, 153--157, 161, 162
\item Coordinate Chart, 116
\item Coordinate Transformation, 62, 68, 77
\item Cotangent, 67, 95, 99, 100, 150
\item Cotangent Bundle, 95, 150
\item Covariant Derivative, 161, 162
\item Curvature, 5, 7, 13--15, 42, 43, 157--160, 163, 173, 174
\indexspace
\item Diffeomorphism, 61, 74, 89, 90, 93, 107, 142
\item Differential Form, 14, 91
\item Differential Forms, 5, 6, 14, 20, 103, 117, 122, 123, 125
\indexspace
\item Einstein, 159, 160, 164, 166
\item Embedding, 175
\item Exponential, 147
\item Exterior Derivative, 6, 109, 110
\indexspace
\item Fiber Bundle, 43, 149, 150
\item Fundamental Group, 33, 34, 37
\indexspace
\item Geodesic, 15, 162--166, 169, 172, 173, 175, 183, 184
\indexspace
\item Hodge, 5, 6, 14, 119, 122, 123
\item Hodge Star, 14, 119, 123
\item Homeomorphism, 29, 55, 149
\item Homology, 183
\indexspace
\item Integration, 113, 114, 182
\indexspace
\item Jacobian, 114, 177--179
\indexspace
\item Laplacian, 132, 133
\item Lie Algebra, 6, 83, 87, 90, 94, 95, 136, 141--143, 146, 147
\item Lie Derivative, 6, 74, 77, 93, 94, 96, 97, 101
\item Lie Group, 58, 135--137, 141, 143, 145--147, 151
\indexspace
\item Manifold, 6, 13, 14, 35, 36, 38, 39, 45, 55, 56, 58, 59, 61--63, 65--67, 72, 73, 75, 77, 79--81, 87, 91, 101, 103, 113, 115, 116, 119, 120, 129, 130, 132, 135, 149, 150, 153, 160, 161, 163
\item Map, 29, 31, 33, 34, 37, 38, 43, 44, 46, 48, 55, 58, 61--63, 74, 77, 79--81, 83, 89, 95, 99, 109, 115, 116, 119, 135, 139, 142, 143, 145, 149--151, 153, 154, 157, 158, 163, 171, 177, 178
\item Metric, 6, 7, 14, 15, 114--119, 125, 131, 160--162, 164--167, 171, 173--175
\indexspace
\item Orientation, 114, 171, 172
\item Oriented, 113, 129, 172
\indexspace
\item Parallel Transport, 162
\item Poisson Bracket, 87
\item Principal Bundle, 42, 46, 150, 151, 157
\item Pullback, 77, 89, 95, 108
\item Pullback Map, 77
\item Pushforward, 14, 77, 78, 81, 89, 90
\indexspace
\item Ricci, 15, 20, 21, 163--166, 173
\item Ricci Tensor, 163--166
\item Riemann, 15, 20, 21, 159, 163, 165, 166, 176, 177
\item Riemann Tensor, 163, 165, 166
\indexspace
\item Stokes Theorem, 5, 7, 14, 129--131
\indexspace
\item Tangent, 5--7, 14, 61--66, 71, 74, 77, 99, 100, 132, 136, 141, 142, 146, 149, 150, 162, 163
\item Tangent Bundle, 71, 74, 149, 150, 163
\item Tangent Space, 6, 7, 14, 61, 62, 65, 136, 141, 142, 150
\item Tensor, 6, 20, 21, 99--101, 103, 115, 117, 160--166, 180
\item Topological, 5, 6, 13, 25--27, 29--31, 33, 37, 38, 41--43, 45--49, 55, 61, 74, 135
\item Topological Space, 25, 27, 29--31, 33, 37, 38, 55, 61, 135
\item Topology, 3, 5, 6, 13, 23, 25--27, 29, 30, 41, 56
\item Torsion, 162, 163
\indexspace
\item Vector, 5--7, 14, 43, 45--47, 59, 62, 63, 65--69, 71--75, 77--81, 83--85, 87, 89--91, 93--96, 99--101, 103--107, 113, 115, 116, 123, 130--132, 141--143, 145, 149--151, 153, 154, 157, 158, 161--163, 174, 175, 177, 179
\item Vector Bundle, 45--47, 150, 151, 157
\item Vector Field, 71--74, 79--81, 83, 89--91, 93--95, 103, 106, 107, 116, 130, 131, 141, 142, 145, 150, 154
\item Vector Fields, 5--7, 14, 71--73, 75, 83, 84, 87, 90, 94--96, 99, 101, 103--105, 107, 115, 116, 141--143, 153, 157, 158, 163, 174, 175, 177
\item Volume Form, 6, 113, 116, 117, 120, 131, 132
\indexspace
\item Wedge, 103, 105--110, 121
\item Wedge Product, 103, 105--107, 110, 121
\end{theindex}